\documentclass[12pt]{report}
\usepackage[titletoc]{appendix}
\usepackage{verbatim}
\usepackage{cite}
\usepackage{graphicx}
\usepackage{color}
\usepackage[fleqn]{amsmath}
\usepackage{amssymb}
\usepackage{slashed}
\usepackage{epsfig}
\usepackage{textcomp}
\usepackage{hyperref}
 \usepackage{geometry}
\usepackage{makecell}
\usepackage{lmodern}
\usepackage{tabularx}
\usepackage{booktabs}
\usepackage{multirow}
\usepackage{bm}
\usepackage{amstext}
\usepackage{lscape}
\usepackage{verbatim}
\usepackage{rotating}
\usepackage{dcolumn}
\begin{document}
\pagenumbering{roman}
\thispagestyle{empty}

\begin{center}
%     {\bf \large One Loop Study of Gluon Fusion Processes at Hadron Colliders within the SM and Beyond }
      {\bf {\Large Implications of nuclear interaction for nuclear structure and astrophysics within the relativistic mean-field model} }
\end{center}

\vspace{0.3 cm}
\begin{center}
    {\it \large By}
\end{center}

\begin{center}
%   {\bf {\Large Ambresh Kumar Shivaji} \\ (Enrollment No. : PHYS07200604037)}
    {\bf {\Large Bharat Kumar } \\ PHYS07201304004}
\end{center}

\begin{center}
\bf {{\large Institute of Physics, Bhubaneswar }}\\
\bf {{\large INDIA}}
%\vskip 1.0 cm
%\epsfig{file=IOP_logo.pdf, width=2.5cm, height= 2.5cm}\\
\end{center}

\vskip 2.0 cm
\begin{center}
%\large
%\emph{
\large{
{ A thesis submitted to the }  \\
 {Board of studies in Physical Sciences }\\
%\vspace{0.2cm}
In partial fulfillment of requirements \\
For the Degree of } \\
{\bf  DOCTOR OF PHILOSOPHY} \\
\emph{of} \\
{\bf HOMI BHABHA NATIONAL INSTITUTE }
\vskip 1.0 cm
\epsfig{file=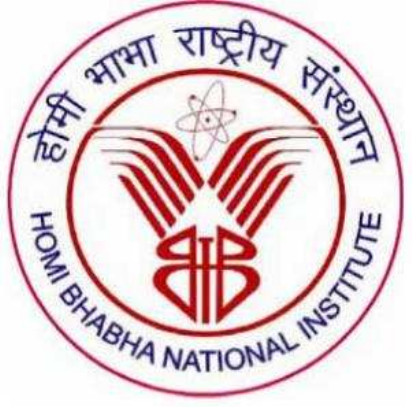, width=4.0cm, height= 4.0cm}\\
\vskip 0.2 cm
{\bf November, 2018}
\end{center}
\newpage
\begin{center}
\large \bf 
STATEMENT BY AUTHOR
% \begin{comment}
\end{center}
This dissertation has been submitted in partial fulfillment of requirements
for an advance degree at Homi Bhabha National Institute (HBNI) and deposited 
in the Library to be made available to borrowers under rules of the HBNI. 

    Brief quotations from this dissertation are allowed without special 
permission, provided that accurate acknowledgement of source is made. 
Requests for permission for extended  quotation from or reproduction    
of this manuscript in whole or in part may be granted by the competent 
Authority of HBNI when in his or her judgment the proposed use of the 
meterial is in the interests of scholarship. In all other instances, 
however, permission must be obtained from the author.\\
\bigskip
\\
\\
\\
\\
\\
  \hspace*{10.5cm} {(Bharat Kumar )}\\
\newpage
\thispagestyle{empty}
\begin{center}
\large \bf
DECLARATION
\end{center}

I, {\bf Bharat Kumar }, hereby declare that the investigations
presented in the thesis have been carried out by me. The matter
embodied in the thesis is original and has not been submitted earlier
as a whole or in part for a degree/diploma at this or any other
Institution/University.

\vskip 3.0cm
\noindent
%\hspace{10.5cm}\underline{Signature}\\
 \hspace*{10.5cm} (Bharat Kumar)\\
%\noindent
%Date :

\newpage
\begin{center}
{\bf List of Publications arising from the thesis}
\end{center}
%\begin{left}
%\hspace {1.0cm}
{\large\bf Journal}\\
	{\large\underline {Published}}
%\end{left}
\begin{enumerate}
%\item{Shape co-existence and parity doublet in Zr isotopes, \\
%{\bf Bharat Kumar}, S. K. Singh and S. K. Patra, \\
% Int. J. Mod. Phys. E {\bf 24}, 1550017 (2015)}.
\item{``Examining the stability of thermally fissile Th and U isotopes'',\\ 
{\bf Bharat Kumar}, S. K. Biswal, S. K. Singh and S. K. Patra,\\ 
Phys. Rev. C {\bf 2015}, 92, 054314-1$-$054314-10}.
\item {``Modes of decay in neutron-rich nuclei'',\\
{\bf Bharat Kumar}, S. K. Biswal, S. K. Singh, Chirashree Lahiri, and 
S. K. Patra,\\
Int. J. Mod. Phys. E {\bf 2016}, 25, 1650062-1$-$1650062-16.}
\item {``Tidal deformability of neutron and hyperon star with relativistic 
mean-field equations of state'', \\
{\bf Bharat Kumar}, S. K. Biswal and S. K. Patra,\\
Phys. Rev. C {\bf 2017}, 95, 015801-1$-$015801-10.}
\item{``Relative mass distributions of neutron-rich thermally fissile nuclei 
within statistical model'',\\
{\bf Bharat Kumar}, M. T. Senthil Kannan, M. Balasubramaniam, B. K. Agrawal,
S. K. Patra, \\
Phys. Rev. C {\bf 2017}, 96, 034623-1$-$034623-10.}
\item {``New parameterization of the effective field theory motivated 
relativistic mean-field model'', \\
{\bf Bharat Kumar}, S. K. Singh, B. K. Agrawal, S. K. Patra,\\
 Nucl. Phys. A {\bf 2017}, 966, 197-207.}
\item {``New relativistic effective interaction for finite nuclei, infinite 
nuclear matter and neutron stars'', \\
{\bf Bharat Kumar}, S. K. Patra, and  B. K. Agrawal\\
		Phys. Rev. C {\bf 2018}, 97, 045806-1$-$045806-16.}
\end{enumerate}
\newpage
{\Large \underline{Communicated}}
\begin{enumerate}
\item{``Structure effects on fission yields'',\\
{\bf Bharat Kumar}, M. T. Senthil Kannan, M. Balasubramaniam, B. K. Agrawal, 
		S. K. Patra, [Communicated to Int. J. Mod. Phys. E]\footnote{A part of the paper will contribute to the thesis}.}
\end{enumerate}
\vskip 0.2mm
{\Large \bf Conferences}
\begin{enumerate}

	\item   {\bf Talk: }''Analysis of parity doublet in medium mass nuclei'', \\
        {\underline{\bf Bharat Kumar}}, S. K. Singh and S. K. Patra, \\
        Proceedings of the DAE Symp. on Nucl. Phys. {\bf 2014}, 59, 96-97.
\item {\bf Poster:} ``$\beta$-decay half-life of Th and U isotopes'',\\
      {\underline{\bf Bharat Kumar}}, S. K.  Biswal, S. K. Singh and  S. K. Patra, \\
      Proceedings of the DAE Symp. on Nucl. Phys. {\bf 2015}, 60, 406-407.
\item {\bf Poster:} ``Evolution of N = 32,34 shell closure in relativistic mean field theory'', \\
{\underline{\bf Bharat Kumar}}, S. K.  Biswal and  S. K. Patra, \\
Proceedings of the DAE-BRNS Symp. on Nucl. Phys. {\bf 2016}, 61, 196-197.
\item {\bf Talk:} ``Tidal effects in equal-mass binary neutron stars'',  \\
{\underline{\bf Bharat Kumar}},S. K. Biswal and S. K. Patra,     \\
Proceedings of the DAE-BRNS Symp. on Nucl. Phys. {\bf 2016}, 61, 868-869.
\item {\bf Poster:} ``Curvature of a neutron star'', \\
{\underline{\bf Bharat Kumar}}, S. K. Biswal and  S. K. Patra,   \\
Proceedings of the DAE-BRNS Symp. on Nucl. Phys. {\bf 2016}, 61, 916-917.
\item {\bf Poster:} ``Search For $\Lambda$ Shell Closures in Multi-$\Lambda$ Hypernuclei'', \\
Asloob A. Rather, M. Ikram, M. Imran, {\underline{\bf Bharat Kumar}}, S. K. Biswal,
S. K. Patra,\\
Proceedings of the DAE-BRNS Symp. on Nucl. Phys. {\bf 2016}, 61, 178-179.
\item {\bf Poster:} ``Competition between $\alpha$, $\beta$ decay and Spontaneous Fission in Z=132 Superheavy Nuclei'', \\
 Asloob A. Rather, M. Ikram, {\underline{\bf Bharat Kumar}}, S. K. Biswal,S. K. Patra\\
Proceedings of the DAE-BRNS Symp. on Nucl. Phys. {\bf 2016}, 61, 202-203.
\item {\bf Poster:} ``Effects of $\delta$-meson on the maximum mass of the hyperon star'',\\
 S. K. Biswal, {\underline{\bf Bharat Kumar}}, S. K. Patra,\\
Proceedings of the DAE-BRNS Symp. on Nucl. Phys. {\bf 2016}, 61, 912-913.
\item {\bf Poster:} ``Effective relativistic mean field model for finite nuclei and neutron stars'', \\ {\underline{\bf Bharat Kumar}}, B. K. Agrawal and
                S. K. Patra, \\
Proceedings of the DAE-BRNS Symp. on Nucl. Phys. {\bf 2017}, 62, 712-713.
\item {\bf Invited talk:} ``Tidal deformability of neutrons and hyperon star'', \\
        {\underline{\bf Bharat Kumar}} and S. K. Patra, \\
Proceedings of the DAE-BRNS Symp. on Nucl. Phys. {\bf 2017}, 62, 21-22.
\end{enumerate}

%{\bf *Papers included in the thesis} \\
%\bigskip
%{\large \bf Conferences }
%\begin{enumerate}
%\item   Analysis of parity doublet in medium mass nuclei, \\
%        {\bf Bharat Kumar}, S. K. Singh and S. K. Patra, \\
%        Proceedings of the DAE Symp. on Nucl. Phys. {\bf 59}, 96 (2014).
%\item $\beta$-decay half-life of Th and U isotopes\\
%{\bf Bharat Kumar}, S. K.  Biswal, S. K. Singh and  S. K. Patra, \\
%      Proceedings of the DAE Symp. on Nucl. Phys. {\bf 60}, 406 (2015).
%\item Evolution of N = 32,34 shell closure in relativistic mean-field theory, \\
%{\bf Bharat Kumar}, S. K.  Biswal and  S. K. Patra, \\
%Proceedings of the DAE-BRNS Symp. on Nucl. Phys. {\bf 61}, 196 (2016).
%\item Tidal effects in equal-mass binary neutron stars,  \\
%{\bf Bharat Kumar},S. K. Biswal and S. K. Patra,\\
%Proceedings of the DAE-BRNS Symp. on Nucl. Phys. {\bf 61}, 868 (2016).
%\item Curvature of a neutron star, \\
%{\bf Bharat Kumar}, S. K. Biswal and  S. K. Patra,   \\
%Proceedings of the DAE-BRNS Symp. on Nucl. Phys. {\bf 61}, 916 (2016).
%\item Effective relativistic mean field model for finite nuclei and
%neutron stars, \\ 
%{\bf Bharat Kumar}, B. K. Agrawal, and  S.K. Patra,\\
%{\it Proceedings of the DAE-BRNS Symp. on Nucl. Phys.}, {\bf 62}, 712 (2017).
%\end{enumerate}
\vskip 2.0cm
\hspace*{9cm} {(Bharat Kumar)}

% Dedications
\newpage
\thispagestyle{empty}
\vspace*{3.0in}
\hspace*{2.0in}
\begin{center}
	{\Large \emph{ Dedicated To My Beloved Mother, Uncle $\&$ Aunti Ji}\\}
\hrule
\end{center}

\newpage
\thispagestyle{empty}
\begin{center}
\large \bf
ACKNOWLEDGMENTS
\end{center}
\vspace*{-0.1in}
{ \noindent
This thesis is no doubt the end of my journey in obtaining my Ph.D but it is 
the first dip in the holy ocean of intense knowledge which I wish to increase 
day by day through my work in nuclear physics as well as astrophysics. My 
thesis is proposed beautifully, shaped well, and put an end with the support 
and encouragement of numerous people including my supervisor, collaborators, 
friends, and family. 

First of all, I would like to express my deep and sincere gratitude to my supervisor Prof. S. K. Patra for his guidance and all the useful discussions and continuous support over the past four years of my PhD research work at the Institute of Physics, Bhubaneswar. I greatly appreciate his patience to bear all my annoying behaviour and frequent arguments during our discussions. Through out my research work he taught me all the tit-bits by appreciating, encouraging and at times by his scoldings too which made me grow as an individual. He is solely responsible for all the positive outcomes of my papers as he fully supported me to have open discussions with researchers around the world. During the collaborative work, his faith in me and his open minded approach helped me in picking up new problems and approaching the right people. His rigorous and tight sessions at the preliminary days helped me in gaining the right momentum and his free and zero restrictions in the final year made me handle problems on my own. I feel very happy to have had the opportunity to work with such a wonderful advisor, and I look forward to our continued collaborations.

My special word of thanks should also go to Dr. Tanja Hinderer. She guided me
very nicely the tidal deformability calculations via Skype while she was a postdoc in Max Plank Institute for Gravitational Physics, Germany. I thoroughly enjoyed the physics discussions with her and the  learnings during the summer school at ICTS-TIFR, Bangalore. Her lecture has helped me a lot  in understanding the basics and also in further calculations. Her help and friendly nature always made me feel at ease with her, and I feel privileged to be associated with a person like her during my PhD.

During my past years, I have collaborated with a fantastic group of people with whom I have discussed various aspects of the work presented here. I take this opportunity to express my deep sense of gratitude and respectful regards to Prof. B. K. Agrawal, for sharing knowledge and taking out his time to review, rectify and suggest on many of my papers. His ideas and concepts have a remarkable influence on my entire work. It was my pleasure to work with him and I truly hope that I will be fortunate again to work with him in future. Also, I thank Dr. M. Balasubramaniam for his comments and suggestions on my paper.

I thank my senior Shailesh bhai for giving me the help I needed as a beginner. He is the person from whom  I learnt the incorporation of ideas into a paper. I had really a good time with another senior of mine Subrat bhai, my best PhD mate, who never treated me as his junior. It was an enjoyable experience to work together in modifying the codes and working on problems. I thank Senthil for his active participations and for showing the enthusiasm to work together starting from the preliminary work to the end of our papers. 

I would also like to extend a huge and warm thanks to Swagatika. She has been 
always beside me during the happy and hard moments to push me through her 
motivations, unconditional support and care for me. I thank her for the 
english corrections and suggestions in two manuscripts of mine. I thank Abdul 
for checking the derivations of the tidal deformability section and his 
valuable inputs in this thesis. It was indeed my pleasure to discuss physics 
with him. I would also like to take this opportunity to convey special thanks 
to Swagatika, Swati, Neha di, Shreyansh bhai, and Dr. P. Landry for carefully 
reading my 
thesis because of which I can see its present shape.  Apart from this, I wish 
to thank my closest friends, Atul, Lakshmi, and Poonam for their moral support. Although, we devote less time together now a days as compared to the time 
spent in Jia Sarai Delhi but the sharing, bonding, affection and care is still 
intact in our hearts and will always be.

Last but not the least, this thesis would be meaningless if I do not mention the backbone of me, my mother. She is the most important person in my life. 
Facing all the hardships she single handedly managed my family and sacrificed 
everything for me and my siblings. She is the one whose strong will and 
determination towards life has given me the inspiration to work hard to the 
best of my capability. Also, my Uncle and Aunty need special attention as they 
are the ones who stood by me during my early struggling days and helped me 
financially, mentally and emotionally. They always treated me like their son 
and loved me immensely all these years. Without my Uncle's blessings and 
support I would not have dared to choose this field as my profession. After 
thanking my living gods; my mother, Uncle and Aunty, I thank the almighty to 
have showered all the blessings and  for providing me all the loving people 
around me whose support and love has made my world the most beautiful place to 
live in.
\vskip 2.0cm
%\vspace{1.5in}
%{\bf Date:}
%\hspace{3.2in} {\bf Bharat Kumar}
 \hspace*{0.32cm}
%\hrule

% \begin{abstract}
% We have studied certain processes, within the Standard Model (SM) and beyond,
% which are initiated by gluon-gluon fusion at Hadron colliders. We consider
% di-vector boson production in association with a jet via gluon fusion in the
% SM. We have taken the model of large extra space-dimensions as an example for the 
% model of new physics. Within this model we consider the associated production of 
% a boson and the KK-gravitons via gluon fusion. All these processes proceed via 
% quark loop diagrams at the leading order itself. Total cross-section calculation
% and various kinematic distributions constitute the main results of these studies. 
% \end{abstract}

% \end{comment}

\numberwithin{equation}{chapter}
\numberwithin{figure}{chapter}
\numberwithin{table}{chapter}
%%%%%%%%%%%%%%%%%%%%%%%%%%%%%%%%%%%%%%%%%%%%%%%%%%%%%%%%%%%%%%%%%%%%%%
\tableofcontents
\addcontentsline{toc}{chapter}{Synopsis}
\newpage
\begin{center}
{\large{\bf Synopsis }}
\end{center}
%\label{chapter:add}
%\begin{flushleft}
%{\large{\bf Synopsis }}\\
%\end{flushleft}
\vskip 2.0mm
Nuclear physicists have been trying to understand systematically the nuclear
systems ranging from finite nuclei to hot and dense nuclear matter within one 
theoretical framework. The nucleus is a many-body system 
constituting of strongly interacting and self-bound ensemble of protons and 
neutrons. Therefore, it becomes very difficult to describe the nucleus in a
transparent way. One of the most successful and widely used methods  
is the relativistic field theory which takes into account the relativistic 
nucleons interacting with each other by exchanging mesons. The simplest 
approach of such a theory is the so-called relativistic mean field (RMF)
approximation or quantum hadrodynamics that describes a nuclear system in the 
form of nucleonic Dirac field interacting with classical meson fields \cite{1}. In this approach, a nuclear system is considered to be composed of 
relativistic nucleons whose self-energy is determined through 
meson fields which are generated by the nuclear density. It has also achieved 
great success in describing a variety of nuclear phenomena both in the 
low-density region and in the high-density region. At the high-density 
region, aLIGO/Virgo detectors and heavy ion colliders have significantly 
extended the window from which the nuclear matter and neutron 
star (NS) can be studied \cite{2}. Such studies help in better understanding
of the RMF model and also give new ideas to improve it further. Since it is a phenomenological model, 
its efficiency can be tested by comparing with the experiment. The purpose of 
this thesis is to study the implications of nuclear interaction for nuclear 
structure and astrophysics within the RMF model. 

Since its discovery in 1896 by Becquerel, the $\alpha$-decay has remained a 
powerful tool to study the nuclear structure. The $\alpha$-decay theory was 
proposed by Gamow, Condon and Gurney in 1928. In simple
quantum mechanical view $\alpha$-decay is a quantum tunneling through the
Coulomb barrier, which is forbidden by classical mechanics. The $\alpha$-decay
is not the only decay mode found in the heavy nuclei, we can also find other 
exotic decay modes like $\beta$-decay, spontaneous fission and $cluster$-decay.
In $cluster$-decay, smaller nuclei  like $^{16}$O, $^{12}$C, $^{20}$Ne~\cite{3} and many other nuclei get emitted from a bigger nucleus. 
In the superheavy region of the nuclear chart, the prominent modes are 
the  $\alpha$-decay and spontaneous fission along the $\beta$-stability line.
In fission, the nucleus splits, either through radioactive decay or 
bombardment of subatomic particles like neutron. Moreover, the center of a 
heavy element spontaneously emits a charged particle as it breaks down into 
smaller nucleus. But it does not occur often, and happens only with the heavier elements like Uranium or Plutonium. Being thermally fissile nature, the two 
Uranium isotopes $^{233}$U and $^{235}$U, and one Plutonium isotope $^{239}$Pu 
in the actinide region are suitable for energy production. 
Among these three radioactive isotopes, $^{235}$U is the most important 
one for production of both nuclear power and nuclear bombs because $^{235}$U 
is readily or more easily fissioned by the absorption of a neutron of low 
energy ($<1$ eV) which is also available on earth crust. 
In chapter \ref{chapter3}, we examine if heavier neutron-rich isotopes of 
$^{216-250}$Th and $^{216-256}$U could exist having thermally fissile 
properties and if so what would be their stability and fission decay 
properties \cite{4}. The potential energy surface (PES) calculation is 
employed which 
is based on the self-consistent Hartree approach. For this purpose, we use 
the so-called quadrupole constrained calculation. The solution without 
constraint will lead to one of the spherical or deformation minimum, but one 
can never get a potential curve. To calculate the complete PES curve, instead 
of minimizing the original Hamiltonian, one has to minimize the following 
constraint Hamiltonian $H'$, defined as:
\begin{eqnarray}
<H'>=<H-\lambda Q>,
\end{eqnarray}
where $\lambda$ is the Lagrangian multiplier, which can be adjusted in such 
a way that the final deformation calculated from quadrupole $<Q>$ is $\beta_2$. 
Here, quadrupole deformation will always have the same angular dependence, 
{\it i e.}, spherical harmonics $Y_{20}(\theta,\phi)$. With the help of PES, we 
determine the fission barrier height which is the difference between the 
ground state and the highest saddle point. The higher the fission barrier, 
the longer is the fission lifetime. Further, the half-lives against the 
$\alpha$ decay of thermally fissile nuclei are obtained with a phenomenological 
formula of Viola and Seaborg, and also by using a quantum mechanical approach 
known as WKB method. A neutron-rich nucleus such as $^{254}$U is stable 
against $\alpha$ decay and spontaneous fission, because the presence of 
a large number of neutrons makes the fission barrier broader. Thus, we can test 
 the $\beta$ decay lifetime to determine the stability of such types of nuclei.

Mass distribution of the fission fragments is one of the crucial 
characteristics in the study of fission theory. Fong \cite{5} has proposed a 
statistical theory to study asymmetric mass division of fission fragments.  
This theory says that the relative fission probability is equal to the 
product of densities of quantum states of the fissioning nuclei at the 
scission point. For the first time, we apply the temperature dependent 
relativistic mean-field model (TRMF) to study the binary mass distributions for 
recently predicted thermally fissile nuclei $^{250}$U and $^{254}$Th using 
level density approach \cite{6}. The probability of yields of a particular 
fragment is obtained within the statistical theory with the inputs from TRMF
like excitation 
energies and level density parameters for the fission fragments at a given 
temperature. The inclusion of temperature dependent BCS pairing is discussed in
chapter \ref{chapter2}, and the level density parameter and its relation with
the relative mass yield is given in chapter \ref{chapter4}.

After discussing the properties of neutron-rich thermally fissile nuclei, 
we shift our focus towards the NS properties, which is a highly neutron-rich 
system of the order of $10^{57}$ neutrons. Nuclear physicists have tried to 
derive an ``equation of state'' (EoS) that describes dense nuclear matter 
which is found inside a NS. Using the EoS of RMF model, the Tolman-Oppenheimer-
Volkof (TOV) equation is solved to compute the structure of the NS.
The study of Love numbers in Newtonian theory dates back 100 years by famous
mathematician A. E. H. Love. In 2008, Flanagan and Hinderer proposed the 
gravitational wave phasing explicitly in terms of the tidal Love number in
general relativity, which was recently confirmed by Earth-based gravitational 
waves (GWs) detectors such as Advanced LIGO, and Virgo \cite{7}. The concept 
of the tides created between the NS is as follows: In NS-NS binary, 
the orbital motion causes the emission of 
GWs. As a result, there is  removal of energy and angular 
momentum from the system, which make the orbits to decrease in radius and 
increase in frequency. Thus, there is the inspiraling motion of the compact 
bodies. After 
that, the GWs enter the frequency band in the inspiral and we 
get a clear picture of the shape and phasing of the waves. When the stars have 
larger orbital separation the tidal interaction between binary is negligible, 
and gravitational frequency is also low because  NSs bodies behave as  
point masses at larger separation. When the star orbits get closer and closer, the orbital frequency increases sufficiently and the influence of the tidal 
interaction becomes significant. Finally, bodies get a tidal deformation thus 
influencing the shape of the bodies and phasing of the GWs.  
In chapter \ref{chapter5}, we have calculated various tidal Love numbers of 
the neutron and hyperon stars such as $k_2,k_3,k_4$, where the RMF EoSs have 
been imported from the hadronic and hyperonic nuclear matter under the 
$\beta$-equilibrium conditions \cite{8}. Apart from these Love numbers, we 
have also reported the shape Love numbers and magnetic Love numbers of the 
stars.

In the literature near about 263 RMF force parameter sets are available, which 
determine the nuclear matter properties around the saturation density \cite{9}.
 It is found that only 7-8$\%$ models satisfy all the experimental and empirical
data. So, we need such type of models which can satisfy not only the nuclear 
matter experimental constraints but also the bulk properties of the finite 
nuclei. In addition, we can also test for EoS that comes from the observation 
of the NS mass. 
The Lagrangian of the G2 parameter set contains most of the self and 
cross-couplings which give better results not only for the finite nuclei 
but also for the nuclear matter systems. Still, the G2 model needs some 
corrections to get more reliable and better results. We have added the extra 
degrees of freedom {\it i.e.,} isovector scalar $\delta$ meson and the cross-coupling of 
$\rho$ mesons to the $\sigma$ and $\omega$ mesons to make the Lagrangian 
more proficient. The contribution of the $\delta$ meson is to take care of the 
large asymmetry of the system. Also, $\delta$ meson results in a stiff 
EoS at high densities due to the increased $\rho$ meson 
nucleon coupling in order to reproduce the empirical symmetry energy at the 
nuclear saturation density.
However, recent experimental results seem to suggest the need of an EoS 
softer even than those of most nonlinear RMF models without the 
inclusion of the $\delta$ meson. For finite nuclei, the inclusion of the 
$\delta$ meson can improve the description of bulk properties of finite 
nuclei, in particular at the drip lines where due to the large isospin 
asymmetry its contribution might be appreciable. 
The contribution of the $\delta$ meson might also be examined by analyzing
the isovector part of the spin-orbit interaction. The $\omega-\rho$ 
cross-coupling plays a crucial role to modify the neutron-skin thickness of 
finite nuclei and the density-dependent symmetry energy of the nuclear matter
at saturation density. 
Therefore, we have also tried to search for a suitable parameter set to 
reproduce all the experimental and empirical constraints.
We have implemented the simulated annealing method to the problem of searching
for global minimum in the hyperspace of the $\chi^2$ function, which depends
on the values of the parameters of a Lagrangian density of the extended RMF 
model. The set of experimental data used in our fitting procedure include 
the binding energy and charge radii of the 8 spherical nuclei, ranging from 
lighter nuclei to heavy ones. While fitting, we have also included the binding 
energy per particle and symmetry energy at saturation density. Finally, we got  
new extended RMF parameter sets named as G3, and IOPB-I, which are applicable for 
finite nuclei, infinite nuclear matter, and neutron stars \cite{9}. In chapter 
\ref{chapter6}, we have discussed more about the G3 and IOPB-I parameter sets.

\addcontentsline{toc}{chapter}{List of Figures}
\listoffigures 
\listoftables
\addcontentsline{toc}{chapter}{List of Tables}

%\chapter{synopsis}
\newpage
\setcounter{page}{1}
\pagenumbering{arabic}

% \pagestyle{headings}
 % Introduction and Overview
%\include{chhabi1}

%\include{synopsis}
%\setcounter{equation}{0}
%\setcounter{figure}{0}

\newpage
\chapter{Introduction }
\label{chapter1}
In 1911, Rutherford and his collaborators discovered nucleus from the famous 
gold-foil scattering experiment. This experiment suggested that mass of the 
atom is concentrated at the center of the atom known as nucleus. After that, 
it was known that the atomic mass number A of a nucleus is a bit more than 
twice the atomic number Z. Since then, many years have passed in pursuit of 
understanding the structure of the nucleus. Interestingly, Landau published a 
seminal paper in 1932 where he showed that the {\it ``density of matter 
becomes so great that atomic nuclei come in contact, forming one gigantic 
nucleus'' \cite{lan32}.} 
This paper of Landau marked the 
first theoretical speculation on the existence of neutron stars(NS). As proposed 
by Landau with his theory, a complete picture of the internal 
configuration of the nucleus in the form of experimental backing came up with the discovery of the neutron 
in 1932 when J. Chadwick used scattering data to calculate the mass of this 
neutral particle \cite{chad32}.
At that time, the nature of nucleon-nucleon interaction was not known because 
of its charge independent nuclear force. In 1935, Yukawa explained the 
nature of nuclear force and suggested that the massive bosons(mesons) mediate 
the interaction between nucleons \cite{yuka35}. He found theoretically that
the $\pi-$meson mass is nearly 140 MeV, which provides an explanation for the residual
strong force between nucleons. The muon has a mass of 105.7 MeV and does not 
participate in the strong nuclear interaction, which was found in 1937 
experimentally in cosmic radiation and interpreted as a particle as suggested 
by Yukawa. But, Yukawa was not satisfied with this finding. Finally, a landmark progress was achieved in 1947 by a team led by the English physicist C. F. 
Powell with the discovery of the $\pi-$meson in cosmic-ray particle 
interaction in Berkeley. Soon after, the heavier mesons $\omega,\rho$ and $\delta$ were found 
in different laboratories, which cover the nature of strong interaction 
\cite{pdg}. Therefore, the concept of immensely strong nuclear force that binds nucleons came into picture, which is why the nuclear process is 
able to release a tremendous amount of energy---such as the thermonuclear fusion
reaction that happen in the sun to produce virtually all elements in a process
known as {\it nucleosynthesis}.

Later, a large number of phenomena and properties of nuclei came to be known,
but many of them lacked in clarity. Nuclear fission is one of the
most important discoveries in nuclear physics by Otto Hahn and Strassmann and
subsequently, the chain reaction by E. Fermi led the foundation of energy 
production from the nucleus. The phenomenon of nuclear fission is a very 
complex 
process producing a huge amount of energy when heavy elements like uranium 
and thorium  are irradiated with slow neutrons. The theory of fission process 
was first given by Meitner in 1939\cite{meit39}. In this process, a parent 
nucleus goes from the  ground state to scission point through a deformation and
then splits into two daughter nuclei. This decay process can be described as
an interplay between the nuclear surface energy coming from the strong 
interaction and the Coulomb repulsion~\cite{bohr39}. 

In this dissertation, we have addressed the answers related to the following 
questions as highlighted in the 2007 Nuclear Science Long Range Plan \cite{sci07}:
\begin{enumerate}
\item {\it What is the nature of the nuclear force that binds protons and 
neutrons into stable nuclei and rare isotopes?}
\item{\it What is the next possibility of thermally fissile neutron-rich nuclei?}
\item {\it What is the nature of neutron stars and dense nuclear matter?}
\end{enumerate}

Theoretically, the above raised questions can be answered with the help of 
various experimental work as follows:

(i) The study of very neutron-rich nuclei at and beyond the drip-line 
employing state-of-the art experimental techniques at leading facilities 
worldwide---such as new radioactive ion beam facilities HIFR at CSR \cite{hir,csr}, FAIR at GSI \cite{fair,gsi}, Spiral at GANIL \cite{ganil}, RIBF at RIKEN \cite{riken} and FRIB at MSU \cite{msu}, respectively. One of the primary goals 
of the above mentioned research facilities on fission barrier and also 
in the production of neutron-rich heavy elements is to allow the study of their 
lifetimes and decays. To have a better understanding on the diversity of 
elements, the nuclear landscape is displayed in Fig. \ref{land} \cite{NDCC}. 
The straight red dotted line shows the N=Z stable light nuclei. The red lines
show the magic numbers. The black squares represent 255 stable nuclei existing
in nature and their half-lives comparable to or longer than the age of the Earth.
The stable nuclei are surrounded by the yellow area which represents a total of 
3225 unstable nuclei, already synthesized in different laboratories in the world. 
The various theoretical models suggest that another 5000 isotopes  
depicting the green area comprises the unknown proton and neutron-rich regions that can be explored experimentally in the near future.

\begin{figure}[!b]
\begin{center}
        \includegraphics[width=0.7\columnwidth]{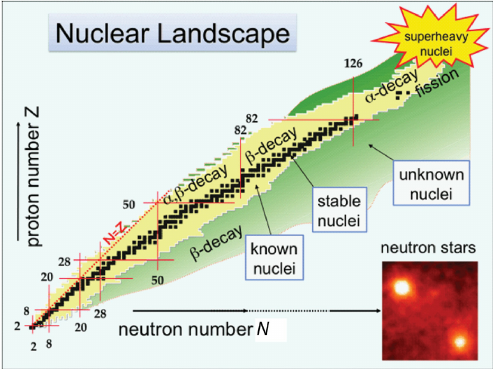}
	\caption{ Nuclear landscape \cite{NDCC}: the black squares represent 
stable plus very long-lived nuclei that define the valley of stability. The 
yellow region assigns the range of unstable nuclei that have been explored by 
reactions employing stable nuclei by addition or removal of neutrons or 
protons. Many thousands of radioactive nuclei far from the valley of 
stability remain to be explored; this "terra incognito" is indicated in green. }
        \label{land}
\end{center}
\end{figure}

(ii) The equation of state (EoS) describes the complex behavior of the dense 
nuclear matter that makes up neutron stars, which is difficult to extract from 
general X-ray astronomy. However, in August 2017, the Advanced Laser 
Interferometer Gravitational-Wave Observatory (aLIGO) and Virgo detectors 
made the first-ever observation of gravitational waves generated by the merger 
of two neutron stars \cite{BNS}. This observation provides new insights 
on the properties of neutron stars, such as its mass and ``tidal deformability''---the stiffness of a star in response to the forces caused 
by its companion's gravitational field. This observation has helped to 
decode many puzzling questions relating to the nature of dense matter, on the 
synthesis of the heavy elements, and to test gravity in the highly-relativistic or supranuclear-density regime \cite{BNS}.
%However, the ground-based gravitational-wave detectors 
%such as advanced Laser Interferometer Gravitational-Wave Observatory (aLIGO), 
%and Virgo detected gravitational waves from the neutron-star merger that 
%provides the new insights into the properties of neutron stars. 
%excitations in the quantum system which may arise due to coupling of continuum 
%states with the loosely bound states, lower matter density and extreme is spin. 
\begin{figure}[!b]
\begin{center}
        \includegraphics[width=0.7\columnwidth]{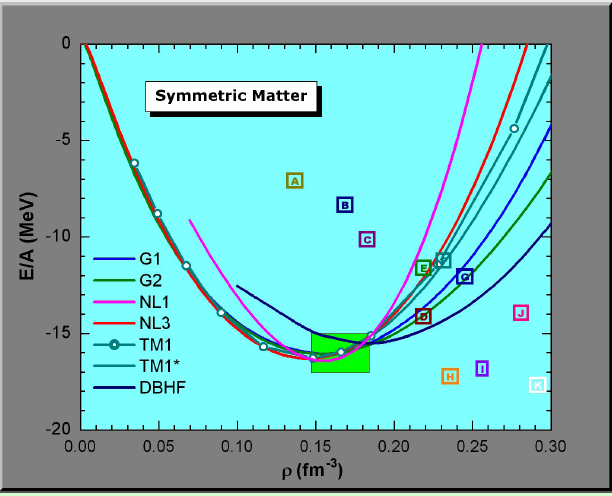}
\caption{ Binding energy (E/A) in MeV for symmetric nuclear matter as a 
function of density. The squares represent various non-relativistic models, 
solid lines represent the relativistic models, and the green box represents the 
empirical data.}
        \label{coster}
\end{center}
\end{figure}

\section{Effective Mean-Field Theory}
At present, nuclear physics and nuclear astrophysics are well described within
the self-consistent effective mean-field models \cite{furnstahl97}. These
effective theories are not only successful to describe the properties of finite
nuclei  but also explain the nuclear matter at supranormal densities \cite{arun05}. Recently, large number of nuclear phenomena are predicted near the nuclear
drip lines within  the relativistic and non-relativistic formalisms
\cite{estal01,chab98,sand20}.  Consequently, several experiments are
planned in various laboratories to probe the deeper side of the unknown
nuclear territories, {\it i.e.}, the neutron and proton drip lines.
Among the effective theories, the relativistic mean-field (RMF) model is
one of the most successful self-consistent formalisms, that is currently
drawing attention to the theoretical studies of such systems.

Although the construction of the energy density functional for the
RMF model is different than those for the  non-relativistic models,
such as Skyrme \cite{doba84,doba96} and Gogny interactions \cite{dech80},
the obtained results for finite nuclei are in general very close to each
other. The same accuracy in prediction is also valid for the properties
of the neutron stars.  At higher densities, the relativistic effects are
accounted appropriately within the RMF model \cite{walecka74}. In the RMF 
model the interactions among nucleons are described through the exchange of 
mesons.  These mesons are collectively taken as effective fields and denoted 
by classical numbers. In brief, the RMF formalism
is the relativistic Hartree or Hartree-Fock approximation to the one-boson
exchange (OBE) theory of nuclear interactions.  In OBE theory, the
nucleons interact with each other by exchange of isovector $\pi$, $\rho$, and
$\delta$ mesons and isoscalars like $\eta$, and $\omega$ mesons. The
$\pi$, and $\eta$ mesons are pseudo-scalar in nature and do not obey the
ground-state parity symmetry. At the mean-field level, they do not contribute to
the ground-state properties of even nuclei.

The first and simplest successful relativistic Lagrangian is formed by
taking only the contribution of the $\sigma$, $\omega$ and $\rho$ mesons
into account without any nonlinear term for the Lagrangian density 
\cite{walecka74}.
This model predicts an unreasonably large incompressibility $K$ of $\sim 550$
MeV for the infinite nuclear matter at saturation \cite{walecka74}. To lower the
value of $K$ to an acceptable range, the self-coupling terms in the $\sigma$
meson are included by Boguta and Bodmer \cite{boguta77}. Based on this
Lagrangian density, a large number of parameter sets, such as NL1 \cite{pg89},
NL2 \cite{pg89}, NL-SH \cite{sharma93}, NL3 \cite{lala97} and NL3$^*$ \cite{nl3s} are  calibrated. The addition of $\sigma$ meson self-couplings improve the
quality of finite nuclei properties and incompressibility remarkably. However,
the equation of states at supranormal densities are quite stiff.
Thus, the addition of vector meson self-coupling is introduced into the
Lagrangian density and different parameter sets are constructed
\cite{bodmer91,gmuca92,sugahara94}. These parameter sets are able to explain the
finite nuclei and nuclear matter properties to a great extent, but the existence
of the Coester band as well as the three-body effects need to be addressed.
Fig. \ref{coster} depicts the binding energy per nucleon as a function of 
baryon density calculated by different relativistic and non-relativistic models 
for symmetric nuclear matter (SNM). It is noticed \cite{coster} that the EoS 
calculated by non-relativistic model do not reach to the  Coester-band or 
empirical saturation point of SNM ($E/A\simeq-16.0$ MeV at $\rho_{0}\simeq0.16$ 
fm$^{-3}$). At higher density regime, all the models are showing different 
nature, which is used for the NS matter and hence questions the reliability of 
these models. Subsequently, nuclear physicists also changed their way of 
thinking and introduced different strategies to improve the result by 
designing the density-dependent coupling constants and effective-field-theory 
motivated relativistic mean field (ERMF) model \cite{furnstahl97,vretenar00}.

Further, motivated by the effective field theory, Furnstahl {\it et al.}
\cite{furnstahl97} used all possible couplings up to fourth order of the
expansion, exploiting the naive dimensional analysis (NDA) and
naturalness, and obtained  the G1 and G2 parameter sets. In the Lagrangian
density, they considered only the contributions of the isoscalar-isovector
cross-coupling, which has a greater implication for the neutron radius and
EoS of asymmetric nuclear matter \cite{pika05}.
Later on it is realized that the contributions of $\delta$ mesons are also
needed to explain certain properties of nuclear phenomena in extreme conditions
\cite{kubis97,G3}. Though the contributions of the $\delta$ mesons to the  bulk
properties are nominal in the normal nuclear matter, the effects are significant
for highly asymmetric dense nuclear matter. The $\delta$ meson splits the
effective masses of proton and neutron, which influences the production of $K^{+,-}$ and
$\pi^{+}/\pi^{-}$ in the heavy-ion collision (HIC)~\cite{ferini05}. Also,
it increases the proton fraction in the $\beta$-stable matter and modifies
the transport properties of the neutron star and heavy-ion reactions
\cite{chiu64,bahc65,lati91}.  The source terms for both the $\rho$ and $\delta$
mesons contain isospin density, but their origins are different. The $\rho$ meson arises from the asymmetry in the number density and the evolution of the $\delta$
meson is from the mass asymmetry of the nucleons.  The inclusion of $\delta$
mesons could influence the certain physical observables like neutron-skin
thickness, isotopic shift, two neutron separation energy $S_{2n}$, symmetry
energy $S(\rho)$, giant dipole resonance, and effective mass of the nucleons,
which are correlated with the isovector channel of the interaction.
The density dependence of symmetry energy is  strongly  correlated
with the neutron-skin thickness in heavy nuclei, but until now experiments
have not fixed the accurate value of the neutron radius, which is under
consideration for verification in parity-violating electron-nucleus
scattering experiments~\cite{horowitz01b,vretenar00}. 

Inspired by all the previous parameter sets we too tried to search for a 
suitable one devoid of the shortcomings mentioned earlier and we finally 
developed new parameter sets G3 and IOPB-I for finite and infinite nuclear 
matter, and neutron stars system within the effective field theory motivated 
RMF theory. The Lagrangian of the G3 set includes all the necessary terms such 
as $\omega-\rho$, and $\sigma-\rho$ cross-couplings, and also $\delta$ meson.  
These cross-couplings modified the nature of the neutron skin-thickness for 
finite nuclei as well as the density-dependent symmetry energy and also 
constrains the equation of state of the pure neutron matter. The detailed 
analysis of the role of each term in the Lagrangian of the G3 and IOPB-I sets 
will be discussed in the next chapter \ref{chapter2}. 
Till there is some discrepancy of the RMF models, which will discuss in the 
next subsection. 

\subsection{Limitations of the model}
It is important to mention a few points about the limitations of the present
approach which are as follows:

(1) In RMF formalism we work in the mean-field approximation of the
meson field. In this approximation, we neglect the vacuum fluctuation,
which is an indispensable part of the relativistic formalism.
While calculating the nucleonic dynamics, we neglect the negative energy
solution which means we work in the no sea approximation~\cite{rufa86}.
It has been discussed that the no-sea approximation and quantum fluctuation
can improve the results upto a maximum of 20\% for very light nuclei \cite{zhu91}. Therefore, the mean-field is not a preferable approach for the
light region of the periodic table. However, for the heavy masses, this
mean-field approach is quite good and can be used for any practical purpose. \\

(2) In order to solve the nuclear many-body system, here we use the Hartee
formalism and neglect Fock term, which corresponds to the exchange correlation.\\

(3) To take care of the pairing correlation, we use a BCS-type pairing
approach. This gives good results for the nuclei near the $\beta$-stability
line, but it fails to incorporate properly the pairing correlation for
the nuclei away from the $\beta$-stability line and superheavy nuclei~\cite{karatzikos10}. Thus, a better approach like Hartree-Fock-Bogoliubov type pairing correlation is more suitable for the present region \cite{ring80,ring96}. \\

(4) Parametrization  plays an important role in improvising the results.
The constants in RMF parametrizations are determined by fixing the
experimental data for few spherical nuclei. We expect that the results may
be improved by refitting the force parameters for more number of nuclei,
including the deformed isotopes.\\

(5) The basic assumption in the RMF theory is that two nucleons interact
with each other through the exchange of various mesons. There is no direct
inclusion of three-body or higher effects. This effect is taken care of
partially by including the self-coupling of mesons, and in recent relativistic
approach various cross-couplings are added because of their importance.\\

(6) Although, there are various mesons observed experimentally, few of them
are taken into consideration in the nucleon-nucleon interaction. The 
contribution of
some of them are prohibited for symmetry reason and many are neglected
due to their negligible contributions, because of their heavy mass. However,
some of them has substantial contribution to the properties of nuclei,
especially when the neutron-proton asymmetry is greater, such as $\delta$
meson~\cite{kubis97,singh14}.

With this, we move on to the following section to introduce the applications related to RMF theory.

\begin{figure}[!b]
\begin{center}
        \includegraphics[width=0.7\columnwidth]{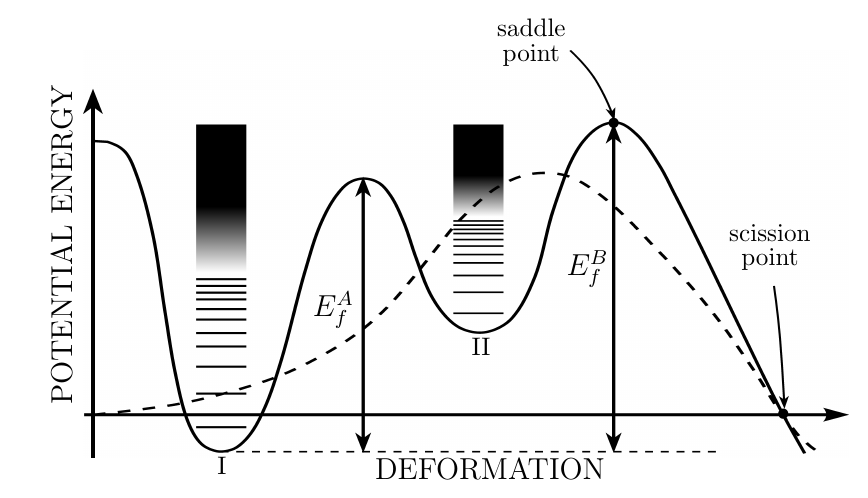}
\caption{Potential energy surface as a function of deformation parameter. The 
	solid line of the fission barrier with ``shell correction''. A dashed line represents the liquid drop fission barrier. The figure is taken from 
Ref. \cite{tpes}.}
        \label{barrier}
\end{center}
\end{figure}

\section{Nuclear fission}
The process of nuclear fission gives rise to many puzzles and complexities, it 
has also proved to be successful in answering many questions of the nuclear
phenomena. But a complete theoretical explanation is still not at hand. 
Initially, after the discovery of nuclear fission N. Bohr and J. Wheeler 
suggested a theoretical explanation by liquid drop model (LDM) of atomic nuclei
\cite{bohr39}. A fissile nucleus is treated as an incompressible liquid drop. 
These drops are uniformly distributed electric charge over the volume of a 
nucleus. The qualitative description of the dependence of the potential 
energy surface on the arbitrary deformation is shown in figure \ref{barrier}. 
The dashed line represents a potential energy surface due to the charged 
liquid drop. In this process, the fissioning nucleus goes through a number of 
intermediate states, and then liquid drops break up into two smaller 
droplets at scission point. Although, the LDM very well explain the general
properties for the actinide nuclei, but it is found that all nuclei in the
ground state show spherical shape. So, it is unable to explain the asymmetric
mass division of the fission fragments which is described  by Strutinsky
in his method of ``shell corrections'' \cite{stru}. The solid line represents 
the inclusion of liquid drop energy with shell correction energy. The ground state (I), first barrier ($E^{A}_f$), second minimum (II), and the second barrier 
($E^{B}_f$) are marked. The single-particle energy of the nucleons are affected 
by the shell correction as marked on the position I and II in figure 
\ref{barrier}.
The gaps in the single-particle spectra have given additional binding to the
nucleus which lower the ground state with respect to the liquid drop energy. 
Therefore, it create a barrier against fission. The height of the fission
barrier ($E^B_{f}$) is calculated by taking the difference between the 
saddle and the ground state energies. The existence of the more massive nuclei 
is found on the saddle point, which makes it quite an exciting and valuable 
quantity to study. Nowadays, many modern approaches of using 
phenomenological nuclear-energy density functionals have come up to obtain 
nuclear ground-state properties in a self-consistent way. The construction of 
potential energy surface as a function of deformation parameter is calculated 
using the Lagrange multiplier to the Hamiltonin and minimize it. 

The study of fission mass distribution is one of the major insights of the 
fission process. Conventionally, there are two different approaches, the 
statistical and the dynamical approaches for the study of fission process 
\cite{gre73,fon56}. The
latter is a collective calculations of the potential energy surface
and the mass asymmetry. Further, the fission fragments are determined
either as the minimum in the potential energy surface or by the maximum in
the WKB penetration probability integral for the fission fragments. In
statistical theory \cite{fon56},
the relative probability of the fission process depends on the density of the
quantum states of the fragments at scission point. The mass and the
charge distribution of the binary and the ternary fission is studied
using the single particle energies of the finite range droplet model
(FRDM) \cite{mrd81,mbs2014}. In FRDM \cite{moller95} formalism, the energy
at a given temperature is calculated using the relation $E(T) = \sum_i
n_i \epsilon_i$  with $n_i$ and $\epsilon_i$, the Fermi-distribution
function  and the single-particle energy corresponding to the ground
state deformation \cite{mbs2014}. The temperature dependence of the
deformations of the fission fragments and  the contributions of the pairing correlations are ignored.
But the self-consistent temperature dependent RMF theory is taken care by the 
quantity as stated above.

\begin{figure}[!b]
\begin{center}
        \includegraphics[width=0.7\columnwidth]{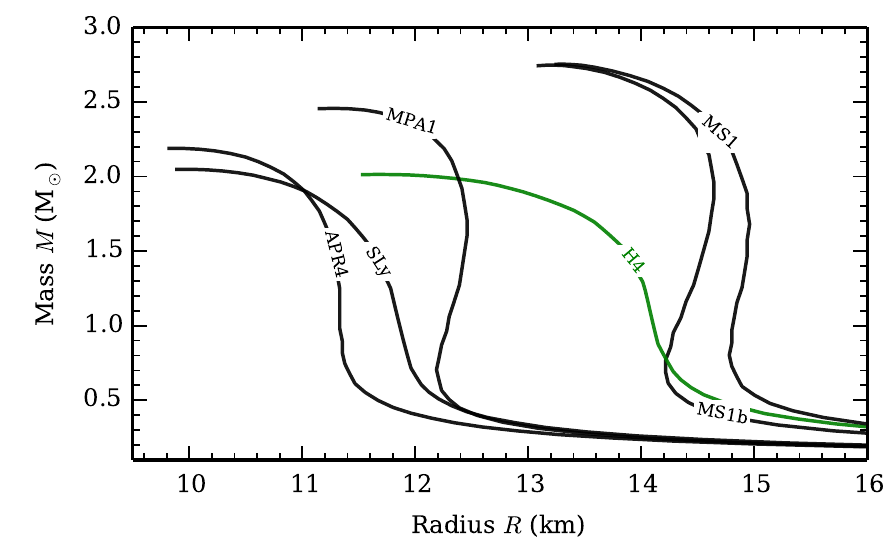}
\caption{The mass-radius plot for the neutron (black curve) and hyperon 
	(green curve) stars \cite{jreadp}.}
        \label{mrplot}
\end{center}
\end{figure}

\section{GW170817:Tidal deformability of neutron stars}
The neutron star is a tiny and compact object in the universe, which is 
formed after a core-collapse supernovae explosion. The mass of a NS is 
precisely measured from the observation of binary pulsar system, and its 
mass is observed to be as large as of 2$M_\odot$ \cite{demo10,antoni13}. 
The size of a 1.4$M_\odot$ NS is about 10 km (or more), and has a central 
density $\rho_c$ as high as 5 to 
10 times larger than the density of normal nuclei $\rho_{nuc}=2.8\times10^{14}$
g/cm$^3$ \cite{steiner10}. Therefore, the NS is one of the densest forms of 
matter in the universe. The nature and composition of such ultra-dense matter 
have remained an essential question in the nuclear physics. At higher density
($2.0\gtrsim \rho/\rho_0\gtrsim 15.0$), various possibilities predict the 
emergence of new phases of matter, condensates of particles such as hyperons, 
kaons, etc.---which actively depend on the theoretical models. The structure 
of NS sensitively depends on the nuclear EoS. The star mass-radius plot is 
shown in Fig. \ref{mrplot}, which is the solution of Tolman-Oppenheimer-Volkoff (TOV) equation, where energy density and pressure are the inputs \cite{tov}. 
A stiff (or hard) EoS tends to have larger pressure gradient for a given 
density. Such an EoS would be harder to compress and offers more support 
against gravity. Conversely, a soft EoS has smaller pressures gradient, and is 
more easily compressed. The figure shows that stars calculated with a stiffer 
EoS yield greater maximum masses and larger radii than stars derived from 
softer EoS. For example, the MS1 and MS1b EoSs suggest larger and massive NS \cite{jreadp}.

\begin{figure}[!b]
\begin{center}
        \includegraphics[width=0.7\columnwidth]{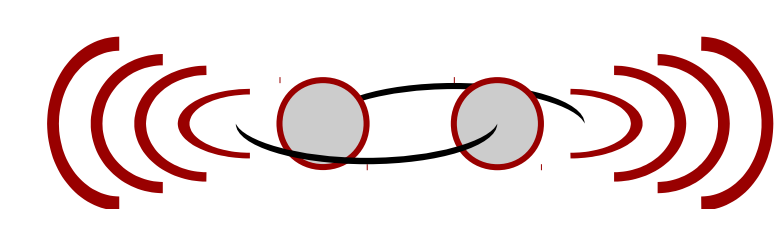}
\caption{The schematic representation of the gravitational waves produced by 
the inspiral of compact bodies in a binary neutron stars system. The figure is
taken from Ref. \cite{philtalk}.}
        \label{tidalphase}
\end{center}
\end{figure}

In 1974, Russell Hulse and Joseph Taylor discovered the first binary neutron 
star system (BNS) called as PSR 1913+16. The most exciting measurement in this 
system is the observation which revealed that BNS orbiting towards
each other and were also shrinking at a rate of 10 mm per year. This shrinkage 
is caused by the loss of orbital energy due to gravitational radiation \cite{hulse}. The measurements provided the first proof that such waves exist. Fig. \ref{tidalphase} is the graphical 
representation of the gravitational waves produced by the BNS \cite{philtalk}. 
In this process, two stars rotate about a common center of mass. As 
they rotate, they send gravitational waves
and in the process, the orbits lose energy and get closer and closer, which is 
called inspiralling. As they get closer they send off more gravitational waves 
and gets even closer, eventually colliding with each other. Just before the 
merger, the star get tidally disturbed by external tidal field because of 
 the other companion star---which produces a small correction in the phase of 
 the gravitational waves. The inspiral phase of 
a NS-NS merger creates an extremely strong tidal gravitational field that deform 
the multipolar structure of the stars. This effect can be specified in terms of 
the so-called tidal deformability $\lambda$ of the stars (or tidal Love number, 
which gives the information about the internal structure of the NS) \cite{tanja,flan,tanja1}.

\begin{figure}[!b]
\begin{center}
        \includegraphics[width=0.7\columnwidth]{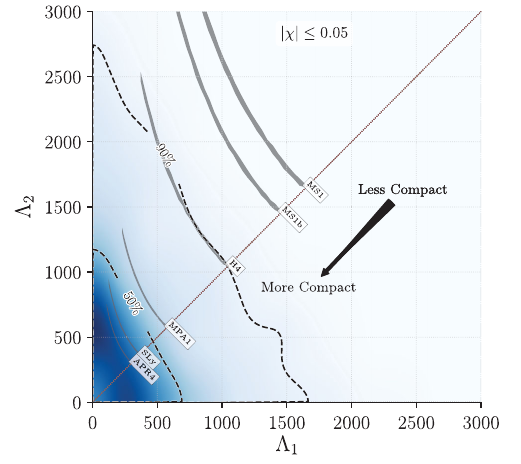}
\caption{The plot for the tidal deformability of binary neutron stars for the
        low-spin prior as given in Fig. 5 of GW170817 \cite{BNS}. The dashed line
        represents the $90\%$ and $50\%$ confidence limit. The diagonal dashed
        line marked the $\Lambda_1=\Lambda_2$ boundary.}
        \label{gw170817}
\end{center}
\end{figure}

The Love numbers directly affect the size of tidal bulging on bodies which
occur due to the non-uniform external gravitational field. To explain this we
take the case of Sun and Earth where Sun is considered as a point mass. It
has been observed that the gravitational field of the Sun is strongest on 
that side of the Earth which is much close to the Sun than the other side.
As a result of which there is relative acceleration leading to the quadrupole
deformation as viewed in the Earth's center-of-mass frame. Thus the overall
result is the formation of the two high tides per day at a given point on
Earth.

Placing a spherical star in a static external
quadrupolar tidal field $\mathcal{E}_{ij}$ results in deformation of the star
along with  quadrupole deformation, which is the leading order perturbation.
Such a deformation is measured by \cite{tanja,tanja1}
\begin{eqnarray}
        \lambda = -\frac {Q_{ij}}{\mathcal{E}_{ij}}=\frac{2}{3} k_2 R^{5} \; ,
        \label{eq40}
\end{eqnarray}
\begin{eqnarray}
        \Lambda=\frac{2k_2}{3C^5},
        \label{eq41}
\end{eqnarray}
where $Q_{ij}$ is the induced quadrupole moment of a star in binary, and
 $\cal{E}$$_{ij}$ is the static external quadrupole tidal field of the companion
 star. $\lambda$ is the tidal deformability parameter depending on the EoS via
both the NS radius and a dimensionless quantity $k_2$, called the second Love
number \cite{flan,tanja1}. $\Lambda$ is the dimensionless
version of $\lambda$, and $C$ is the compactness parameter ($C=M/R$). However,
in general relativity we have to distinguish $k_2$ between gravitational
fields generated by masses (electric type), and those generated by the motion 
of masses, i.e., mass currents (magnetic type)  that has no analogue in
Newtonian gravity \cite{phil,thib1}.
Equation \eqref{eq40} indicates that $\lambda$ strongly depends on the radius 
of the NS as well as on the value of $k_2$. Moreover, $k_2$ depends on the 
internal structure of the constituent body and directly enters into the 
gravitational wave phase of inspiraling BNS which in turn conveys information 
about the EoS. As the radii of the NS increases, the deformation by the 
external field becomes large as there will be an increase in gravitational 
gradient with the simultaneous increase in radius. In other words, 
stiff (soft) EoS  yields large (small) deformation in the BNS system. 
Since the force of attraction between stars becomes
more and more important in the course of time, because of the reduction of the
orbital distance between them. The orbital distance between the binary decreases
as the companion star emits gravitational radiation. As a result, the binary
accelerates and finally merges with each other and possibly turns to a black
hole. Before the merger, the estimation of the leading order quadrupole 
electric tidal Love number $k_{2}$ along with other higher order Love numbers 
$k_{3}$ and $k_{4}$ are very important for the detection of gravitational wave.

Recently, advanced LIGO and Virgo detectors informed first time the
direct detection of gravitational waves from inspiralling NS-NS binary, which
is referred as GW170817 \cite{BNS}. The chirp mass of binary is 
found to be 1.188$^{+0.004}_{-0.002}M_\odot$ for the 90$\%$ credible intervals, 
which is precisely measured from data analysis of GW170817.
The dimensionless tidal deformabilities $\Lambda_1$ and $\Lambda_2$ with 
$90\%$ and $50\%$ confidence limit are shown in Fig. \ref{gw170817}, that 
obtained for two stars in the BNS merger observed by GW170817. The  
measurement is reported in the form of a limit given for the average 
dimensionless tidal deformability$^{1}$\footnotetext[1]{The definition of the 
dimensionless tidal deformability $\tilde{\Lambda}$ is given by Eq. 
\eqref{lambar}. } $\tilde{\Lambda}\le800$ for low-spin prior \ref{gw170817}.
According to GW170817, the stiffer EoSs are ruled out such as MS1, and MS1b,
respectively. 

\section{Plan of the thesis}
The main aim of this thesis is to study the implications of nuclear interaction
for nuclear structure and astrophysics within the RMF model. Also, we extend
the version of RMF models which is successful in the finite and infinite
nuclear matter regime. 

The thesis is organized as follows. After the introduction, we outline the ERMF Lagrangian including $\omega-\rho$ cross-coupling and $\delta$ meson in 
chapter \ref{chapter2}. The equation of motion of the fields 
are derived for different fields $\phi,\sigma,\omega,\rho,\delta$, and 
electromagnetic field. The temperature dependent BCS and Quasi-BCS pairing 
correlation for the open shell nuclei are also discussed in this chapter. 
I would outline briefly the EoS for infinite nuclear matter and its 
properties. These derivations are the building blocks of different 
calculations presented in the forthcoming chapters. 

In chapter \ref{chapter3}, we use the axially deformed RMF formalism to 
calculate the bulk properties of the thermally fissile nuclei including the 
potential energy surface and single-particle energy levels. For this purpose,
we describe the selection of the basis space for exotic nuclei which 
require a large model space to get a proper convergence solution of the
system. Finally, various decay modes are calculated using either empirical 
formula or by using the well-known double folding formalism with M3Y 
nucleon-nucleon potential.

In chapter \ref{chapter4}, we use the temperature-dependent axially deformed
RMF formalism. To calculate the total binding energy of the system, we use the
axially symmetric harmonic oscillator basis space for Fermion and Boson. 
The excitation energy, the level density parameter, and inverse level density 
parameters are calculated within the TRMF formalism. The relative mass 
distributions of neutron-rich thermally fissile nuclei $^{254}$Th and 
$^{250}$U are studied using a statistical model. The calculated results are 
compared with the finite range droplet models (FRDM) predictions.

The Newtonian and relativistic tidal Love number mathematical derivations are 
given in the chapter \ref{chapter5}. In particular,we will derive the 
important expression for the various tidal Love numbers, average tidal deformability, orbital frequency, gravitational energy flux, orbital decay and gravitational wave phase of the binary neutron star.
Next, we discuss the state of matter in neutron and hyperon star by using the 
condition of hydrostatic beta equilibrium. 
Then we present the mass-radius results with the help of famous TOV equation.
The various tidal Love numbers and tidal deformability of neutron and hyperon 
stars are calculated. The cut-off frequency of the neutron and hyperon stars 
are also discuss in this chapter. Throughout the thesis, we have taken the value of $G=c=1$.

In chapter \ref{chapter6}, following the derivation in chapter \ref{chapter2} of binding energy, charge radius, and analytical expression for the symmetry 
energy and incompressibility coefficient of the symmetric nuclear matter at 
saturation we discuss the strategy of the parameter fitting using the 
simulated annealing method in chapter \ref{chapter6}. After getting the 
new parameter sets G3, and IOPB-I, the results on binding energy, two-neutron 
separation energy, isotopic shift, and neutron-skin thickness of finite nuclei 
are discussed thoroughly. The mass-radius and tidal deformability of neutron 
star obtained by new parameter sets are also discussed in chapter \ref{chapter6}. 

Finally,the summary and concluding remarks are given in chapter \ref{chapter7}.

\setcounter{equation}{0}
\setcounter{figure}{0}

\newpage
\chapter{Relativistic mean-field theory }
\label{chapter2}

%The complex nuclear many-body problem can not be solved correctly. Therefore, 
%the relativistic quantum field theory for the nuclear many-body problem 
%developed by J. D. Walecka in 1974, 
Quantum hadrodynamics is a quantum field theory where nucleons and mesons are 
treated as elementary degrees of freedom. The credit for origin of relativistic nuclear model goes to the work of D$\ddot{u}$err \cite{duerr} who revised the 
non-relativistic field theoretical nuclear model of Johnson and Teller \cite{teller}, and was re-introduced by Green and Miller in \cite{miller}. 
Finally, the simple $\sigma-\omega$ model was introduced by J. D. Walecka 
in 1974, who mentioned the main features of the nucleon-nucleon 
interaction \cite{walecka74}. This model has been renormalizable. Unfortunately, renormalizable has encountered difficulties due to substantial effects from 
loop integrals that incorporate the dynamics of the quantum vacuum. The 
effective theory is an alternative. In this chapter, we first added the 
isovector part into the effective Lagrangian in which the coupling of nucleons 
to the $\delta$ and $\rho$ mesons and the cross-coupling of the $\rho$ mesons 
to the $\sigma$ and $\omega$ mesons along with their interactions to 
$\sigma,$ $\omega,$ and $\rho$ mesons are included. Then, we have explained how 
to use it to compute ground-state properties of finite nuclei.  Next, we will 
discuss the 
pairing correlations for open-shell nuclei. Finally, we move to the discussion 
related to the equation of states for an infinite nuclear matter, which is 
very important nowadays after the detection of gravitational waves from 
binary neutron stars. 
%Finally, we will discuss the tidal interaction of the
%binary neutron star system.

\section {Energy density functional and equations of motion} {\label{EDF}}

The beauty of an effective Lagrangian is that one can ignore the basic 
difficulties of the formalism, like renormalization and divergence of the 
system \cite{furnstahl97}. The model can be used directly by fitting the 
coupling constants and some masses of the mesons.
The ERMF Lagrangian has an infinite number of terms with all types of self- 
and cross-couplings. It is necessary to develop a truncation procedure for
practical use.  
Generally, the meson fields constructed in the Lagrangian are smaller than the 
mass of the nucleon. Their ratio could be used as a truncation scheme as  
is done in Refs. \cite{furnstahl97,muller96,serot97,estal01} along with the NDA 
and naturalness properties. The basic nucleon-meson ERMF Lagrangian 
(with $\delta$ meson, $W R$)  up to fourth order with exchange mesons like 
$\sigma$, $\omega$, $\rho$ mesons and photon $A$ is given as \cite{furnstahl97,singh14}: 
%
%\begin{widetext}
\begin{eqnarray}
{\cal E}({r}) & = &  \sum_\alpha \varphi_\alpha^\dagger({r})
\Bigg\{ -i \mbox{\boldmath$\alpha$} \!\cdot\! \mbox{\boldmath$\nabla$}
+ \beta \left[M - \Phi (r) - \tau_3 D(r)\right] + W({r})
+ \frac{1}{2}\tau_3 R({r})
+ \frac{1+\tau_3}{2} A ({r})
\nonumber \\[3mm]
& &
- \frac{i \beta\mbox{\boldmath$\alpha$}}{2M}\!\cdot\!
  \left (f_\omega \mbox{\boldmath$\nabla$} W({r})
  + \frac{1}{2}f_\rho\tau_3 \mbox{\boldmath$\nabla$} R({r}) \right)
  \Bigg\} \varphi_\alpha (r)
%\nonumber \\[3mm]
%& & \null
  + \left ( \frac{1}{2}
  + \frac{\kappa_3}{3!}\frac{\Phi({r})}{M}
  + \frac{\kappa_4}{4!}\frac{\Phi^2({r})}{M^2}\right )
   \frac{m_s^2}{g_s^2} \Phi^2({r})
\nonumber \\[3mm]
& & \null 
-  \frac{\zeta_0}{4!} \frac{1}{ g_\omega^2 } W^4 ({r})
+ \frac{1}{2g_s^2}\left( 1 +
\alpha_1\frac{\Phi({r})}{M}\right) \left(
\mbox{\boldmath $\nabla$}\Phi({r})\right)^2
 - \frac{1}{2g_\omega^2}\left( 1 +\alpha_2\frac{\Phi({r})}{M}\right)
 \nonumber \\[3mm]
 & &  \null 
 \left( \mbox{\boldmath $\nabla$} W({r})  \right)^2
% \nonumber \\[3mm]
% & &  \null 
 - \frac{1}{2}\left(1 + \eta_1 \frac{\Phi({r})}{M} +
 \frac{\eta_2}{2} \frac{\Phi^2 ({r})}{M^2} \right)
  \frac{m_\omega^2}{g_\omega^2} W^2 ({r})
   - \frac{1}{2e^2} \left( \mbox{\boldmath $\nabla$} A({r})\right)^2
   \nonumber \\[3mm]
 & & \null
   - \frac{1}{2g_\rho^2} \left( \mbox{\boldmath $\nabla$} R({r})\right)^2
%    \nonumber \\[3mm]
%  & & \null
   - \frac{1}{2} \left( 1 + \eta_\rho \frac{\Phi({r})}{M} \right)
   \frac{m_\rho^2}{g_\rho^2} R^2({r})
%nonumber\\[3mm]
% &
   -\Lambda_{\omega}\left(R^{2}(r) W^{2}(r)\right)
   \nonumber\\
   & &
   +\frac{1}{2 g_{\delta}^{2}}\left( \mbox{\boldmath $\nabla$} D({r})\right)^2
   +\frac{1}{2}\frac{ {  m_{\delta}}^2}{g_{\delta}^{2}}\left(D^{2}(r)\right)\;,
\label{eqq1}
\end{eqnarray}
%\end{widetext}
%
where $\Phi$, $W$, $R$, $D$, and $A$ are the fields$^{2}$\footnotetext[2]{
$W(r)=g_{\omega} V_{0}(r), \Phi(r)=g_{\sigma}\Phi_{0}(r), R(r)=g_{\rho} b_{0}(r),$ and $D(r)=g_{\delta}D_{0}(r)$ are the scaled mean-fields with different 
coupling constants.}; $g_\sigma$, $g_\omega$, 
$g_\rho$, $g_\delta$, and $\frac{e^2}{4\pi}$  are the coupling constants;
and $m_\sigma$, $m_\omega$, $m_\rho$, and $m_\delta$ are the masses for 
$\sigma$, $\omega$, $\rho$, and $\delta$ mesons and photon, respectively.
The parameters, such as $\eta_{1}, \eta_{2}, \Lambda_{\omega}, \alpha_{1},
\alpha_{2}, f_{\omega}$ have their own importance to explain various
properties of finite nuclei and nuclear matter. For instance, the surface
properties of finite nuclei is analyzed through non-linear interactions
of $\eta_{1}$ and $\eta_{2}$ as discussed in Ref. \cite{estal01}.

Now, our aim is to solve the field equations for the baryons and mesons 
(nucleon, $\sigma$, $\omega$, $\rho$, and $\delta$) using the variational 
principle. We obtained the meson equation of motion using the equation 
%%
%\begin{eqnarray}
	$\left(\frac{\partial\mathcal{E}}{\partial\phi_i}\right)_{\rho=const}=0.$
%\end{eqnarray}
%
The single-particle energy for the nucleons is obtained by using the Lagrange
multiplier $\varepsilon_i$, which is the energy eigenvalue of the
Dirac equation constraining the normalization condition 
$\sum_i\varphi_i^\dagger({r}) \varphi_i({r})=1 $ \cite{furn87}.  
The Dirac equation for the wave function $\varphi_i({r})$ becomes
\begin{eqnarray}
\frac{\partial}{\partial\varphi_i^\dagger({r})}\Bigg[{\cal E}{({r})}-
\sum_i\varphi_i^\dagger({r}) \varphi_i({r})\Bigg]= 0,
\end{eqnarray}
i.e.
%
%\begin{widetext}
\begin{eqnarray}
\Bigg\{-i \mbox{\boldmath$\alpha$} \!\cdot\! \mbox{\boldmath$\nabla$}
& + & \beta [M - \Phi(r) - \tau_3 D(r)] + W(r)
  +\frac{1}{2} \tau_3 R(r) + \frac{1 +\tau_3}{2}A(r)
\nonumber \\[3mm]
& & \null  
\left.
   -  \frac{i\beta \mbox{\boldmath$\alpha$}}{2M}
   \!\cdot\! \left [ f_\omega \mbox{\boldmath$\nabla$} W(r)
    + \frac{1}{2}f_{\rho} \tau_3 \mbox{\boldmath$\nabla$} R(r) \right]
     \right\} \varphi_\alpha (r) =
     \varepsilon_\alpha \, \varphi_\alpha (r) \,.
     \label{eq2}
	\label{Dirac}
\end{eqnarray}
%\end{widetext}
%
The mean-field equations for $\Phi$, $W$, $R$, $D$, and $A$ are given by
%
%\begin{widetext}
\begin{eqnarray}
-\Delta \Phi(r) + m_s^2 \Phi(r)  & = &
g_s^2 \rho_s(r)
-{m_s^2\over M}\Phi^2 (r)
\left({\kappa_3\over 2}+{\kappa_4\over 3!}{\Phi(r)\over M}
\right )
\nonumber  \\[3mm]
& & \null
+{g_s^2 \over 2M}
\left(\eta_1+\eta_2{\Phi(r)\over M}\right)
{ m_\omega^2\over  g_\omega^2} W^2 (r)
+{\eta_{\rho} \over 2M}{g_s^2 \over g{_\rho}^2}
{ m_\rho^2 } R^2 (r)
\nonumber  \\[3mm]
& & \null
+{\alpha_1 \over 2M}[
(\mbox{\boldmath $\nabla$}\Phi(r))^2
+2\Phi(r)\Delta \Phi(r) ]
+ {\alpha_2 \over 2M} {g_s^2\over g_\omega^2}
(\mbox{\boldmath $\nabla$}W(r))^2 ,\;\;\;\;\;\;\;
 \label{eq3}  \\[3mm]
-\Delta W(r) +  m_\omega^2 W(r)  & = &
g_\omega^2 \left( \rho(r) + \frac{f_\omega}{2} \rho_{\rm T}(r) \right)
-\left( \eta_1+{\eta_2\over 2}{\Phi(r)\over M} \right ){\Phi(r)
\over M} m_\omega^2 W(r)
\nonumber  \\[3mm]
& & \null
-{1\over 3!}\zeta_0 W^3(r)
+{\alpha_2 \over M} [\mbox{\boldmath $\nabla$}\Phi(r)
\cdot\mbox{\boldmath $\nabla$}W(r)
%                    +\Phi(r)\Delta W(r)] \,,
+\Phi(r)\Delta W(r)]
\nonumber\\
& &
-2\;\Lambda_{\omega}{g_{\omega}}^2 {R^{2}(r)} W(r) \,,
\label{eq4}  \\[3mm]
-\Delta R(r) +  m_{\rho}^2 R(r)  & = &
{1 \over 2 }g_{\rho}^2 \left (\rho_{3}(r) +
{1 \over 2 }f_{\rho}\rho_{\rm T,3}(r)
\right ) -  \eta_\rho {\Phi (r) \over M }m_{\rho}^2 R(r)
\nonumber\\
& &
-2\;\Lambda_{\omega}{g_{\rho}}^{2} R(r) {W^{2}(r)} \,,
\label{eq5}  \\[3mm]
-\Delta A(r)   & = &
e^2 \rho_{\rm p}(r)    \,, 
\label{eq6} \\[3mm]
-\Delta D(r)+{m_{\delta}}^{2} D(r) & = &
g_{\delta}^{2}\rho_{s3} \;,
\label{eq7}
\end{eqnarray}
%\end{widetext}
%
%
where the baryon, scalar, isovector, proton, and tensor densities are
\begin{eqnarray}
\rho(r) & = &
\sum_i n_i \varphi_i^\dagger(r) \varphi_i(r) \,
\nonumber\\
	&=& \rho_{p}(r)+\rho_{n}(r)
\nonumber\\
&= &
\frac{2}{(2\pi)^{3}}\int_{0}^{k_{p}}d^{3}k
+\frac{2}{(2\pi)^{3}}\int_{0}^{k_{n}}d^{3}k,
\label{den_nuc} 
\end{eqnarray}

\begin{eqnarray}
\rho_s(r) & = &
\sum_i n_i \varphi_i^\dagger(r) \beta \varphi_i(r) 
\nonumber \\ 
	&=&\rho_{s p}(r) +  \rho_{s n}(r)
\nonumber \\
&=&\sum_{i} \frac{2}{(2\pi)^3}\int_{0}^{k_{i}} d^{3}k 
\frac{M_{i}^{\ast}} {(k^{2}_i+M_{i}^{\ast 2})^{\frac{1}{2}}},
\label{scalar_density} 
\end{eqnarray}

\begin{eqnarray}
\rho_3 (r) & = &
\sum_i n_i \varphi_i^\dagger(r) \tau_3 \varphi_i(r) \nonumber \\
&=& \rho_{p} (r) -  \rho_{n} (r),
\label{isovect_density}
\end{eqnarray}
\begin{eqnarray}
\rho_{s3} (r) & = &
\sum_i n_i \varphi_i^\dagger(r) \tau_3 \beta \varphi_i(r) 
\nonumber \\
&=& \rho_{ps} (r) -  \rho_{ns} (r)
\label{isoscal_density}
\end{eqnarray}
\begin{eqnarray}
\rho_{\rm p}(r) & = &
\sum_i n_i \varphi_i^\dagger(r) \left (\frac{1 +\tau_3}{2}
\right)  \varphi_i(r),
\label{charge_density} 
\end{eqnarray}
\begin{eqnarray}
\rho_{\rm T}(r) & = &
\sum_i n_i \frac{i}{M} \mbox{\boldmath$\nabla$} \!\cdot\!
\left[ \varphi_i^\dagger(r) \beta \mbox{\boldmath$\alpha$}
\varphi_i(r) \right] \,,
\label{tensor_dens} 
\end{eqnarray}
and
\begin{eqnarray}
\rho_{\rm T,3}(r) & = &
\sum_i n_i \frac{i}{M} \mbox{\boldmath$\nabla$} \!\cdot\!
\left[ \varphi_i^\dagger(r) \beta \mbox{\boldmath$\alpha$}
\tau_3 \varphi_i(r) \right].  
\label{isotensor_dens} 
\end{eqnarray}
%$\qquad$     
Here $k_{i}$ is the nucleon's  Fermi momentum and the summation is 
over all the occupied states. The qualitative structure of the fields (such as
V and S) for the finite nucleus are shown in Fig. \ref{nosea}. The nucleons 
and mesons are composite particles and their vacuum polarization effects have been neglected. Hence, the 
negative-energy states do not contribute to the densities and current 
\cite{pg89}. In the 
fitting process, the coupling constants of the effective Lagrangian are
 determined from a set of experimental data which takes into account the large 
part of the vacuum polarization effects in the {\it no-sea approximation}. 
It is clear that the {\it no-sea approximation} is essential to determine 
the stationary solutions of the relativistic mean-field equations which 
describe the ground-state properties of the nucleus. The Dirac sea holds the 
negative-energy eigenvectors of the Dirac Hamiltonian, which is different 
for different nuclei. Thus, it depends on the specific solution of the set of 
nonlinear RMF equations. The Dirac spinors can be expanded in terms of vacuum 
solutions which form a complete set of plane wave functions in spinor space. 
This set will be complete when the states with negative energies are the part 
of the positive energy states and create the Dirac sea of the vacuum.

\begin{figure}[!b]
\begin{center}
\includegraphics[width=0.6\columnwidth]{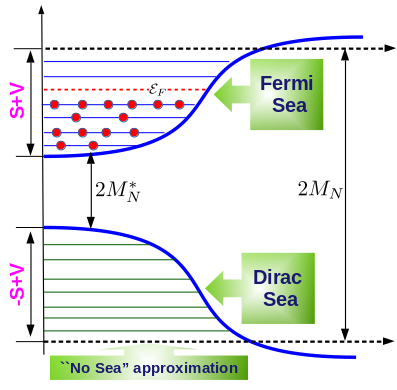}
\caption{ The qualitative structure of the Lorentz scalar field S and vector 
	field V in finite nuclei. $M_N$, and $M^{*}_N$ are the mass of the 
	nucleons and effective mass of the nucleons, respectively. Dirac 
	positive energy S+V and negative energy -S+V states are shown. }
        \label{nosea}
\end{center}
\end{figure}

The effective masses of proton, $M_{p}^{\ast}$, and neutron, $M_{n}^{\ast}$ are 
written as 
\begin{equation}
M_{p}^{\ast}=M- \Phi (r) - D({r}),
\end{equation}
\begin{equation} 
M_{n}^{\ast}=M- \Phi (r) + D({r}).
\end{equation}
%fulfill the following condition:
The vector potential is
\begin{equation}
V(r)=g_{\omega}V_{0}(r)+\frac{1}{2} g_{\rho}\tau_{3}b_{0}(r)
+e\frac{(1-\tau_3)}{2}A_0(r). 
\end{equation}
and the scalar potential is 
\begin{equation}
	S(r)= g_{\sigma}\Phi_0(r)+g_\delta \tau_3 \delta_0(r)
\end{equation}
where, $\Phi_0(r)$, $\delta_0(r)$ are scalar fields by $\sigma$ and $\delta$
mesons, respectively. 
The set of coupled differential equations is solved self-consistently to 
describe the ground-state properties of finite nuclei. In the fitting procedure, we used the experimental data of binding energy (BE) and charge radius $r_{ch}$ 
for a set of spherical nuclei ($^{16}$O, $^{40}$Ca, $^{48}$Ca, $^{68}$Ni, $^{90}$Zr, $^{100,132}$Sn, and  $^{208}$Pb). 
The total binding energy is obtained by 
\begin{eqnarray}
E_{total} &=& E_{part}+E_{\sigma}+E_{\omega}+E_{\rho}
\nonumber\\   
&&
	+E_{\delta}+E_{\omega\rho}+E_{c}+E_{pair}+E_{c.m.},
	\label{TBE}
\end{eqnarray}
where $E_{part}$ is the sum of the single-particle energies of the nucleons and 
$E_{\sigma}$, $E_{\omega}$, $E_{\rho}$, $E_{\delta}$, and $E_{c}$ are the 
contributions of the respective mesons and Coulomb fields. The pairing 
$E_{pair}$ and the center of mass motion $E_{c.m.}=\frac{3}{4}\times41A^{-1/3}$ 
MeV energies are also taken into account \cite{elliott55,negele70,estal01}.

\section{Temperature dependent BCS pairing}{\label{TBCS}}

The pairing correlation plays a distinct role in open-shell nuclei.
The effect of pairing correlation is markedly seen with increase
in mass number A. Moreover it helps in understanding the deformation
of medium and heavy nuclei. It has a lean effect on
both bulk and single particles properties of lighter mass nuclei
because of the availability of limited pairs near the Fermi surface.
We take the case of T=1 channel of pairing correlation i.e,
pairing between proton- proton and neutron-neutron. In this case, a nucleon of
quantum states $\vert jm_z\rangle$ pairs with
another nucleons having same $I_z$ value with quantum states $\vert j-m_z
\rangle$,
 since it is the time reversal partner of the other. In both nuclear and atomic
domain the ideology of BCS pairing is the same. The even-odd mass staggering
of isotopes was the first evidence of its kind for the pairing energy.
Considering the mean-field formalism, the violation of the particle number is
seen only due to the pairing correlation. We find terms like $\varphi^{\dagger} \varphi$ (density) in the RMF Lagrangian density but we put an embargo on terms
of the form $\varphi^{\dagger}\varphi^{\dagger}$ or $\varphi\varphi$ since it 
violates the particle number conservation.
We apply externally the BCS pairing approximation for our calculation to take
the pairing correlation into account. The pairing interaction energy in terms of occupation
probabilities $v_i^2$ and $u_i^2=1-v_i^2$ is written
as~\cite{pres82,Patra93}:

\begin{equation}
E_{pair}=-G\left[\sum_{i>0}u_{i}v_{i}\right]^2,
\end{equation}
with $G$ is the pairing force constant.
The variational approach with respect to the occupation number $v_i^2$ gives the BCS equation
\cite{pres82}:
\begin{equation}
2\epsilon_iu_iv_i-\triangle(u_i^2-v_i^2)=0,
\label{eqn:bcs}
\end{equation}
with the pairing gap $\triangle=G\sum_{i>0}u_{i}v_{i}$. The pairing gap ($\triangle$) of proton and neutron is taken from the empirical formula \cite{va73,gam90}:
\begin{equation}
\triangle = 12 \times A^{-1/2}.
\end{equation}

To calculate the properties of nuclei at finite temperature one has to include 
the temperature in the set of coupled equations. For this, the temperature is
introduced in the partial occupancies in the BCS approximation is given by,
\begin{equation}
n_i=v_i^2=\frac{1}{2}\left[1-\frac{\epsilon_i-\lambda}{\tilde{\epsilon_i}}[1-2 f(\tilde{\epsilon_i},T)]\right],
\end{equation}
with
\begin{eqnarray}
f(\tilde{\epsilon_i},T) = \frac{1}{(1+exp[{\tilde{\epsilon_i}/T}])} & and \nonumber \\[3mm]
	\tilde{\epsilon_i} = \sqrt{(\epsilon_i-\lambda)^2+\triangle^2}.&
\end{eqnarray}

The function $f(\tilde{\epsilon_i},T)$ represents the Fermi Dirac
distribution for quasi particle energy $\tilde{\epsilon_i}$. The chemical potential $\lambda_p (\lambda_n)$
for protons (neutrons) is obtained from the constraints of particle number
equations

\begin{eqnarray}
\sum_i n_i^{Z}  = Z, \nonumber \\
\sum_i n_ i^{N} =  N.
\end{eqnarray}
The sum is taken over all proton and neutron states. The entropy is obtained by,
\begin{equation}
S = - \sum_i \left[n_i\, ln(n_i) + (1 - n_i)\, ln (1- n_i)\right].
\end{equation}
% energy equation
The total energy and the gap parameter are obtained by minimizing the free energy,
\begin{equation}
F = E - TS.
\end{equation}
In constant pairing gap calculations, for a particular value of pairing gap $\triangle$
and force constant $G$, the pairing energy $E_{pair}$ diverges, if
it is extended to an infinite configuration space. In fact, in all
realistic calculations with finite range forces, $\triangle$ is not
constant, but decreases with large angular momenta states above the Fermi
surface. Therefore, a pairing window in all the equations are extended
up-to the level $|\epsilon_i-\lambda|\leq 2(41A^{-1/3})$ as a function
of the single particle energy. The factor 2 has been determined so as
to reproduce the pairing correlation energy for neutrons in $^{118}$Sn
using Gogny force \cite{gam90,Patra93,dech80}.

\section{Quasi-BCS pairing}
For fitting, we consider a seniority-type interaction as a tool by taking a 
constant value of G for the active pair shell.
The BCS approach does not go well for nuclei away from the stability line
because, in the present case, with the increase in the number of neutrons or 
protons the corresponding Fermi level goes to zero and the number of available 
levels above it minimizes. To complement this situation we see that the 
particle-hole and pair excitation's reach the continuum. In Ref. \cite{doba84} we 
notice that if we make the BCS calculation using the quasi-particle state as in 
Hartree-Fock-Bogoliubov (HFB) calculation, then the BCS binding 
energies are coming out to be very close to the HFB, but rms radii (i.e the single-particle wave functions) greatly depend on the size of the box where the 
calculation is done. This is because of the unphysical neutron (proton) gas
in the continuum where wave-functions are not confined in a region.
The above shortcomings of the BCS approach can be improved by 
means of the so-called quasi-bound states,  i.e, states bound because of 
their own centrifugal barrier (centrifugal-plus-Coulomb barrier for protons)
\cite{estal01,chab98,sand20}. Our calculations are done by confining the 
available space to one harmonic oscillator shell each above and below the Fermi 
level to exclude the unrealistic pairing of highly excited states in the continuum \cite{estal01}.

%\section{Results and discussions} \label{sec3}

\subsection{The Nuclear Equation of State}
The nuclear equation of state plays vital role in explaining many 
properties of nuclear matter and neutron star etc. Most importantly, the EoS 
very well explains the complex behaviour of the dense matter that makes up 
neutron star. In static, infinite, uniform, and isotropic nuclear matter, all 
the gradients of the fields in Eqs. \eqref{eq3}--\eqref{eq7} vanish. By the 
definition of infinite nuclear matter, the electromagnetic interaction is also 
neglected. The expressions for energy density and pressure for such a system 
are obtained from the energy-momentum tensor \cite{walecka86}:
\begin{eqnarray}
T_{\mu \nu}=\sum_{i}{\partial_{\nu} \phi_{i}} 
\frac{ {\partial \cal L}} {\partial \left({\partial^{\mu} \phi_{i}}\right)} 
-g_{\mu \nu}{\cal L}.
\end{eqnarray}
The zeroth component of the energy-momentum tensor $<T_{00}>$ gives the 
energy density and the third component $<T_{ii}>$ determine the pressure of 
the system \cite{singh14}:
%
%\begin{widetext}
\begin{eqnarray}\label{eqn:eos1}
{\cal{E}} & = &  \frac{2}{(2\pi)^{3}}\int d^{3}k E_{i}^\ast (k)+\rho  W+
\frac{ m_{s}^2\Phi^{2}}{g_{s}^2}\Bigg(\frac{1}{2}+\frac{\kappa_{3}}{3!}
\frac{\Phi }{M} + \frac{\kappa_4}{4!}\frac{\Phi^2}{M^2}\Bigg)
\nonumber\\
&&
 -\frac{1}{2}m_{\omega}^2\frac{W^{2}}{g_{\omega}^2}\Bigg(1+\eta_{1}\frac{\Phi}{M}+\frac{\eta_{2}}{2}\frac{\Phi ^2}{M^2}\Bigg)-\frac{1}{4!}\frac{\zeta_{0}W^{4}}
	{g_{\omega}^2}+\frac{1}{2}\rho_{3} R
 \nonumber\\
 &&
-\frac{1}{2}\Bigg(1+\frac{\eta_{\rho}\Phi}{M}\Bigg)\frac{m_{\rho}^2}{g_{\rho}^2}R^{2}-\Lambda_{\omega}  (R^{2} W^{2})
+\frac{1}{2}\frac{m_{\delta}^2}{g_{\delta}^{2}}\left(D^{2} \right),
	\label{eq20}
\end{eqnarray}
%\end{widetext}

%The pressure is given by

%\begin{widetext}
\begin{eqnarray}\label{eqn:eos2}
P & = &  \frac{2}{3 (2\pi)^{3}}\int d^{3}k \frac{k^2}{E_{i}^\ast (k)}-
\frac{ m_{s}^2\Phi^{2}}{g_{s}^2}\Bigg(\frac{1}{2}+\frac{\kappa_{3}}{3!}
\frac{\Phi }{M}+ \frac{\kappa_4}{4!}\frac{\Phi^2}{M^2}  \Bigg)
\nonumber\\
& &
 +\frac{1}{2}m_{\omega}^2\frac{W^{2}}{g_{\omega}^2}\Bigg(1+\eta_{1}\frac{\Phi}{M}+\frac{\eta_{2}}{2}\frac{\Phi ^2}{M^2}\Bigg)+\frac{1}{4!}\frac{\zeta_{0}W^{4}}{g_{\omega}^2}
  \nonumber\\
& &
+\frac{1}{2}\Bigg(1+\frac{\eta_{\rho}\Phi}{M}\Bigg)\frac{m_{\rho}^2}{g_{\rho}^2}R^{2}+\Lambda_{\omega} (R^{2} W^{2})
-\frac{1}{2}\frac{m_{\delta}^2}{g_{\delta}^{2}}\left(D^{2}\right),
	\label{eq21}
\end{eqnarray}
%\end{widetext}
%
where $E_{i}^\ast(k)$=$\sqrt {k^2+{M_{i}^\ast}^2} \qquad  (i= p,n)$
is the energy and $k$ is the momentum of the nucleon. In the context of 
density functional theory, it is possible to parametrize the exchange and 
correlation effects through local potentials (Kohn--Sham potentials), as long as 
those contributions are small enough \cite{Ko65}. The Hartree values control the 
dynamics in the relativistic Dirac-Br\"uckner-Hartree-Fock calculations. Therefore, the local meson fields in the RMF formalism can be interpreted as 
Kohn-Sham potentials and in this sense Eqs. (\ref{eq2}--\ref{eq7}) include 
effects beyond the Hartree approach through the nonlinear couplings \cite{furnstahl97}.

\subsection{Nuclear Matter Properties}{\label{nuclear-matter}}
In 1936, the semi-empirical mass formula was developed by Bethe-Weizsacker
based on the liquid drop model. This mass model
is quite successful in describing some bulk properties of finite nuclei.
In the liquid drop model, the binding energy per particle of a nucleus can 
be constructed as follows:
\begin{eqnarray}
	\frac{E(Z,N)}{A}=M-a_v+a_s A^{-1/3}+a_c \frac{Z^2}{A^{2/3}}+a_{asym} 
	\frac{(N-Z)^2}{A^2}+\dots
\end{eqnarray}
where $M$ is the nucleon mass, and $A=Z+N$ is the total number of nucleons.
$a_v,a_s,a_c,$ and $a_{asym}$ are the volume, surface, Coulomb, and asymmetry
term, respectively. The liquid drop model can be extended to introduce the
infinite nuclear matter (means $Z,N,$ and $V$ go to infinity) by switching
off the Coulomb term {\it i.e} $a_c=0$, and neglecting the surface 
contribution. Therefore, the binding energy per nucleon of the system is given 
by

\begin{eqnarray}
	\frac{E(Z,N)}{A}-M &=&-a_v+a_{asym}\frac{(N-Z)^2}{A^2} \nonumber\\
e(\rho,\alpha)&=&\frac{{\cal E}}{\rho_{B}} - M\equiv-a_v+a_{asym}\alpha^2 ,
\end{eqnarray}
where $\alpha=(N-Z)/A=\frac{\rho_n-\rho_p}{\rho_n+\rho_p}$ is known as the 
neutron excess of the infinite nuclear matter. As suggested above the infinite 
nuclear matter is incompressible. Conventionally, we can expand binding energy 
per particle for compressible nuclear matter around $\alpha=0$:
\begin{eqnarray}
	e(\rho,\alpha)&=&e(\rho,\alpha=0)+\Big(\frac{\partial e(\rho,\alpha)}
	{\partial\alpha}\Big)_{\alpha=0}\alpha+\frac{1}{2}\Big(\frac{\partial^2
e(\rho,\alpha)}{\partial\alpha^2}\Big)_{\alpha=0}\alpha^2+\dots\nonumber\\
	&=&{e}(\rho) + S(\rho) \alpha^2 +{\cal O}(\alpha^4) ,
\end{eqnarray}
where the linear term in $\alpha$ vanishes due to the charge symmetry of 
nuclear force, ${e}(\rho)$ is energy density of the symmetric nuclear matter 
(SNM) ($\alpha$ = 0), and $\alpha=1$ corresponds to pure neutron matter. The 
symmetry energy $S(\rho)$ of the system defined as:
\begin{eqnarray}
S(\rho)=\frac{1}{2}\left[\frac{\partial^2 {e}(\rho, \alpha)}
{\partial \alpha^2}\right]_{\alpha=0}.
\end{eqnarray}
The isospin asymmetry arises due to the difference in densities and masses 
of the neutron and proton. The density-type isospin asymmetry is taken care by 
$\rho$ meson (isovector-vector meson) and mass asymmetry by the $\delta$ meson 
(isovector - scalar meson). The general expression for symmetry energy $S(\rho)$ 
is a combined expression of $\rho$ and $\delta$ mesons, which is defined 
as~\cite{matsui81,kubis97,estal01,roca11}
\begin{eqnarray}
S(\rho)=S^{kin}(\rho) + S^{\rho}({\rho})+S^{\delta}
(\rho),
\end{eqnarray}
with
\begin{eqnarray}
S^{kin}(\rho)=\frac{k_F^2}{6E_F^*},\;
S^{\rho}({\rho})=\frac{g_{\rho}^2\rho}{8m_{\rho}^{*2}}
\end{eqnarray}
and
\begin{eqnarray}
S^{\delta}(\rho)=-\frac{1}{2}\rho \frac{g_\delta^2}{m_\delta^2}\left(
\frac{M^*}{E_F}\right)^2 u_\delta \left(\rho,M^*\right).
\end{eqnarray}
The last function $u_\delta$ is from the discreteness of the Fermi momentum. 
This momentum is quite large in nuclear matter and can be treated as a 
continuum and continuous system.
The function $u_\delta$ is defined as
\begin{eqnarray} 
u_\delta \left(\rho,M^*\right)=\frac{1}{ 1+ 3 \frac{g_\delta^2}{m_\delta^2}
\left(\frac{\rho^s}{M^*}-\frac{\rho}{E_F}\right)}.
\end{eqnarray}
In the limit of continuum, the function $u_\delta \approx 1$. 
The whole symmetry energy ($S^{kin}+S^{pot}$) arises from 
$\rho$ and $\delta$ mesons and is given as
\begin{eqnarray}{\label{eqn:sym}}
S(\rho)=\frac{k_F^2}{6E_F^*}+\frac{g_{\rho}^2\rho}{8{m_{\rho}^{*2}}}  
-\frac{1}{2}\rho \frac{g_\delta^2}{m_\delta^2}\left(\frac{M^*}{E_F}\right)^2,
\end{eqnarray}
where $E_F^*$ is the Fermi energy and $k_F$ is the
Fermi momentum. 
The mass of the $\rho$ meson modified, because of the cross-coupling of 
$\rho$-$\omega$ fields and is given by
\begin{eqnarray}
m_{\rho}^{*2}=\left(1+\eta_{\rho}\frac{\Phi}{M}\right)m_{\rho}^2
+2g_{\rho}^2(\Lambda_\omega W^2).
\end{eqnarray}
The cross-coupling of isoscalar-isovector mesons ($\Lambda_\omega$)  
modifies the density dependence of $S(\rho)$ without affecting the 
saturation properties of the SNM \cite{singh13,horowitz01b,horowitz01a}.
In the numerical calculation, the coefficient of symmetry energy $S(\rho)$ 
is obtained by the energy difference of symmetric and pure neutron matter 
at saturation. In our calculation, we have taken the isovector channel into 
account to make the new parameters, which incorporate the currently existing 
experimental observations, and predictions are made keeping in mind some future 
aspects of the model. The symmetry energy can be expanded as a Taylor series 
around the saturation density $\rho_0$ as
\begin{eqnarray}
S(\rho)=J + L{\cal Y} 
+ \frac{1}{2}K_{sym}{\cal Y}^2 +\frac{1}{6}Q_{sym}{\cal Y}^3 + {\cal O}[{\cal Y}^4],
	\label{eq30}
\end{eqnarray} 
where $J=S(\rho_0)$ is the symmetry energy at saturation and 
${\cal Y} = \frac{\rho-\rho_0}{3\rho_0}$. The coefficients $L(\rho_0)$, 
$K_{sym}(\rho_0)$, and $Q_{sym}$ are defined as:
\begin{eqnarray}
	L=3\rho\frac{\partial S(\rho)}{\partial {\rho}}\bigg{|}_{ \rho=\rho_0},\;
\end{eqnarray}
\begin{eqnarray}
	K_{sym}=9\rho^2\frac{\partial^2 S(\rho)}{\partial {\rho}^2}\bigg{|}_{ {\rho=\rho_0}},\;
\end{eqnarray}
\begin{eqnarray}
        Q_{sym}=27\rho^3\frac{\partial^3 S(\rho)}{\partial {\rho}^3}\bigg{|}_{ {\rho=\rho_0}}.\;
\end{eqnarray}
Similarly, we obtain the asymmetric nuclear matter incompressibility as $K(\alpha)=K+K_\tau \alpha^2+{\cal O}(\alpha^4)$ and $K_\tau$ is given by \cite{chen09}
\begin{eqnarray}
	K_{\tau}=K_{sym}-6L-\frac{Q_{0}L}{K},\;
	\label{eq34}
\end{eqnarray}
where $Q_{0}=27\rho^3\frac{\partial^3 (\cal{E}/\rho)}{\partial {\rho}^3}$ in SNM.

%r
Here, $L$ is the slope and $K_{sym}$ represents the curvature of $S(\rho)$ at 
saturation density. A large number of investigations have been made to fix the 
values of $J$, $L$, and $K_{sym}$ \cite{singh13,tsang12,dutra12,xu10,newton11,steiner12,fattoyev12}. The density dependence of symmetry energy is a key quantity 
to control the properties of both finite nuclei and infinite nuclear matter 
\cite{roca09}. Currently, the available information on symmetry energy 
$J=31.6\pm2.66$ MeV and its slope $L=58.9\pm16$ MeV at saturation density are 
obtained by various astrophysical observations \cite{li13}.  
To date, the precise values of $J,L,$ and the neutron radii for finite
nuclei are not known experimentally; it is essential to discuss the behavior 
of the symmetry energy as a function of density in our new parameter sets 
G3 and IOPB-I (see chapter \ref{chapter6}). 
\newpage

\setcounter{equation}{0}
\setcounter{figure}{0}

\newpage
\chapter{Decay modes of Th and U isotopes}
\label{chapter3} 
The properties of recently predicted thermally fissile Th and U isotopes
are studied within the framework of relativistic mean-field approach using 
axially deformed harmonic oscillator basis. We calculated the ground, first 
intrinsic excited 
state for highly neutron-rich thorium and uranium isotopes. The possible modes 
of decay such as $\alpha$ decay and $\beta$ decay are analyzed. We found that 
the neutron-rich isotopes are stable against $\alpha$ decay, however they are 
very unstable against $\beta$-decay. The life time of these nuclei is 
predicted to be tens of seconds against $\beta$ decay. If these nuclei utilized
before their decay time, a lot of energy can be produced with the help of 
multifragmentation fission \cite{sat08}. Also, these nuclei have great 
implications from the astrophysical point of view. In specific case of 
$^{228-230}$Th and $^{228-234}$U isotopes, we found isomeric states having 
energy range of 2 to 3 MeV and three maxima in the potential energy surface.

\section{Introduction}\label{sec1}

Nowadays uranium and thorium isotopes have attracted  great attention 
in nuclear physics due to the thermally fissile nature of some of 
them \cite{sat08}. These thermally fissile materials have tremendous 
importance in energy production. To date, the known thermally fissile nuclei 
are $^{233}$U, $^{235}$U and $^{239}$Pu; of these, only $^{235}$U has a 
long lifetime, and it is the only thermally fissile isotope available in 
nature \cite{sat08}. Thus, presently an important area of research is the 
search for any other thermally fissile nuclei apart from $^{233}$U,
 $^{235}$U and $^{239}$Pu. Recently, Satpathy {\it et al.} \cite{sat08} 
showed that uranium and thorium isotopes with neutron number N=154-172 have 
a thermally fissile property. They performed a calculation with a typical 
example of $^{250}$U as this nucleus has a low fission barrier with a 
significantly large barrier width, which makes it stable against spontaneous 
fission. Apart from the thermally fissile nature, these nuclei also play an 
important role in nucleosynthesis in stellar evolution. As these nuclei are 
stable against spontaneous fission, the prominent decay modes may be the 
emission of $\alpha$, $\beta$, and $cluster$ particles from the neutron-rich 
thermally fissile (uranium and thorium) isotopes.

To measure the stability of these neutron-rich U and Th isotopes,  we 
investigate the $\alpha$ and $\beta$ decay properties of these nuclei. Also, 
we extend our calculations to estimate the binding energy, root mean square 
radii, quadrupole moments and other structural properties. 

From the last three decades, the  relativistic mean field (RMF) formalism
has been  a formidable theory in describing finite nuclear properties 
throughout the periodic chart and infinite nuclear matter properties 
concerned with the dense cosmic objects such as neutron star. Along the same
line RMF theory is also good enough for study the clusterization \cite{aru05}, 
$\alpha$ decay \cite{bidhu11}, and $\beta$ decay of nuclei. The presence of 
cluster in  heavy nuclei like  $^{222}$Ra, $^{232}$U, $^{239}$Pu, and 
$^{242}$Cm has been studied using RMF formalism \cite{bk06,patra07}. 
It gives a clear prediction of $\alpha$ like (N=Z) matter in the central part 
for heavy nuclei and a $cluster$-like structure (N=Z and $N\neq Z$) for 
light-mass nuclei \cite{aru05}. The proton emission as well as cluster 
decay phenomena is well studied using RMF formalism with M3Y~\cite{love79},  
LR3Y~\cite{bir12}, and NLR3Y\cite{bidhu14} nucleon-nucleon potentials in 
the framework of single- and double-folding models, respectively. 
Here, we used the RMF formalism with the well known NL3 parameter 
set \cite{lala97}  for all our calculations.

\section{RMF Formalism}\label{sec2}
In the present chapter, we use the axially deformed RMF formalism 
to calculate various nuclear phenomena. The meson-nucleon interaction is given 
by the Lagrangian density ~\cite{patra91,walecka74,walecka86,horo 81,bogu77,pric87}
\begin{eqnarray}
{\cal L}&=&\overline{\psi_{i}}\{i\gamma^{\mu}
\partial_{\mu}-M\}\psi_{i}
+{\frac12}\partial^{\mu}\sigma\partial_{\mu}\sigma
-{\frac12}m_{\sigma}^{2}\sigma^{2}\nonumber\\
&& -{\frac13}g_{2}\sigma^{3} -{\frac14}g_{3}\sigma^{4}
-g_{s}\overline{\psi_{i}}\psi_{i}\sigma-{\frac14}\Omega^{\mu\nu}
\Omega_{\mu\nu}\nonumber\\
&&+{\frac12}m_{w}^{2}V^{\mu}V_{\mu}
+{\frac14}c_{3}(V_{\mu}V^{\mu})^{2} -g_{w}\overline\psi_{i}
\gamma^{\mu}\psi_{i}
V_{\mu}\nonumber\\
&&-{\frac14}\vec{B}^{\mu\nu}.\vec{B}_{\mu\nu}+{\frac12}m_{\rho}^{2}{\vec
R^{\mu}} .{\vec{R}_{\mu}}
-g_{\rho}\overline\psi_{i}\gamma^{\mu}\vec{\tau}\psi_{i}.\vec
{R^{\mu}}\nonumber\\
&&-{\frac14}F^{\mu\nu}F_{\mu\nu}-e\overline\psi_{i}
\gamma^{\mu}\frac{\left(1-\tau_{3i}\right)}{2}\psi_{i}A_{\mu} ,
\label{eq:7}
\end{eqnarray}
where, $\psi$ is the single-particle Dirac spinor and meson fields are 
denoted as $\sigma$, $V^\mu$, and $R^\mu$ for $\sigma$, $\omega$, and $\rho$ 
mesons, respectively. 
The electromagnetic interaction between protons is denoted by photon 
field $A^\mu$. $g_s$, $g_\omega$, $g_\rho$, and $\frac{e^2}{4\pi}$ are the 
coupling constants for the $\sigma$,$\omega$, and $\rho$ meson and photon 
fields, respectively. The strengths of the self-coupling $\sigma$ meson 
($\sigma^3$ and $\sigma^4$) are denoted by $g_2$ and $g_3$, with $c_3$ as the 
non-linear coupling constant for $\omega$ meson. The nucleon mass is denoted
M, where the $\sigma$, $\omega$, and $\rho-$ meson masses are $m_s$, 
$m_\omega$, and $m_\rho$, respectively. 
The field tensors of the isovector mesons and the photon are given by,
\begin{eqnarray}
\Omega^{\mu\nu} & = & \partial^{\mu} V^{\nu} - \partial^{\nu} V^{\mu}, \, \\[3mm]
\vec{B}^{\mu\nu} & = & \partial^{\mu} \vec{R}^{\nu} - \partial^{\nu} \vec{R}^{\mu}, \\%  - g_{\rho} (\vec{R}^{\mu}\times\vec{R}^{\nu}), \,  \\[3mm]
F^{\mu\nu} & = & \partial^{\mu} A^{\nu} - \partial^{\nu} A^{\mu}.       \, 
\end{eqnarray}

From the classical Euler-Lagrangian equation, we get the Dirac equation and 
Klein- Gordan equation for the nucleon and meson fields, respectively. The 
Dirac equation for the nucleon is solved by expanding the Dirac spinor into 
lower and upper components, while the mean-field equation for bosons is 
solved in the deformed harmonic oscillator basis with $\beta_0$ as the 
deformation parameter. The nucleon equation along with different meson 
equations form a set of coupled equations, which can be solved by iterative 
method. 

%Various types of densities such as the baryon (vector), scalar, 
%isovector and proton (charge) densities are given as
%\begin{eqnarray}
%\rho(r) & = &
%\sum_i \psi_i^\dagger(r) \psi_i(r) \,,
%\label{eqFN6} \\[3mm]
%\rho_s(r) & = &
%\sum_i \psi_i^\dagger(r) \gamma_0 \psi_i(r) \,,
%\label{eqFN7} \\[3mm]
%\rho_3 (r) & = &
%\sum_i \psi_i^\dagger(r) \tau_3 \psi_i(r) \,,
%\label{eqFN8} \\[3mm]
%\rho_{\rm p}(r) & = &
%\sum_i \psi_i^\dagger(r) \left (\frac{1 -\tau_3}{2}
%\right)  \psi_i(r) \,.
%\label{eqFN9} % \\[3mm]
%\end{eqnarray}
The calculations are simplified under the shadow of various symmetries like 
conservation of parity, no-sea approximation  and time reversal symmetry, 
which kills all spatial components of the meson fields and the 
antiparticle-state contribution to the nuclear observable. 
%The center-of-mass correction is calculated with the nonrelativistic 
%approximation, which gives $E_{c.m}=\frac{3}{4}  41A^{-1/3}$ (in MeV). 

The quadrupole deformation parameter $\beta_2$ is calculated from the 
resulting quadropole moments $Q$ of the proton $Q_p$ and neutron $Q_n$ through 
the following relation:
\begin{eqnarray}
Q = Q_n + Q_p = \sqrt{\frac{16\pi}5} (\frac3{4\pi} AR^2\beta_2),
\end{eqnarray}
where $R=1.2 A^{1/3}$. The root-mean-square charge radius ($r_{ch}$), proton
radius ($r_{p}$), neutron radius ($r_{n}$), and matter radius ($r_{m}$) are
given as \cite{patra91}:
\begin{eqnarray}
<r_p^2>=\frac{1}{Z}\int\rho_p(r_{\bot},z)r^2_pd\tau_p,
\end{eqnarray}
\begin{eqnarray}
<r_n^2>=\frac{1}{N}\int\rho_n(r_{\bot},z)r_n^2d\tau_n,
\end{eqnarray}
\begin{eqnarray}
r_{ch}=\sqrt{r_p^2+0.64},
\end{eqnarray}
\begin{eqnarray}
<r_m^2>=\frac{1}{A}\int\rho(r_{\bot},z)r^2d\tau,
\end{eqnarray}
here all symbols have their own usual meaning. The total energy of the system 
is calculated by Eq. \eqref{TBE}.

\subsection{Pauli blocking approximation}\label{block}
It is a tough task to compute the BE and quadrupole moment of odd-N or odd-Z
or both odd-N and odd-Z numbers (odd-even, even-odd, or odd-odd) nuclei.
To do this, one needs to include the additional time-odd term, as done in the
SHF Hamiltonian~\cite{stone07}, or empirically the pairing force in order
to take care the effect of an odd neutron  or odd-proton~\cite{onsi20}. In an
odd-even or odd-odd nucleus, the time reversal symmetry is violated in
mean-field models. In our RMF calculations, we neglect the space components
of the vector fields, which are odd under time reversal and parity. These are
important in the determination of magnetic moments~\cite{holf88} but have a
very small effect on bulk properties such as BEs or quadrupole deformations,
and they can be neglected\cite{lala99} in the present context.
Here, for the odd-Z or odd-N calculations, we employ the Pauli blocking
approximation, which restores the time-reversal symmetry. In this approach,
one pair of conjugate states, $\pm{m}$, is taken out of the pairing scheme.
The odd particle stays in one of these states, and its corresponding conjugate
state remains empty. In principle, one has to block different states around
the Fermi level in turn to find the one that gives the lowest energy
configuration of the odd nucleus. For odd-odd nuclei, one needs to block both
the odd neutron and odd proton.

\section{Calculations and results}\label{sec4}
In this section, we evaluate our results for the BE, rms radius, quadrupole 
deformation parameter for recently predicted thermally fissile isotopes of 
Th and U. These nuclei are quite heavy and require a large number of 
oscillator bases, which takes considerable time for computation. In the first 
subsection here we describe how to select the basis space, and the results and 
discussion follow.

\subsection{Selection of basis space}\label{sec5}

The Dirac equation for fermions (proton and neutron) and the equation of 
motion for bosons ($\sigma-$, $\omega-$, $\rho-$ and $A_0$) obtained from
the RMF Lagrangian are solved self-consistently using an iterative method.
These equations are solved in an axially deformed harmonic oscillator 
expansion basis, $N_F$ and $N_B$ for fermionic and bosonic wave functions,
respectively.

For heavy nuclei, a large number of basis space $N_F$ and $N_B$ are needed
to get a converged solution. To reduce the computational time without
compromising the convergence of the solution, we have to choose an optimal
number of model spaces for both fermion and boson fields.
To choose optimal values for $N_F$ and $N_B$, we select $^{240}$Th as a
test case and increase the basis quanta from 8 to 20 step by step. The
obtained results for BE, charge radius and quadrupole deformation
parameter are shown in Fig.~\ref{bea1}. In our calculations, we notice an 
increment of 200 MeV in BE upon going from $N_F=N_B=$8 to 10. This increment
in energy decreases upon going to a higher oscillator basis. For example,
change in energy is $\sim 0.2$ MeV  with a change in $N_F=N_B$ from
14 to 20, and the increment in $r_c$ values is 0.12 fm, respectively.  
Keeping in mind the increase in convergence time for larger quanta as well as
the size of the nuclei considered, we finally use $N_F=N_B=20$ in our 
calculations to get suitable convergence results, which is the current accuracy 
of the present RMF models.

\begin{figure}[ht]
	\begin{center}
\includegraphics[width=0.7\columnwidth]{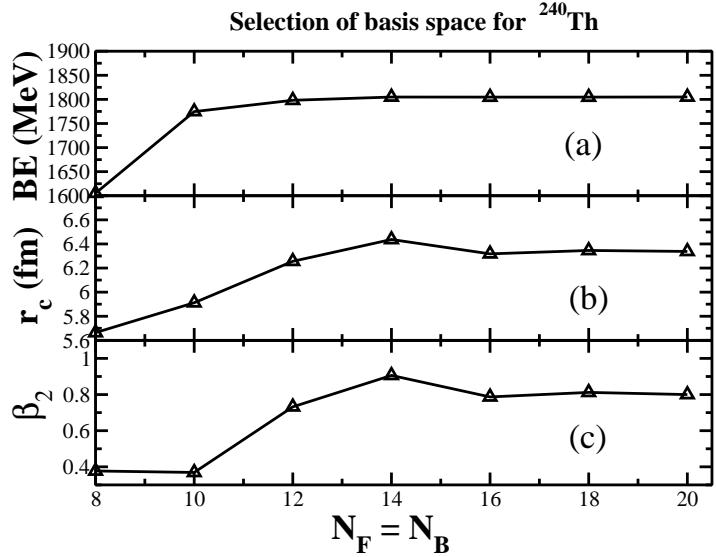}
	\end{center}
\caption{Variation of the (a) calculated binding energy (BE), (b) charge radii 
($r_{c}$), and quadrupole deformation parameter ($\beta_{2}$) are given with 
the bosonic and fermionic basis. 
}
\label{bea1}
\end{figure}

\subsection{Binding energies, charge radii and quadrupole deformation 
parameters}\label{sec6}

To be sure about the predictivity of our model, first we calculate the BE,
charge radii $r_c$, and quadrupole deformation parameter $\beta_2$ for some 
of the known cases. We have compared our results with the experimental data 
wherever available or with the Finite Range Droplet Model (FRDM) of 
M\"oller {\it et al.} \cite{NDCC,moller97,moller95,angel13}. 
The results are listed in Tables~\ref{Tab1} and~\ref{Tab2}. From the tables, 
it is obvious that the calculated BEs are comparable with the FRDM as well as 
experimental values. Further inspection of the tables reveals that the FRDM 
results are closer to the data. This may be due to the fitting of the FRDM 
parameters for almost all known data. However, in the case of most RMF 
parametrizations, the constants are determined by using a few spherical 
nuclei data along with certain nuclear matter properties. Thus the prediction 
of the RMF results are considered to be reasonable, but not excellent.

Ren {\it et al.}~\cite{ren01,ren03} have reported that the ground states of 
several superheavy nuclei are highly deformed states. Since, these are 
very heavy isotopes, the general assumption is that the ground state 
probably remains in deformed configuration (liquid drop picture). When these 
nuclei are excited either by a thermal neutron or by any other means, its 
intrinsic excited state becomes extraordinarily deformed and attains the 
scission point before it goes to fission. This can also be easily realized 
from the PES curve. Our calculations agree with the prediction of Ren
{\it et al.} for other superheavy regions of the mass table. However, this 
conclusion is contradicted in~\cite{soba91}, according to which the ground 
state of superheavy nuclei is either spherical or normally deformed. 

\begin{table}
\hspace{0.1 cm}
\caption{The calculated binding energies BE, quadrupole deformation parameter
$\beta_{2}$, rms radii for the ground states and few selective intrinsic 
excited state of U isotopes, using  RMF formalism with NL3 parameter set. The experimental and FRDM data~\cite{NDCC,moller97,moller95,angel13} are also 
included in the table. See the text for more details.}
\scalebox{0.8}{
{\begin{tabular}{ccccccccccccccccccccccc}
\Xhline{3\arrayrulewidth}
& \multicolumn{6}{ c }{RMF (NL3)} &\multicolumn{2}{c }{FRDM}& \multicolumn{3}{c }{Experiment}  \\
	\cmidrule(lr){2-7}\cmidrule(lr){8-9}\cmidrule{10-12}
Nucleus &       $r_n$   &       $r_p$   &       $r_{rms}$       &       $r_{ch}$        &       $\beta_2$       &        BE (MeV)       &        BE(MeV)        &$\beta_2$&     $r_{ch}$        &       $\beta_2$       &        BE (MeV)       \\
\Xhline{3\arrayrulewidth}
%	&&&&&&&&&&&\\
$^{216}$U       &       5.762   &       5.616   &       5.700     &       5.673   &       0       &       1660.5  &       1649.0  &-0.052&                &               &               \\
        &       6.054   &       5.946   &       6.008   &       5.999   &       0.608     &       1650.8  &&              &               &               &               \\
%        &               &               &               &               &               &               &       &       &               &               &               \\
$^{218}$U       &       5.789   &       5.625   &       5.721   &       5.682   &       0       &       1678.0  &       1666.7  &0.008&         &               &       1665.6  \\
        &       6.081   &       5.957   &       6.029   &       6.011   &       0.606     &       1666.9  &       &       &               &               &               \\
%        &               &               &               &               &               &               &       &       &               &               &               \\
$^{220}$U       &       5.819   &       5.641   &       5.745   &       5.698   &       0       &       1692.2  &       1681.2  &0.008&         &               &       1680.8  \\
        &       6.109   &       5.971   &       6.052   &       6.025   &       0.605     &       1682.6  &&              &               &               &               \\
%        &               &               &               &               &               &               &&              &               &               &               \\
$^{222}$U       &       5.849   &       5.661   &       5.772   &       5.717   &       0       &       1705.1  &       1695.7  &0.048&         &               &       1695.6  \\
        &       6.142   &       5.990    &       6.079   &       6.043   &       0.611    &       1697.9  &&              &               &               &               \\
%        &               &               &               &               &               &               &&              &               &               &               \\

$^{224}$U       &       5.878   &       5.681   &       5.798   &       5.737   &       0       &       1717.9  &       1710.8  &0.146&         &               &       1710.3  \\
        &       6.198   &       6.032   &       6.131   &       6.085   &       0.645    &       1712.8  &&              &               &               &               \\
%        &               &               &               &               &               &               &&              &               &               &               \\
$^{226}$U       &       5.907   &       5.701   &       5.824   &       5.757   &       0       &       1730.8  &       1724.7  &0.172&         &               &       1724.8  \\
        &       6.232   &       6.053   &       6.160    &       6.106   &       0.652    &       1727.4  &&              &               &               &               \\
%        &               &               &               &               &               &               &&              &               &               &               \\
        &       5.935   &       5.721   &       5.850    &       5.776   &       0       &       1743.6  &&              &               &               &               \\
$^{228}$U       &       5.966   &       5.743   &       5.877   &       5.798   &       0.210    &       1741.7  &       1739.0  &0.191&         &               &       1739    \\
        &       6.259   &       6.068   &       6.182   &       6.120    &       0.651    &       1741.3  &&              &               &               &               \\
%        &               &               &               &               &               &               &&              &               &               &               \\
%        &          &          &          &          &          &         &&              &               &               &               \\
$^{230}$U       &       5.964   &       5.739   &       5.875   &       5.795   &       0       &       1756.0  &       1752.6  &0.199&         &       0.260    &       1752.8  \\
        &       6.000       &       5.765   &       5.907   &       5.821   &       0.234    &       1755.4  &&              &               &               &               \\
        &       6.293   &       6.091   &       6.213   &       6.143   &       0.658    &       1753.7  &&              &               &               &               \\
%        &               &               &               &               &               &               &&              &               &               &               \\
%        &          &          &          &          &          &         &&              &               &               &               \\
$^{232}$U       &       5.994   &       5.755   &       5.900     &       5.810    &       0       &       1766.8  &       1765.7  &0.207&         &       0.267    &       1765.9  \\
        &       6.033   &       5.785   &       5.935   &       5.840    &       0.251    &       1768.2  &&              &               &               &               \\
        &       6.364   &       6.167   &       6.286   &       6.218   &       0.712    &       1766.8  &&              &               &               &               \\
%        &               &               &               &               &               &               &&              &               &               &               \\
%        &          &          &          &           &          &         &&              &               &               &               \\
$^{234}$U       &       6.021   &       5.767   &       5.923   &       5.823   &       0       &       1776.4  &       1778.2  &0.215& 5.829   &       0.265    &       1778.6  \\
        &       6.065   &       5.803   &       5.963   &       5.858   &       0.267    &       1780.3  &&              &               &               &               \\
        &       6.415    &       6.209   &       6.334   &       6.260   &       0.738    &       1778.2  &&              &               &               &               \\
%        &               &               &               &               &               &               &&              &               &               &               \\
%        &          &          &          &          &          &         &&              &               &               &               \\
$^{236}$U       &       6.092   &       5.819   &       5.987   &       5.874   &       0.276    &       1791.7  &       1790.0  &0.215& 5.843   &       0.272    &       1790.4  \\
        &       6.446   &       6.230   &       6.363   &       6.281   &       0.744    &       1789.4  &&              &               &               &               \\
%        &               &               &               &               &               &               &&              &               &               &               \\
%        &          &          &          &          &          &         &&              &               &               &               \\
$^{238}$U       &       6.124   &       5.838   &       6.015   &       5.892   &       0.283    &       1802.5  &       1801.2  &0.215& 5.857   &       0.272    &       1801.7  \\
        &       6.488   &       6.263   &       6.402   &       6.314   &       0.763    &       1800.4  &&              &               &               &               \\
%        &               &               &               &               &               &               &&              &               &               &               \\
\Xhline{3\arrayrulewidth}
\end{tabular}\label{Tab1} }}
\end{table}

%\end{document}

\begin{table}
\hspace{0.1 cm}
\caption{Same as Table I,  but for Th isotopes.}
\scalebox{0.8}{
{\begin{tabular}{cccccccccccccccccccccc}
\Xhline{3\arrayrulewidth}
& \multicolumn{6}{ c }{RMF (NL3)} &\multicolumn{2}{ c }{FRDM}& \multicolumn{3}{ c }{Experiment}  \\
	\cmidrule(lr){2-7}\cmidrule(lr){8-9}\cmidrule(lr){10-12}
Nucleus &       $r_n$   &       $r_p$   &       $r_{rms}$       &       $r_{ch}$        &       $\beta_2$       &       BE (MeV)        &       BE (MeV)        &$\beta_2$&     $r_{ch}$        &       $\beta_2$       &       BE(MeV) \\
\Xhline{3\arrayrulewidth}
%&&&&&&&&&&&\\
$^{216}$Th      &       5.781   &       5.594   &       5.704   &       5.651   &       0       &       1673.5  &       1663.6  &0.008&         &               &       1662.7  \\
        &       6.034    &       5.897   &       5.977   &       5.951   &       0.567    &       1663.8  &&              &               &               &               \\
%        &               &               &               &               &               &               &&              &               &               &               \\
$^{218}$Th      &       5.812   &       5.611   &       5.730    &       5.667   &       0       &       1686.5  &       1677.2  &0.008&         &               &       1676.7  \\
        &       6.105   &       5.959   &       6.045   &       6.013   &       0.616    &       1678.2  &&              &               &               &               \\
%        &               &               &               &               &               &               &&              &               &               &               \\
$^{220}$Th      &       5.842   &       5.631   &       5.757   &       5.687   &       0       &       1698.1  &       1690.2  &0.030&         &               &       1690.6  \\
        &       6.140   &       5.983   &       6.076   &       6.036   &       0.624    &       1692.8  &&              &               &               &               \\
%        &               &               &               &               &               &               &&              &               &               &               \\
$^{222}$Th      &       5.873   &       5.651   &       5.784   &       5.707   &       0       &       1709.7  &       1704.6  &0.111&         &       0.151   &       1704.2  \\
        &       6.174   &       6.007   &       6.107   &       6.060   &       0.631    &       1706.1  &&              &               &               &               \\
%        &               &               &               &               &               &               &&              &               &               &               \\

$^{224}$Th      &       5.902   &       5.672   &       5.81    &       5.728   &       0       &       1721.4  &       1717.4  &0.164&         &       0.173   &       1717.6  \\
        &       6.222   &       6.021   &       6.142   &       6.074   &       0.640    &       1718.9  &&              &               &               &               \\
%        &               &               &               &               &               &               &&              &               &               &               \\
%        &          &          &          &          &         &        &         &&         &         &        \\
$^{226}$Th      &       5.931   &       5.692   &       5.837   &       5.748   &       0       &       1733.0  &1729.9& 0.173             &             &    0.225           &       1730.5        \\
        &       6.25    &       6.036   &       6.166   &       6.089   &       0.642    &       1731.9  &&              &               &               &               \\
%        &               &               &               &               &               &               &&              &               &               &               \\
$^{228}$Th         &       5.955    &       5.710   &       5.859   &       5.766    &       0  &       1743.9  &1742.5&0.182              &  5.748             &    0.229            &  1743.0             \\
     &       5.989   &       5.729   &       5.888   &       5.785   &       0.227       &       1744.5  &         &&    &         &         \\
        &       6.292   &       6.065   &       6.203   &       6.118   &       0.661    &       1743.4  &&              &               &               &               \\
%        &               &               &               &               &               &               &&              &               &               &               \\
%        &       &          &          &          &          &         &&              &               &               &               \\
$^{230}$Th      &       5.990    &       5.727   &       5.888   &       5.783   &       0       &       1754.2  &       1754.6  &0.198& 5.767   &       0.246   &       1755.1  \\
        &       6.026   &       5.751   &       5.920    &       5.807   &       0.232    &       1756.0  &&              &               &               &               \\
        &       6.315   &       6.111    &       6.236   &       6.163   &       0.671    &       1753.1  &&              &               &               &               \\
%        &               &               &               &               &               &               &&              &               &               &               \\
%        &          &         &          &          &         &         &&              &               &               &               \\
$^{232}$Th      & 6.060      &  5.773       & 5.950      & 5.828   &0.251        &1767.0        &1766.2& 0.207             & 5.784              &   0.248             &  1766.7             \\
        &   6.240        &     6.010     &    6.151      &   6.063       &  0.681         & 1765.0        &         &&    &         &         \\
%        &          &          &          &          &           &         &&              &               &               &               \\
%        &               &               &               &               &               &               &&              &               &               &               \\
%        &           &          &          &           &          &         &&              &               &               &               \\
%      &      &      &     &      &      &     &&              &               &               &               \\
$^{234}$Th        &       6.093   &       5.793   &       5.979   &       5.848   &       0.269    &       1777.5  &       1777.2  &0.215&         &       0.238   &       1777.6  \\
%        &          &          &          &          &           &         &&              &               &               &               \\
%        &               &               &               &               &               &               &&              &               &               &               \\
%        &          &          &           &          &          &         &&              &               &               &               \\
$^{236}$Th      &6.122       &5.812     &6.006     &5.866      & 0.272   & 1787.6    &       1787.6  &0.215&         &               &       1788.1  \\
%        &          &          &           &          &           &          &&              &               &               &               \\
%        &          &          &         &          &           &        &&              &               &               &               \\
%        &               &               &               &               &               &               &&              &               &               &               \\
%        &          &         &          &          &           &         &&              &               &               &               \\
$^{238}$Th      &6.152       & 5.832      &6.033    & 5.887     & 0.281      & 1797.5  &       1797.7  &0.224&         &               &       1797.8  \\
%        &           &           &          &          &           &          &&              &               &               &               \\
%        &         &         &          &          &           &        &&              &               &               &               \\
%        &               &               &               &               &               &               &&              &               &               &               \\
%        &          &          &          &          &           &         &         &  &       &               &               \\
$^{240}$Th      & 6.180        &5.846     & 6.057   &5.901        &0.292         & 1806.6       &1807.2& 0.224             &               &               &               \\
%        &           &          &          &         &          &        &&              &               &               &               \\
%        &           &          &          &          &            &         &&              &               &               &               \\
%        &               &               &               &               &               &               &&              &               &               &               \\
\Xhline{3\arrayrulewidth}
\end{tabular}\label{Tab2} }}
\end{table}
In some cases of U and Th isotopes, we get more than one solution. The 
solution corresponding to the maximum BE is the ground state configuration 
and all other solutions are the intrinsic excited states. In some cases, the 
ground state BE does not match the experimental data. However, the BE, whose 
quadrupole deformation parameter $\beta_2$ is closer to the experimental data 
or to the FRDM value match well with each other. For example, the BEs 
of $^{236}$U are 1791.7, 1790.0, and 1790.4 MeV with RMF, FRDM, and 
experimental data, respectively, and the corresponding $\beta_2$ are 0.276, 
0.215, and 0.272. Similar to the BE, we get $\beta_2$ and charge radius $r_c$
 RMF results comparable with the FRDM and experimental values.
% In some cases, we have also 
%noticed oblate and super-deformed states, which are slightly above the ground state configuration.   

\begin{figure}[ht]
        \begin{center}
\includegraphics[width=0.7\columnwidth]{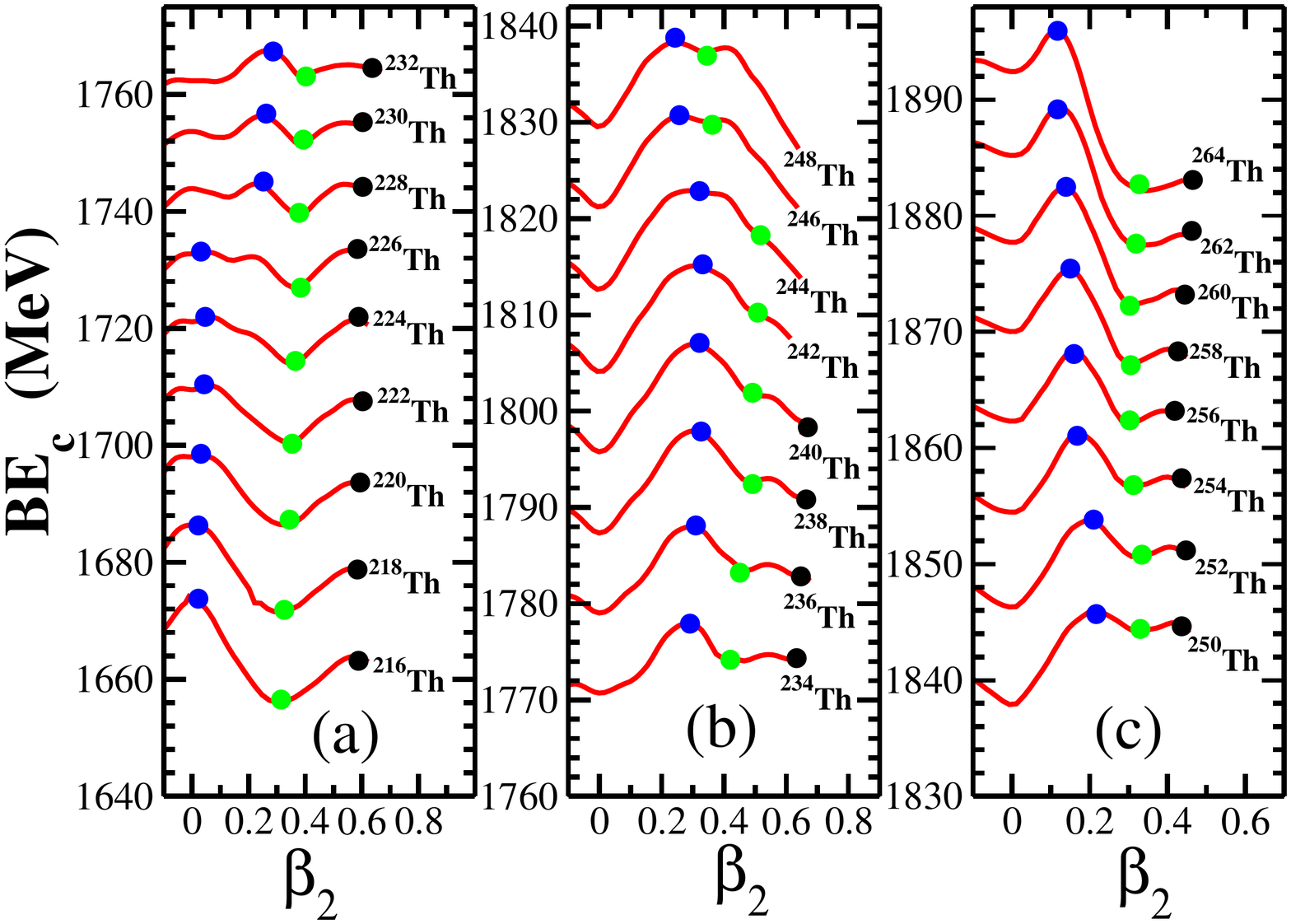}
        \end{center}
\caption{The potential energy surface is a function of the quadrupole 
deformation parameter ($\beta_{2}$) for Th isotopes. The difference between 
the leftmost (blue) and the middle (green) circles represents the first 
fission barrier heights B$_{f}$ (in MeV). See text for details.}
\label{bea2}
\end{figure}

\begin{figure}[ht]
        \begin{center}
\includegraphics[width=0.7\columnwidth]{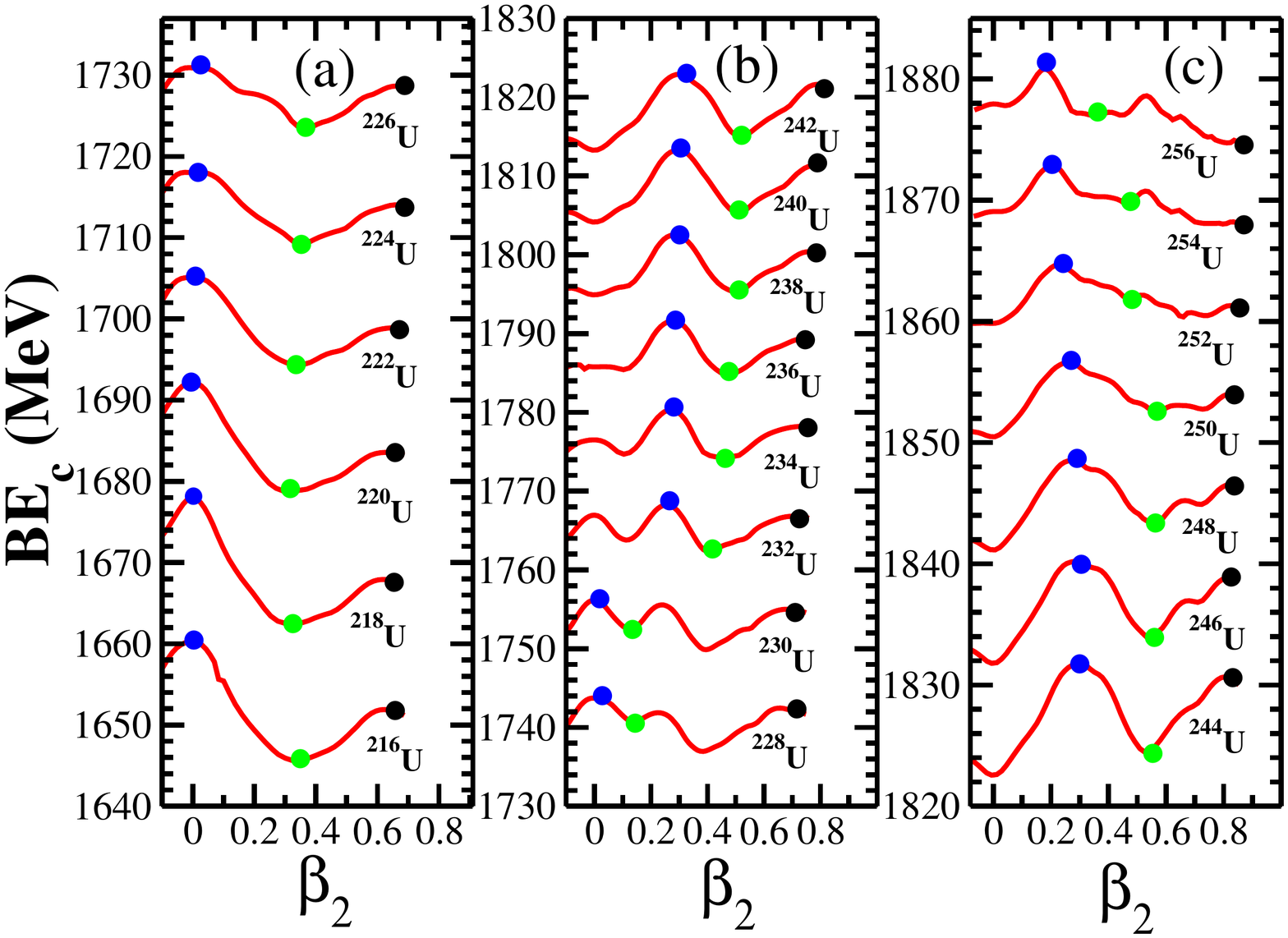}
        \end{center}
\caption{Same as Fig. 3.2, but for U isotopes.}
\label{bea3}
\end{figure}

\subsection{Potential energy surface}\label{sec7}

In late 1960's, the structure of PES found renewed interest for its role in 
the nuclear fission process. In the majority of PESs for actinide nuclei, 
there exist a second maximum, which splits the fission barrier into inner and 
outer segments ~\cite{lynn80}.
It also has a crucial role for the characterization of ground state, intrinsic
excited state, occurrence of shape coexistence, radioactivity, and spontaneous 
and induced fission. The structure of the PES is defined mainly from the shell 
structure, which is strongly related to the distance between the mass centers 
of the nascent fragments. The macroscopic-microscopic liquid drop theory has 
been given a key concept of fission, where the surface energy is in the form 
of collective deformation of the nucleus. 

In Figs.~\ref{bea2} and ~\ref{bea3} we have plotted the PES for some selected 
isotopes of Th and U nuclei. The constraint binding energy BE$_{c}$ versus the 
quadrupole deformation parameter $\beta_2$ are shown. 
A nucleus undergoes fission process when the nucleus becomes highly elongated
along an axis. This can be done most simply by modifying the single-particle 
potential with the help of a constraint, i.e., the Lagrangian multiplier 
$\lambda$. Then, the system becomes more or less compressed depending
on the Lagrangian multiplier $\lambda$. In other words, in a constraint 
calculation, we minimize the expectation value of the Hamiltonian $<H'>$ 
instead of $<H>$, which are related to each other by the relation~\cite{patra09,flocard73,koepf88,fink89,hirata93}:
\begin{eqnarray}
H^{'}=H-\lambda Q,\qquad {with} \qquad Q=r^{2}Y_{20}(\theta, \phi),
\end{eqnarray}
where $\lambda$ is fixed by the condition $<Q>_\lambda$ = $Q_{0}$.

\begin{table}
\begin{center}
\hspace{0.1 cm}
\caption{First fission barrier heights B$_{f}$ (in MeV) of some
even-even actinide nuclei from RMF(NL3) calculations compared
with FRDM and experimental data~\cite{moller95}.}
\renewcommand{\tabcolsep}{0.2 cm}
\renewcommand{\arraystretch}{1.2}
{\begin{tabular}{ccccc}
\hline
\hline
 \multicolumn{1}{ c }{Nucleus} &\multicolumn{1}{ c }{B$^{cal.}_{f}$ }& 
\multicolumn{1}{ c }{B$^{FRDM}_{f}$~\cite{moller95}}&\multicolumn{1}{ c }{B$^{exp.}_{f}$~\cite{moller95}} \\ 
\cline{1-3}
\hline
$^{228}$Th & 5.69 &  7.43  &6.50\\
$^{230}$Th & 5.25 &  7.57  &7.0\\
$^{232}$Th & 4.85 &  7.63  &6.30\\
$^{234}$Th & 4.34 &  7.44  &6.65\\
$^{232}$U & 5.65 &  6.61  &5.40\\
$^{234}$U & 6.30 &  6.79  &5.80\\
$^{236}$U & 6.64 &  6.65  &5.75\\
$^{238}$U & 7.15 &  4.89  &5.90\\
$^{240}$U & 7.66 &  5.59  &5.80\\
\hline
\hline
\end{tabular}\label{Tab3}}
\end{center}
\end{table}

\begin{figure}[ht]
        \begin{center}
\includegraphics[width=0.7\columnwidth]{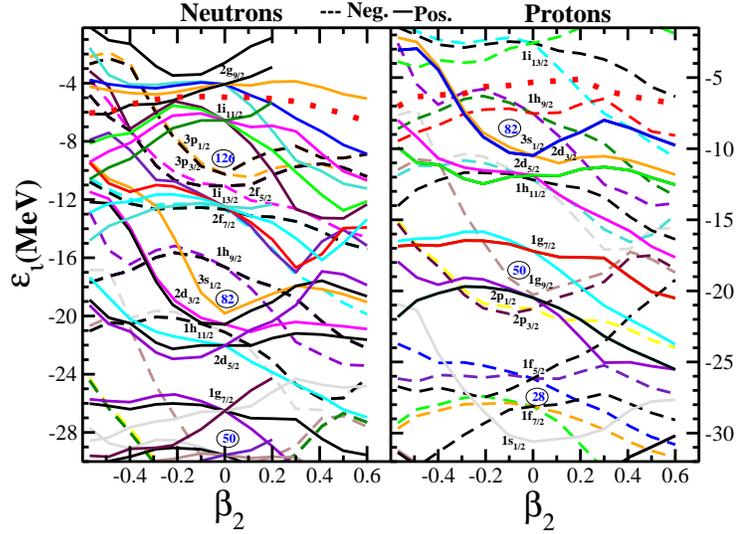}
        \end{center}
\caption{Single-particle energy levels for $^{232}$Th as a function of 
the quadrupole deformation parameter $\beta_{2}$. The Fermi levels are denoted 
by the dotted(red) curve.}
\label{bea4}
\end{figure}

\begin{figure}[ht]
        \begin{center}
\includegraphics[width=0.7\columnwidth]{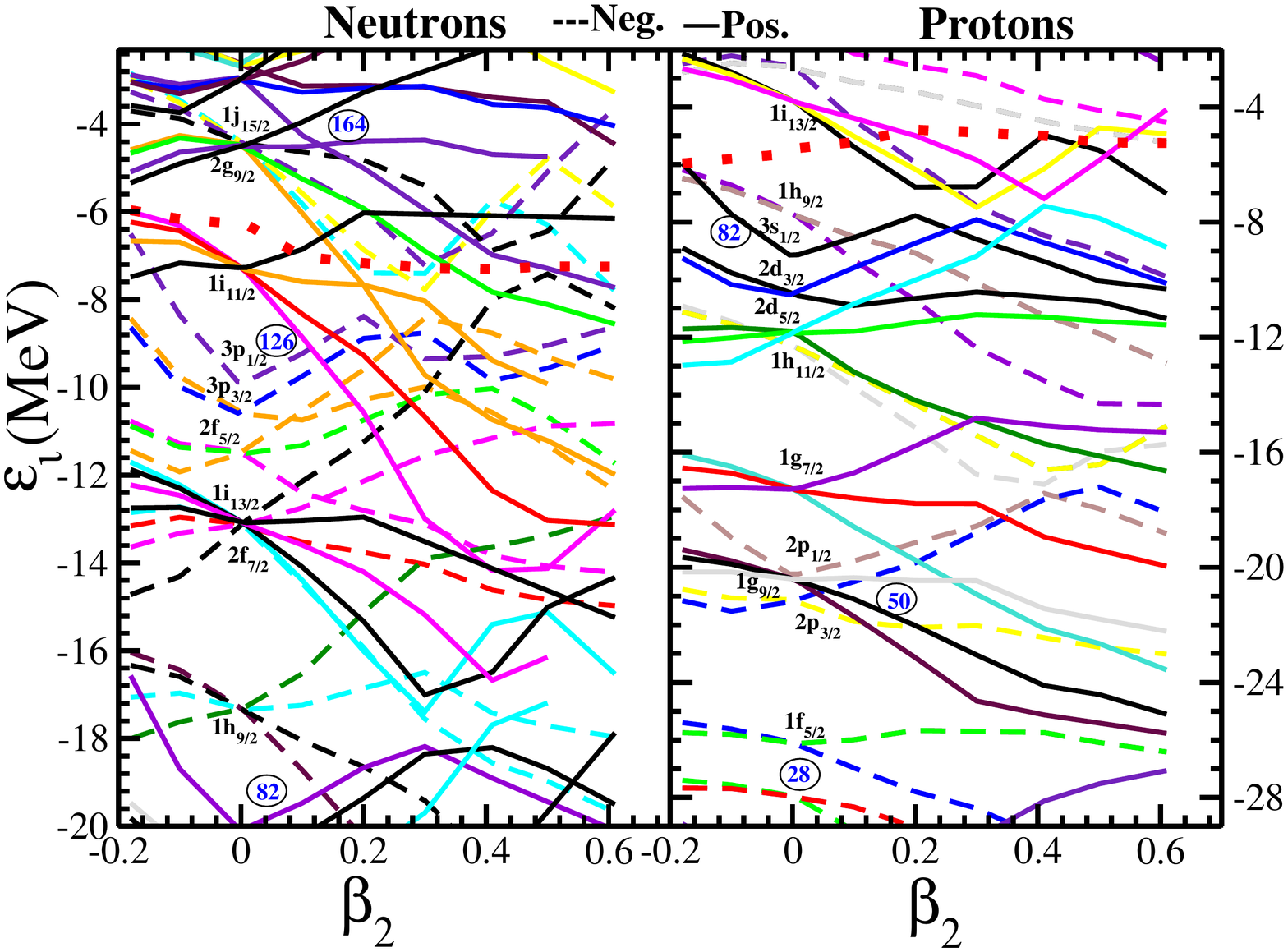}
        \end{center}
\caption{Same as Fig. 3.4 but for the $^{236}$U nucleus.
}
\label{bea5}
\end{figure}

Usually, in an axially deformed constraint calculation for a nucleus,
we see two maxima in the PES diagram: (i) prolate and (ii) oblate or
spherical. However, in some cases, more than two maxima are seen.
 If the ground state energy is distinctly more than other maxima, then
the nucleus has a well-defined ground-state configuration. On the other
hand, if the difference in BEs between two or three maxima is negligible, then 
the nucleus is in the shape coexistence configuration. In this case, a 
configuration mixing calculation is needed to determine the ground-state 
solution of the nucleus, which is beyond the scope of the present calculation. 
It is to be noted here that in a constraint calculation, the maximum BE
(major peak in the PES diagram) corresponds to the ground-state configuration 
and all other solutions (minor peaks in the PES curve) are the intrinsic 
excited states. 

The fission barrier $B_f$ is an important quantity for study the properties of 
fission reaction.  We calculate the fission barrier from the PES curve for 
some selected even-even nuclei, which are listed in Table~\ref{Tab3}. From the 
table, it can be seen that the fission barrier for $^{228}$Th turns out to be 
5.69 MeV, comparable to the FRDM and experimental values of $B_f$ = 7.43  and 
6.50 MeV, respectively. Similarly, the calculated $B_f$ of $^{232}$U is 
5.65 MeV, which also agrees well with the experimental data, 5.40 MeV. In 
some cases, the fission barrier height is 1$-$2 MeV lower or higher than 
the experimental data. The double-humped fission barrier is reproduced in all 
these cases. Similar types of calculations are done in Refs.~\cite{meng06,bing12,nan14,zhao15}. 

In nuclei like $^{228-230}$Th and $^{228-234}$U, we find three maxima. Among 
these maxima, two are found at normal deformation (spherical and normal 
prolate), but the third one is situated far away, i.e., at a relatively large 
quadrupole deformation. Upon careful inspection, one can also see that one of 
them (mostly the peak nearer to the spherical region) is not strongly 
pronounced and can be ignored in certain cases. This third maximum 
separates the second barrier by a depth of  1$-$2 MeV, responsible for the
formation of resonance state, which has been observed experimentally
\cite{back72}. For some of the uranium isotopes $^{216-230}$U, the ground 
states are predicted to be spherical in the RMF formalism, agreeing with the 
FRDM results. The other isotopes in the series $^{232-256}$U are found to be 
in the prolate ground state, matching the experimental data.
Similarly, the thorium nuclei $^{216-226}$Th are spherical in shape and 
$^{228-264}$Th are in the prolate ground configuration.
In addition to these shapes, we also note sallow regions in the PES curves of 
both Th and U isotopes. These fluctuation in the PES curves could be due to 
the limitation of mean field approximation and one needs a theory beyond 
mean-field to overcome such fluctuations. For example, the Generator 
Coordinate Method or Random Phase Approximation could be an improved 
formalism to take care of such effects~\cite{brink68}. Beyond the second 
hump, we find the PES curve goes down and down, and never up again. This is 
the process wherein the liquid drop gets more and more elongated and reaches 
the fission stage. The PES curve from which it starts to go down is marked by
the scission points, which are shown by the rightmost, black dot on the PES 
curves of Figs.~\ref{bea2} and ~\ref{bea3}.

\subsection{Evolution of single-particle energy with deformation}\label{sec8}

In this subsection, we evaluate the neutron and proton single-particle 
energy levels for some selected Nilssion orbits with different values of 
deformation parameter $\beta_{2}$ using the constraint calculations. The 
results are shown in Figs.~\ref{bea4} and ~\ref{bea5}, which explain the
origin of the shape change along the $\alpha$ decay chains of the thorium
and uranium isotopes. The positive-parity orbit is shown by solid line, 
negative-parity orbit by dashed lines and the dotted (red) line indicates 
the Fermi energy for $^{232}$Th and $^{236}$U.

For small-Z nuclei, the electrostatic repulsion is very weak, but at a 
higher value of Z (superheavy nuclei), the electrostatic repulsion is much 
stronger so that the nuclear liquid drop becomes unstable to surface 
distortion ~\cite{mayer66} and fission. In such a nucleus, the single-particle 
density is very high and the energy separation is small, which determines 
that the shell stabilizes the unstable Coulomb repulsion. This effect is clear 
for heavy elements approaching N=126, with a gap between 3p$_{1/2}$ and 
1i$_{11/2}$ of about 2$-$3 MeV, in a neutron single-particle of $^{236}$U 
and $^{232}$Th. In both Figs.\ref{bea4} and ~\ref{bea5}, the neutron 
single-particle energy level 1i$_{13/2}$ lies between 2f$_{7/2}$ and 2f$_{5/2}$,
 creating a distinct shell gap at N=114. In $^{232}$Th and $^{236}$U, with 
increasing deformation the opposite-parity levels of 2g$_{9/2}$ and 1j$_{15/2}$,which are far apart in the spherical solution, come close to each other. This 
gives rise to the parity doublet phenomena \cite{singh14,kumar15,hab02}.

\section {Mode of decays}\label{sec9}

In this section, we will discuss various mode of decays encounter by these 
thermally fissile nuclei both in the $\beta$-stability line and away from it. 
This is important, because the utility of heavy, and mostly, nuclei which are 
away from stability lines depends very much on their lifetime. For example, 
we do not get $^{233}$U and $^{239}$Pu in nature, because of their short 
lifetimes, although these two nuclei are extremely useful for energy 
production. That is why $^{235}$U is the most needed isotope in the uranium 
series, for its thermally fissile nature in energy production in the fission 
process, both for civilian and for military use. The common mode of 
instability for such heavy nuclei are spontaneous fission, $\alpha$, $\beta$,
and $cluster$ decays. All these decays depend on the neutron-to-proton ratio 
as well as the number of nucleons present in the nucleus.

\subsection{$\alpha$- and $\beta$-decays half-lives}\label{sec10}

In some previous work \cite{bk06,patra07}, it has been analyzed the densities 
of nuclei in a more detailed manner. From that analysis, we concluded that 
there is no visible cluster either in the ground or in the excited intrinsic 
states. The possible clusterizations are the $\alpha$-like matter in the 
interior and neutron-rich matter in the exterior region of normal and 
neutron-rich superheavy nuclei, respectively. Thus, the possible mode of 
decays are $\alpha$ decay for $\beta$-stable nuclei and $\beta^-$ decay for 
neutron-rich isotopes. To estimate the stability of such nuclei, we have to 
calculate the $\alpha$-decay $T_{1/2}(\alpha)$ and the $\beta$-decay 
$T_{1/2}(\beta)$ half-lives times.

\subsubsection { The Q$_\alpha$ energy and $\alpha$-decay half-life 
              T$_{1/2}^{\alpha}$ }\label{sec11}

To calculate the $\alpha$ decay half-life $T_{1/2}^{\alpha}$, one has to know 
the $Q_{\alpha}$ energies of the nucleus. This can be estimated by knowing 
the BEs of the parents and daughter and the BE of the $\alpha$ particle, i.e., 
the BE of $^{4}$He. The BEs are obtained from experimental data wherever 
available and from other mass formulas as well as the RMF Lagrangian as 
have been discussed earlier in Ref. \cite{patra23}. The $Q_{\alpha}$
energy is evaluated by using the relation:
\begin{eqnarray}
Q_{\alpha}(N,Z)&=& BE(N, Z)- BE(N-2,Z-2)\nonumber\\
&-& BE(2,2).
\end{eqnarray}
Here, $BE(N, Z)$, $BE(N-2,Z-2)$ and $BE(2,2)$ are the BEs of the parent, 
daughter, and $^4$He nuclei (BE= 28.296 MeV) with neutron number N and proton 
number Z.
\begin{figure}[ht]
	\begin{center}
\includegraphics[width=0.7\columnwidth]{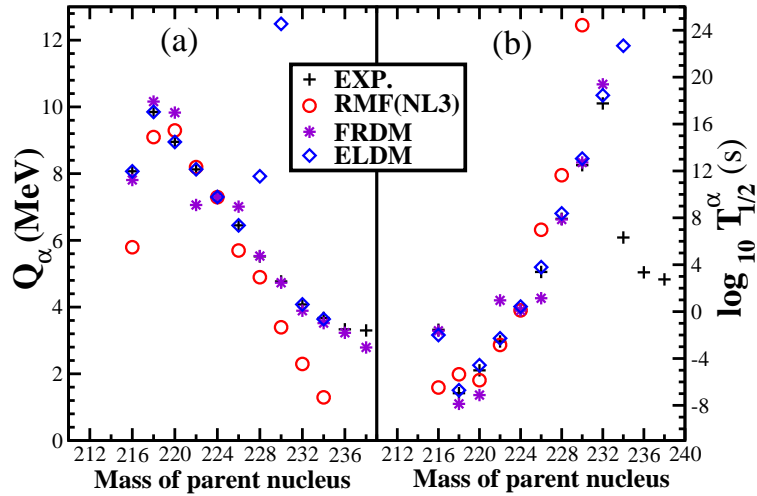}
	\end{center}
\caption{The Q$_\alpha$ and half-life time $T^{\alpha}_{1/2}$ of the 
$\alpha$ decay chain for Th isotopes are calculated using RMF, 
FRMD~\cite{moller97,moller95}, and ELDM~\cite{sb02} and compared with the 
experiment (EXP.)~\cite{NDCC}.}
\label{bea11}
\end{figure}

\begin{figure}[ht]
	\begin{center}
\includegraphics[width=0.7\columnwidth]{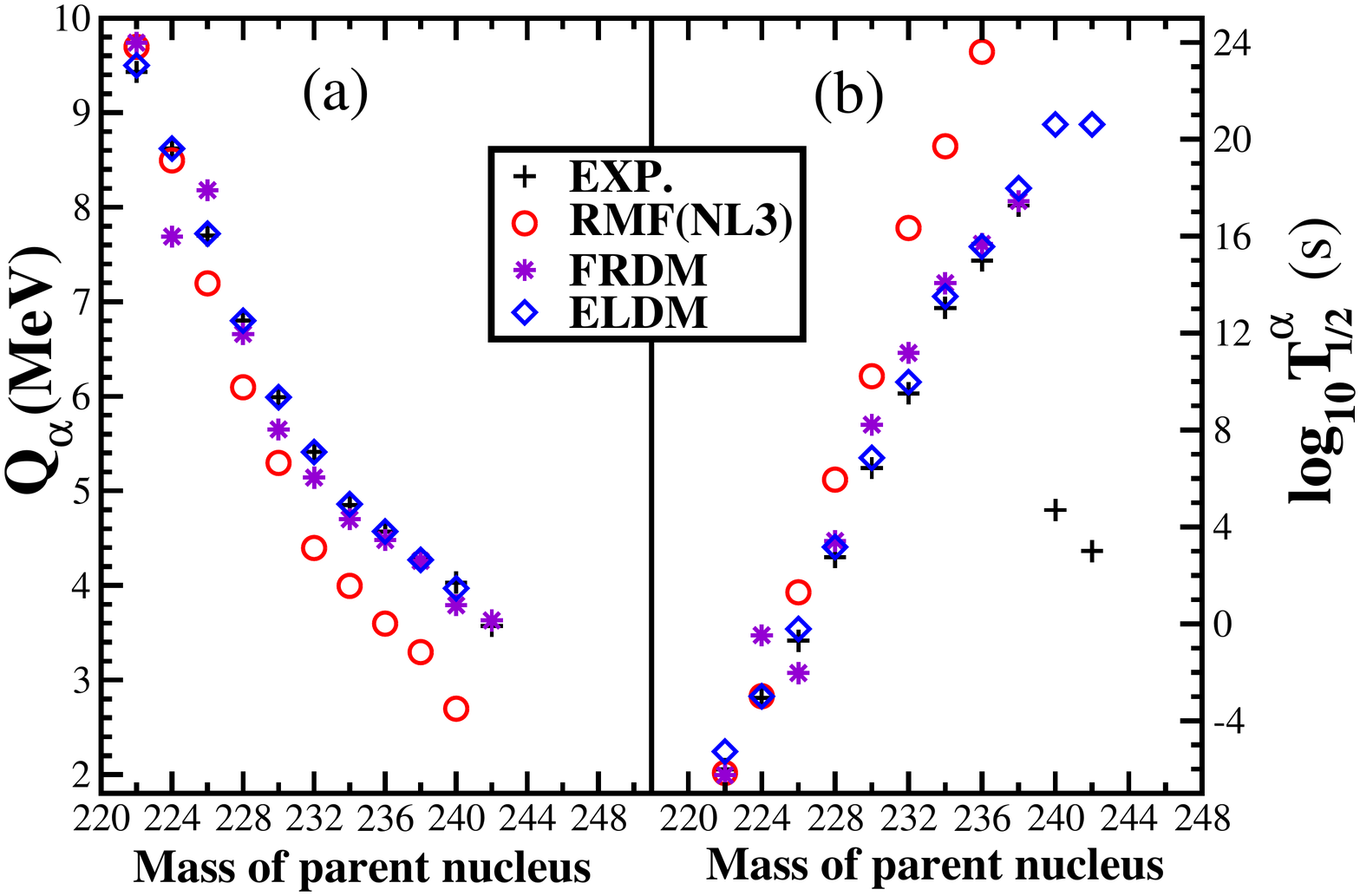}
	\end{center}
\caption{Same as Fig. 3.6, but for U.}
\label{bea12}
\end{figure}
Knowing the $Q_{\alpha}$ values of nuclei, we roughly estimate the $\alpha$ 
decay half-lives $log_{10}T_{1/2}^{\alpha} (s)$ of various isotopes 
using the phenomenological formula of Viola and Seaborg \cite{viol01}:
\begin{equation}
log_{10}T^{\alpha}_{1/2}(s)=\frac {(aZ-b)}{\sqrt{Q_{\alpha}}}-(cZ+d) + 
h_{log}.
\end{equation}
The value of the parameters $a$, $b$, $c$, and $d$ are taken from the
recent modified parametrizations of Sobiczewski {\it et al.} \cite{sobi89}, 
which are $a$ = 1.66175, $b$ = 8.5166, $c$ = 0.20228, $d$ = 33.9069.
The quantity $h_{log}$ accounts for the hindrances associated with the odd 
proton and neutron numbers as given by Viola and Seaborg~\cite{viol01}, namely

$h_{log}=\begin{array}{ll}
           0, &  \mbox{ Z and N even}\\
        0.772,&  \mbox{ Z odd and N even}\\ 
        1.066,&  \mbox{ Z even and N odd}\\ 
        1.114,&  \mbox{ Z and N odd}. 
\end{array}$

The $Q_{\alpha}-$values obtained from RMF calculations for Th and U isotopes 
are shown in Figs.~\ref{bea11} and~\ref{bea12}. Our results also compared 
with other theoretical predictions \cite{moller97,sb02} and experimental 
data \cite{angel13}. The agreement of RMF results with others as well as 
with experiment is pretty well. Although the agreement in $Q_{\alpha}$ value 
is quite good, one must note that the $T^{\alpha}_{1/2}(s)$ values may vary a 
lot, because of the exponential factor in the equation. That is why it is 
better to compare $log_{10}T^{\alpha}_{1/2}(s)$ instead of 
$T^{\alpha}_{1/2}(s)$ values. These are compared in the right panel in 
Figs.~\ref{bea11} and \ref{bea12}. We note that our prediction matches well 
with other calculations as well as experimental data.

Further, a careful analysis of $log_{10}T^{\alpha}_{1/2}$ (in seconds) for 
even-even thorium reveals that the $Q_{\alpha}$ value decreases with an 
increase in the mass number A of the parent nucleus. The $Q_{\alpha}$ energy 
of Th isotopes given by Duarte {\it et al.} \cite{sb02} deviates a lot when 
mass of the parent nucleus reaches A=230. The corresponding $log_{10}T^{\alpha}_{1/2}$ increases almost monotonically linearly with an increase in mass number 
of the same nucleus. The experimental values of $log_{10}T^{\alpha}_{1/2}$ 
deviate a lot in the heavy mass region (with parent nuclei 234-238). A similar 
situation is found in the case of uranium isotopes also, which are shown in 
Fig.~\ref{bea12}.
It is noteworthy that the origin of $\alpha$ decay or $cluster$ decay
phenomena are purely quantum mechanical process. Thus the quantum tunneling
plays an important role in such decay processes. The deviation of experimental
$\alpha$ decay lifetime from the calculated results obtained by the empirical
formula may not be suitable for such heavy nuclei, which are away from
the stability line, and more involved quantum mechanical treatment is needed
for such cases.

\subsection{Quantum mechanical calculation of $\alpha$-decay half-life 
              $T_{1/2}^{WKB}$}

In this subsection, we do an approximate evaluation of the $\alpha$ decay 
half-life using a quantum mechanical approach.  This approach is used 
recently ~\cite{bidhu11} for the evaluation of proton-emission as well as 
cluster decay. The obtained results satisfactorily match with known 
experimental data. Since these nuclei are prone to $\alpha$ decay or 
spontaneous fission, we need to calculate these decays to examine the 
stability. It is shown by Satpathy {\it et al.} \cite{sat08} that by addition 
of neutrons to Th and U isotopes, the neutron-rich nuclei become surprisingly 
stable against spontaneous fission. Thus, the possible modes of decay may be 
$\alpha$ and $\beta$ emission. To estimate the $\alpha$ decay, one needs 
the optical potential of the $\alpha$ and daughter nuclei, where a bare 
nucleon-nucleon potential, such as M3Y~\cite{love79}, LR3Y~\cite{bir12}, 
NLR3Y\cite{bidhu14} or DD-M3Y~\cite{kobos84} interaction is essential.  
In our calculations, we have taken the widely used M3Y interaction for this 
purpose. For simplicity, we use spherical densities of the cluster and 
daughter. Here, $\rho_{c}$ is the cluster density of the $\alpha$ particle 
and the daughter nucleus $\rho_{d}$ are obtained from RMF(NL3) formalism 
\cite{lala97}. Then, the nucleus-nucleus optical potential is calculated by 
using the well-known double-folding procedure to the M3Y~\cite{love79} 
nucleon-nucleon interaction, supplemented by a zero-range pseudopotential 
representing the single-nucleon exchange effects (EX). 
The Coulomb potential $V_{c}$ is added to obtain the total interaction 
potential $V(R) = V_{n}(R) + V_{c}(R)$. When the $\alpha$ particle tunnels 
through the potential barrier between two turning points, the probability of 
emission of the $\alpha$ particle is obtained by the WKB approximation. 
\begin{figure}[ht]
	\begin{center}
\includegraphics[width=0.8\columnwidth]{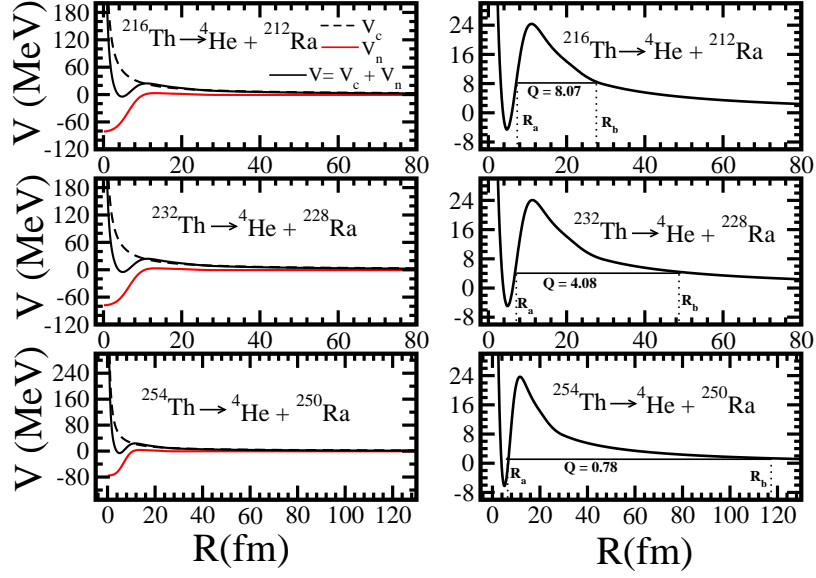}
\caption{In the left panel of the plot Coulomb potential $V_c (R)$, total 
interaction potential V(R) and folded potential $V_{n}(R)$(M3Y+EX) for Th 
isotopes are given. In the right panel of the figure the penetration path with 
an energy equal to the $Q$(MeV) value of the $\alpha$-decay shown by 
horizontal line.}
\label{bea9}
	\end{center}
\end{figure}
\begin{figure}[ht]
	\begin{center}
\includegraphics[width=0.8\columnwidth]{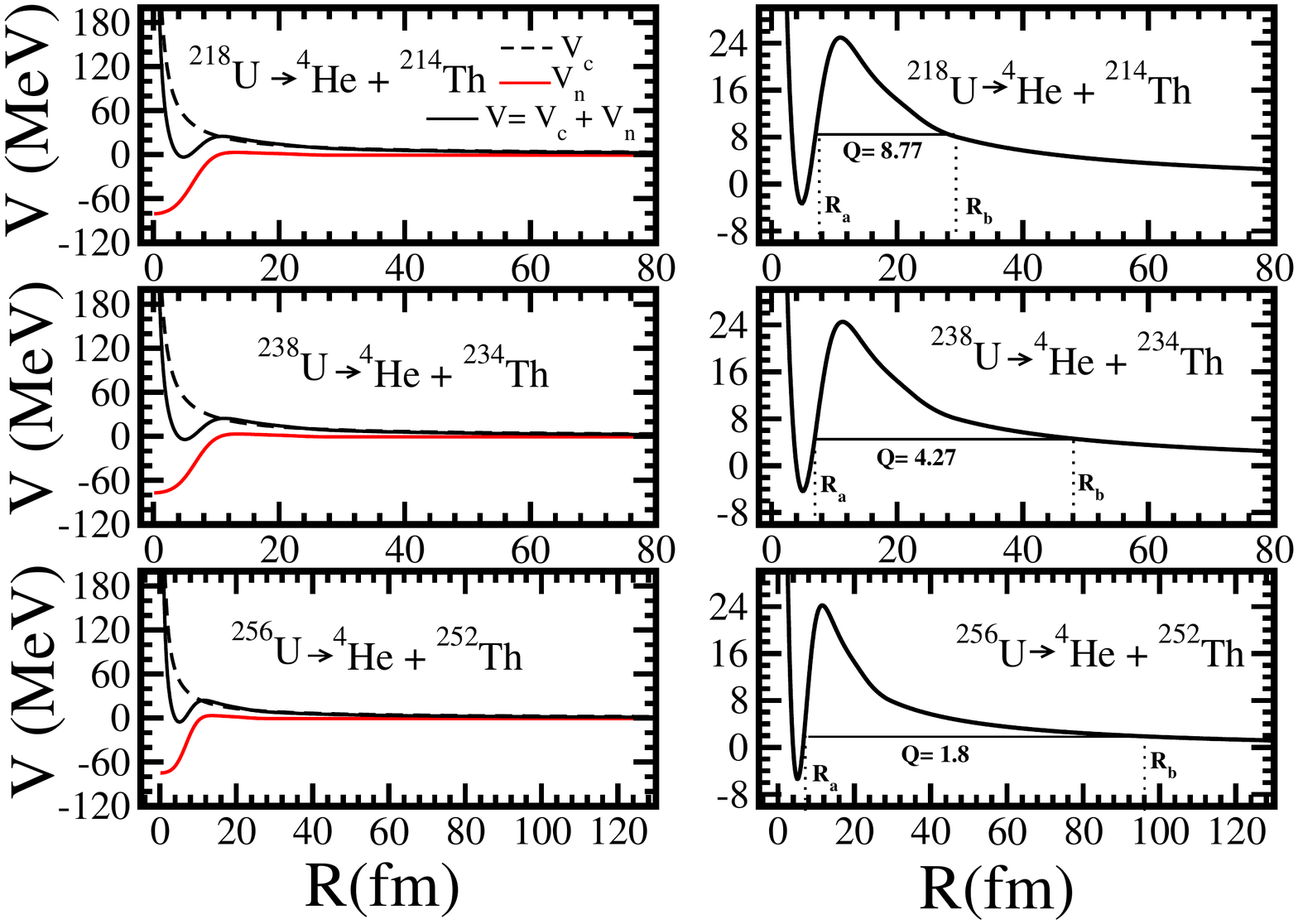}
\caption{Same as Fig. 3.8, but for U.}
\label{bea10}
	\end{center}
\end{figure}
Using this approximation, we have made an attempt to investigate the 
$\alpha$ decay of the neutron-rich thorium and uranium isotopes.
\begin{table*}
\hspace{0.1 cm}
\caption{
The penetrability P is evaluated using WKB approximation. The half-lives 
T$_{1/2}^{WKB}$ and T$_{1/2}^{\alpha}$ are calculated by quantum mechanical 
tunneling processes and Viola - Seaborg formula\cite{viol01}. Experimental
Q$_\alpha$ values are used for known masses and for unknown nuclei, Q$_\alpha$ 
obtained from RMF.}
{\begin{tabular}{cccccccccccc}
\Xhline{3\arrayrulewidth}
	\multicolumn{1}{ c }{Parent}& &\multicolumn{1}{ c }{Q$_\alpha$(MeV)}&& 
	\multicolumn{1}{ c }{P}&&\multicolumn{1}{ c }{T$_{1/2}^{WKB.}$(s.)}
	&&\multicolumn{1}{ c }{T$_{1/2}^{\alpha}$(s.)} \\ 
\cline{1-4}
\hline
$^{216}$Th  && 8.072~\cite{NDCC} &&  2.109$\times$10$^{-21}$&&3.28$\times10^{-1}$&&3.0$\times10^{-3}$ \\
$^{232}$Th  && 4.081~\cite{NDCC} &&  6.17$\times$10$^{-37}$&&1.12$\times10^{17}$&&5.07$\times10^{17}$ \\
$^{254}$Th  && 0.78 [RMF] &&  1.90$\times$10$^{-55}$&&3.6$\times$10$^{35}$ &&
1$\times$10$^{107}$\\
$^{218}$U   && 8.773~\cite{NDCC} &&  1.26$\times$10$^{-20}$&&5.5$\times10^{-2}$&&1.0$\times10^{-4}$ \\
$^{238}$U  && 4.27~\cite{NDCC} &&  1.163$\times$ 10$^{-37}$&&5.95$\times$ 10$^{15}$&&2.219$\times$10$^{17}$ \\
$^{256}$U   && 1.8 [RMF]&& 3.783$\times$10$^{-44}$&&1.83$\times$10$^{22}$&&
1.218$\times$10$^{55}$ \\
\Xhline{3\arrayrulewidth}
\end{tabular}\label{wkb}}
\end{table*}
The double folded~\cite{love79,khoa94} interaction potential V$_n$(M3Y+EX)
between the alpha cluster and daughter nucleus having densities $\rho_{c}$ 
and $\rho_{d}$ is 
\begin{eqnarray}
 V_n(\vec R)=\int {\rho_c}(\vec {r_c}){\rho_d}(\vec {r_d}) v|{\vec {r_c}
             -\vec {r_d}+\vec {R}=s}| d^{3}{r_c} d^{3}{r_d},
\end{eqnarray}
where $v(s)$ is the zero-range pseudopotential representing the single-nucleon 
exchange effects,\\
\begin{eqnarray}
v(s)=7999 \frac {e^{-4s}}{4s} - 2134\frac{e^{-2.5s}}{2.5s}+J_{00}(E) \delta(s),
\end{eqnarray}
with the exchange term~\cite{basu03} 
\begin{eqnarray}
      J_{00}(E)=-276(1-0.005E/A_\alpha(c))   {MeV fm^{3}}.
\end{eqnarray}
Here, A$_\alpha(c)$ is the mass of the cluster, i.e., the $\alpha$ particle 
mass and E is energy measured in the center of mass of the $\alpha$ particle 
or the cluster-daughter nucleus system, equal to the released $Q-$value.
The Coulomb potential between the $\alpha$ and daughter nucleus is
\begin{eqnarray}
           V{_c}(R)=Z{_c}Z{_d}e^{2}/R, 
\end{eqnarray}
and total interaction potential  is
\begin{eqnarray}
            V(R) = V{_n}(R)+V{_c}(R).
\end{eqnarray}
When the $\alpha$ particle tunnels through a potential barrier of two turning
points R$_a$ and R$_b$, then the probability of the $\alpha$ decay is given by
\begin{eqnarray}
P = exp \Bigg[ -\frac{2}{\hbar} \int_{R{_a}}^{R{_b}} {2\mu[V(R)-Q]}^{1/2} 
dR\Bigg].
\end{eqnarray}
The decay rate is defined as
\begin{eqnarray}
\lambda = \nu P,
\end{eqnarray}
with the assault frequency $\nu$ = 10$^{21}$ s$^{-1}$. The half-life is
calculated as:
\begin{eqnarray}
T^{WKB}_{1/2}= \frac {0.693}{\lambda}.
\end{eqnarray}
The total interaction potential (black curve) of the daughter and $\alpha$
nuclei are shown in Figs.~\ref{bea9} and~\ref{bea10}. The central well is due 
to the average nuclear attraction of all the nucleons and the hill-like
structure is due to the electric repulsion of the protons. The $\alpha$ 
particle with Q-value trapped inside the two turning points R$_a$ and R$_b$ of 
the barrier. The penetration probability P is given in the third column of 
Table~\ref{wkb} for some selected cases of thorium and uranium isotopes. The 
probabilities for $^{216}$Th and $^{218}$U are relatively high, because of the 
small width as compared to its barrier height. So, it is easier for the 
$\alpha$ particle to escape from these two turning points. 

As we increase the number of neutron in a nucleus, the Coulomb force becomes 
weak due to the hindrance of repulsion among the protons. In such cases, the 
width of the two turning points is very large and the barrier height is small.
Thus, the probability of $\alpha$ decay for $^{254}$Th and $^{256}$U is 
almost infinity and this type of nuclei are stable against $\alpha$ or
$cluster$ decays. These neutron-rich thermally fissile Th and U isotopes are
also stable against spontaneous fission. Since these are thermally fissile 
nuclei, a feather touch deposition of energy with thermal neutron, it
undergoes fission. Thus, half-life of the spontaneous fission is nearly 
infinity. We have compared our quantum mechanical tunneling results with the 
empirical formula of  Viola and Seaborg~\cite{viol01} in Table~\ref{wkb}.
For known nuclei, like $^{232}$Th and $^{238}$U both the results match
well, however, it deviates enormously from each other for unknown nuclei both 
in neutron-rich and neutron-deficient region. In general, independent of the
formula or model used, the $\alpha$ decay mode is rare for ultra-asymmetric
nuclei. In such isotopes, the possible decay mode is the $\beta$ decay.

\subsection{$\beta$-decay}\label{sec12}

As we have discussed, the prominent mode of instability of neutron-rich
Th and U nuclei is the $\beta$ decay, we have given an estimation of such
decay in this subsection. Actually, the $\beta$-decay lifetime should be
evaluated on the microscopic level, but that is beyond the present thesis.
Here we have used the empirical formula of Fiset and Nix \cite{Fiset72}, which
is defined as:
\begin{figure}[ht]
\begin{center}
\includegraphics[width=0.7\columnwidth]{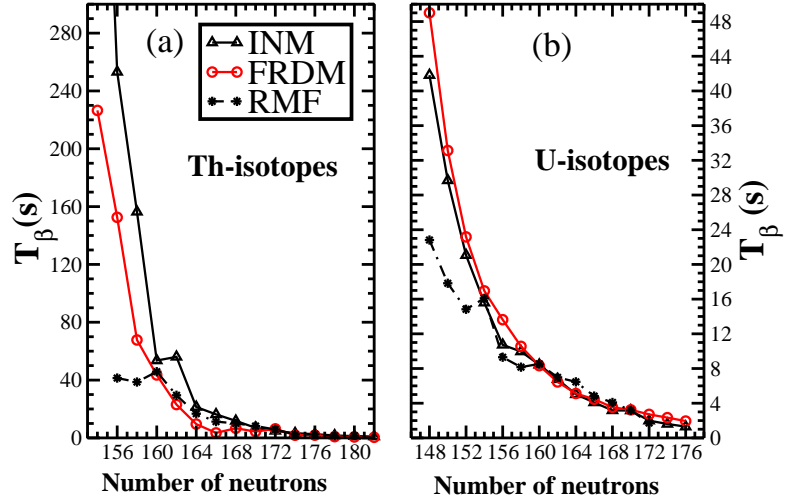}
\caption{The $\beta$-decay half-lives for Th and U isotopes are calculated 
using the formula of Fiset and Nix~\cite{Fiset72} [eq. (24)]. Ground-state 
binding energies are taken from FRDM~\cite{moller97}, INM~\cite{nayak99}, and 
RMF models.}
\label{bea15}
\end{center}
\end{figure}
\begin{eqnarray}
T{_\beta} = (540 \times 10^{5.0})\frac {m{_e}^{5}} {\rho_{d.s.} (W{_\beta}^{6}-m{_e}^{6})}   s.  
\end{eqnarray}

Similar to the $\alpha$ decay, we evaluate the $Q_{\beta}$ value for Th and U 
series using the relation $Q{_\beta} = BE(Z+1,A) - B(Z,A)$ and $W{_\beta} = Q{_\beta} + m{_e}^{2}$. Here, $\rho_{d.s.}$ is the average density of states in 
the daughter nucleus (e$^{-A/290}$ $\times$ number of states within 1 MeV of 
the ground state).
To evaluate the bulk properties, such as BE of odd-Z nuclei, we used the Pauli 
blocking prescription as discussed in Sec.~\ref{block}. The obtained results 
are displayed in Fig.~\ref{bea15} for both Th and U isotopes.
From the figure, it is clear that for neutron-rich Th and U nuclei, the
prominent mode of decay is $\beta$ decay. This means that once a neutron-rich
thermally fissile isotope is formed by some artificial means in the laboratory
or naturally in supernovae explosion, it immediately undergoes $\beta$ decay.
In our rough estimation, the lifetime of $^{254}$Th and $^{256}$U, which
are the nuclei of interest, is tens of seconds. If this prediction of
time period is acceptable, then on the nuclear physics scale, it is reasonably 
a good time for further use of the nuclei.  It is worth to mention here that
thermally fissile isotopes in the Th and U series have neutron number
N=154-172 keeping N=164 in the middle of the island. So, in the case of the
short lifetime of $^{254}$Th and $^{256}$U, one can choose a lighter isotope
of the series for practical utility.  

\section{Conclusions}\label{sec14}
In summary, we have done a thorough structural study of recently predicted 
thermally fissile isotopes in the Th and U series in the framework of 
RMF theory. Although there are certain limitations of the present approach, 
the qualitative results will remain unchanged even when the drawbacks of the 
model taken into account. The heavier isotopes of these two nuclei show 
various shapes including very large prolate deformations in highly excited 
configurations. The change in single-particle orbits along the line of 
quadrupole deformation are analyzed and parity doublet states found in some 
cases. Using an empirical estimation, we find that the neutron-rich isotopes 
of these thermally fissile nuclei are predicted to be stable against $\alpha$
and $cluster$ decays. Spontaneous fission also does not occur, because the 
presence of a large number of neutrons makes the fission barrier broader.  
However, these nuclei are highly $\beta$ unstable. Our calculation predicts 
that the $\beta$ lifetime is tens of seconds for $^{254}$Th and $^{256}$U
and this time increases for nuclei that have lower neutron number but thermally 
fissile. This finite lifetime of these thermally fissile isotopes could be
very useful for energy production with nuclear reactor technology. If these
neutron-rich nuclei are used as nuclear fuel, the reactor will achieve critical
conditions much more rapidly than with normal nuclear fuel, because of the 
release of a large number of neutrons during the fission process.

\setcounter{equation}{0}
\setcounter{figure}{0}

\newpage
\chapter{Relative mass distributions of neutron-rich thermally fissile nuclei } 
\label{chapter4}

In this chapter, we study the binary mass distribution for the recently 
predicted thermally fissile neutron-rich uranium and thorium nuclei using 
statistical model.  The level density parameters needed for the study are 
evaluated from the excitation energies of the temperature dependent 
relativistic mean field formalism. The excitation energy and the level density 
parameter for a given temperature are employed in the convolution integral 
method to obtain the probability of the particular fragmentation. As 
representative cases, we present the results for the binary yields of 
$^{250}$U and $^{254}$Th. The relative yields are  presented for three 
different temperatures: $T =$ 1, 2 and 3 MeV.  

\section{Introduction}
The fission phenomenon is one of the most interesting subjects in the field of 
nuclear physics. To study fission properties, a large number of models have 
been proposed. The fissioning of a nucleus is successfully explained by the 
liquid drop model, and the semi-empirical mass formula is the best and simple 
oldest tool to get a rough estimation of the energy released in a fission 
process. The pioneering work of Vautherin and Brink \cite{vat70}, who applied 
the Skyrme interaction in a self-consistent method for the calculation of 
ground-state properties of finite nuclei, opened a new dimension in the 
quantitative estimation of nuclear properties. Subsequently, the Hartree-Fock 
and time dependent Hartree-Fock formalisms \cite{pal} were also implemented to 
study the properties of fission. Most recently, the microscopic relativistic 
mean field approximation, which is another successful theory in nuclear 
physics, is also used for the study of nuclear fission \cite{skp10}.

In the last few decades, the availability of neutron-rich nuclei in various 
laboratories across the globe opened up new research in the field of nuclear 
physics, because of their exotic decay properties. The effort towards the 
synthesis of superheavy nuclei in laboratories such as Dubna (Russia), 
GSI (Germany), RIKEN (Japan) and BNL (USA) is also quite remarkable. Due to 
all these, the periodic table has been extended, to date, upto atomic number 
$Z = 118$ \cite{audi12}. The decay modes of these superheavy nuclei are very 
different than the usual modes. Mostly, we understand that a neutron-rich 
nucleus has a large number of neutron than nuclei in the light or medium mass 
region of the periodic table. The study of these neutron-rich superheavy 
nuclei is very interesting because of their ground-state structures and 
various mode of decays, including multifragment fission (more than two fragments) \cite{skp10}. Another interesting feature of some neutron-rich uranium and 
thorium nuclei is that, similar to $^{233}$U, $^{235}$U and $^{239}$Pu, the 
nuclei $^{246-264}$U and $^{244-262}$Th are also thermally fissile, which is 
extremely important for energy production in the fission process. If the 
neutron-rich uranium and thorium nuclei are viable sources, then these 
nuclei will be more effective for achieving the critical condition in a 
controlled fission reaction. 

Now the question arises, how we can get a reasonable estimation of the mass 
yield in the spallation reaction of these neutron-rich thermally fissile 
nuclei? As mentioned earlier in this section, there are many formalisms 
available in the literature to study these cases. Here, we adopt the 
statistical model developed by Fong \cite{fon56}. The calculation is further 
extended by Rajasekaran and Devanathan \cite{mrd81} to study the binary mass 
distributions using the single-particle energies of the Nilsson model. The 
obtained results are in good agreement with the experimental data.
In the present study, we would like to replace the single-particle energies 
with the excitation energies of a successful microscopic approach: the 
relativistic mean field (RMF) formalism. 

For last few decades, the RMF formalism \cite{walecka74,gam90,patra91,gam90} 
with various parameter sets has successfully reproduced the bulk 
properties, such as binding energies, root-mean-square radii, quadrupole 
deformation, etc., not only for nuclei near the $\beta-$stability line but 
also for nuclei away from it. Further, the RMF formalism has been successfully 
applied to the study of clusterization of known cluster emitting heavy nucleus
\cite{aru05,bks06,skp07} and the fission of hyper-hyper deformed $^{56}$Ni
nucleus \cite{rkg08}. Rutz \textit{et al.} \cite{rutz95} reproduced the double, and triple humped fission barrier of $^{240}$Pu, $^{232}$Th and the asymmetric 
ground-states of $^{226}$Ra using the RMF formalism. Moreover, the symmetric 
and asymmetric fission modes are also successfully reproduced. 
Patra \textit{et al.} \cite{skp10} studied the neck configuration in the 
fission decay of neutron-rich U and Th isotopes. 
The main goal of this present chapter is to understand the binary fragmentation yields of such neutron-rich thermally fissile superheavy nuclei. $^{250}$U and 
$^{254}$Th are taken for further calculations as the representative cases. 
We have used the temperature dependent relativistic mean field (TRMF) as 
described in chapter \ref{chapter2}, which modifies the BCS pairing (in 
Sec. \ref{TBCS}). 

\section{Formalism} \label{sec2}
The possible binary fragments of the considered nucleus are obtained by 
equating the charge-to-mass ratio of the parent nucleus to the fission 
fragments as \cite{mbs2014}:
\begin{equation}
\frac{Z_P}{A_P} \approx \frac{Z_i}{A_i}, \label{eq1}
\end{equation}
with $A_P$, $Z_P$ and $A_i$, $Z_i$ ($i$ = 1 and 2) correspond to mass and 
charge numbers of the parent nucleus and the fission fragments \cite{mrd81}. 
The constraints, $A_1 + A_2 = A$, $Z_1 + Z_2 = Z$ and $ A_1 \ge A_2$ are 
imposed to satisfy the conservation of charge and mass number in a nuclear 
fission process and to avoid the repetition of fission fragments. 
Another constraint i.e., the binary charge numbers from $ Z_2 \ge $ 26 to 
$Z_1 \le$ 66, is also taken into consideration from the experimental yield 
\cite{berg71} to generate the combinations, assuming that the fission 
fragments lie within these charge range.
\subsection{Statistical theory}
The statistical theory \cite{fon56,cole} assumes that the probability of the 
particular fragmentation is directly proportional to the folded level 
density $\rho_{12}$ of that fragments with the total excitation energy 
$E^*$, i.e., $P(A_j,Z_j) \propto \rho_{12}(E^*)$. Here,
\begin{equation}
\rho_{12}(E^*) = \int_{0}^{E^*} \rho_{1}(E^*_{1})\,\rho_{2}(E^* - E_1^*)\, dE^*_1, \label{eq22}
\end{equation} 
and $\rho_i$ is the level density of two fragments ($i$ = 1, 2). The nuclear 
level density \cite{bet37,mor72} is expressed as a function of fragment 
excitation energy $E^{\ast}_i$ and the single-particle level density 
parameter $a_i$:
\begin{equation}
\rho_i\left(E^{\ast}_i\right) = \dfrac{1}{12} \left(\dfrac{\pi^{2}}{a_i}\right)^{1/4} E^{\ast (-5/4)}_i\exp\left(2\sqrt{a_iE^{\ast}_i}\right).\label{eq33}
\end{equation}
In Refs. \cite{mbs2014,sen17}, the excitation energies of the fragments using 
the ground state single-particle energies of finite range droplet model (FRDM) 
\cite{moller97} at a given temperature $T$ keeping the total number of proton 
and neutron fixed is calculated. In the present study, we apply the 
self-consistent TRMF theory to calculate the $E^*$ of the fragments. 
The excitation energy is calculated as,
\begin{equation}
E^{*}_{i}(T) = E_{i}(T) - E_{i}(T = 0). \label{eq4}
\end{equation}
The level density parameter $a_i$ is given as,
\begin{equation}
a_i = \frac{E^*_i}{T^{2}}. \label{eq5}
\end{equation}
The relative yield is calculated as the ratio of the probability of a given 
binary fragmentation to the sum of the probabilities of all the possible 
binary fragmentations:
\begin{equation}
Y(A_j,Z_j)=\frac{P(A_j,Z_j)}{\sum_{j} P(A_j,Z_j)}, \label{eq6}
\end{equation}
where $A_j$ and $Z_j$ refer to the binary fragmentations involving two 
fragments with mass and charge numbers $A_1$, $A_2$ and $Z_1$, $Z_2$ obtained 
from Eq. \eqref{eq1}. The competing basic decay modes such as neutron/proton 
emission, $\alpha$ decay, and ternary fragmentation are not considered. In 
addition to these approximations, we have also not included the dynamics of 
the fission reaction, which are really important to get a quantitative 
comparison with the experimental measurements. The presented results are the 
prompt disintegration of a parent nucleus into two fragments (democratic 
breakup). The resulting excitation energy would be liberated as prompt 
particle emission or delayed emission, but such secondary emissions are also 
ignored.

\section{Results and discussions} \label{sec3}

In our very recent work \cite{mts16}, we have calculated the ternary mass 
distributions for $^{252}$Cf, $^{242}$Pu and $^{236}$U with the fixed third 
fragments A$_3 =~^{48}$Ca, $^{20}$O and $^{16}$O respectively for the three 
different temperatures $T =$ 1, 2 and 3 MeV within the TRMF formalism. The 
structure effects of binary fragments are also reported in Ref. \cite{cinthol}. 
In this article, we study the mass distribution of $^{250}$U and $^{254}$Th 
as representative cases from the range of neutron-rich thermally 
fissile nuclei $^{246-264}$U and $^{244-262}$Th. Because of the neutron-rich 
nature of these nuclei, a large number of neutrons are emitted during the 
fission process. These nucleons help to achieve the critical condition much 
sooner than the normal fissile nuclei. 

To assure the predictability of the statistical model, we also study the 
binary fragmentation of naturally occurring $^{236}$U and $^{232}$Th nuclei. 
The possible binary fragments are obtained using Eq. \eqref{eq1}. To 
calculate the total binding energy at a given temperature, we use the axially 
symmetric harmonic  oscillator basis expansion N$_F$ and N$_B$ for the Fermion 
and Boson wave-functions to solve the Dirac and the Klein Gordon Eqs.
(as described in chapter \ref{chapter2}, Sec. \ref{EDF}) iteratively. It is 
reported \cite{bhar15} that the effect of basis space on the calculated 
binding energy, quadrupole deformation parameter ($\beta_2$) and the rms radii 
of nucleus are  almost equal for the basis set  N$_F =$ N$_B =$ 12 to 20 in 
the  mass region A $\sim$ 200 . Thus, we use the basis space N$_F =$ 12 and 
N$_B=$ 20 to study the binary fragments up to mass number A $\sim$ 182.
The binding energy is obtained by minimizing the free energy, which gives the
most probable quadrupole deformation parameter $\beta_2$ and the proton 
(neutron) pairing gaps $\triangle_p$ ($\triangle_n$) for the given 
temperature. At finite temperature, the continuum corrections due to the
excitation of nucleons need to be considered. The level density in the 
continuum depends on the basis space N$_F$ and N$_B$ \cite{niu09}. It is 
shown that the continuum corrections need not be included in the calculations 
of level densities up to the temperature T $\sim$ 3 MeV \cite{bka00,bka98}. 

\begin{figure}[t]
	\begin{center}
\includegraphics[width=0.8\columnwidth]{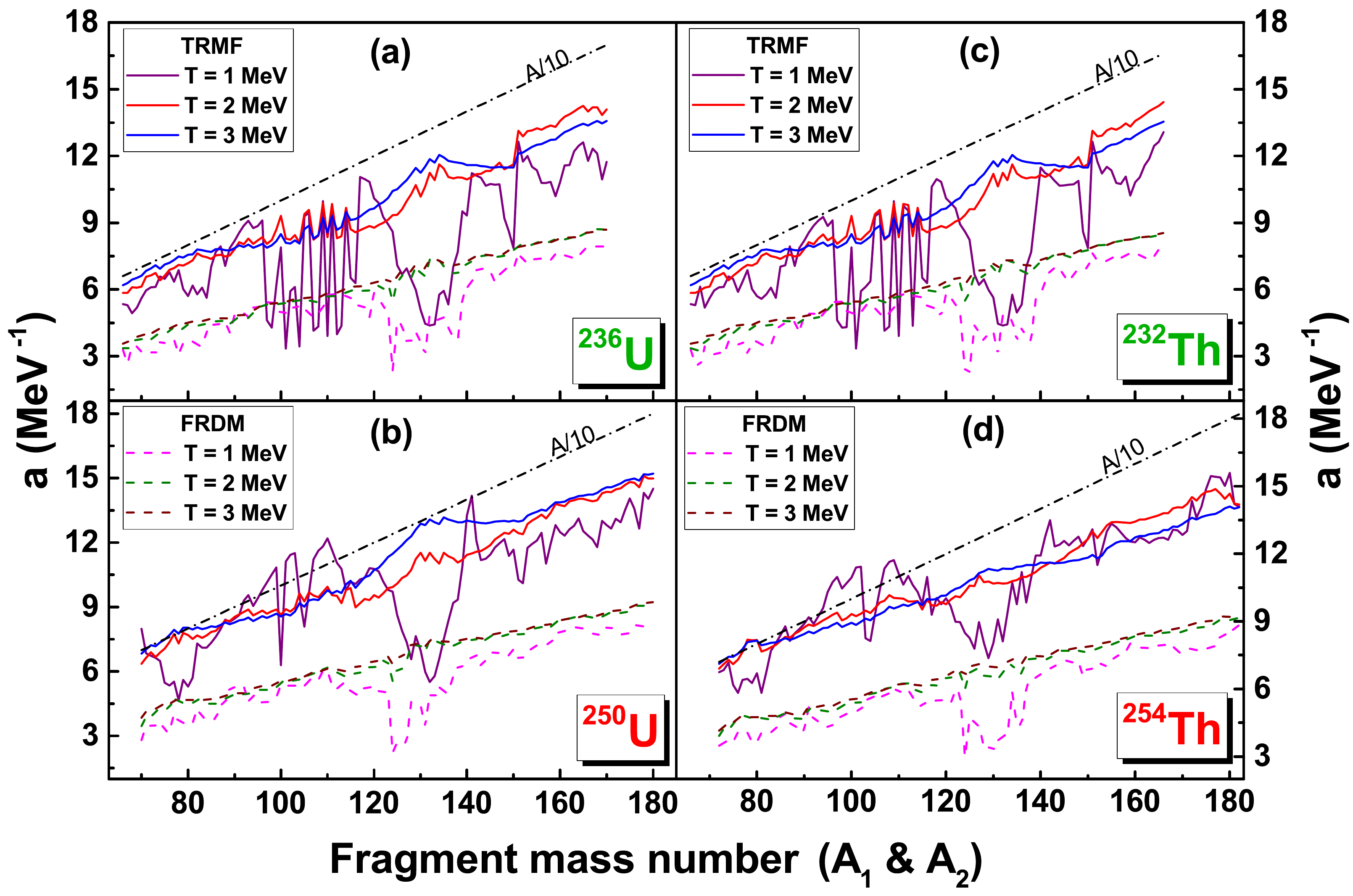}
	\end{center}
\caption{The level density parameter $a$ for the binary fragmentation of 
$^{236}$U, $^{250}$U, $^{232}$Th, and $^{254}$Th at temperature $T =$ 1, 2 and 
3 MeV within the TRMF (solid lines) and FRDM (dashed lines) formalisms.}
\label{U_ldp}
\end{figure}

\subsection{Level density parameter and level density within TRMF and FRDM formalisms }\label{ldp}

In TRMF, the excitation energies $E^*$ and the level density parameters 
$a_i$ of the fragments are obtained self consistently from Eqns. \eqref{eq4} 
to \eqref{eq5}. The FRDM calculations are also done for comparison. In this 
case, level density of the fragments are evaluated from the ground-state 
single-particle energies of the FRDM of M\"{o}ller \textit{et al.} \cite{moller95} which are retrieved from the Reference Input Parameter Library (RIPL-3) 
\cite{ripl3}. 
The total energy at a given temperature is calculated as 
$E(T) = \sum n_i \epsilon_i$; $\epsilon_i$ are the ground-state single-particle energies and $n_i$ are the Fermi-Dirac distribution function. The 
$T$-dependent energies are obtained by varying the occupation numbers at a 
fixed particle number for a given temperature and given fragment. The level 
density parameter $a$ is a crucial quantity in the statistical theory for the 
estimation of yields. These values of $a$ for the binary fragments of 
$^{236}$U, $^{250}$U, $^{232}$Th and $^{254}$Th obtained from TRMF and FRDM 
are depicted in Fig. \ref{U_ldp}. The empirical estimations $a=A/K$ are also
given for comparison, with $K$ being the inverse level density parameter. In 
general, the $K$ value varies from 8 to 13 with the increasing temperature. 
However, the level density parameter is considered to be constant up to 
$T\approx$4 MeV. Hence, we take the practical value of $K = 10$ as mentioned 
in Ref. \cite{ner2002}. The $a$ values of TRMF are close to the empirical 
level density parameter. The FRDM level density parameters are appreciably 
lower than the referenced $a$. Further, in both models at $T =$ 1 MeV, there 
are more fluctuations in the level density parameter due to the shell effects 
of the fragments. At $T =$ 2 and 3 MeV, the variations are small. This may be 
due to the fact that the shell become degenerate at the higher temperatures. 
All fragments becomes spherical at temperature $T\approx$3 MeV as shown in 
Ref. \cite{cinthol}. 
\begin{figure}[t]
	\begin{center}
\includegraphics[width=0.8\columnwidth]{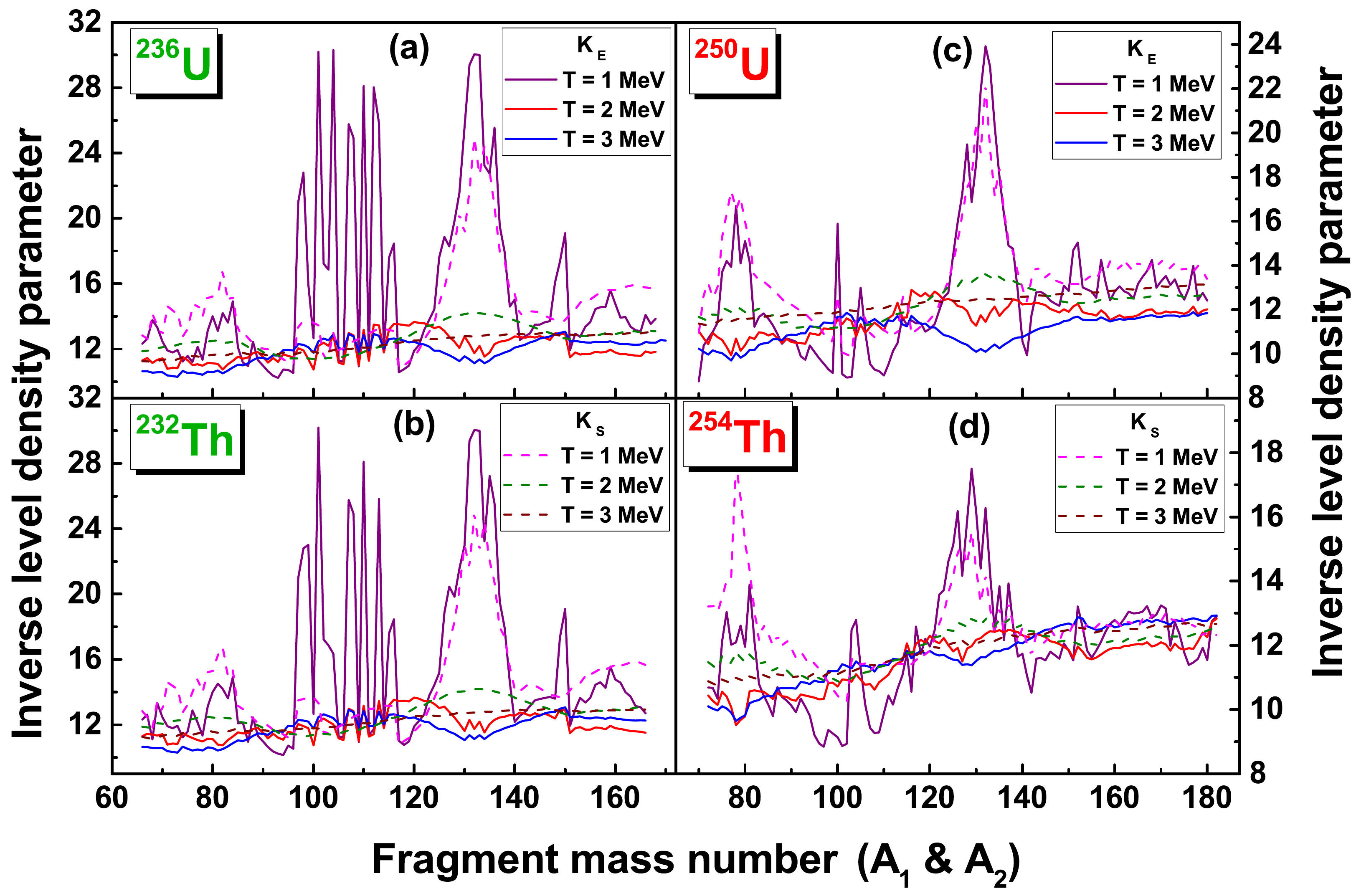}
	\end{center}
\caption{The inverse level density parameters $K_E$ (solid lines) and $K_S$ 
(dashed lines) are obtained for $^{236}$U, $^{250}$U, $^{232}$Th, and 
$^{254}$Th at temperatures $T =$ 1, 2 and 3 MeV.}
\label{keks}
\end{figure}
\begin{figure}[t]
	\begin{center}
\includegraphics[width=0.8\columnwidth]{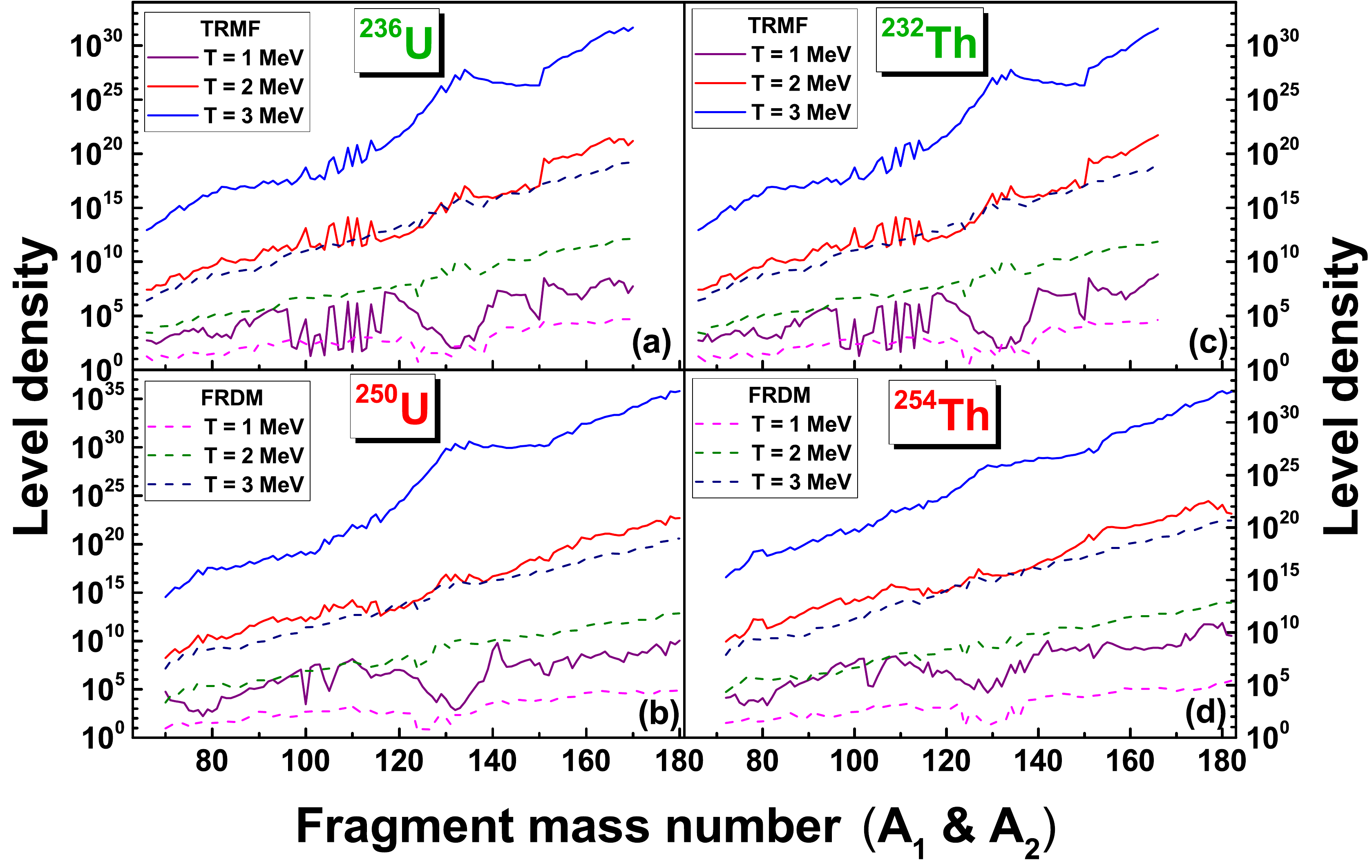}
	\end{center}
\caption{The level density of the binary fragmentations of 
$^{236}$U, $^{250}$U, $^{232}$Th, and $^{254}$Th at temperature $T =$ 1, 2 
and 3 MeV within the TRMF (solid lines) and FRDM (dashed lines) formalisms.}
\label{LD}
\end{figure}
The level density parameter $a$ is evaluated in two different ways using excitation energy and the entropy of 
the system as:
\begin{eqnarray}
a_E = \dfrac{E^*}{T^2}, \\ \nonumber
a_S = \dfrac{S}{2T}. 
\end{eqnarray} 
 For instance, the inverse level density parameters $K_E$ and $K_S$ of 
 $^{236}$U, $^{250}$U, $^{232}$Th, and $^{254}$Th within TRMF formalism are 
 depicted in Fig. \ref{keks}. Both $K_S$ and $K_E$ have maximum fluctuation 
 upto 30 MeV at $T = $1 MeV. These values reduce to $10-13$ MeV at temperature 
 $T = 2$ MeV or above. It is to be noted that at $T =$ 3 MeV, the inverse 
 level density parameter substantially lower around the mass number 
 $A\sim$130 in all cases. This may be due to the neutron closed shell 
 ($N =$ 82) in the fission fragments of $^{236}$U and $^{232}$Th and the 
 neutron-rich nuclei $^{250}$U and $^{254}$Th. The level density for the 
 fission fragments of $^{236}$U, $^{250}$U, $^{232}$Th and $^{254}$Th are 
 plotted as a function of mass number in Fig. \ref{LD} within the TRMF and 
 FRDM formalisms at three different temperatures, $T =$ 1, 2 and 3 MeV. 

The level density $\rho$ has maximum fluctuations at $T =$ 1 MeV for all 
considered nuclei in TRMF model, similar to the level density parameter $a$. 
The $\rho$ values are substantially lower at mass number $A \sim 130$ for all 
nuclei. In Fig. \ref{LD}, one can notice that the level density has small 
kinks in the mass region $A \sim 71-81$ of $^{236}$U  and $A \sim 77-91$ of 
$^{250}$U, compared with the neighboring nuclei at temperature $T =$ 2 MeV. 
Consequently, the corresponding partner fragments have also higher $\rho$ 
values. A further inspection reveals that the level density of the closed 
shell nucleus around $A \sim$ 130 has higher value than the neighboring nuclei 
for both $^{236,250}$U, but it has lower yield due to the smaller level 
density of the corresponding partners. At $T =$ 3 MeV, the level density of 
the fragments around mass number $A \sim $72 and  130 have larger values 
compared to other fragments of $^{236}$U. On the other hand, the level 
density in the vicinity of neutron number $N = 82$ and proton number $Z = 50$ 
for the fragments of the neutron-rich $^{250}$U nucleus is quite high, because 
of the close shell of the fragments. This is evident from the small kink in 
the level density of $^{130}$Cd ($N =$ 82), $^{132}$In ($N \sim$ 82) and $^{135}$Sn ($Z =$ 50). Again, for $^{232}$Th, the level densities are found to be 
maximum at around mass number  $A \sim $81 and 100 for $T =$ 2 MeV. In case 
of $^{254}$Th, the $\rho$ values are found to be large for the fragments 
around $A \sim$ 78 and 97 at $T =$ 2 MeV. Their corresponding partners have 
also similar behavior. For higher temperature $T =$ 3 MeV, the higher $\rho$ 
values of $^{232}$Th fragments are notable around  mass number $A \sim 130$. 
Similarly, for $^{254}$Th, the fission fragments around $A \sim$ 78 have 
higher level density at $T =$ 3 MeV. In general, the level density increases 
towards the neutron closed shell ($N =$ 82) nucleus.

%\subsection{Fission mass distribution}
\begin{figure}[t]
	\begin{center}
\includegraphics[width=0.8\columnwidth]{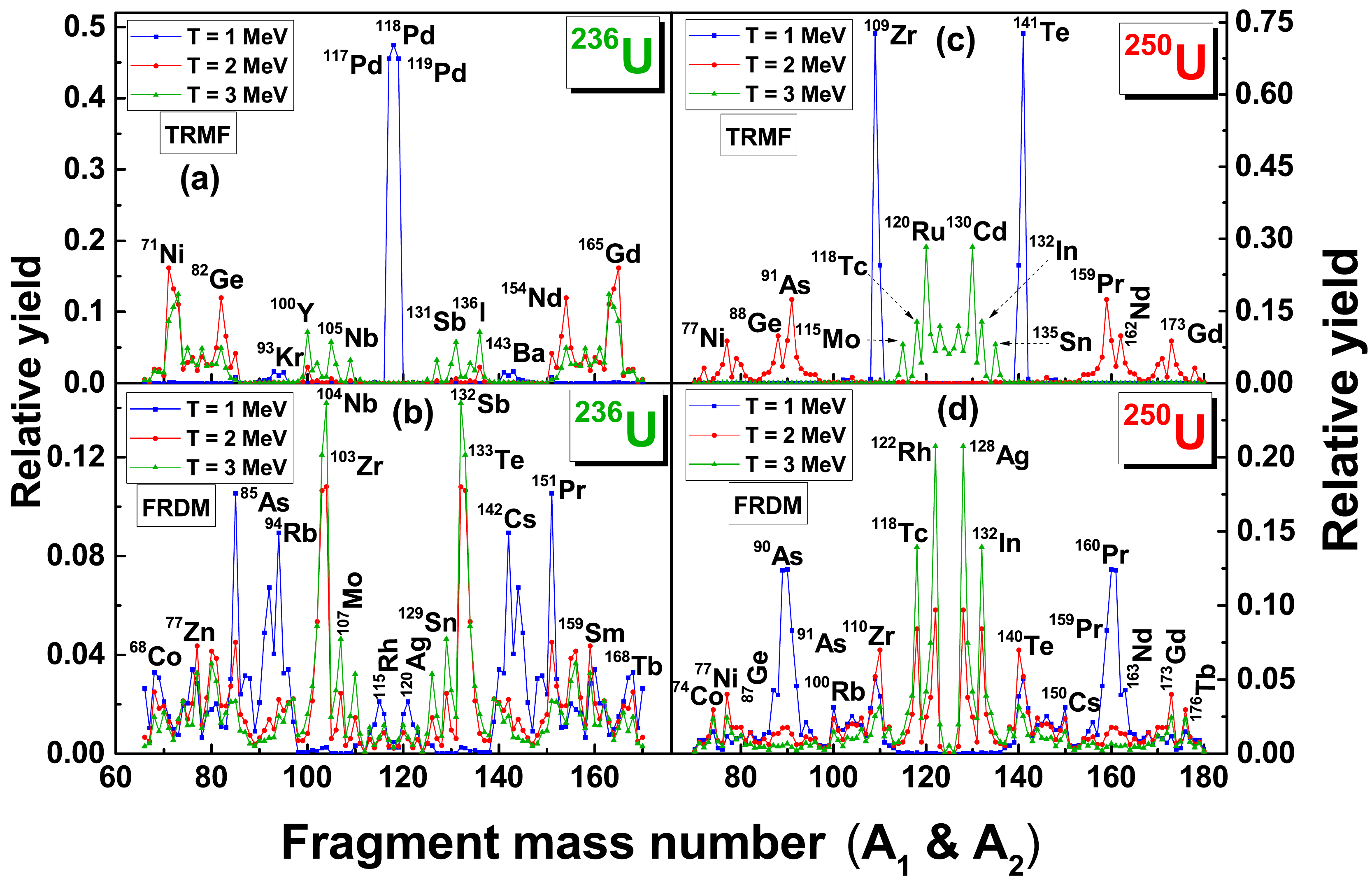}
	\end{center}
\caption{Mass distribution of $^{236}$U and $^{250}$U at temperatures $T =$ 1, 
2 and 3 MeV. The total yield values are normalized to the scale 2.}
\label{U_yield}
\end{figure}
\begin{figure}[t]
	\begin{center}
\includegraphics[width=0.8\columnwidth]{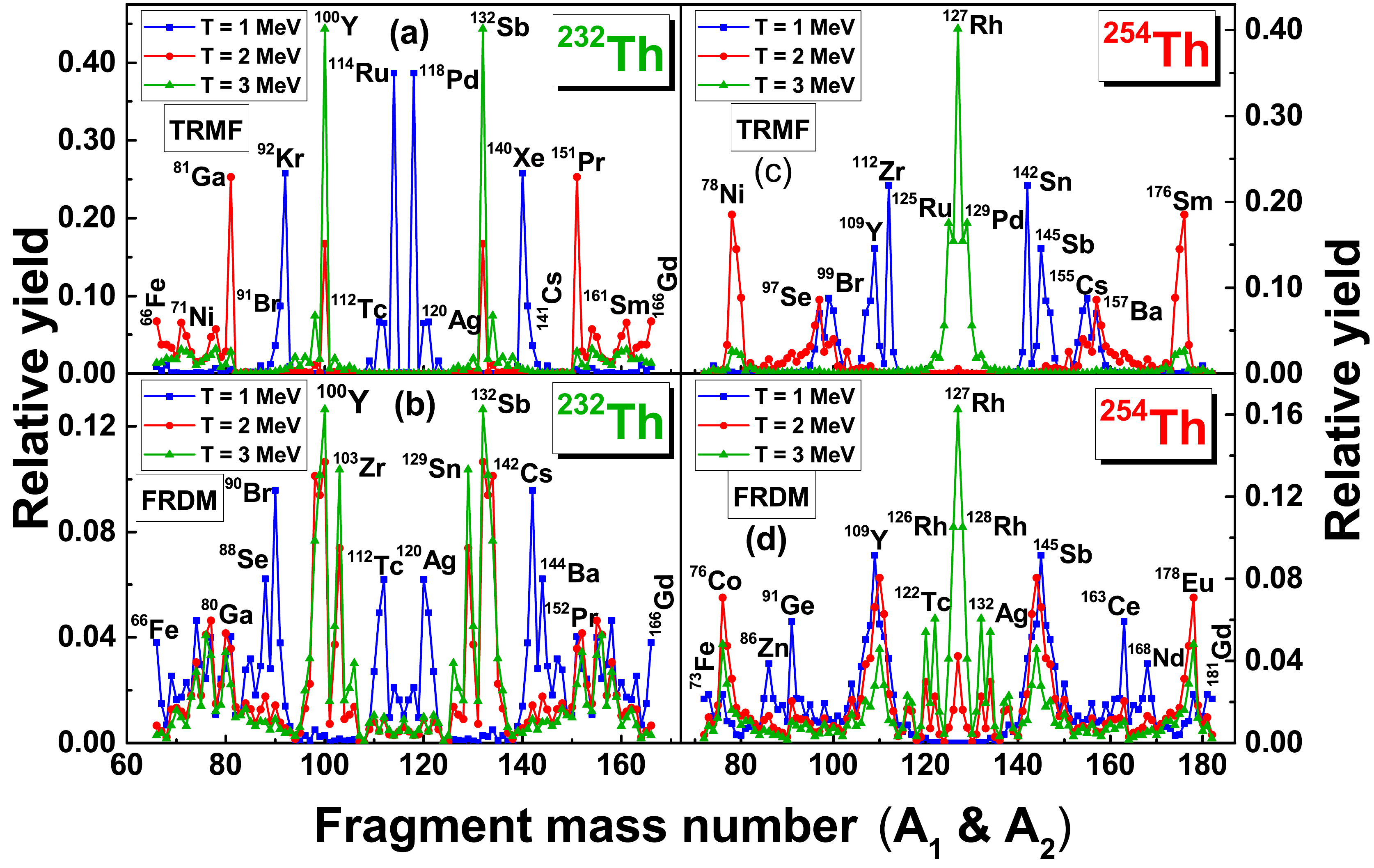}
	\end{center}
\caption{Mass distribution of $^{232}$Th and $^{254}$Th at temperatures T = 1, 2 and 3 MeV.  The total yield values are normalized to the scale 2.}
\label{Th_yield}
\end{figure} 

\subsection{Relative fragmentation distribution in binary systems }
In this section, the mass distributions of $^{236}$U, $^{232}$Th and the 
neutron-rich nuclei $^{250}$U and $^{254}$Th are calculated at temperatures 
$T =$ 1, 2 and 3 MeV using TRMF and FRDM excitation energies and the level 
density parameters $a$ as explained in Sec. \ref{sec2}. The binary mass 
distributions of $^{236,250}$U and $^{232,254}$Th are plotted in 
Figs. \ref{U_yield} and \ref{Th_yield}. The total energy at finite 
temperature and ground-state energy are calculated using the TRMF formalism as 
discussed in Sec. \ref{ldp}. From the excitation energy E$^{\ast}$ and the 
temperature $T$, the level density parameter $a$ and the level density $\rho$ 
of the fragments are calculated using Eq. \eqref{eq33}. From the fragment level 
densities $\rho_i$, the folding density $\rho_{12}$ is calculated using the 
convolution integral as in Eq. \eqref{eq22} and the relative yields are 
calculated using Eq. \eqref{eq6}. The total yields are normalized to the scale 
2. 
 
The mass yield of normal nuclei $^{236}$U and $^{232}$Th are briefly explained 
first, followed by a detailed description of the neutron-rich nuclei. The 
results of most favorable fragment yields of $^{236,250}$U and $^{232,254}$Th 
are listed in Table \ref{tab1} at three different temperatures $T =$ 1, 2 and 
3 MeV, for both TRMF and FRDM formalisms.
From Figs. \ref{U_yield} and \ref{Th_yield}, it is shown that the mass 
distributions for $^{236}$U and $^{232}$Th are quite different from those of 
the neutron-rich $^{250}$U and $^{254}$Th isotopes. 
 
The symmetric binary fragmentation $^{118}$Pd $+^{118}$Pd for $^{236}$U is the 
most favorable combination. In TRMF, the fragments with close shell 
($N =$ 100 and $Z =$ 28) combinations are more probable at the temperature 
$T =$ 2 MeV. The blend region of neutron and proton close shell 
($N \approx$ 82 and $Z \approx$ 50) has the considerable yield values at 
$T =$ 3 MeV. The fragmentations $^{151}$Pr $+^{85}$As, $^{142}$Cs $+^{94}$Rb 
and $^{144}$Ba $+^{92}$Kr are the favorable combinations at temperature 
$T =$ 1 MeV in FRDM formalism. For higher temperatures $T =$ 2 and 3 MeV, the 
closed shell or near closed shell fragments ($N =$  82, 50 and $Z =$ 28) 
have larger yields. From Fig. \ref{Th_yield} in TRMF formalism, the 
combinations $^{118}$Pd $+^{114}$Ru and $^{140}$Xe $+^{92}$Kr are the possible 
fragments at $T =$ 1 MeV for the nucleus $^{232}$Th. At $T =$ 2 MeV, we find 
maximum yields for the fragments with the close shell or near close shell 
combinations ($N =$ 82, 50). For higher temperature $T =$ 3 MeV, near the 
neutron close shell ($N \sim$ 82), $^{132}$Sb $+^{100}$Y is the most favorable 
fragmentation pair compared with all other yields. Similar fragmentations are 
found in the FRDM formalism at $T =$ 2 and 3 MeV. In addition, the probability 
of the evaluation of $^{129}$Sn $+^{103}$Zr is also quite substantial in the 
fission process. For $T =$ 1 MeV, the yield is more or less similar with the 
TRMF model.

From Fig. \ref{U_yield}, for $^{250}$U the fragment combinations $^{140,141}$Te $+^{110,109}$Zr have the maximum yields at $T =$ 1 MeV in TRMF. This is also 
consistent with the evolution of the subclosed proton shell $Z = 40$ in $Zr$ 
isotopes \cite{blatt}. Contrary to this almost symmetric binary yield, the 
mass distribution of this nucleus in FRDM formalism an asymmetric evolution of fragment combinations such as $^{160,159}$Pr $+^{90,91}$As, $^{163,162}$Nd$+^{87,88}$Ge and $^{150}$Cs $+^{100}$Rb. Interestingly, at $T =$ 2 and 3 MeV, 
the more favorable fragment combinations have one of the closed shell 
nuclei. At $T =$ 2 MeV, $^{159}$Pr $+^{91}$As, $^{162}$Nd $+^{88}$Ge and 
$^{173}$Gd $+^{77}$Ni are the more probable fragmentations (see Fig. \ref{U_yield}(c)). It is reported by Satpathy {\it et al.} \cite{satpathy} and 
experimentally verified by Patel {\it et al.} \cite{patel} 
that $N =$ 100 is a neutron close shell for the deformed region, 
where $Z = 62$ acts like a magic number. In FRDM, $^{128}$Ag $+^{122}$Rh, 
$^{132}$In $+^{118}$Tc, $^{140}$Te $+^{110}$Zr and $^{173}$Gd $+^{77}$Ni have 
larger yield at temperature $T =$ 2 MeV. With the TRMF method, the most 
favorable fragments are confined in the single region ($A \approx 114-136$) 
which is a blend of vicinity of neutron ($N =$ 82) and proton ($Z =$ 50) 
closed shell nuclei at $T =$ 3 MeV. 
The fragment combinations $^{130}$Cd $+^{120}$Ru, $^{132}$In $+^{118}$Tc 
and $^{135}$Sn $+^{115}$Mo are the major yields for $^{250}$U at $T =$ 3 MeV 
in TRMF calculations. With the FRDM method, at $T =$ 3 MeV, more probable 
fragments are similar those at $T =$ 2 MeV. A comparison between 
Fig. \ref{U_yield}(c) and \ref{U_yield}(d) clarifies that, although the 
prediction of FRDM and TRMF at $T =$ 3 MeV are qualitatively similar, they are
quantitatively very different at $T =$ 2 MeV in both the predictions.
Also, from Fig. \ref{U_yield}, it is inferred that the yields of the fragment 
combinations in blend region increases and in other regions decreases at 
$T =$ 2 MeV.

%\subsection{Relative fragmentation in binary ststems }

In the present study, the total energy of the parent nucleus A is more than 
the sum of the energies of the  daughters $A_1$ and $A_2$. Here, the dynamics 
of entire process starting from the initial stage up to the scission are 
ignored. As a result, the energy conservation in the spallation reaction is 
not taken into account. The fragment yield can be regarded as the relative 
fragmentation probability, which is obtained from Eq. \eqref{eq6}. Now we analyze the fragmentation yields for Th isotopes and the results are depicted in Fig. 
\ref{Th_yield} and Table \ref{tab1}. In this case, one can see that the mass 
distribution broadly spreads through out the region $A_i=66-166$. Again,
the most concentrated yields can be divided into two regions I
($A_1 =$ 141-148 and $A_2 =$ 106-113) and II ($A_1 =$ 152-158 and $A_2 =$ 102-96) for $^{254}$Th in TRMF formalism at the temperature $T =$ 1 MeV. The most 
favorable fragmentation $^{142}$Sn $+^{112}$Zr is obtained from region I. The 
other combinations in that region have also considerable yields. In region II, 
the isotopes of Ba and Cs appears, curiously, along with their corresponding 
partners. Categorically, in FRDM predictions, region I has larger yields at 
$T =$ 1 MeV. The other possible fragmentations are $^{163}$Ce $+^{91}$Ge, 
$^{168}$Nd $+^{86}$Zn and $^{181}$Gd $+^{73}$Fe (See Fig. \ref{Th_yield} 
(b,d)). The mass distribution is different with different temperature, and the 
maximum yields at $T =$ 2 MeV in TRMF formalism are $^{174,175,176}$Sm $+^{80,79,78}$Ni. Apart from these combinations, there are other considerable yields 
can be seen in Fig. \ref{Th_yield} for region II.
The prediction of maximum probability of the fragments production in FRDM 
method are  $^{144}$Sb $+^{110}$Y, $^{178}$Eu $+^{76}$Co and $^{127}$Rh $+^{127}$Rh at $T =$ 2 MeV. Besides these yields, one can find other notable evolution 
of masses in region I due to the vicinity of the proton close shell. 
Interestingly, at $T =$ 3 MeV, the symmetric binary combination 
$^{127}$Rh $+^{127}$Rh has the largest  yield due to the neutron close shell 
($N = $ 82) of the fragment $^{127}$Rh. The other yield fragments have 
exactly or nearly a magic nucleon combination, mostly neutron ($N =$ 82) as 
one of the fragment. A considerable yield is also seen for the proton close 
shell ($Z =$ 28) Ni or/and ($Z=$62) Sm isotopes supporting our earlier prediction \cite{cinthol}. This confirms the prediction of Sm as a deformed magic 
nucleus \cite{satpathy,patel}. 
Another observation of the present calculations show that the yields of the 
neutron-rich nuclei agree with the symmetric mass distribution of 
Chaudhuri \textit{et al.} \cite{cha2015} at large excitation energy, which 
contradicts the recent prediction of large asymmetric mass distribution of 
neutron-deficient Th isotopes \cite{pasca}. These two results \cite{cha2015,pasca} along with our present calculations confirm that the symmetric or 
asymmetric mass distribution at different temperature depends on the proton 
and neutron combination of the parent nucleus.
In general, both TRMF and FRDM predict maximum yields for both symmetric and
asymmetric binary fragmentations followed by other secondary fragmentations 
emissions, depending on the temperature as well as the mass number of the 
parent nucleus. Thus, the binary fragments have larger level density $\rho$ 
comparing with other nuclei because of neutron/proton close shell fragment 
combinations at $T =$ 2 and 3 MeV. This results consistent with the fact that 
most favorable fragments have larger phase space than the neighboring nuclei 
as reported earlier \cite{mts16,cinthol}. 

\begin{table}
	\caption{The relative fragmentation yield (R.Y.) = $Y(A_j,Z_j)=\dfrac{P(A_j,Z_j)}{\sum P(A_j,Z_j)}$ for $^{236}$U, $^{250}$U, $^{232}$Th and $^{254}$Th, obtained with TRMF at the temperatures $T=$ 1, 2 and 3 MeV are 
		compared with the FRDM prediction (The yield values are normalized to 2).}  
	
	\centering
	\scalebox{0.65}{	
	\begin{tabular}{cccccccccccc} 
	\Xhline{3\arrayrulewidth}

		\multirow{2}{1cm}{Parent} &\multirow{2}{*}{T (MeV)}  & \multicolumn{2}{c}{TRMF} & \multicolumn{2}{c}{FRDM} 
		&\multirow{2}{1cm}{Parent} &\multirow{2}{*}{T (MeV)}   & \multicolumn{2}{c}{TRMF} & \multicolumn{2}{c}{FRDM } \\
 \cmidrule(lr){3-4} \cmidrule(lr){5-6}\cmidrule(lr){9-10} \cmidrule(lr){11-12}
		
		&&Fragment &R.Y.& Fragment & R.Y.& &&Fragment &R.Y.& Fragment & R.Y.\\
		%	\vspace{0.05cm}
	\Xhline{3\arrayrulewidth}	
		%	&&\\
		\multirow{11}{*}{$^{236}$U}&\multirow{3}{*}{1}& $^{118}$Pd + $^{118}$Pd  &0.949&  $^{151}$Pr + $^{85}$As  &0.210 & \multirow{11}{*}{$^{250}$U}&\multirow{3}{*}{1}& $^{141}$Te + $^{109}$Zr &1.454&  $^{160}$Pr + $^{90}$As  &0.248 \\
		
		& 	& $^{119}$Pd + $^{117}$Pd  &0.910& $^{142}$Cs + $^{94}$Rb  &0.178&& 	&  $^{140}$Te + $^{110}$Zr &0.491&  $^{161}$Pr + $^{89}$As  &0.247\\
		
		&& $^{143}$Ba + $^{93}$Kr  &0.032& $^{144}$Ba + $^{92}$Kr  &0.134&& 	&  $^{148}$Xe + $^{102}$Sr &0.014&  $^{159}$Pr + $^{91}$As  &0.166\\
%		\cmidrule(lr){2-6} \cmidrule(lr){8-12}
		
		&\multirow{3}{*}{2}& $^{165}$Gd + $^{71}$Ni  &0.323& $^{132}$Sb + $^{104}$Nb   &0.216&&\multirow{4}{*}{2}& $^{159}$Pr + $^{91}$As  &0.348& $^{128}$Ag + $^{122}$Rh   &0.193\\	
		
		& &	$^{164}$Gd + $^{72}$Ni  &0.264& $^{133}$Te + $^{103}$Zr  &0.213&	& 	& $^{162}$Nd + $^{88}$Ge  &0.197& $^{132}$In + $^{118}$Tc  &0.168\\	
		
		&&$^{163}$Gd + $^{73}$Ni  &0.0.221 &$^{151}$Pr + $^{85}$As  &0.210&& & $^{160}$Pr + $^{90}$As  &0.176& $^{140}$Te + $^{110}$Zn &0.140 \\
		
		&& $^{154}$Nd + $^{82}$Ge  &0.240& $^{159}$Sb + $^{77}$Zn   &0.087&& & $^{173}$Gd + $^{77}$Ni  &0.175& $^{141}$Te + $^{109}$Zn   &0.100\\
		
%		\cmidrule(lr){2-6} \cmidrule{8-12}

		&\multirow{4}{*}{3}& $^{163}$Gd + $^{73}$Ni  &0.249& $^{132}$Sb + $^{104}$Nb   &0.283&&\multirow{4}{*}{3}& $^{130}$Cd + $^{120}$Ru  &0.565& $^{128}$Ag + $^{122}$Rh   &0.414\\	
		
		& 	& $^{164}$Gd + $^{72}$Ni &0.214& $^{133}$Te + $^{103}$Zr  &0.242&	& & $^{132}$In + $^{118}$Tc  &0.255& $^{132}$In + $^{118}$Tc  &0.278\\
		
		&&	$^{136}$I + $^{100}$Y  &0.143& $^{134}$Te + $^{102}$Zr  &0.102&& & $^{127}$Ag + $^{123}$Rh  &0.236& $^{129}$Ag + $^{121}$Rh   &0.149\\
		
		&& $^{131}$Sb + $^{105}$Nb  &0.114& $^{129}$Sn + $^{107}$Mo &0.092& & & $^{135}$Sn + $^{115}$Mo  &0.161& $^{130}$Cd + $^{120}$Ru  &0.083\\\hline	
		
		\hline %\hline
		%%%%%%%%%%%%%%%%%%%%%%%%%%%%%%%%%%%%%%%%%%%%%%%%%%%%%%%%%%%%%%%%%%%%%%%%%%%%
		\multirow{10}{*}{$^{232}$Th}&\multirow{4}{*}{1}& $^{118}$Pd + $^{114}$Ru  &0.773&  $^{142}$Cs + $^{90}$Br  &0.190&\multirow{10}{*}{$^{254}$Th}&\multirow{4}{*}{1}& $^{142}$Sn + $^{112}$Zr  &0.439&  $^{145}$Sb + $^{109}$Y  &0.183\\
		
		& 	& $^{140}$Xe + $^{92}$Kr  &0.515& $^{144}$Ba + $^{88}$Se  &0.124&& 	& $^{145}$Sb + $^{109}$Y  &0.291& $^{163}$Ce + $^{91}$Ge  &0.118\\
		
		&& $^{141}$Cs + $^{91}$Br  &0.174& $^{120}$Ag + $^{112}$Tc  &0.123&& & $^{155}$Cs + $^{99}$Br  &0.176& $^{144}$Sb + $^{110}$Y   &0.115\\
		
		&& $^{120}$Ag + $^{112}$Tc  &0.129& $^{158}$Pm + $^{74}$Cu  &0.092&& 	& $^{157}$Ba + $^{97}$Se  &0.139& $^{168}$Nd + $^{86}$Zn  &0.077\\	
%		\cmidrule(lr){2-6} \cmidrule(lr){8-12}
		
		&\multirow{3}{*}{2}& $^{151}$Pr + $^{81}$Ga  &0.505& $^{132}$Sb + $^{100}$Y   &0.213&&\multirow{3}{*}{2}& $^{176}$Sm + $^{78}$Ni  &0.370& $^{144}$Sb + $^{110}$Y   &0.161\\
		
		& &	$^{132}$Sb + $^{100}$Y   &0.334& $^{134}$Te + $^{98}$Sr  &0.202&& 	& $^{175}$Sm + $^{79}$Ni  &0.290& $^{178}$Eu + $^{76}$Co   &0.141\\
		
		&& $^{166}$Gd + $^{66}$Fe  &0.134& $^{129}$Sn + $^{103}$Zr   &0.146&&  & $^{157}$Ba + $^{97}$Se  &0.172& $^{144}$Sb + $^{110}$Y   &0.132\\
		
%		\cmidrule(lr){2-6} \cmidrule(lr){8-12}

		&\multirow{3}{*}{3}&  $^{132}$Sb + $^{100}$Y  &0.886&  $^{132}$Sb + $^{100}$Y   &0.252&&\multirow{3}{*}{3}& $^{127}$Rh + $^{127}$Rh  &0.803& $^{127}$Rh + $^{127}$Rh  &0.325\\	
		
		& 	& $^{134}$Te + $^{98}$Sr &0.148& $^{129}$Sn + $^{103}$Zr  &0.207&	& 	& $^{129}$Pd + $^{125}$Ru  &0.350& $^{127}$Rh + $^{127}$Rh  &0.210\\
		
		&& $^{155}$Nd + $^{77}$Zn  &0.063& $^{134}$Te + $^{98}$Sr &0.153&& & $^{128}$Rh + $^{126}$Rh  &0.307& $^{132}$Ag + $^{122}$Tc  &0.120\\	
	\Xhline{3\arrayrulewidth}	
		\label{tab1}
	\end{tabular}}
\end{table}

To this end,  it may be mentioned that the differences in the mass 
distributions or the relative yields calculated using TRMF and FRDM approaches  mainly arise due to the differences in the level densities associated with 
these approaches. The mean values and the fluctuations in the level density 
parameter and the corresponding level density are even qualitatively different 
in both the approaches considered. This possibly stems from the fact that the 
single-particle energies in the FRDM are temperature independent. The 
temperature dependence of the excitation energy, required to calculate the 
level density parameter, comes only from the modification of the 
single-particle occupancy due to the Fermi distribution. In the TRMF approach, 
the excitation energy for each fragment at a given temperature is calculated
self-consistently. Therefore, the deformation and the single-particle energies 
changes with temperature.

For the neutron-rich nuclei, the fragments having neutron/proton close shell 
$N =$ 50, 82 and 100 have maximum possibility of emission at $T =$ 2 and 3 MeV (for both nuclei $^{250}$U and $^{254}$Th). This is a general trend we could 
expect for all neutron-rich nuclei. It is worthwhile to mention some of the 
recent reports and predictions of multifragment fission for neutron-rich 
uranium and thorium nuclei. When such a neutron-rich nucleus breaks into 
nearly two fragments, the products exceed the drip-line, leaving few nucleons 
(or light nuclei) free. As a result, these free particles along with the 
scission neutrons enhance the chain reaction in a thermonuclear device. These 
additional particles (nucleons or light nuclei) responsible for reaching the 
critical condition much faster than in the usual fission for normal thermally 
fissile nucleus. Thus, the neutron-rich thermally fissile nuclei, such as 
$^{246-264}$U and $^{244-262}$Th, will be very useful for energy production.

\section{Summary and conclusions} \label{se4}
The fission mass distributions of $\beta-$stable nuclei $^{236}$U and $^{232}$Th and the neutron-rich thermally fissile nuclei $^{250}$U and $^{254}$Th are 
studied within a statistical model. The possible combinations are obtained by 
equating the charge-to-mass ratio of the parents to that of the fragments. 
The excitation energies of fragments are evaluated from the temperature 
dependent self-consistent binding energies at the given $T$ and the 
ground-state binding energies which are calculated from the RMF model. The 
level densities and the yields combinations are manipulated using the 
convolution integral approach. The fission mass distributions of the 
aforementioned nuclei are also evaluated from the FRDM formalism for 
comparison. The level density parameter $a$ and inverse level density 
parameter $K$ are also studied to see the difference between results with 
these two methods. Besides fission fragments, the level densities are also 
discussed in the present chapter. For $^{236}$U and $^{232}$Th, the symmetric 
and nearly symmetric fragmentations are more favorable at temperature $T =$ 1 
MeV. Interestingly, in most of the cases we find one of the favorable fragment 
has a close shell or near close shell configuration ($N =$ 82,50 and $Z =$ 28) 
at temperature $T =$ 2 and 3 MeV. Further, Zr isotopes has larger yield values 
for $^{250}$U and 
$^{254}$Th with their accompanying possible fragments at $T =$ 1 MeV. The Ba 
and Cs isotopes with their partners are also more possible for $^{254}$Th. 
This could be due to the deformed close shell in the region $Z=52-66$ of the 
periodic table \cite{jonh}. The Ni isotopes and the neutron close shell 
($N \sim$ 100) nuclei are some of the prominent yields for both $^{250}$U and 
$^{254}$Th at temperature $T =$ 2 MeV. At $T =$ 3 MeV, the neutron close shell 
($N =$ 82) is one of the largest yield fragments. The symmetric fragmentation 
 $^{127}$Rh $+^{127}$Rh is possible for $^{254}$Th due to the $N = $82 close 
 shell occurs in binary fragmentation. For $^{250}$U, the larger yield values 
 are confined to the junction of neutron and proton closed shell nuclei.

\setcounter{equation}{0}
\setcounter{figure}{0}

\newpage
\chapter{Tidal deformability}
\label{chapter5} 

In this chapter$^{3}$\footnotetext[3]{This paper was published before the 
detection of gravitational waves from binary neutron star merger.}, we 
systematically study the tidal deformability for neutron 
and hyperon stars using relativistic mean field equations of state (EoSs). The 
tidal effect plays an important role during the early part of the evolution 
of compact binaries. Although, the deformability associated with the EoSs has 
a small correction, it gives a clean gravitational wave signature in binary 
inspiral. These are characterized by various Love numbers $k_l$ ($l$=2, 3, 4), 
that depend on the EoS of a star for a given mass and radius. The tidal effect 
of star could be efficiently measured through advanced LIGO detector from the 
final stages of inspiraling binary neutron star merger.

%%%%%%%%%%%%%%%%%%%%%%%%%%%%%%%%%%%%%%%%%%%%%%%%%%%%%%%%%%%%%%%%%%%%
%\section{Tides in binary neutron star systems}
\section{Introduction}

%\section{Two body system}
\begin{figure}[!b]
\begin{center}
        \includegraphics[width=0.8\columnwidth]{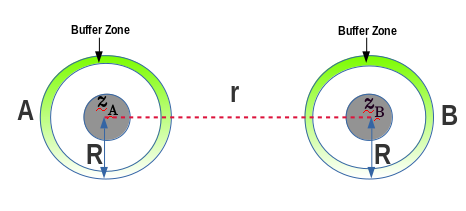}
\caption{Plot for the two body system . The light green shaded region 
represents the weak-field buffer zone. The center-of-mass of each body is 
$z_A$, and $z_B$, respectively. The space-time is decomposed into a weak-field 
region and a strong-field region surrounding each compact object (gray). }
        \label{body2}
        \end{center}
\end{figure}
 When two neutron stars approach, they come under the influence of each 
other via gravity and then get distorted. As an after effect is that tides are 
raised exactly the same way as tides are created on Earth due to Moon. 
The newly formed tides pick the energy out of the orbit resulting in the 
speedy motion of the inspiral. This can be detected and measured in the form 
of gravitational waves. Larger are the size of the neutron stars, bigger are 
the tides formed. 
 From the equation of state we can determine the size of neutron stars 
 alongwith its tidal deformation. Moreover, in the experimental front, from
 the measurements of the neutron star masses and extent of tidal deformation 
their size and equation of state can be calculated.

Consider two bodies A, and B as shown in the Fig \ref{body2}, which are 
separated by a distance $r$ and both the bodies are having a typical size 
of radius $R$ \cite{flen1,will}. We assume that $R<r$, which indicates that 
the binaries are entirely isolated. Also, each body is surrounded by a vacuum region which we 
call as the buffer zone. When the objects are very close to each other, they 
will share their outer envelopes. As the bodies are isolated from one another
neither any ejection of matter has taken place nor accreted. The orbital 
motion of each body will get affected when the binaries start interacting and 
transfer their masses.

The dynamics of the external and internal body is governed by the mutual 
gravitational interaction, which yields the orbital timescale given by 
$T_{orb}\sim(r^3/ m)^{1/2}$, where m is the mass of the typical body. Similarly, the internal dynamics of the inner body is controlled by the hydrodynamical 
process, that gives the internal times scale $T_{int}\sim(R^3/ m)^{1/2}$. 
Subsequently, the internal and external dynamics take place over the widely 
separated time scale ($T_{int}<<T_{orb}$), which means that both the dynamics 
have decoupled mostly from one another. Notice that the star's dynamics are not 
fully decoupled. For example, Moon is responsible for formation of tides on 
Earth and it is because of the orbital position of the Moon. Then the tidal 
deformation of the Earth modifies its gravitational potential. As a result, it 
affects the orbit of the Moon.

\subsection{Newtonian tidal interactions}
Newton's law of universal gravitation predicts that the gravitational force 
between two objects is proportional to the objects masses and inversely 
proportional to the square of the distance between them. The law, which 
applies to weakly interacting objects traveling at speed much slower than 
that of light. 
As we know that the Newtonian gravity is used as a limit of weak-field 
relativistic gravity, based on this Einstien developed the concept of 
curvature space-time and general relativity. Newtonian gravity is conveniently 
formulated in a fixed rectilinear coordinate system in terms of an absolute 
time coordinate. 

We consider each body is constituted of N number of arbitrary points as shown 
in Fig. \ref{body2}. Then, the total potential $u$ is a linear superposition of 
the individual potentials created by each body. Hence we can write the 
gravitational potential at position $\vec{\it x}$ due to the point mass 
\cite{vines,vines1,icts}

\begin{eqnarray}
	{\it u}=- \sum_A{\frac{M}{|\vec{\it x}-\vec{\it x'}|}},
\end{eqnarray}
and also due to a system of the extended object with density $\rho$ 

\begin{eqnarray}
	{\it u}=- \int{d^3{\it x'}{\frac{\rho({\vec{\it x'}})}{|\vec{\it x}-\vec{\it x'}|}}}\;\;.
	\label{eb}
\end{eqnarray}

To calculate the dynamics of the extended objects, we determine the 
external and internal potentials of each body. The total potential is 
represented as
        ${\it u=u^{int}_A+u^{ext}_A}$.

Using Taylor's series expansion, we expand a field $\phi(\vec{\it x})$ around 
$\vec{\it z_A}$ as

\begin{eqnarray}
	\phi(\vec{\it x})&=&\phi(\vec{\it z_A})+({\vec{\it x}}-\vec{\it z_A})
	\frac{\partial\phi(\vec{\it x})}{\partial \vec{\it x}}\Big|_{\vec{\it z_A}}+\frac{({\vec{\it x}}-\vec{\it z_A})^2}{2}
\frac{\partial^2\phi(\vec{\it x})}{\partial^2 \vec{\it x}}\Big|_{\vec{\it z_A}}+\dots\nonumber\\
&=&\phi(\vec{\it z_A})+({{\it x}}-{\it z_A})^i\partial_i\phi(\vec{\it x})\Big|_{\vec{\it z_A}}
+({{\it x}}-{\it z_A})^i({{\it x}}-{\it z_A})^j \partial_i\partial_j \phi(\vec{\it x})\Big|_{\vec{\it z_A}}+\dots\nonumber\\
&=&\sum_{l=0}\frac{1}{l!}{(\it x-z_A)^L} \partial_L\phi(\vec{\it x})\Big|_{\vec{\it z_A}}.
\end{eqnarray}
where L is a multi-index representing the $l$ indices such as
$(\it{x-z_A})^L=(\it x-z_A)^i (\it x-z_A)^j\dots(\it x-z_A)^l$ and
$\partial_L=\partial/\partial{\it x}^i\; \partial/\partial{\it x}^j\dots
\partial/\partial{\it x}^l$. 
Now, we write the internal potential at a point $\vec{\it z_A}$ using 
Eq. \eqref{eb}
\begin{eqnarray}
{\it u^{int}_A({\vec{\it x}}})&=&-\sum^{\infty}_{l=0} \frac{(-1)^l}{l!}
\int{d^3{\vec{\it x'}}\rho_A(\vec{\it x'})({\it x'-z_A})^L \partial_L \Big(\frac{1}{|\vec{\it x}-\vec{\it x'}|}\Big)\Big|_{\vec{\it x'}=\vec{\it z_A}}}\nonumber\\
&=&-\sum^{\infty}_{l=0} \frac{(-1)^l}{l!}\int{d^3{\vec{\it x'}}\rho_A(\vec{\it x'})({\it x'-z_A})^L \partial_L \Big(\frac{1}{r}}\Big)\Big|_{\vec{\it x'}=\vec{\it z_A}},\;\;
\end{eqnarray}
where $r=|\vec{\it x}-\vec{\it x'}|=\sqrt{\delta_{ij}{\it x^i x^j}}$, and $n^i=\frac{{\it x^i}}{r}$.
We note that $\partial_L\frac{1}{|\vec{\it x}-\vec{\it x'}|}$ is a symmetric
trace free(STF) tensor for $\vec{\it x}\neq \vec{\it x'}$ because the trace of
any pair of indices $\partial_i \partial_i\frac{1}{|\vec{\it x}-\vec{\it x'}|}
=\Delta^2\frac{1}{|\vec{\it x}-\vec{\it x'}|}=-4\pi\delta^{(3)}(\vec{\it x}-\vec{\it x'})$ is zero unless $\vec{\it x}$ coincides with the center-of-mass
\cite{thorn}. 
Now, we use the following identities \cite{will}:\\
\begin{eqnarray}
(i)\;\;\partial_j r&=&\partial_j\sqrt{\delta_{ik}{\it x^i x^k}}=\frac{1}{2r}\delta_{ik}\Big[\delta_{ij}{\it x^k}+\delta_{jk}{\it x^i}\Big]\nonumber\\
&=&\frac{1}{2r}\Big[\delta_{jk}{\it x^k}+\delta_{ij}{\it x^i}\Big]=\frac{1}{2r}\Big[2{\it x^j}\Big]=n^j\;\;\;\;
\label{nj}
\end{eqnarray}
\begin{eqnarray}
(ii)\;\;\partial_j{n_k}&=&\partial_j\Big(\frac{x^k}{r}\Big)=\frac{\delta_{jk}}{r}-\frac{1}{r^2}{\it x^k}{\partial_j r}\nonumber\\
&=&\frac{\delta_{jk}}{r}-\frac{1}{r^2}x^k n^j=\frac{1}{r}(\delta_{jk}-n_j n_k)
	\label{dnj}
\end{eqnarray}
\begin{eqnarray}
	(iii)\;\;\partial_j\Big(\frac{1}{r}\Big)&=&-\frac{1}{r^2}\partial_j r=-\frac{1}{r^2}
n_j
\end{eqnarray}
\begin{eqnarray}
(iv)\;\;\partial_L\Big(\frac{1}{r}\Big)&=&\sum^{\infty}_{l=0}(-1)^l(2l-1)!!
	\frac{n^{<L>}}{r^{l+1}}
\label{prl}
\end{eqnarray}
The internal potential at a point $\vec{\it x}$ can now be written as
\begin{eqnarray}
	{\it u^{int}_A}&=&-\sum^{\infty}_{l=0} \frac{(-1)^l}{l!}
        \partial_L \Big(\frac{1}{r}\Big)\Big|_{\vec{\it x'}=\vec{\it z_A}}
	\int{d^3{\vec{\it x'}}\rho_A(\vec{\it x'})({\it x'-z_A})^L}\nonumber\\
	&=&-\sum^{\infty}_{l=0} \frac{(-1)^l}{l!}
        \partial_L \Big(\frac{1}{r}\Big)\Big|_{\vec{\it x'}=\vec{\it z_A}}
	M^{L}_A\nonumber\\
	&=&-\sum^{\infty}_{l=0} \frac{(-1)^l}{l!}(-1)^l(2l-1)!!\frac{n^{<L>}}{r^{l+1}}M^{L}_A \nonumber\\
&=&-\sum^{\infty}_{l=0} \frac{(2l-1)!!}{l!}\frac{n^{<L>}}{r^{l+1}}M^{L}_A,
	\end{eqnarray}
where $M^{L}_A$ is the $l^{th}$ multipole moment of the body defined by
\begin{eqnarray}
	M^{L}_A=\int{d^3{\vec{\it x'}}\rho_A(\vec{\it x'})({\it x'-z_A})^L}
	\end{eqnarray}
$M_A=\int{d^3{\it x}\;\rho}\;,\;\;\;  l=0$ for monopole\\
$M^{i}_A=\int{d^3{\it x}\;\rho ({\it x-z_A})^i}=0\;,\;\;\;  l=1$ for dipole\\
$M^{ij}_A=\int{d^3{\it x}\;\rho ({\it x-z_A})^{<ij>}}=Q^{ij}_A\;,\;\;\;  l=2$ 
for quadrupole\\
where $({\it x-z_A})^{<ij>}=({\it x-z_A})^i ({\it x-z_A})^j-\frac{1}{3}
\delta_{ij}|\vec{\it x}-\vec{\it z_A}|^2$\\
\begin{eqnarray}
	n^{<ij>}=n^i n^j-\frac{1}{3}\delta^{ij},
\end{eqnarray}
Similarly, we can expand the external potential about the center-of-mass of
body A
\begin{eqnarray}
	u^{ext}_A(\vec{\it x})&=&\sum_{l=0}\frac{1}{l!}(\it x-z_A)^L \partial_L\phi^{ext}_A(\vec{\it x})\Big|_{\vec{\it x}=\vec{\it z_A}}\nonumber\\
	&=&\sum_{l=0}\frac{1}{l!}(\it x-z_A)^L \mathcal{E}^{L}_A .
\end{eqnarray}
Where $\mathcal{E}^{L}_A$ is the external tidal field due to the potential
from object B
\begin{eqnarray}
	\mathcal{E}^{L}_A&=&-\partial_L u^{ext}_A(\vec{\it x})\Big|_{\vec{\it x}=\vec{\it z_A}}\nonumber\\
	&=&-\partial_L u^{int}_B(\vec{\it x})\Big|_{\vec{\it x}=\vec{\it z_A}} \nonumber\\
	&=&-\partial_L\Big(-\sum^{\infty}_{l=0}\frac{(-1)^l}{l!} M^{L}_B \partial_L\frac{1}{|\vec{\it x}-\vec{\it z_B}|}\Big)_{\vec{\it x}=\vec{\it z_A}}\nonumber\\
	&=&\sum^{\infty}_{l=0}\frac{(-1)^l}{l!} M^{L}_B \partial_L\frac{1}{|\vec{\it z_A}-\vec{\it z_B}|}=\sum^{\infty}_{l=0}\frac{(-1)^l}{l!} M^{L}_B \partial_L\frac{1}{r_{AB}}.
\end{eqnarray}
Here, $r_{AB}=|\vec{\it z_A}-\vec{\it z_B}|$ is the separation between two body.
Some useful identities are:
\begin{eqnarray}
	\mathcal{E}^{L}=-\partial_L\Big(\frac{M_B}{r_{AB}}\Big).
\end{eqnarray}
for $L=2$
\begin{eqnarray}
	\mathcal{E}^{ij}&=&-\partial_i\partial_j\Big(\frac{M_B}{r_{AB}}\Big)\nonumber\\
	&=&-\frac{M_B}{r^3_{AB}}(3n_i n_j-\delta_{ij})_{AB}\nonumber\\
	&=&-\frac{3M_B}{r^3_{AB}}(n_i n_j-\frac{1}{3}\delta_{ij})_{AB}\nonumber\\
	&=&-\frac{3M_B}{r^3_{AB}} n^{<ij>}_{AB} .
\end{eqnarray}
\subsection{Constructing the Lagrangian and energy describing a binary system 
of extended objects}
The Lagrangian is constructed from the kinetic and potential energies of 
the system as $L=T-V$. Using the Lagrangian we can determine the equation of
motion via the Euler-Lagrange equations.

First we consider the bodies to be inside their respective "buffer" zones as 
shown in Fig. \ref{body2}, i.e. when the distances are large compared to the 
size of the body but small compared to the distance of the companion. 
The total kinetic energy for each object can be decomposed into the 
contributions from the center-of-mass of each body as well as from the 
internal kinetic energy about the center-of-mass of each body
\cite{icts}:
\begin{eqnarray}
	T&=&T_A+T_B ;\nonumber\\
	T&=&\frac{1}{2}\int_A{d^3{\it x}\;\rho \vec{\dot{\it z_A}^2}}+\frac{1}{2}M_B \vec{\dot{\it z_B}^2}+T^{int}\nonumber\\
	&=&\frac{1}{2} v^2 \Big[\frac{M_A M^2_{B}}{M^2}+\frac{M_B (-M^2_{A})}{M^2}\Big]+T^{int}\nonumber\\
	&=&\frac{1}{2} v^2 M_A M_B \Big[\frac{M_A+M_B}{M^2}\Big]+T^{int}\nonumber\\
	&=&\frac{1}{2} v^2 \frac{M_A M_B}{M}+T^{int}\nonumber\\
	&=&\frac{1}{2}\mu v^2+T^{int},
\end{eqnarray}
where $\mu=\frac{M_A M_B}{M}$ is the reduced mass of the body.

Similarly, For the extended object A, the point-mass companion's potential is
$M_B/ |\vec{\it x}-\vec{\it z_B}|$. When this potential is expanded about A's
center of mass it becomes
\begin{eqnarray}
	u^{companion}_A=\sum^{\infty}_{l=0} \frac{1}{l!}(x-z_A)^L \mathcal{E}_L
	.
\end{eqnarray}
The potential energy $V_A$ due to the center of mass motion $\vec{\it z_A}$ is 
\begin{eqnarray}
	V_A&=&-\frac{1}{2}\int_A{d^3{\it x}\; \rho_A\; u^{companion}_A}\nonumber\\
	&=&-\frac{1}{2}\int_A{d^3{\it x}\; \Big(-\sum^{\infty}_{l=0}\frac{1}{l!}
        (\vec{\it x}-\vec{\it z_A})^L\mathcal{E}_L\Big)\rho_A}\nonumber\\
	&=&-\frac{M_A M_B}{2\;r_{AB}}+\frac{1}{2}\sum^{\infty}_{l=1}\frac{1}{l!}
 \mathcal{E}_L\int_A{d^3{\it x}\;(\vec{\it x}-\vec{\it z_A})^L \rho_A}\nonumber\\
	&=&-\frac{M_A M_B}{2\;r_{AB}}+\frac{1}{4}\mathcal{E}_{ij} Q^{ij} .
\end{eqnarray}

Similarly, the potential energy $V_B$ of the point mass influenced by the 
field of the extended body is
\begin{eqnarray}
	V_B&=&-\frac{1}{2}M_B u^{companion}_B \nonumber\\
	&=&-\frac{1}{2}M_B \sum^{\infty}_{l=0}\frac{(2l-1)!!}{l!}\frac{n^{<L>}_B}{r^{l+1}_{BA}}M_L\nonumber\\
	&=&-\frac{M_A M_B}{2\;r_{BA}}+\sum^{\infty}_{l=1}\frac{(2l-1)!!}{l!}\frac{n^{<L>}_B}{r^{l+1}_{BA}}M_L\nonumber\\
	&=&-\frac{M_A M_B}{2\;r}-\frac{1}{2}\frac{3}{2} Q_{ij}M_{B}\frac{n^{<ij>}_B}{r^3}\nonumber\\
	&=&-\frac{M_A M_B}{2\;r_{BA}}+\frac{1}{4}\mathcal{E}_{ij} Q^{ij} .
\end{eqnarray}
Now, we want to calculate the total potential energy upto $l=2$ and put 
$r_{AB}=r_{BA}=r$. Thus,
\begin{eqnarray}
	V&=&V_A+V_B+V^{int}\, ,\nonumber\\
	&=&-\frac{M_A M_B}{r}+\frac{1}{2}\mathcal{E}_{ij} Q^{ij}+V^{int} .
\end{eqnarray}
Hence, the total Lagrangian of the system can be written as
\begin{eqnarray}
	L&=&\frac{1}{2}\mu v^2+T^{int}-(-\frac{M_A M_B}{r}+\frac{1}{2}\mathcal{E}_{ij} Q^{ij}+V^{int})\nonumber\\
	&=&\frac{1}{2}\mu v^2+\frac{M_A M_B}{r}-\frac{1}{2}\mathcal{E}_{ij} Q^{ij}+T^{int}-V^{int}\nonumber\\
	&=&\frac{1}{2}\mu v^2+\frac{M_A M_B}{r}-\frac{1}{2}\mathcal{E}_{ij} Q^{ij}+L_{int}(Q,\dot{Q},...) \; .
	\label{lg}
\end{eqnarray}

The internal Lagrangian describes only the elastic potential energy associated 
with the tidal deformation \cite{flan}:
\begin{eqnarray}
        L_{int}=-\frac{1}{4 \lambda}Q_{ij}Q^{ij} .
\end{eqnarray}
Where $\lambda$ is the tidal deformability parameter. We will assume that the 
tidal effects are small and can be treated as a linear perturbation. The 
tidally induced adiabatic quadrupole moment is given by

\begin{eqnarray}
	Q_{ij}=-\lambda \mathcal{E}_{ij} .
	\label{lam}
\end{eqnarray}
Noted that Eq.\eqref{lam} can be translated to a relativistic result within the
same approximation and interpreting $\mathcal{E}$ regarding the curvature 
tensor, i.e. $\mathcal{E}_{ij}=R_{0i0j}$ call as Riemann tensor.

Eq. \eqref{lg} can be written as
\begin{eqnarray}
	L&=&\frac{1}{2}\mu v^2+\frac{M_A M_B}{r}-\frac{1}{2}\mathcal{E}_{ij} Q^{ij}-\frac{1}{4\lambda}Q_{ij}Q^{ij}\nonumber\\
	&=&\frac{1}{2}\mu v^2+\frac{M_A M_B}{r}-\frac{1}{2}\mathcal{E}_{ij} 
	(-\lambda \mathcal{E}^{ij})-\frac{1}{4} (-\lambda \mathcal{E}_{ij}) (-\lambda \mathcal{E}^{ij})\nonumber\\
	&=&\frac{1}{2}\mu v^2+\frac{M_A M_B}{r}+\frac{1}{2} (\lambda \mathcal{E}_{ij}) 
(\mathcal{E}^{ij})-\frac{1}{4}(\lambda \mathcal{E}_{ij})(\mathcal{E}^{ij})\nonumber\\
	&=&\frac{1}{2}\mu v^2+\frac{M_A M_B}{r}+\frac{1}{4} \lambda\; \mathcal{E}_{ij}\mathcal{E}^{ij}\nonumber\\
	&=&\frac{1}{2}\mu v^2+\frac{M_A M_B}{r}+\frac{\lambda}{4}  6\frac{M^{2}_B}{r^6} \nonumber\\
	&=&\frac{1}{2}\mu v^2+\frac{M_A M_B}{r}+\frac{3}{2} \lambda  \frac{M^{2}_B}{r^6} .
\end{eqnarray}
For the Lagrangian above, we can calculate the energy associated with the 
system by 
\begin{eqnarray}
E=\frac{1}{2}\mu v^2-\frac{M_A M_B}{r}-\frac{3}{2} \lambda  \frac{M^{2}_B}
{r^6}  .
\end{eqnarray}
Now, we need to transform Cartesian to spherical polar coordinates, and the
motion takes place in a fixed orbital plane with $\theta=\pi/2$, and 
$\dot{\theta}=0$. We can define finally energy of the two body system comes out to be \cite{vines}: 
\begin{eqnarray}
E=\frac{1}{2}\mu (\dot{r}^2+r^2\dot{\phi}^2)-\frac{\mu M}{r}-\frac{3 M_B^2 \lambda}{2\;r^6} .
	\label{enr}
\end{eqnarray}
For the linear tidal correction $r=r_0(1+\delta r)$, where $\delta r =dr/r_0<<1$ is the dimensionless quantity. 
Consider a circular orbit, $\dot{r}=0=\ddot{r}$,$\dot{\phi}=\Omega$ is the 
orbital angular frequency, and $r_0=M^{1/3}\Omega^{-2/3}$ is the orbital 
radius, respectively. At linear order tidal effects we 
obtain $\delta r=\frac{3M_B^2 \lambda \Omega^{10/3}}{\mu M^{8/3}}$. Hence, the 
Eq. \eqref{enr} can be written as
\begin{eqnarray}
E(\Omega)&=& \frac{1}{2} \mu r_0^2 (1+2\delta r)\Omega^2-\frac{\mu M}{r_0 (1+\delta r)}-\frac{3 M_B^2\lambda}{2\;r_0^6(1+\delta r)^6}\nonumber\\
	&=&-\frac{1}{2}\mu (M\Omega)^{2/3}\Big[1-\frac{9 M_B^2\lambda\Omega^{10/3}}{\mu M^{8/3}}+\dots\Big] .
	\label{en}
\end{eqnarray}
Tidal contribution to the energy $E$ of the binary stars in terms of the 
post-Newtonian dimensionless parameter $x$ is given by 
\begin{eqnarray}
E(x)=-\frac{1}{2} M \eta x\Big[1+(PN-PP\;\; corr.)-9\frac{M_B}{M_A}
\frac{\lambda_A}{M^5}x^5+A\leftrightarrow B\Big] .
\end{eqnarray}
where, $x=(\Omega M)^{2/3}$, and $\eta=M_A M_B/M^2$ is the symmetric mass 
relation. Second term represents the leading order terms to the post-Newtonian
point-particle corrections.
\subsection{Gravitational wave energy flux}
The rate of energy loss due to gravitational radiation can be found from the
quadrupole formula $\dot{E}_{GW}=-\frac{1}{5}\Big<\dddot{Q}^T_{ij}\;\dddot{Q}^{T}_{ij}\Big>$. The quantity $Q^{T}_{ij}$ is the total quadrupole of the system, 
and can be written as
\begin{eqnarray}
Q^{T}_{ij}&=&\mu r^2 n^{<ij>}+\frac{3\lambda M_B}{r^3}n^{<ij>}\nonumber\\
	&=&n^{<ij>}\Big[\mu r^2 +\frac{3\lambda M_B}{r^3}\Big]\nonumber\\
	&=&n^{<ij>} \zeta ,
	\end{eqnarray}
where $\zeta=\mu r^2 +\frac{3\lambda M_B}{r^3}$, and $n^{<ij>}=n_i n_j-\frac{1}{3}\delta_{ij}$. For circular orbit $\dot{r}=0$, $n_i=(cos \phi,sin \phi,0), \phi_i=(-sin \phi, cos \phi,0), \dot{n_i}=\Omega \phi_i,\dot{\phi}=-\Omega n_i, n_i \phi^i =1, n^i n_i=1$, and $\phi_i \phi^i=1$, respectively. \\
\begin{eqnarray}
	\frac{dn^{<ij>}}{dt}&=&\dot{n_i} n_j+n_i \dot{n_j}\nonumber\\
	&=&\Omega\phi_i n_j+n_i\Omega \phi_j\nonumber\\
	&=&\Omega (\phi_i n_j+n_i\phi_j), \nonumber\\
\frac{d^2n^{<ij>}}{dt^2}&=&\Omega\frac{d}{dt}(\phi_i n_j+n_i\phi_j)\nonumber\\
	&=&\Omega(\dot{\phi_i} n_j+\phi_i \dot{n_j}+\dot{n_i}\phi_j+n_i\dot{\phi_j})\nonumber\\
	&=&\Omega(-\Omega n_i n_j+\phi_i\Omega \phi_j+\Omega \phi_i\phi_j-n_i\Omega n_j)\nonumber\\
	&=&\Omega^2(-2n_i n_j+2\phi_i\phi_j)\nonumber\\
	&=&2\Omega^2(-n_i n_j+\phi_i\phi_j) ,\nonumber\\
	\frac{d^3n^{<ij>}}{dt^3}&=&2\Omega^2\frac{d}{dt}(-n_i n_j+\phi_i\phi_j)\nonumber\\
	&=&2\Omega^2(-\dot{n_i}n_j-n_i\dot{n_j}+\dot{\phi_i}\phi_j+\phi_i\dot{\phi_j})\nonumber\\
	&=&2\Omega^2(-\Omega \phi_i n_j-n_i\Omega\phi_j-\Omega n_i \phi_j-\phi_i \Omega n_j)\nonumber\\
	&=&-2\Omega^3(2\phi_i n_j+2 n_i \phi_j)\nonumber\\
	&=&-4 \Omega^3 (\phi_i n_j+n_i \phi_j) ,
\end{eqnarray}
So, the flux can be written as
\begin{eqnarray}
	\dot{E}_{GW}&=&-\frac{1}{5}\Big<\dddot{Q}^T_{ij}\;\dddot{Q}^{T}_{ij}\Big>\nonumber\\
	&=&-\frac{1}{5} 4 \zeta^2\Omega^3 (\phi_i n_j+n_i \phi_j) 4 \Omega^3 (\phi_i n_j+n_i \phi_j) \nonumber\\
	&=&-\frac{1}{5} 16 \zeta^2\Omega^6 (\phi_i n_j \phi_i n_j+\phi_i n_j n_i \phi_j+n_i \phi_j \phi_i n_j+n_i\phi_j n_i \phi_j)\nonumber \\
	&=&-\frac{1}{5} 16 \zeta^2\Omega^6(1+0+0+1)
	=-\frac{32}{5}\zeta^2\Omega^6\nonumber\\
	&=&-\frac{32}{5}\Omega^6\Big[\mu r^2+\frac{3\lambda M_B}{r^3}\Big]^2 \nonumber\\
	&=&-\frac{32}{5}\Omega^6\Big[\mu^2 r^4+\frac{6\mu r^2\lambda M_B}{r^3}
+\frac{9\lambda^2 M_B^2}{r^6}\Big]\nonumber\\
	&=&-\frac{32}{5}\Omega^6\Big[\mu^2 r^4+\frac{6\mu \lambda M_B}{r}+\dots\Big]\nonumber\\
	&=&-\frac{32}{5}\Omega^6\Big[\mu^2 r_0^4 (1+\delta r)^4+\frac{6\mu \lambda M_B}{r_0 (1+\delta r)} \Big]\nonumber\\
	&=&-\frac{32}{5}\Omega^6\Big[\mu^2 r_0^4 (1+4\delta r)+\frac{6\mu \lambda M_B}{r_0 } \Big]\nonumber\\
	&=&-\frac{32}{5}\Omega^6 \mu^2 r_0^4\Big[1+4\delta r+\frac{6\lambda M_B}{\mu r_0^5 }\Big]\nonumber\\
	&=&-\frac{32}{5}\Omega^6 \mu^2 (M^{1/3}\Omega^{-2/3})^4\Big[1+4\frac{3\lambda M_B^2 \Omega^{10/3}}{\mu M^{8/3}}+\frac{6\lambda M_B}{\mu(M^{1/3}\Omega^{-2/3})^5}\Big]\nonumber\\
	&=&-\frac{32}{5}\Omega^{10/3} \mu^2 M^{4/3}\Big[1+\frac{6\lambda M_B\Omega^{10/3}}{\mu M^{5/3}}\Big(1+\frac{2M_B}{M}\Big)\Big]\nonumber\\
	&=&-\frac{32}{5}\Omega^{10/3} \mu^2 M^{4/3} \Big[1+\delta\dot{E}\Big],
\end{eqnarray}
where $\dot{E}$ is the tidal correction on the gravitational waves.
\begin{eqnarray}
\delta \dot{E}=\frac{6\lambda M_B\Omega^{10/3}}{\mu M^{5/3}}\Big(1+\frac{2M_B}{M}\Big) ,
\end{eqnarray}
The energy flux for a quasicircular inspiral for binary stars is given by 
\cite{vines}:
\begin{eqnarray}
\dot{E}_{GW}=-\frac{32}{5}\eta^2 x^5 \Big[1+(PN-PP\;\;corr.)+6\frac{M_A+3M_B}{M_A} \frac{\lambda_A}{M^5}x^5+A\leftrightarrow B\Big] .
\end{eqnarray}
\subsection{Orbital decay due to gravitational waves}
To estimate the orbital decay, we consider the tidal contribution to the 
energy $E$ and energy flux $\dot{E}_{GW}$ for quasicircular inspiral 
\cite{vines,icts}. Therefore, one can write
\begin{eqnarray}
\frac{d\Omega}{dt}=\frac{d\Omega}{dE}\;\frac{dE}{dt}=\frac{\dot{E}_{GW}}{(\frac{dE}{d\Omega})}\;\; .
\label{ot}
 \end{eqnarray}
Differentiating Eq. \eqref{en} with respect to $\Omega$, we get
\begin{eqnarray}
\frac{dE}{d\Omega}&=&-\frac{1}{3} \mu M^{2/3} \Omega^{-1/3}\Big[1-\frac{54 \lambda M_B^2 \Omega^{10/3}}{\mu M^{8/3}}\Big] \nonumber\\
	&=&-\frac{1}{3} \mu M^{2/3} \Omega^{-1/3}[1+\Delta] ,
\end{eqnarray}
where, $\Delta=-\frac{54 \lambda M_B^2 \Omega^{10/3}}{\mu M^{8/3}}$, and
Eq. \eqref{ot} can be written for an extended object with a point mass. Now,
\begin{eqnarray}
	\frac{d\Omega}{dt}&=& \frac{-\frac{32}{5}\Omega^{10/3} \mu^2 M^{4/3} [1+\delta\dot{E}]}{-\frac{1}{3} \mu M^{2/3} \Omega^{-1/3}[1+\Delta]}\nonumber\\
	&=&\frac{96}{5} \mu M^{2/3} \Omega^{11/3}[1+\delta\dot{E}-\Delta]\nonumber\\
	&=&\frac{96}{5} \mu M^{2/3} \Omega^{11/3}\Big[1+\frac{6\lambda M_B\Omega^{10/3}}{\mu M^{5/3}}\Big(1+\frac{2M_B}{M}\Big)+\frac{54 \lambda M_B^2 \Omega^{10/3}}{\mu M^{8/3}}\Big]\nonumber\\
	&=&\frac{96}{5} \mu M^{2/3} \Omega^{11/3}\Big[1+\frac{6\lambda M_B\Omega^{10/3}}{\mu M^{8/3}}\Big(9 M_B+M+2M_B\Big)\Big]\nonumber\\
	&=&\frac{96}{5} \mu M^{2/3} \Omega^{11/3}\Big[1+\frac{6\lambda M_B\Omega^{10/3}}{\mu M^{8/3}}\Big(M+11M_B\Big)\Big]\nonumber\\
	&=&\frac{96}{5} \mathcal{M}^{5/3}_{ch}\Omega^{11/3}\Big[1+\frac{6\lambda M_B\Omega^{10/3}}{\mu M^{8/3}}\Big(M+11M_B\Big)\Big] ,
\end{eqnarray}
where, $\mathcal{M}_{ch}=\mu^{3/5} M^{2/5}$ is the chirp mass of the binary 
system, and similarly, we can write orbital decay for two extended objects 
\begin{eqnarray}
	\frac{d\Omega}{dt}&=&\frac{96}{5} \mathcal{M}^{5/3}_{ch}\Omega^{11/3}\Big[1+\frac{6\lambda_A M_B\Omega^{10/3}}{\mu M^{8/3}}\Big(M+11M_B\Big)\Big]+(A\leftrightarrow B)\nonumber\\
	&=&\frac{96}{5} \mathcal{M}^{5/3}_{ch}\Omega^{11/3}\Big[1+\frac{6\Omega^{10/3}}{\mu M^{8/3}}\Big(\lambda_A M_B (M+11M_B)+\lambda_B M_A (M+11M_A)\Big)\Big]\nonumber\\
	&=&\frac{96}{5} \mathcal{M}^{5/3}_{ch}\Omega^{11/3}\Big[1+\frac{156 \Omega^{10/3}}{ M^{5/3}}\frac{1}{26}\Big(\frac{M_A+12M_B}{M_A}\lambda_A+\frac{M_B+12M_A}{M_B}
\lambda_B\Big)\Big]\nonumber\\
	&=&\frac{96}{5} \mathcal{M}^{5/3}_{ch}\Omega^{11/3}\Big[1+\frac{156 \Omega^{10/3}}{ M^{5/3}}\tilde{\lambda}\Big],
\end{eqnarray}
where, $\tilde{\lambda}$ is the average tidal deformability of the binary stars
\cite{tanja}.
\begin{eqnarray}
\tilde{\lambda}=\frac{1}{26}\Big(\frac{M_A+12M_B}{M_A}\lambda_A+\frac{M_B+12M_A}{M_B}\lambda_B \Big) .
	\label{avtide}
\end{eqnarray}
When the masses of the stars are equal $M_A=M_B$ then $\tilde{\lambda}=\lambda_A
=\lambda_B$. The evolution of the gravitational-wave frequency is determined by
\cite{flan}
\begin{eqnarray}
	\frac{dx}{dt}&=&\frac{d}{dt}(\Omega M)^{2/3}=\frac{2}{3}M^{2/3} \Omega^{-1/3}
\frac{d\Omega}{dt}\nonumber\\
	&=&\frac{2}{3}M^{2/3} \Omega^{-1/3} \frac{96}{5} \mu M^{2/3} \Omega^{11/3}
\Big[1+\frac{6\lambda_A M_B\Omega^{10/3}}{\mu M^{8/3}}\Big(11 M_B+M\Big)+A\leftrightarrow B\Big]\nonumber\\
	&=&\frac{64}{5} \mu M^{4/3} \Omega^{10/3}\Big[1+(PN-PP\;\;corr.)+\frac{156 \Omega^{10/3}}{ M^{5/3}}\tilde{\lambda}\Big]\nonumber\\
	&=&\frac{64}{5} \frac{\eta x^5}{M} \Big[1+(PN-PP\;\;corr.)+156\frac{\tilde{\lambda}}{M^5}x^5\Big]\nonumber\\
	&=&\frac{64}{5} \frac{\eta x^5}{M} \Big[1+(PN-PP\;\;corr.)+156\tilde{\Lambda}\; x^5\Big] .
\end{eqnarray}
Where $\tilde{\Lambda}$ is the dimensionless tidal deformability, and the 
quadrupole gravitational wave phase $\Phi$ can be calculated by 
\begin{eqnarray}
\frac{d\Phi}{dt}=2 \Omega=2 x^{3/2}/M\; .
\end{eqnarray} 
%%%%%%%%%%%%%%%%%%%%%%%%%%%%%%%%%%%%%%%%%%%%%%%%%%%%%%%%%%%%
\subsection{Relativistic tidal interactions}
We will now derive the $l= 2,3,4$ tidal love numbers of a 
star which depend on the equation of state.
The star's quadrupole moment $Q_{ ij}$, and the external tidal 
field $\mathcal{E}_{ij}$ come in the asymptotic expansion coefficients of the 
total metric at a large distance $r$ from the star. This expansion includes the 
metric component $g_{tt}$ in asymptotically Cartesian, mass-centered 
coordinates \cite{thorn1}, and we have 
\begin{eqnarray}
	-\frac{(1+g_{tt})}{2}&=&\phi^{int}+\phi^{ext}  \\
&=&-\frac{M}{r}-\frac{3Q_{ij}}{2r^3}\Big(n_i n_j-\frac{1}{3}\delta_{ij}\Big)
+\dots+\frac{\mathcal{E}_{ij}}{2} r^2 n_i n_j \;.
\end{eqnarray}
where $n^i=x^i/r$, and both $Q_{ij}$ and $\mathcal{E}_{ij}$ are symmetric and 
traceless. The first term represents the monopole piece of the standard 
gravitational potential, which depends on the object's mass $M$. The second 
term is the body's response decaying with $r^{-3}$ that measured regarding the 
tidal Love number $k_2$. The last term is describing an external tidal field 
growing with $r^2$. Now, our aim is to derive the relativistic Love numbers 
within the context of the linear perturbation theory. To this, the initially 
spherical object is perturbed slightly by an applied tidal field. Therefore, we 
start with the equilibrium of a static, spherically symmetric, and the 
relativistic star  which is described by a space-time metric 
$g^{(0)}_{\alpha\beta}$ given by \cite{misner}:

\begin{eqnarray}
g^{(0)}_{\alpha\beta}=diag\Big[-e^{\nu(r)},e^{\lambda(r)},r^2,r^2 sin^2{\theta}\Big].
\end{eqnarray}
Static perturbations of the system is considered to study the tidal
deformation of the star. Also, we focus on the even-parity perturbations in the 
Regge-Wheeler gauge which is the angular dependence of the components of a
linearized metric perturbation into spherical harmonics \cite{wheeler}. The 
perturbed metric can be written as 
\begin{eqnarray}
	h_{\alpha\beta}=diag\Big[-e^{\nu(r)} H(r),-e^{\lambda(r)}H(r),r^2 K(r),r^2 sin^2{\theta}\;K(r)\Big]Y_{20}(\theta,\phi) ,
\end{eqnarray}
where $Y_{20}(\theta,\phi)=\frac{1}{4}\sqrt{(\frac{5}{\pi}}(-1+3\;cos^2{\theta})$\;\;,$Y_{20}(\theta,\phi)=\frac{1}{4}\sqrt{(\frac{5}{\pi}}(-1+3\;x^2)$. Here,
$x=cos \theta,\; sin^2{\theta}=1-cos^2{\theta}=1-x^2$.
\begin{eqnarray}
	h_{\alpha\beta}=diag\Big[-e^{\nu(r)} H(r),-e^{\lambda(r)}H(r),r^2 K(r),r^2 (1-x^2)\;K(r)\Big]\frac{1}{4}\sqrt{\frac{5}{\pi}}(-1+3\;x^2)\; .\;\;\;\;\;
\end{eqnarray}
The full metric of the space-time is given by
\begin{eqnarray}
	g_{\alpha\beta}=g^{0}_{\alpha\beta}+h_{\alpha\beta}\; .
	\label{matric}
\end{eqnarray}

Eq. \eqref{matric} can be solved with the following conditions:\\
(i) First, the perturbed Einstein equations (for $l\ge2$) for the perfect 
fluid is to be determined. Then, the derivative of the stress-energy-momentum 
tensor is to be found out by taking 
\begin{eqnarray}
\delta T_{\mu}^{\nu}=(\delta{\mathcal{E}}+\delta P)u_{\mu}u^{\nu}
+(P+\mathcal{E})(u_{\mu}\delta u^{\nu}+u_{\nu}\delta u^{\mu})+\delta P \delta_{\mu}^{\nu} \; .
\end{eqnarray}
The nonvanishing components of the stress-energy moments tensor are  
$\delta T_{0}^{0}=-\delta\mathcal{E}=-(dP/d\mathcal{E})^{-1}\delta P$ and 
$\delta T_{i}^{i}=\delta P$. However, all off-diagonal components vanish 
identically, since $\delta u^{i}=0$.\\
(ii) $ T_{\mu}^{\nu}$ of the Einstein equations \eqref{EIN} can be expressed 
concerning the perturbed metric and inserting previous result (i). Therefore, 
the linearized Einstein equation with all the combined components is given
by 
\begin{eqnarray}
\delta G_{\mu}^{\nu}=8 \pi \delta T_{\mu}^{\nu} .
\end{eqnarray}
Now, we need to write different components of the Einstein tensors $\delta G_{\mu}^{\nu}$. 

For $tt-$component:
\begin{eqnarray}
	\delta{G_{t}^{t}}&=&\frac{e^{-\lambda(r)}}{8r^2}\sqrt{\frac{5}{\pi}}(-1+3x^2)(-4e^{\lambda(r)}K(r)+H(r)(2+6e^{\lambda(r)}\nonumber\\
	&-&2r\lambda'(r))+r(2H'(r)+K'(r)(6-r\lambda'(r))+2r K''(r))) .
\end{eqnarray}
For $rr-$component:
\begin{eqnarray}
\delta{G_{r}^{r}}&=&\frac{e^{-\lambda(r)}}{8r^2}\sqrt{\frac{5}{\pi}}(-1+3x^2)
(4e^{\lambda(r)}K(r)+H(r)(-2+6e^{\lambda(r)}\nonumber\\
	&-&2r\nu'(r))-r(2H'(r)+K'(r)(2+r\nu'(r)))) .
\end{eqnarray}

For $\theta\theta-$component:
\begin{eqnarray}
\delta{G_{\theta}^{\theta}}&=&-\frac{1}{16r} e^{-\lambda (r)} \sqrt{\frac{5}{\pi}}(-1+3x^2)(2 H(r)\lambda'(r)-2 H(r)\nu'(r)+r H(r)\lambda'(r)\nu'(r)\nonumber\\
&-&r H(r)\nu'(r)^2+H'(r)(-4+r\lambda'(r)-3r\nu'(r))+K'(r)(-4+r\lambda'(r)
\nonumber\\
&-&r \nu'(r))-2 r H''(r)-2 r K''(r)-2rH(r)\nu''(r)).\;\;\;\;
\end{eqnarray}
For $\phi\phi-$component is same as the $\theta\theta-$component. Hence, $\delta{G_{\theta}^{\theta}}-\delta{G_{\phi}^{\phi}}=0$.\\
For $r\theta-$component:
\begin{eqnarray}
	\delta{G_{\theta}^{r}}&=&-\frac{3}{4}e^{-\lambda(r)}\sqrt{\frac{5}{\pi}}\;\;x (H'(r)+K'(r)+H(r)\nu'(r))=0 ,\nonumber\\
	K'(r)&=&-H'(r)-H(r) \nu'(r) ,\nonumber\\
	K''(r)&=&-H''(r)-H'(r)\nu'(r)-H'(r)\nu''(r) .
\end{eqnarray}
\begin{eqnarray}
	\delta P&=&(\delta{G_{\theta}^{\theta}}+\delta{G_{\phi}^{\phi}})/{16\pi}\nonumber\\
	&=&-\frac{1}{128\pi^{3/2}r} e^{-\lambda (r)} \sqrt{5}(-1+3x^2)(2 H(r)\lambda'(r)-2 H(r)\nu'(r)+r H(r)\lambda'(r)\nu'(r)\nonumber\\
&-&r H(r)\nu'(r)^2+H'(r)(-4+r\lambda'(r)-3r\nu'(r))+K'(r)(-4+r\lambda'(r)
	\nonumber\\
	&-&r \nu'(r))-2 r H''(r)-2 r K''(r)-2rH(r)\nu''(r)).\;\;\;\;
\end{eqnarray}

To obtain the differential equation for $H$, we need to subtract the 
$rr$-component of the Einstein equation from the $tt$-component.
\begin{eqnarray}
	\delta G_{\alpha}^{\beta}=8\pi \delta T^{\beta}_{\alpha},\;\;\;\;\;\;
	\nonumber\\
\delta G_{t}^{t}-\delta G_{r}^{r}=8 \pi (T^{0}_{0}+T^{i}_{i})\;\;\;\;\;\;
	\nonumber\\
	\delta G_{t}^{t}-\delta G_{r}^{r}+8 \pi(\frac{d\mathcal{E}}{dP} \delta P+\delta P)=0 \;\;\;\;\;\;\;
	\nonumber\\
-\frac{1}{4r^2}\Big(H(r)(24 e^{\lambda(r)}+r(2(-1+\frac{d\mathcal{E}}{dP})\nu'(r)+(1+\frac{d\mathcal{E}}{dP})r\nu'(r)^2-\lambda'(r)(2(3+\frac{d\mathcal{E}}{dP})\nonumber\\
+(1+\frac{d\mathcal{E}}{dP})r \nu'(r))+2(1+\frac{d\mathcal{E}}{dP})r\nu''(r)))
+r(-(1+\frac{d\mathcal{E}}{dP})H'(r)(-4+r\lambda'(r)-3r\nu'(r))\nonumber\\
+(-(3+\frac{d\mathcal{E}}{dP})(-4+r\lambda'(r))+(-1+\frac{d\mathcal{E}}{dP})r
\nu'(r))(-H'(r)-H(r)\nu'(r))\nonumber\\
+2r((1+\frac{d\mathcal{E}}{dP})H''(r)+(3+\frac{d\mathcal{E}}{dP})(-H'(r)\nu'(r)-H''(r)-H(r)\nu''(r))))\Big)=0 .\;\;\;\;\;\;	
	\label{Hr}
\end{eqnarray}
We can write the second-order radial differential equation for the metric 
variable H in the form
\begin{eqnarray}
	H''(r)+C_1 H'(r)+C_0 H=0 ,
	\label{HPP}
\end{eqnarray}
where $C_1$, and $C_0$ are the coefficients of the $H'(r)$, and $H(r)$, 
which are extracted from the Eq. \eqref{Hr}.
\begin{eqnarray}
C_1=\frac{2}{r}+\frac{1}{2}(\nu'(r)-\lambda'(r))=\frac{2}{r}+e^{\lambda(r)}
\Big[\frac{2 m(r)}{r^2}+4\pi r(P-\mathcal{E})\Big],
\end{eqnarray}
and
\begin{eqnarray}
	C_0&=&e^{\lambda(r)}\Big[-\frac{l(l+1)}{r^2}+4\pi(\mathcal{E}+P)\frac{d\mathcal{E}}{dP}+4\pi(\mathcal{E}+P)\Big]\nonumber\\
	&+&\nu''(r)+(\nu'(r))^2+\frac{1}{2r}(2-r\nu')(3\nu'+\lambda')\nonumber\\
	&=& e^{\lambda(r)}\Big[-\frac{l(l+1)}{r^2}+4\pi(\mathcal{E}+P)\frac{d\mathcal{E}}{dP}+4\pi(5\mathcal{E}+9P)\Big]-(\nu'(r))^2,
\end{eqnarray}
where the TOV (as derived in appendix \ref{tov}) equations are used to 
rewrite $C_1$ and $C_0$. We also consider the multipolar order $l$, which 
taken the leading quadrupolar even-parity tide \cite{tanja}.
Therefore, the second-order differential equation for H can be rewritten as
\begin{eqnarray}
	&&	H''(r)+\Big[\frac{2}{r}+e^{\lambda(r)}\Big(\frac{2 m(r)}{r^2}+4\pi r(P-\mathcal{E})\Big)\Big]H'(r)+\Big[e^{\lambda(r)}\Big(-\frac{l(l+1)}{r^2}
\nonumber\\
	&+&4\pi(\mathcal{E}+P)\frac{d\mathcal{E}}{dP}+4\pi(5\mathcal{E}+9P)\Big)-(\nu'(r))^2\Big]H(r)=0 .
\end{eqnarray}
The second-order differential equation for H is converted into two first order 
differential equation for $\beta$ as 
\begin{eqnarray}
	\frac{dH(r)}{dr}=\beta(r) ,
\end{eqnarray}
and 
\begin{eqnarray}
	\frac{d\beta(r)}{dr}&=& e^{\lambda(r)}H(r)\Big[-4\pi\Big(5\mathcal{E}+9P+\frac{(\mathcal{E}+P)}{dP/d\mathcal{E}}\Big)+\frac{l(l+1)}{r^2}+4e^{\lambda(r)} \Big(\frac{m(r)}{r^2}+4\pi r P\Big)^2\;\Big]\nonumber\\
&+&\frac{2\beta(r)}{r}e^{\lambda(r)}\Big[-1+\frac{m(r)}{r}+2\pi r^2(\mathcal{E}-P)\Big]\nonumber\\
&=&\Big(1-\frac{2m(r)}{r}\Big)^{-1}H(r)\Big[-4\pi(5\mathcal{E}+9P+\frac{(\mathcal{E}+P)}{dP/d\mathcal{E}})+\frac{l(l+1)}{r^2}+2\Big(1-\frac{2m(r)}{r}\Big)^{-1}\nonumber\\
&& \Big(\frac{m(r)}{r^2}+4\pi r P\Big)^2\Big]+\frac{2\beta(r)}{r}\Big(1-\frac{2m(r)}{r}\Big)^{-1}\Big[-1+\frac{m(r)}{r}+2\pi r^2(\mathcal{E}-P)\Big], \nonumber\\
\end{eqnarray}
Defining the quantity $y(r)=r \beta(r)/H(r)$ for the internal solution, we can 
differentiate $y(r)$ with respect to $r$. The equations are
\begin{eqnarray}
	y'(r)H(r)+y H'(r)-\beta(r)-r\beta'(r)&=&0\nonumber\\
	y'(r)H(r)+(y-1)\beta(r)-r\beta'(r)&=&0\nonumber\\
	\beta(r)(y(r)-1)+\frac{2H(r)y(r)\Big(r-m(r)+2\pi r^3(P-\mathcal{E})\Big)}{r(r-2m(r))}-\frac{r^2 H(r)}{r-2m(r)}\Big(\frac{l(l+1)}{r^2}&&\nonumber\\
	+\frac{4(m(r)+4\pi r^3 P)^2}{r^3(r-2m)}-\frac{4\pi(P+\mathcal{E}+(9P+5\mathcal{E})/dP/d\mathcal{E}}{dP/d\mathcal{E}}\Big)+H(r)y'(r)&=&0\nonumber\\
	\frac{y(r)H(r)}{r}(y(r)-1)+\frac{2H(r)y(r)\Big(r-m(r)+2\pi r^3(P-\mathcal{E})\Big)}{r(r-2m(r))}-\frac{r^2 H(r)}{r-2m(r)}\Big(\frac{l(l+1)}{r^2}&&\nonumber\\
	+\frac{4(m(r)+4\pi r^3 P)^2}{r^3(r-2m)}-\frac{4\pi(P+\mathcal{E}+(9P+5\mathcal{E})/dP/d\mathcal{E}}{dP/d\mathcal{E}}\Big)+H(r)y'(r)&=&0 ,\;\;\;\;
\nonumber\\
\end{eqnarray}
Multiplying $r/H(r)$ in both sides, we can get 
\begin{eqnarray}
	y(r)(y(r)-1)+\frac{2y(r)\Big(r-m(r)+2\pi r^3(P-\mathcal{E})\Big)}{(r-2m(r))}-\frac{r^3}{r-2m(r)}\Big(\frac{l(l+1)}{r^2}&&\nonumber\\
	+\frac{4(m(r)+4\pi r^3 P)^2}{r^3(r-2m)}-\frac{4\pi(P+\mathcal{E}+(9P+5\mathcal{E})/dP/d\mathcal{E}}{dP/d\mathcal{E}}\Big)+ry'(r)&=&0\;\;\;\;
\nonumber\\
	ry'(r)+y(r)^2+y(r)\Big(-1+\frac{2(r-m(r)+2\pi r^3(P-\mathcal{E}))}{(r-2m(r))}\Big)-\frac{r^3}{r-2m(r)}\Big(\frac{l(l+1)}{r^2}&&\nonumber\\
	+\frac{4(m(r)+4\pi r^3 P)^2}{r^3(r-2m)}-\frac{4\pi(P+\mathcal{E}+(9P+5\mathcal{E})/(dP/d\mathcal{E})}{dP/d\mathcal{E}}\Big) &=&0\;\;\;\;\nonumber\\
ry'(r)+y(r)^2+y(r)\Big(\frac{r-4\pi r^3(\mathcal{E}-P)}{(r-2m(r))}\Big)+
	r^2 \Big\{\frac{4\pi r \Big(\frac{P+\mathcal{E}}{dP/d\mathcal{E}}+9P+5\mathcal{E}-\frac{l(l+1)}{r^2}\Big)}{r-2m(r)}&&\nonumber\\
	-4\Big[\frac{m(r)+4\pi r^3 P}{r^2(1-2m(r)/r}\Big]^2\Big\} &=&0\;\;\;\;\nonumber\\
	ry'(r)+y(r)^2+y(r)F(r)+r^2 Q(r)&=&0 .\nonumber\\
	&&
	\label{yeq}
\end{eqnarray}
Here, $F(r)=\frac{r-4\pi r^3(\mathcal{E}-P)}{(r-2m(r))}$, and 
$Q(r)=\frac{4\pi r \Big(\frac{P+\mathcal{E}}{dP/d\mathcal{E}}+9P+5\mathcal{E}-\frac{l(l+1)}{r^2}\Big)}{r-2m(r)}-4\Big[\frac{m(r)+4\pi r^3 P}{r^2(1-2m(r)/r}\Big]^2 $.

Again, we can write the second-order differential equation \eqref{HPP} for 
outside the star $r=R$ regarding associated Legendre Legendre equation with 
m=2 \cite{thorn2,thib1}:
\begin{eqnarray}
	(x^2-1)H''(x)+2xH'(x)-\Big(l(l+1)+\frac{4}{x^2-1}\Big)H(x)=0 ,
	\label{HX}
\end{eqnarray}
where the prime stands for $d/dx$, $m(r=R)=M$, and $x=R/M-1$ is the 
independent variable. Therefore, the general solution of the Eq. \eqref{HX}
can be written as
\begin{eqnarray}
	H(x)=a_P \hat{P}_{l2}(x)+a_Q \hat{Q}_{l2}(x) .
	\label{hxx}
\end{eqnarray}
Where $a_P$, and $a_Q$ are the integration constants which determines through 
matching to the internal solution. Substituting the value of $y(r)$ into 
Eq. \eqref{hxx}, we get
\begin{eqnarray}
	y(x)=(1+x)\frac{\hat{P}'_{l2}(x)+a_l \hat{Q}'_{l2}(x)}{\hat{P}_{l2}(x)+a_l \hat{Q}_{l2}(x)},
\end{eqnarray}
where $a_l=a_Q/a_P$ determines by matching to the internal solution concerning the compactness $C=M/R$ of the star
\begin{eqnarray}
	a_l=-\frac{\hat{P}'_{l2}(x)-C y_l\; \hat{P}_{l2}(x)}{\hat{Q}'_{l2}(x)-C y_l\; \hat{Q}_{l2}(x)}\Big|_{x=R/M-1}.
\end{eqnarray}
On the other hand, $a_l$ can be related to dimensionless tidal Love number
$k_l$ \cite{thib1}. One can define as
\begin{eqnarray}
	k_l=\frac{1}{2}C^{2l+1} a_l ,
	\label{kl}
\end{eqnarray}
with $l=2,3,4,\dots$ are the quadrupole, octupole, and hexadecapole numbers.
For quadrupole Love number $k_2$, we have 
\begin{eqnarray}
	k_2&=&\frac{1}{2}C^{5} a_2\nonumber\\
&=& \frac{8}{5}(1-2C)^{2}C^{5}[2C(y_{2}-1)-y_{2}+2]\Big\{ 2C(4(y_{2}+1)C^{4}+(6y_{2}-4)C^{3}\nonumber\\
&+&(26-22y_{2})C^{2}+3(5y_{2}-8)C-3y_{2}+6)-3(1-2C)^{2}\nonumber\\
	&&(2C(y_{2}-1)-y_{2}+2)log\Big( \frac{1}{1-2C}\Big)\Big\}^{-1},   
	\label{love2}
\end{eqnarray}
%\end{widetext}
for octupole Love number $k_3$
%\begin{widetext}
\begin{eqnarray}
k_{3} &=& \frac{8}{7}(1-2C)^{2}C^{7}[2(y_{3}-1)C^{2}-3(y_{3}-2)C+y_{3}-3]
	\Big\{  2C[4(y_{3}+1) C^{5}+2(9y_{3}-2)C^{4}\nonumber\\
&-&20(7y_{3}-9)C^{3}+5(37y_{3}-72)C^{2}-45(2y_{3}-5)C
+15(y_{3}-3)]-15(1-2C)^{2}\nonumber\\
&&(2(y_{3}-1)C^{2}-3(y_{3}-2)C+y_{3}-3)log\Big(\frac{1}{1-2C}\Big)\Big\}^{-1},
	\label{love3}
\end{eqnarray}
%\end{widetext}
and for hexadecapole Love number $k_4$ can be written as
\begin{eqnarray}
k_{4} &=& \frac{32}{147}(1-2C)^{2}C^{9}[12(y_{4}-1)C^{3}-34(y_{4}-2)C^{2}
+28(y_{4}-3)C-7(y_{4}-4)]\nonumber \\
&&\Big\{ 2C[8(y_{4}+1)C^{6}+(68y_{4}-8)C^{5}+(1284-996y_{4})C^{4}+40(55y_{4}-116)C^{3}\nonumber\\
	&+&(5360-1910y_{4})C^{2}+105(7y_{4}-24)C-105(y_{4}-4)]-15(1-2C)^{2}[12(y_{4}-1)C^{3}\nonumber \\
	&-&34(y_{4}-2)C^{2}+28(y_{4}-3)C-7(y_{4}-4)] log\Big(\frac{1}{1-2C}\Big)\Big\}^{-1}.
	\label{love4}
\end{eqnarray}
\begin{center}
	{\bf \underline {\large Tidal deformability of neutron and hyperon star}}
\end{center}
%\section{Introduction}\label{sec1}
The detection of gravitational wave is a major breakthrough in astrophysics/cosmology which is detected for the first time by advanced Laser Interferometer 
Gravitational-wave Observatory (aLIGO) detector \cite{black}. Inspiraling and 
coalescing of binary black-hole results in the emission of gravitational waves.
We may expect that in a few years the forthcoming aLIGO \cite{black}, VIRGO 
\cite{virgo}, and KAGRA \cite{kagra} detectors will also detect gravitational 
waves emitted by binary neutron star (NSs). This detection is certainly posed 
to be a valuable guide and will help in a better understanding of a highly 
compressed baryonic system. Flanagan and Hinderer \cite{flan,tanja,tanja1} 
have recently pointed out that tidal effects are also potentially measurable 
during the early part of the evolution when the waveform is relatively clean. 
It is understood that the late inspiral signal may be influenced by tidal 
interaction  between binary stars (NS-NS), which gives the important 
information about the equation-of-state (EoS).
The study of Refs. \cite{baio,baio1,vines,pann,bd,dam,read,vines1,bd1,fava} 
inferred that the tidal effects could be measured using the recent generation 
of gravitational wave (GW) detectors.\\

In 1911, the mathematician A. E. H. Love \cite{Love} introduced the 
dimensionless parameter in Newtonian theory that is related to the tidal 
deformation of the Earth is due to the gravitational attraction between the 
Moon and the Sun on it. 
This Newtonian theory of tides has been imported to the general relativity 
\cite{dam,pois}, where it shows that the electric and magnetic type 
dimensionless gravitational Love number is a part of the tidal field associated with the gravitoelectric and gravitomagnetic interactions. The tidal 
interaction in a compact binary system has been encoded in the Love number 
and is associated with the induced deformation responded by changing shapes 
of the massive body. We are particularly interested for a NS in a close binary 
system, focusing on the various Love numbers $k_l$ ($l$=2, 3, 4) due to the 
shape changes (like quadrupole, octupole and hexadecapole in the presence of 
an external gravitational field). Although higher Love numbers ($l$=3, 4) give 
negligible effect, still these Love numbers ($k_{l}$) can have vital 
importance in future gravitational wave astronomy.  However, geophysicists are 
interested to calculate the surficial Love number $h_{l}$, which describes 
the deformation of the body's surface in a multipole expansion \cite{dam,pois,
phil}. \\

We have used the EoS from the relativistic mean field (RMF) \cite{walecka74,
pg89,ring96} and the newly developed effective field theory motivated RMF (E-RMF) \cite{furn96,furn97} approximation in our present calculations. Here, the 
degrees of freedoms are nucleon, $\sigma-,$ $\omega-$, $\rho-$, $\pi-$mesons 
and photon. This theory fairly explains the observed massive neutron star, 
like PSR J1614-2230 with mass $M=1.97\pm 0.04M_\odot$ \cite{demo10} and 
 PSR J0348+0432 ($M=2.01\pm 0.04 M_\odot$) \cite{antoni13}. 

The baryon octets are also introduced as the stellar system is in extraordinary
condition such as highly asymmetric or extremely high density medium 
\cite{bharat07}. The coupling constants for nucleon-mesons are fitted to 
reproduce the properties of a finite number of nuclei, which then predict not 
only the observables of $\beta-$stable nuclei, but also of drip-lines and 
superheavy regions \cite{walecka74,boguta77,rein86,gam90,rein88,sharma93,lala97}. 
The hyperon-meson couplings are obtained from the quark model \cite{oka,naka,naka1}. Recently, however, the couplings are improved by taking into 
consideration some other properties of strange nuclear matter \cite{naka2}.

\begin{table}
%	\hspace {0.1cm}
\begin{center}
\caption{Parameters and saturation properties for NL3 \cite{lala97},
G2 \cite{furn97}, FSUGold \cite{pieka05}, and FSUGold2 \cite{fsu2}. The
parameters $g_\sigma$, $g_\omega$, $g_\rho$, $k_{3}$, and $k_{4}$ are
calculated from nuclear matter the given saturation  properties using
the relations suggested by the authors of Ref. \cite {glen20}. }
\scalebox{0.7}{
{\begin{tabular}{ccccc}
\hline
\cline{1-5}
\hline
\Xhline{3\arrayrulewidth}
\multicolumn{1}{c}{Parameters}
&\multicolumn{1}{c}{NL3}
&\multicolumn{1}{c}{G2}
&\multicolumn{1}{c}{FSUGold}
&\multicolumn{1}{c}{FSUGold2}\\
\hline
$m_{n}$(MeV)&939&939&939&939\\
$m_\sigma$(MeV)&508.194&520.206&491.5&497.479\\
$m_\omega$(MeV)&782.501&782&783&782.5\\
$m_\rho$(MeV)&763&770&763&763\\
$g_\sigma$&10.1756&10.5088&10.5924&10.3968\\
$g_\omega$&12.7885&12.7864&14.3020&13.5568\\
$g_\rho$&8.9849&9.5108&11.7673&8.970\\
$k_{3}$(MeV)&1.4841&3.2376&0.6194&1.2315\\
$k_{4}$&-5.6596&0.6939&9.7466&-0.2052\\
$\eta_{1}$&0&0.65&0&0\\
$\eta_{2}$&0&0.11&0&0\\
$\eta_\rho$&0&0.390&0&0\\
$\zeta_{0}$&0&2.642&12.273&4.705\\
$\Lambda_{\omega}$&0&0&0.03&0.000823\\
\hline
	&&Nuclear matter properties&& \\
\hline
$\rho_{0}$(fm$^{-3})$&0.148&0.153&0.148&0.1505$\pm$0.00078\\
E/A(MeV)&-16.299&-16.07&-16.3&-16.28$\pm$0.02\\
K$_\infty$(MeV)&271.76&215&230&238.0$\pm$ 2.8\\
J(MeV)&37.4&36.4&32.59&37.62$\pm$1.11\\
L(MeV)&118.2&101.2&60.5&112.8$\pm$ 16.1\\
$m_{n}^*$/$m_{n}$&0.6&0.664&0.610&0.593$\pm$0.004\\
\Xhline{3\arrayrulewidth}
\end{tabular}\label{paratab1} }}
\end{center}
\end{table}

\section{Results and Discussions:}
In this section, we present the results of our calculations in Figs. 1-9 and
Table \ref{tidetab2}. Our calculated results of the EoS and related outcomes 
are discussed in the subsequent subsections.

\subsection{EoSs of neutron and hyperon star}
For a given Lagrangian density Eq. \eqref{EDF}, one can solve the equations 
of motion in the mean-field level, {\it i.e.} the exchange of mesons create a 
uniform meson field, where the nucleons moves in a simple harmonic motion. 
Then we calculate the energy-momentum tensor within the mean field 
approximation (i.e. the meson fields are replaced by their classical number) 
and get the EoS as a function of baryon density (Eqns. \eqref{eq20}, and 
\eqref{eq21} are derived in chapter \ref{chapter2}). 
The EoS remains uncertain at density larger than the saturation density of 
nuclear matter, $\rho_{n}\sim$3 $\times$ 10$^{14}$ g cm$^{-3}$.
 At these densities, neutrons can no longer be considered, which may consists 
mainly of heavy baryons (mass greater than nucleon) and several other species 
of particles expected to appear due to the rapid rise of the baryon chemical 
potentials \cite{baldo}. 
The $\beta$-equilibrium and charge neutrality are two important conditions for 
the neutron/hyperon rich-matter. Both these conditions force the stars to have 
$\sim$90$\%$ of neutrons and $\sim$10$\%$ protons. With the inclusion of baryons, the $\beta-$equilibrium conditions between chemical potentials for different 
particles are 
\begin{eqnarray}\label{beta}
\mu_p = \mu_{\Sigma^+}=\mu_n-\mu_{e}, \nonumber\\
\mu_n=\mu_{\Sigma^0}=\mu_{\Xi^0}=\mu_{n},\nonumber \\
\mu_{\Sigma^-}=\mu_{\Xi^-}=\mu_n+\mu_{e},\nonumber \\
\mu_{\mu}=\mu_{e}, \nonumber\\
\end{eqnarray}
and the charge neutrality condition is satisfied by  
\begin{eqnarray}\label{charge}
n_p+n_{\Sigma^+}=n_e+n_{\mu^-}+n_{\Sigma^-}+n_{\Xi^-}.
\end{eqnarray}
The EoS then gives the corresponding pressure and energy density (as discussed 
in chapter \ref{chapter2}) of the charge-neutral $\beta$-equilibrated NS matter 
(which includes the lowest lying octet of baryons). 
\begin{figure}[ht]
\begin{center}
\includegraphics[width=0.7\columnwidth]{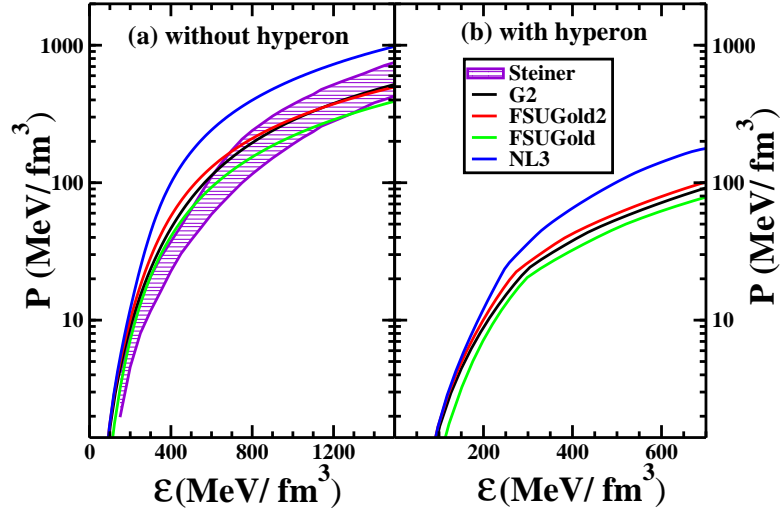}
\end{center}
\caption{The equations of state obtained for nuclear and hyper-nuclear matter 
under charge neutrality as well as the $\beta-$equilibrium condition for 
G2 \cite{furn97}, FSUGold2 \cite{fsu2}, FSUGold \cite{pieka05}, and 
NL3 \cite{lala97} force parameters are compared with the empirical data 
\cite{steiner10} (shaded area in the graph) for $r_{ph}=R$ with the 
uncertainty of $2\sigma$. Here, $R$ and $r_{ph}$ are the neutron radius and 
the photospheric radius, respectively. }
\label{fig1}
\end{figure}

Fig.\ref{fig1} displays the EoSs for G2 \cite{furn97}, FSUGold2 \cite{fsu2}, 
FSUGold \cite{pieka05} and NL3 \cite{lala97} parameter sets. From Fig. 
\ref{fig1}(a), it is obvious that all the EoSs follow a similar trend. Among 
these four, the celebrity NL3 set gives the stiffest EoS and the relatively 
new FSUGold represents the softer character. This is because of the large and 
positive $k_{4}$ value as well as the introduction of isoscalar-isovector 
coupling ($\Lambda_{\omega}$) in the FSUGold set \cite{pieka05}. To have an 
understanding on the softer and stiffer EoSs by various parametrizations, we 
compared their coupling constants and other parameters of the sets in Table 
\ref{paratab1}. We notice a large variation in their effective masses, 
incompressibilities and other nuclear matter properties at saturation. For 
higher energy density $\mathcal{E} \sim 500-1400$ MeV fm$^{-3}$, except 
NL3 set, which has the lowest nucleon effective mass, all other sets are 
found in the region of empirical data with the uncertainty of 2$\sigma$ 
\cite{steiner10}.

Fig. \ref{fig1}(b) shows a hump type structure on the nucleon-hyperon 
EoS at $\mathcal{E}$ around 400-500 MeV fm$^{-3}$. This kink 
($\mathcal{E}\sim$ 200-300 MeV) shows the presence of hyperons in the dense 
system. Here, the repulsive component of the vector potential becomes more 
important than the attractive part of the scalar interaction. As a result the 
coupling of the hyperon-nucleon strength gets weak. At a given baryon density, 
the inclusion of hyperons lower significantly the pressure compared to the 
EoS without hyperons. This is possible due to the higher energy of the 
hyperons, as the neutrons are replaced by the low-energy hyperons. The hyperon 
couplings are expressed as the ratio of the meson-hyperon and meson-nucleon 
couplings:
\begin{eqnarray}
\chi_{\sigma} = \frac {g_{Y\sigma}}{g_{N\sigma}}, \chi_{\omega} = \frac {g_{Y\omega}}{g_{N\omega}},\chi_{\rho} = \frac {g_{Y\rho}}{g_{N\rho}}.
\end{eqnarray}
In the present calculations, we have taken $\chi_{\sigma}$ = 0.7 and 
$\chi_{\omega}=\chi_{\rho} $ = 0.783. One can find similar calculations for 
stellar mass in Refs. \cite{glen20,mene,weis12,lope14}.

\subsection{Mass and radius of neutron and hyperon star}

Once the equations of state for various relativistic forces are fixed, then
we extend our study for the evaluation of the mass and radius of the isolated
neutron star. The Tolmann-Oppenheimer-Volkov (TOV) equations (as derived in
Appendix \ref{tov}) have to be solved for this purpose, where EoSs are the 
inputs. 

For a given EoS, the TOV equations must be integrated from the boundary
conditions $P(0) = P_{c}$, and $M(0)=0$, where $P_{c}$ and $M(0)$ are the 
pressure and mass of the star at $r=0$ and the value of $r$(= $R$), where the 
pressure vanish defines the surface of the star. Thus, at each central density 
we can uniquely determine a mass $M$ and a radius $R$ of the static neutron 
and hyperon stars using the four chosen EoSs. The estimated result for the 
maximum mass as a function of radius compared with the highly precise 
measurements of two massive ($\sim 2 M_\odot$) neutron stars \cite{demo10,antoni13} and extraction of stellar radii from x-ray observation \cite{steiner10}, 
are shown in Figs. \ref{fig2}(a) and \ref{fig2} (b) . From recent observations 
 \cite{demo10,antoni13}, it is clearly illustrated that the maximum mass 
predicted by any theoretical model should reach the limit $\sim 2.0 M_\odot$, 
which is consistent with our present prediction from the G2 EoS of nucleonic 
matter compact star with mass 1.99$M_\odot$ and radius 11.25 km. From x-ray 
observation, Steiner {\it et al.} \cite{steiner10} predicted that the 
most-probable neutron star radii lie in the range 11-12 km with neutron star 
masses$\sim$1.4$M_\odot$ and predicted that the EoS is relatively soft in the 
density range 1-3 times the nuclear saturation density. As explained earlier, 
stiff EoS like NL3 predicts a larger stellar radius 13.23 km and a 
maximum mass 2.81 $M_\odot$. Though FSUGold and FSUGold2 are from the same 
RMF model with similar terms in the Lagrangian, their results for NS are quite 
different with FSUGold2 suggesting a larger and heavier NS with mass  
2.12$M_\odot$ and radius 12.12 km compare to mass and radius (1.75$M_\odot$ 
and 10.76 km) of the FSUGold. Because in FSUGold2 EoS at high densities, the 
impact comes from the quartic vector coupling constant $\zeta_{0}$ and also 
the large value of the slope parameter $L=112.8\pm16.1$ MeV (see Table \ref{paratab1}) tend to predict the NS with large radius \cite{horowitz01a}. From the 
observational point of view, there are large uncertainties in the determination of the radius of the star \cite{rute,gend,cott}, which is a hindrance to get 
a precise knowledge on the composition of the star atmosphere. One can see 
that G2 parameter is able to reproduce the recent observation of 2.0$M_\odot$ 
NS. But the presence of hyperon matter under $\beta$-equilibrium soften the 
EoS, because they are more massive than nucleons and when they start to fill 
their Fermi sea slowly replacing the highest energy nucleons. Hence, the 
maximum mass of NS is reduced by $\sim$0.5 unit solar mass due to the high 
baryon density. For example, the stiffer NL3 EoS gives the maximum NS mass 
$\sim$2.81$M_\odot$ and the presence of hyperon matter reduces the mass to 
$\sim$2.25$M_\odot$ as shown in Fig. \ref{fig2}(b).

\begin{figure}[ht]
\begin{center}
\includegraphics[width=0.7\columnwidth]{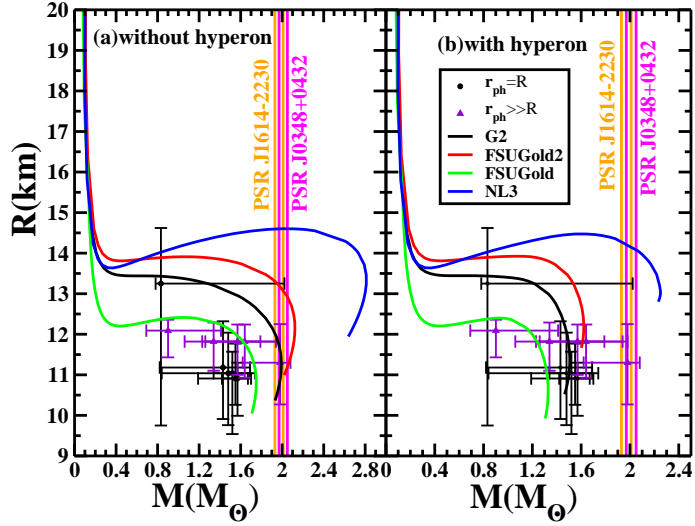}
\end{center}
\caption{The mass-radius profile for the force parameters like G2 \cite{furn97}, FSUGold2 \cite{fsu2}, FSUGold \cite{pieka05} and NL3 \cite{lala97} used. The 
solid circles($r_{ph}=R$) and triangles($r\gg R$) are represent the 
observational constraints \cite{steiner10}, where $r_{ph}$ is the photospheric 
radius. The verticle sheded region corresponds to the recent observation \cite{demo10,antoni13}.}
\label{fig2}
\end{figure}
These results give us warning that most of the present sets of hyperon 
couplings are unable to reproduce the recently observed mass of NS like 
PSR J1614-2230 with mass  $M = 1.97\pm 0.04M_{\odot}$ \cite{demo10} and 
the PSR J0348+0432 with  $M = 2.01\pm 0.04M_{\odot}$ \cite{antoni13}. 
Probably, this suggest us to modify the coupling constants and get the 
EoS properly, so that one can explain all the mass-radius observations to date. Further, one can see that in Fig. \ref{fig2}(b) mass-radius curve of G2, 
FSUGold, FSUGold2 with hyperon lies in the range of the predicted EoS between 
the $r_{ph}$=R and $r_{ph}\gg$R cases is the high density behavior \cite{steiner10}.

\subsection{ Various tidal Love number of compact star}\label{Love}

To estimate the Love numbers $k_l$ ($l$=2, 3, 4), along with the evaluation
of the TOV equations, we have to compute $y = y_{l}(R)$ with initial
boundary condition $y(0)=l$ from the following first order differential 
equation \eqref{yeq} iteratively \cite{tanja,tanja1,thib1,pois}: 
Once, we know the value of $y= y_{l}(R)$, the electric tidal Love numbers 
$k_{l}$ are found from the Eq. \eqref{kl}.

\begin{figure}[ht]
\begin{center}
\includegraphics[width=0.7\columnwidth]{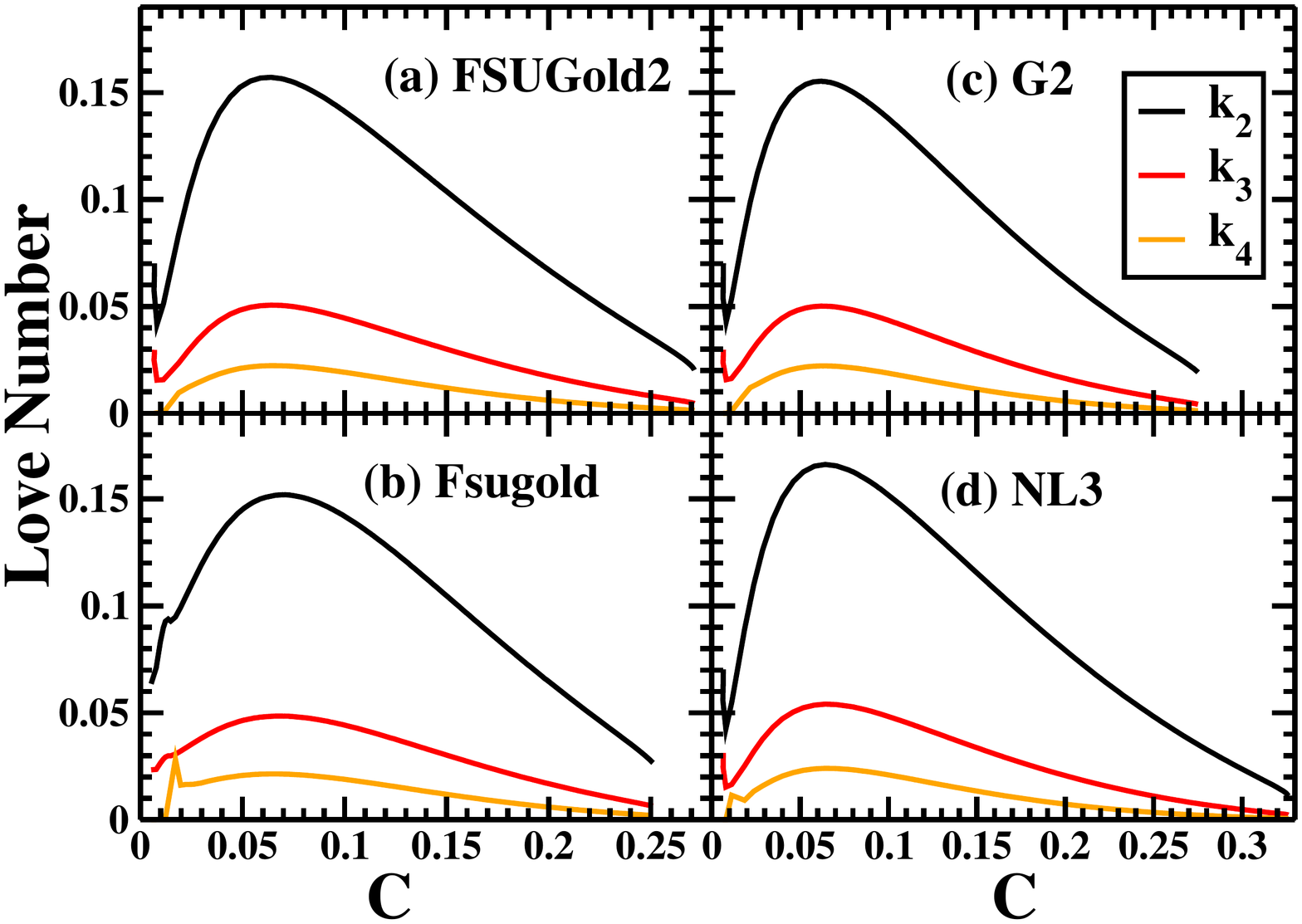}
\end{center}
\caption{The tidal Love numbers $k_{2},k_{3},k_{4}$ as a function of the mass 
of the four selected EoSs of the neutron star.}
\label{fig3}
\end{figure}
\begin{figure}[ht]
\begin{center}
\includegraphics[width=0.7\columnwidth]{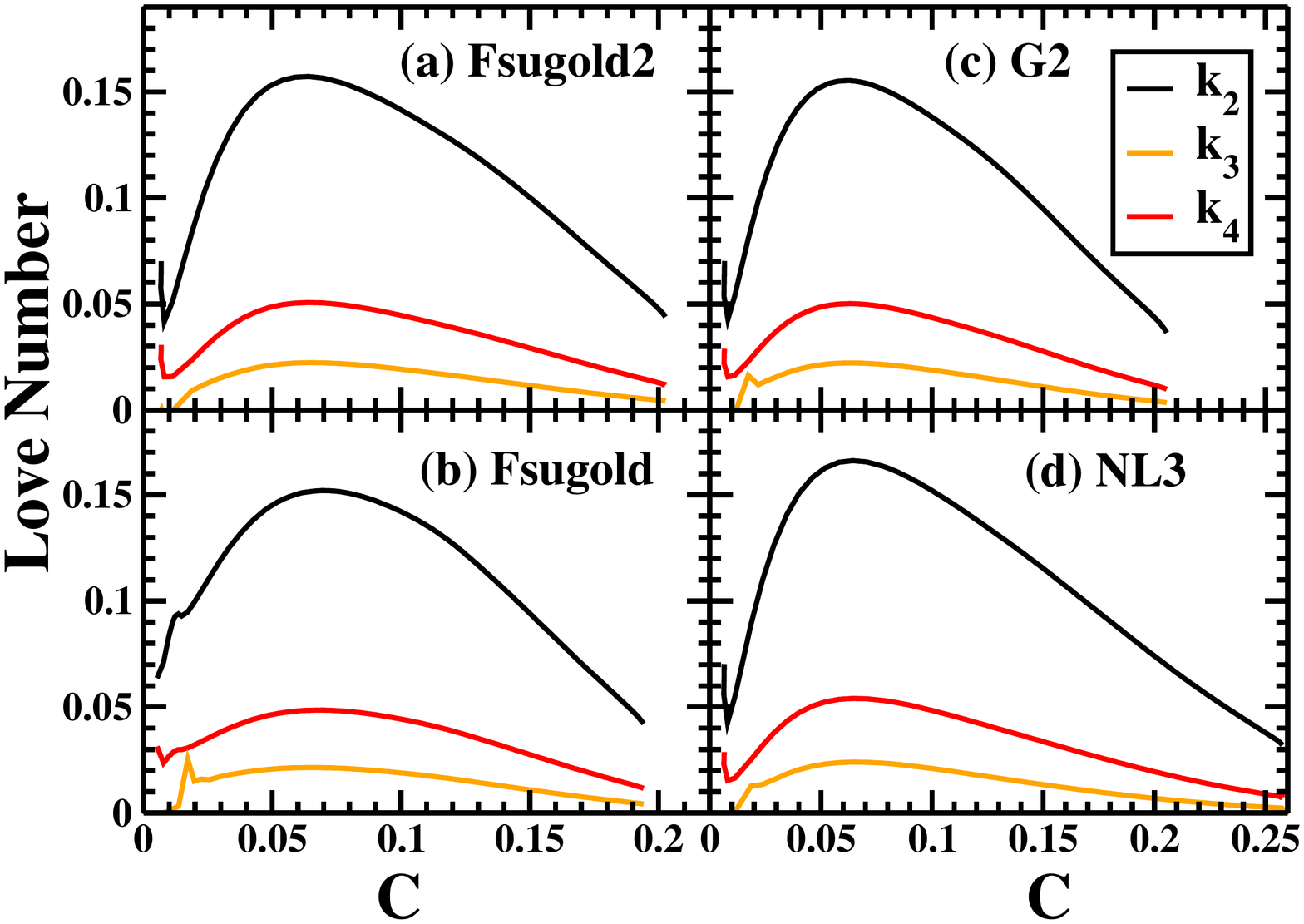}
\end{center}
\caption{Same as Fig. \ref{fig3} but for hyperon star.}
\label{fig4}
\end{figure}
As we have emphasized earlier, the dimensionless Love number $k_{l}$ 
($l=2, 3, 4$) is an important quantity to measure the internal structure of 
the constituent body. These quantities directly enter into the gravitational 
wave phase of inspiralling binary neutron stars (BNS) and extract the 
information of the EoS. Notice that Eqns. \eqref{love2}-\eqref{love4} contain 
an overall factor $(1- 2C)^{2}$, which tends to zero when the compactness 
approaches the compactness of the black hole, {\it i.e.} $C^{BH}$=1/2 \cite{thib}. 
Also, it is to be pointed out that with the presence of multiplication order 
factor $C$ with $(1- 2C)^{2}$ in the expression of $k_l$ that the value of the 
Love number of a black hole simply becomes zero, {\it i.e.}, $k^{BH}_{l}$=0.

\begin{table}
\hspace{0.1 cm}
\caption{Properties of a 1.4$M_\odot$ neutron  and hyperon star for 
different  classes of the EoS. The quadrupolar tidal deformability $\lambda$ 
and uncertainty error $\Delta\tilde{\lambda}$ in (10$^{36}$g cm$^{2}$ s$^{2}$).}
\scalebox{0.7}{
{\begin{tabular}{ccccccccccccccccccccccc}
\hline
\hline
 \multicolumn{13}{ c }{Neutron Star}\\
%\cline{1-13}%\cline{6-7}\cline{7-10}
\Xhline{3\arrayrulewidth}
EoS &       $R(km)$   &       $C$ &$f_{c}$(Hz)  &       $k_{2}$       &       $k_{3}$        &       $k_{4}$       &  $h_{2}$ & $h_{3}$ &$h_{4}$&     $\lambda$
&$\Delta\tilde{\lambda}$&$\Lambda$\\
\Xhline{3\arrayrulewidth}
%&&&&&&&&&&&&\\
NL3&14.422&0.144&1256.7&0.1197&0.0353&0.0142&0.9775&0.6519&0.5074&7.466&2.027
&1288.81\\
G2&13.148&0.157&1440.9&0.0934&0.0265&0.0103&0.8879&0.5951&0.4596&3.668&1.486
&652.76\\
FSUGold2&13.850&0.149&1332.4&0.1040&0.0301&0.0119&0.9275&0.6237&0.4854&5.299&1.763&944.08\\
FSUGold&12.236&0.170&1608.0&0.0882&0.0244&0.0071&0.8589&0.5634&0.4268&2.418&1.178&414.13\\
\multicolumn{13}{ c }{Hyperon Star}\\
%\cline{1-13}
%&&&&&&&&&&&&\\
NL3&14.430&0.143&1252.9&0.1203&0.0355&0.0143&0.9800&0.6541&0.5096&7.527&2.018&1341.20\\
G2&12.686&0.163&1520.6&0.0804&0.0229&0.0088&0.8434&0.5707&0.4399&2.641&1.321&465.83\\
FSUGold2&13.690&0.151&1355.9&0.0988&0.0287&0.0113&0.9108&0.6154&0.4789&4.750&1.696&839.04\\
(FSUGold)$_{1.3M_\odot}$&9.922&0.194&2119.0&0.0421&0.0116&0.0042&0.6884&0.4683&0.3518&0.4048&0.530
&102.14\\
\Xhline{3\arrayrulewidth}
\end{tabular}\label{tidetab2}}}
\end{table}

Fig. \ref{fig3} shows the tidal Love numbers $k_{l}$ ($l$=2, 3, 4) as a 
function of compactness parameter $C$ for the NS with four selected EoSs. The 
result of $k_{l}$ suddenly deceases with increasing compactness 
($C = 0.06-0.25$). For each EoS, the value of $k_{2}$ appears to be maximum 
between $C=0.06-0.07$. However, we are mainly interested in the NS 
masses at $\sim$1.4$M_\odot$.  Because of the tidal interactions in the NS 
binary, the shape of the star acquires quadrupole, octupole, hexadecapole and 
other higher order deformations. The value of the Love numbers for 
corresponding shapes are shown in Table \ref{tidetab2}. The values of $k_l$ 
decreases gradually with increase of multipole moments. Thus, the quadrupole 
deformibility has maximum effects on the binary star merger.   
Similarly, in Fig. \ref{fig4}, the dimensionless Love number $k_{l}$ is shown 
as a function of compactness for the hyperon star. With the inclusion of 
hyperons, the effect of the core is negligible due to the softness of the 
EoSs. The values of $k_{l}$ is different for a typical neutron-hyperon star 
with mass 1.4 $M_{\odot}$ for various sets are listed in the lower portion of
Table \ref{tidetab2}. The radius and respective mass-radius ratio is also 
given in the Table \ref{tidetab2}. The table also reflects that the Love 
numbers decrease slightly or remains unchanged with the addition of hyperon 
in the NS. The NS surface or solid crust is not responsible for any tidal
effects, but instead it is the matter mainly in the outer core that gives the 
largest contribution to the tidal Love numbers. It is relatively unaffected by 
changing the composition of the core. Thus instigate the calculation for the 
surficial Love number $h_l$ for both neutron and hyperon star binary. 
  
\begin{figure}[ht]
\begin{center}
\includegraphics[width=0.7\columnwidth]{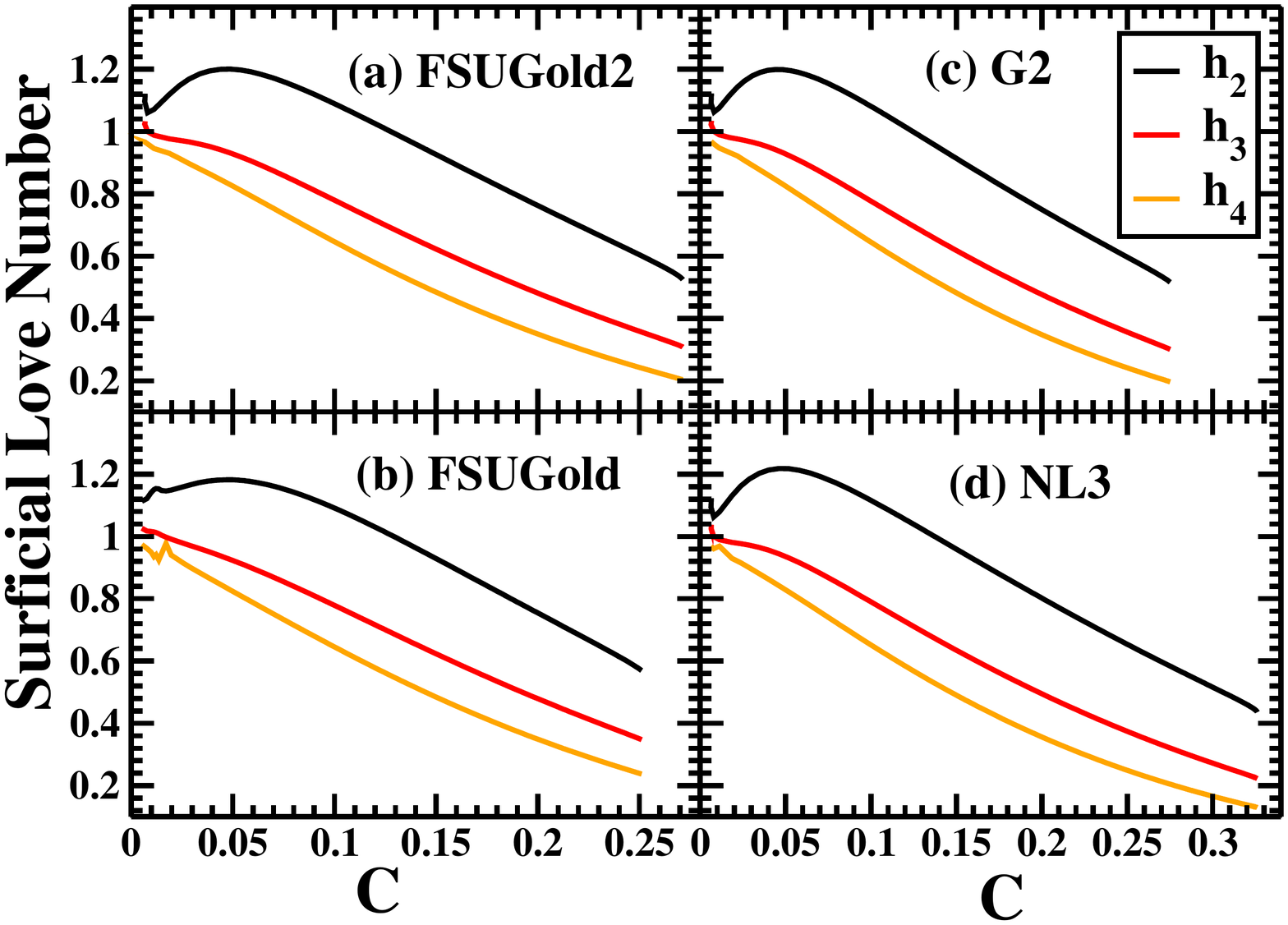}
\end{center}
\caption{Surficial Love number $h_{l}$ as a function of compactness $C$ of a 
neutron star, for selected values of $l$.}
\label{fig5}
\end{figure}
\begin{figure}[ht]
\begin{center}
\includegraphics[width=0.7\columnwidth]{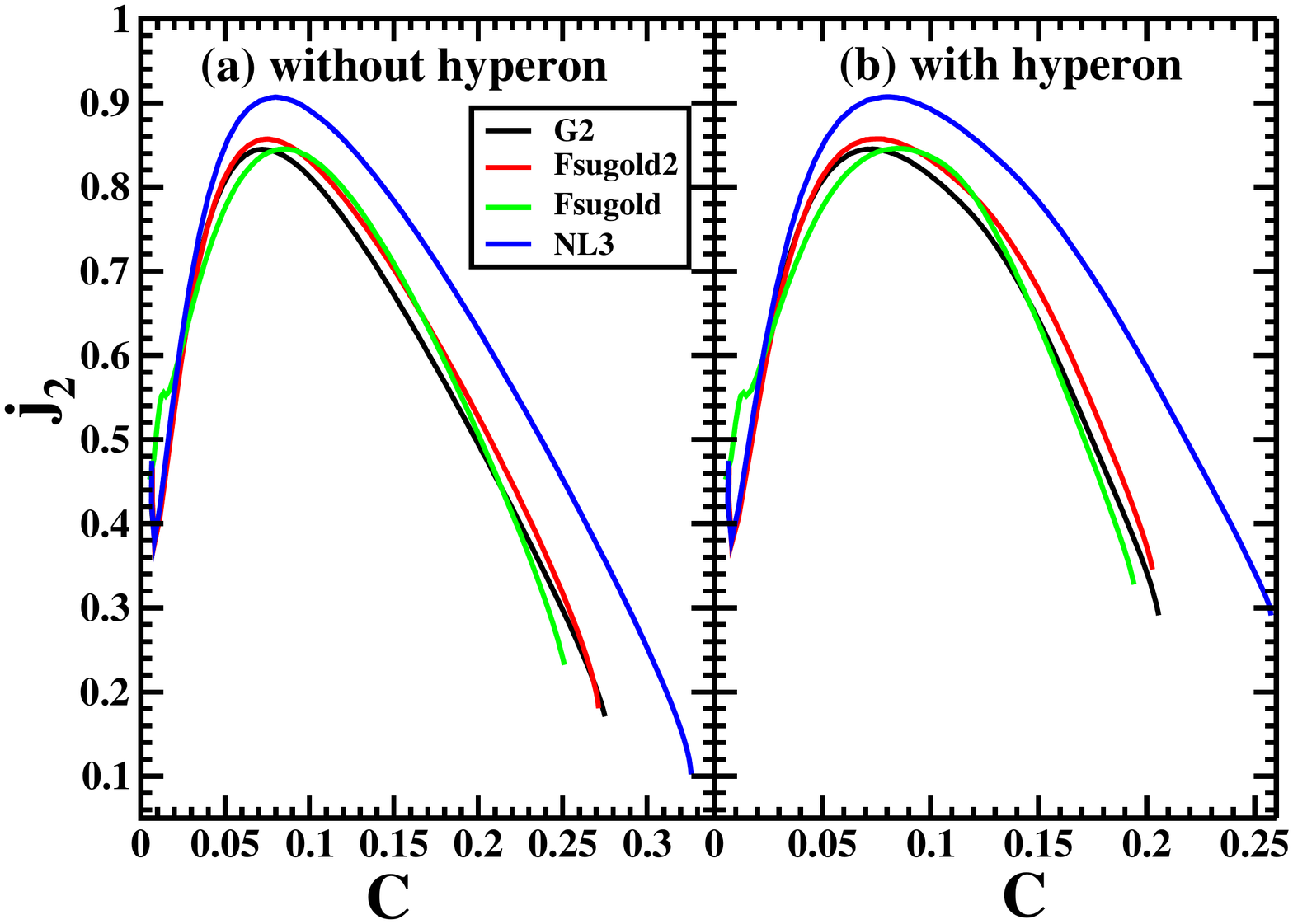}
\end{center}
\caption{The magnetic tidal Love number for selected EoSs. }
\label{fig6}
\end{figure}
Next, we calculate the surficial Love number $h_{l}$ which describes the 
deformation of the body's surface in a multipole expansion. Recently, Damour 
and Nagar \cite{thib} have given the surficial Love number
(also known as shape Love number) $h_{l}$ for the coordinate displacement 
$\delta R$ of the body's surface under the external tidal force. Alternatively, Landry and Poisson \cite{phil} have proposed the definition of Newtonian Love 
number in terms of a curvature perturbation $\delta \mathcal{R}$ instead of a 
surface displacement $\delta R$. For a perfect fluid, the relation between the 
surficial Love number $h_{l}$ and tidal Love number $k_{l}$ is given as
\begin{eqnarray}
  h_{l}=\Gamma_{1}+2 \Gamma_{2} k_{l},
\end{eqnarray}
where
\begin{eqnarray}
 \Gamma_{1}=\frac{l+1}{l-1}(1-C)F(-l,-l,-2l;2C)-\frac{2}{l-1}F(-l,-l-1,-2l;2C), \nonumber 
\end{eqnarray}
and
\begin{eqnarray}
\Gamma_{2}=\frac{l}{l+2}(1-C)F(l+1,l+1,2l+2;2C)\nonumber
+\frac{2}{l+2}F(l+1,l,2l+2;2C).
\end{eqnarray}
where F(a,b,c;z) is the hypergeometric function. Figure \ref{fig5} shows the 
results of surficial Love number $h_l$ of a NS as a function of compactness 
parameter C. Unlike the initially increasing and then decreasing trend of the 
tidal Love number $k_l$, the surficial Love number $h_l$ decreases almost 
exponentially with the compactness parameter. At the minimum value of the 
compactness parameter, the maximum value of the shape Love number of each 
multipole moment approaches 1. Thus, we zero in on to the Newtonian relation 
i.e., $h_{l}=1+2k_{l}$. Again one can compute from Table \ref{tidetab2} 
that the surficial Love number $h_l$ decreases $\sim 20 \%$ from one moment 
to another. For example, $h_2=0.9775$ and $h_3=0.6519$ and $h_4=0.5074$ for 
NL3 parameter sets.

Furthermore, we also calculate the ``magnetic'' tidal Love number $j_l$. Here,
we give only the quadrupolar case ($l=2$), which is expressed as: 
\begin{eqnarray}
j_{2}=\Big \{96 C^{5}(2C-1)(y_{2}-3)\Big\}\Big\{5(2C(12(y_{2}+1)C^{4}
\nonumber \\+2(y_{2}-3)C^{3}+2(y_{2}-3)C^{2}+3(y_{2}-3)C-3y_{2}+9)
\nonumber \\+3(2C-1)(y_{2}-3)log(1-2C))\Big\}^{-1}. 
\end{eqnarray}
After inserting the value of $y_{2}$ in eq. \eqref{love2}, we compute the  
magnetic tidal Love number $j_{2}$ in a hydrostatic equilibrium condition for 
a nonrotating NS. This gives important information about the internal structure \cite{pois} without changing the tidal Love number $k_{2}$. At $C$=0.01, the 
magnetic Love number $j_{2}$ is nearly 0.4. In both cases (with and without 
hyperons), $j_{2}$ is maximum within the compactness 0.06 to 0.07 for all the 
four EoSs (See Fig. \ref{fig6}). Then the value of $j_2$ decreases sharply 
with an increase of compactness. The NL3 parameter set gives a maximum 
$j_{2}$ in both the systems, while the rest of the three sets predict 
comparable $j_2$.

\begin{figure}[ht]
\begin{center}
\includegraphics[width=0.7\columnwidth]{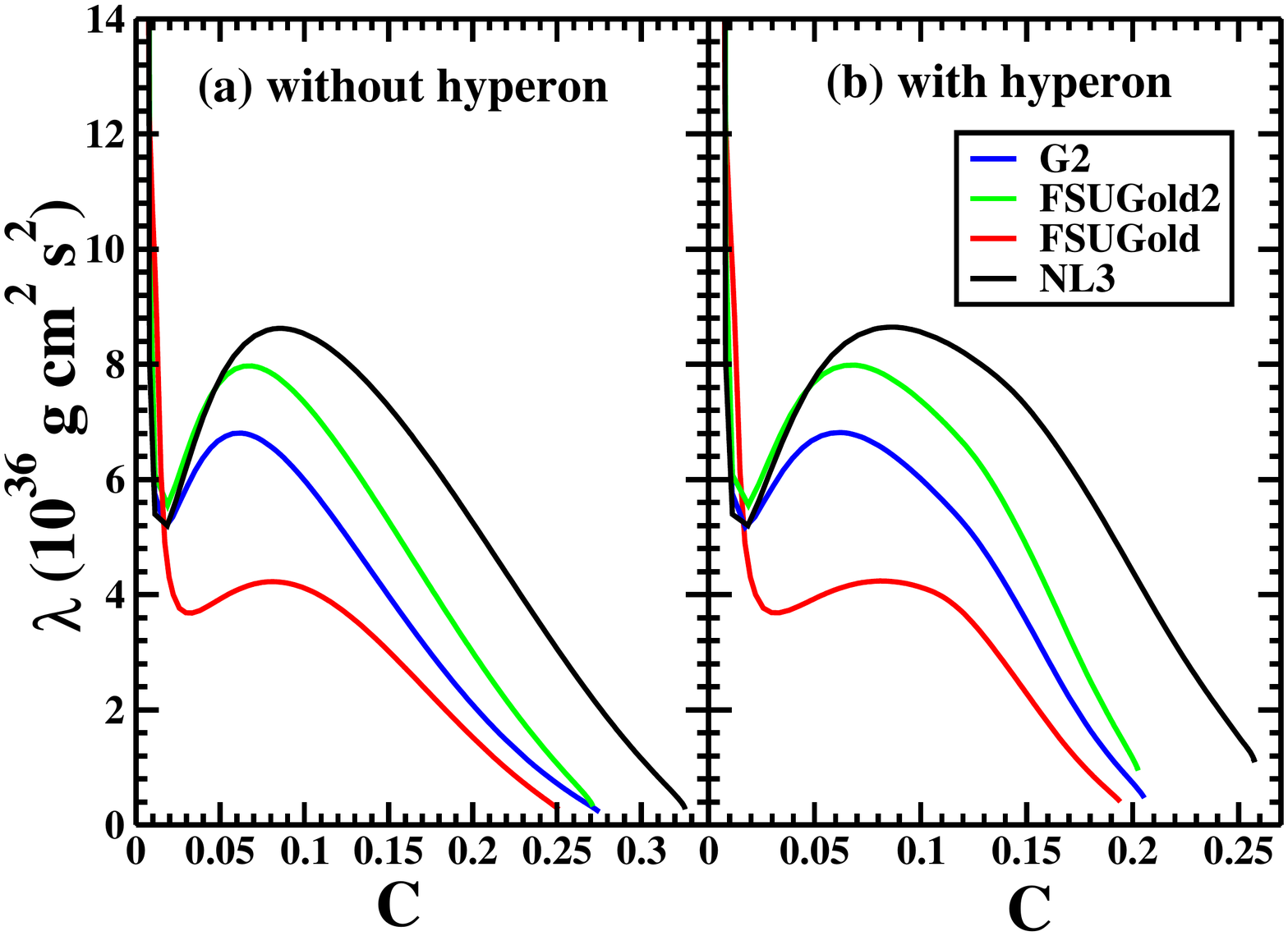}
\end{center}
\caption{The tidal deformability $\lambda$ as a function of the compactness $C$ for the four EoS with and without hyperon. }
\label{fig7}
\end{figure}
\begin{figure}[ht]
\begin{center}
\includegraphics[width=0.7\columnwidth]{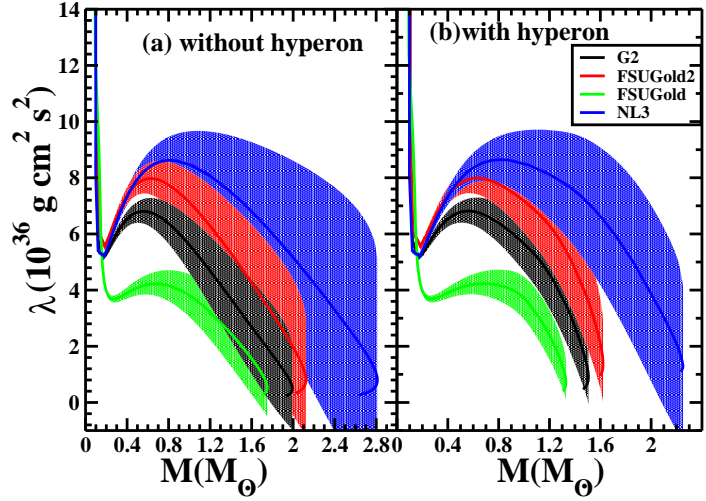}
\end{center}
\caption{Tidal deformability $\lambda$ of a single NS as a function of the 
neutron-star mass for a range of EoSs. The estimate of uncertainties in 
measuring $\lambda$ for equal mass binaries at a distance of $D = 100$ Mpc is 
shown for the aLIGO detector in shaded area. (b) Same as (a), but for hyperon 
star.}
\label{fig8}
\end{figure}
\begin{figure}[ht]
\begin{center}
\includegraphics[width=0.7\columnwidth]{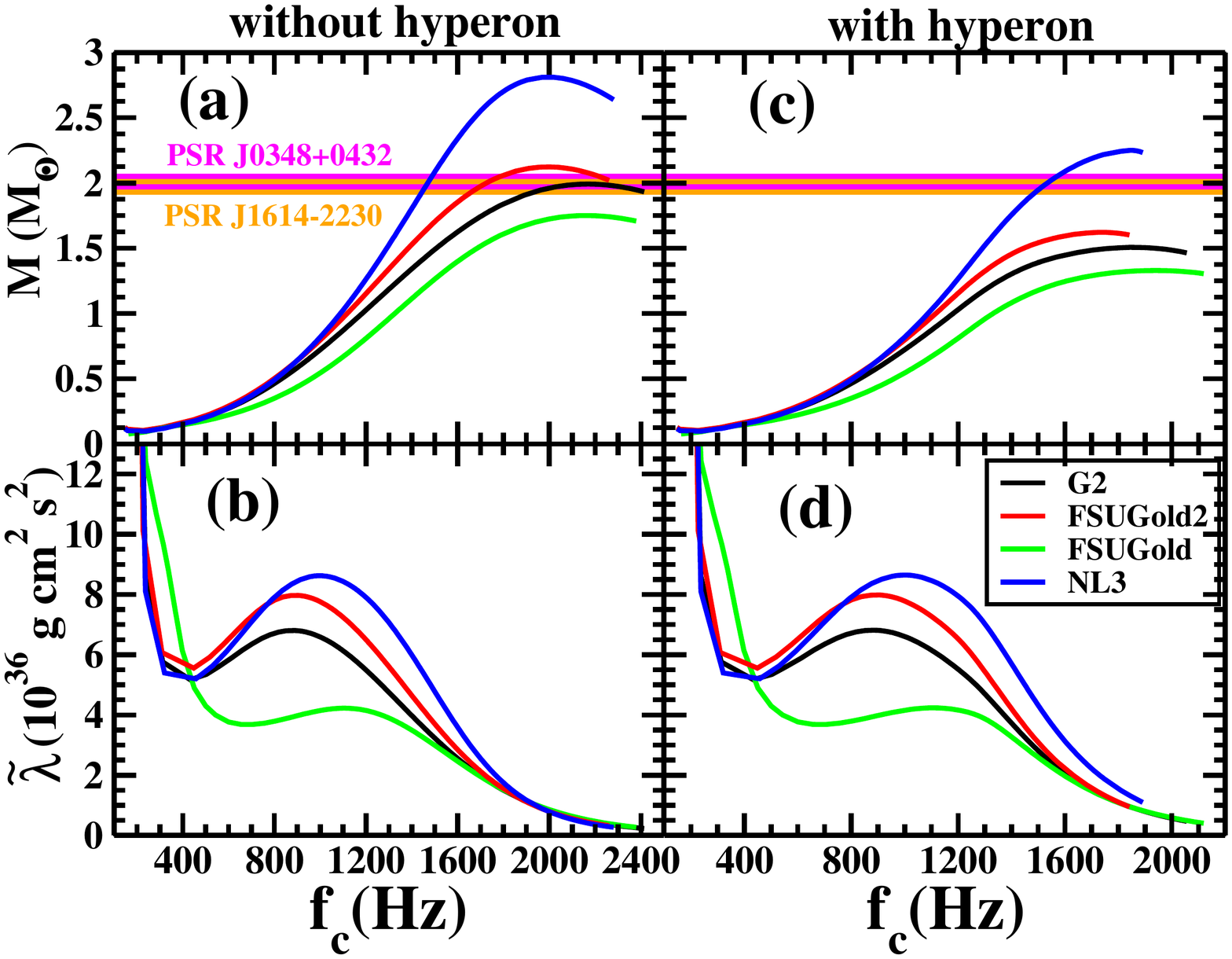}
\end{center}
\caption{(a),(c) The mass-cut-off frequency $f_{c}$ profile of normal and 
hyperon stars using four EoSs. (b) The tidal deformabilty with cut-off 
frequency plot of the neutron star. (d) Same as (b), but for hyperon star.} 
\label{fig9}
\end{figure}
\subsection{Tidal deformability and cut-off frequency of compact star}\label{tidal}
To examine the results of tidal deformability with and without hyperons, we 
have shown the tidal deformability $\lambda$ as a function of $C$ in 
Fig. \ref{fig7}, where we have considered a single NS under the influence of 
an external tidal field with adiabatic approximation using the four EoSs. In 
this case, the orbital evolution time 
scale is much larger than the time scale needed to assume the star as a 
stationary configuration. From the very beginning, we mark an infinitely large 
$\lambda$ corresponding to a small compactness i.e., $C\sim$0.02. Further, the 
$\lambda$ value falls to a minimum that rises again resulting in a hump like 
pattern for each EoS. It is noteworthy that in Fig. \ref{fig7}(b) by 
introducing the NL3 case with hyperon, there is remarkable but mere deviation 
in $\lambda$ value i.e 7.527 g cm$^{2}$ s$^{2}$ ( without hyperon $\lambda$= 7.466 g cm$^{2}$ s$^{2}$). Since, the tidal deformability $\lambda$ is a surface 
phenomenon, it is very much affected by the radius of the star in both neutron 
star and hyperon star. Thus, the tidal deformability $\lambda$ becomes highly 
sensitive on the radius $R$ even though $k_2$ is small. We estimate the radii 
to be within 12.236$-$14.422 km for a NS of mass 1.4$M_\odot$ and the range is 
13.690$-$14.430 km for neutron-hyperon star for all the four stiff or soft 
EoS (see Table \ref{tidetab2}).
 
Figure \ref{fig8}, shows the tidal deformability for both neutron and hyperon 
stars. We have a large radius for a smaller stellar mass of $\sim 0.1 M_\odot$ 
in both cases. At this value of mass and radius, the tidal deformability 
$\lambda$ becomes maximum, because for a large radius with smaller mass, the 
force of attraction within the star is weak and when another star comes closure, the gravitational pull over ride maximum at the surface part of the star. 
This phenomenon is true for both neutron as well as hyperon stars \cite{tanja,tanja1}. Then, suddenly the tidal deformibility decreases and again increases as 
shown in the figure making a broad peak at around $M=0.7-0.8 M_\odot$ and then 
decreases smoothly with increase in the mass of the star. Since, the tidal 
deformibility depends a lot on both mass and radius of a NS, it is imperative 
to measure the radius of the star precisely, as the mass is already measured 
with very good precession. Recently, Steiner {\it et al.}, predicted the most 
extreme limit for the tidal deformabilities between $0.6\times$ 10$^{36}$ and 
$6\times$ 10$^{36}$ g cm$^{2}$ s$^{2}$ for 1.4$M_\odot$ with 95$\%$ confidence. This range can be constrained on high dense matter of any measurements 
\cite{ste15}. Mostly, the 
binaries masses are about 1.4$M_\odot$, so in particular we are interested in 
studying the phenomena within this mass range and the results are summarize in 
Table \ref{tidetab2}. Comparing the results, we notice that the tidal 
deformability $\lambda$ is quite sensitive to the EoS. It is more for stiffer 
EoS, because of the high-density behavior of the symmetry energy \cite{fattev}.

Finally, we calculate the weighted tidal deformability (Eq. \eqref{avtide}}) of 
the binary neutron star of mass $m_{1}$ and $m_{2}$ and the root mean square 
(rms)  measurement uncertainty $\Delta\tilde{\lambda}$ can be calculated following approximate formula \cite{tanja,tanja1}:
\begin{eqnarray}
\Delta\tilde{\lambda}\approx \alpha \left(\frac{M}{M_\odot}\right)^{2.5}
\left(\frac{m_{2}}{m_{1}}\right)^{0.1}\left(\frac{f_{cut}}{Hz}\right)^{-2.2}
\left(\frac{D}{100 Mpc}\right), 
\end{eqnarray}
where $\alpha$ = 1.0 $\times$ 10$^{42}$ g cm$^2$ s$^2$ is the tidal 
deformability for a single Advanced LIGO detector and $f_{cut}$ ($f_{end}$) 
cutoff frequency \cite{dam} for the end stage of the inspiral binary neutron 
stars. D denotes the luminosity distance from the source to observer.

The weighted tidal deformibility for neutron and hyperon stars and their 
corresponding masses as cut-off frequency $f_{cut}$ is shown in Fig. 
\ref{fig9}. The cut-off frequency is a stopping criterion to estimate when the 
tidal model no longer describes the binary. Here, we take the cut-off to be 
approximately when the two neutron stars come into contact, estimated as in 
Eq.36 of Ref. \cite{dam}. Specifically, we use $f_{cut} = 2 f_{orb.}^{N(R_{1}+
R_{2})}$, where $f_{orb.}^{N(R_{1}+R_{2})}$ is the Newtonian orbital frequency
corresponding to the orbital separation where two unperturbed neutron stars
with radii $R_{1}$ and  $R_{2}$ would touch. In the upper panel 
Fig. \ref{fig9}(a), \ref{fig9}(c), it shows the variation of mass of the 
binary as a function of cut-off frequency $f_{cut}$. Here, we considered 
$m_1=m_2$, i.e., both the masses of the binary are equal. Initially, the 
masses of the stars 0.2$M_\odot$ remain almost constant up to 
$f_{cut}\approx 400$ Hz. Then the mass increases nearly exponentially up to
a maximum mass of $\approx$1.75$-$2.81$M_\odot$ (for NS) and  $\approx$1.33$-$2.25$M_\odot$ (for hyperon star) and then decreases. By this time, the cut-off 
frequency $f_{cut}$ attains quite large value. When the individual mass of the 
binary is 1.4$M_\odot$, the NL3 set weighted tidal deformibility achieves the 
cut-off frequency $f_{cut}\approx 1256.7$ Hz as the minimum contrary to the 
$f_{cut}\approx 1608.0$ Hz of FSUGold at the same mass of the single NS. 
It is also clear from the figure that the weighted tidal deformability 
of the NS for the four models are 7.466, 3.668, 5.229 and 2.418 for NL3, G2, 
FSUGold2 and FSUGold, respectively, with the corresponding frequencies 1256.7,
1440.9, 1332.4 and 1608.0 Hz. 

Using the cut-off frequency, we calculate the uncertainty in the measurement 
of the tidal deformability ($\Delta\tilde{\lambda}$) obtained from these four 
EoSs for an equal-mass binary star inspiral at 100 Mpc from aLIGO detector 
(shaded region in Fig. \ref{fig8}). The uncertainty in the lower mass region 
(0.4$-$1.0$M_\odot$) of the NS $\Delta\tilde{\lambda}$ is smaller. Similar 
results are found in the case of hyperon star also. Interestingly, the error 
($\Delta\tilde{\lambda}$) increases with increase the mass of the binary for 
all the EoSs. From Table \ref{tidetab2}, by comparing the $\Delta\tilde{\lambda}$ obtained from all the EoSs, we find that predicted errors are greater than 
the measured value for a star of mass 1.4$M_\odot$.

\section{Summary and Conclusions}\label{conc}
In summary, four different models have been extensively applied which are
obtained from effective field theory motivated relativistic mean field 
formalism. This effective interaction model satisfies the nuclear 
saturation properties and reproduce the bulk properties of finite nuclei with 
a very good accuracy. We used these four forces of interaction and calculate 
the EoS for neutron and hyperon stars matter. It is noteworthy that each term 
of the interaction has its own meaning and has specific character. The 
inclusion of extra terms (nucleons replaced by baryons octet) in the 
Lagrangian contribute to soften the EoS and the matter becomes less 
compressible. Hence, there is a decrease in the maximum mass by 
$\sim 0.5M_{\odot}$ than the pure neutron star. 

We have extended our calculations to various tidal responses both for 
electric-type (even-parity) and magnetic-type (odd-parity) of neutron and 
hyperon stars in the influence of an external gravitational tidal field. The 
Love numbers are directly connected with surficial Love number $h_{l}$ 
associated with the surface properties of the stars. Subsequently, we study 
the quadrupolar tidal deformability $\lambda$ of normal neutron star and 
hyperon star using different set of equations of state. These tidal 
deformabilities particularly depend on the quadrupole Love number $k_{2}$ and 
radius ($R$) of the isolated star. Although the maximum value of $k_{2}$ is 
not very sensitive to the EoS for neutron and hyperon stars lying in the range 
$k_{2}\approx 0.144-0.170$ and $0.143-0.194$ for neutron and hyperon stars, 
respectively, but it is very much sensitive to the radius of the star. 

We find that aLIGO can constrain on the existence of hyperon star, i.e., the 
inner core of the NS has hyperons, but detecting them can be much harder. 
However, it should be able to constraint the neutron star deformability to 
$\lambda\leq$10 $\times$ 10$^{36}$ g cm$^{2}$ s$^{2}$ for a binary of 
1.4$M_\odot$ neutron stars at a distance 100 Mpc from the detector. In future, 
we expect that aLIGO should be able to measure $\lambda$ even for neutron 
stars masses up to 2.0$M_\odot$ and consequently  constrain the stiffness of 
the equations of state. It is worth mentioning that the present calculations 
are based on the extrapolation of the formula $\Delta \tilde \lambda$ given in 
Ref. \cite{tanja,wade,lsc}. Here, the systematic uncertainties in the model 
that was used to obtain the measurability estimates are neglected.

 \setcounter{equation}{0}
 \setcounter{figure}{0}

\newpage
\chapter{New parameter sets G3, and IOPB-I }
\label{chapter6}

In this chapter, we have discussed the newly developed relativistic forces, G3, 
and the Institute of Physics Bhubaneswar-I (IOPB-I), which are broadly 
applicable for the study of finite nuclei, infinite nuclear matter, and 
neutron star. Using these forces, we calculate the binding energies, charge 
radii, isotopic shift, and neutron-skin thickness for some selected nuclei. 
From the ground-state properties of superheavy nuclei (Z=120), it is 
noticed that considerable shell gaps appear at neutron numbers N=172, 184, and 
198, manifesting the magicity at these numbers. The low-density behavior of 
the equation of state for pure neutron matter is compatible with other 
microscopic models. Along with the nuclear symmetry energy, its slope and 
curvature parameters at the saturation density are consistent with those 
extracted from various experimental data. We calculate the neutron star 
properties such as mass-radius, and tidal deformability using the equation of 
state composed of nucleons and leptons in $\beta$-equilibrium condition.

\section{Introduction}
The nuclear physics inputs are essential in understanding the properties
of dense objects like neutron stars. The relativistic mean field models
based on the effective field theory (ERMF) motivated Lagrangian density have
been instrumental in describing the neutron star properties, since,
the ERMF models enables one to readily include the contributions from various
degrees of freedoms such as hyperons, kaons and Bose condensates. The model 
parameters are obtained by adjusting them to reproduce the experimental data 
on the bulk properties for a selected set of finite nuclei. However, these 
parameterizations give remarkable results for bulk properties such as binding 
energy, quadrupole moment, root mean square radius not only for beta stable 
nuclei, but also for nuclei away from the stability line \cite{gam90,estal01}.   However, the same model, sometimes does not appropriately reproduce the 
behavior of the symmetric nuclear matter and pure neutron matter at 
supranormal densities as well as those for the pure neutron matter at the
subsaturation densities.

The ERMF model usually includes the contributions from the self and 
cross-couplings of isoscalar-scalar $\sigma$, isoscalar-vector $\omega$ and 
isovector-vector $\rho$ mesons. The inclusion of various self and 
cross-couplings makes the model flexible to accommodate various phenomena
associated with the finite nuclei and neutron stars adequately without 
compromising the quality of the fit to those data considered a priory.
For example, the self-coupling of $\sigma$ mesons remarkably reduces the
nuclear matter incompressibility to the desired  values \cite{boguta77}.
The cross-coupling of $\rho$ mesons with $\sigma$ or $\omega$ allows one to 
vary the neutron-skin thickness in a heavy nucleus like $^{208}$Pb  over a 
wide range \cite{pika05,pika12}. These cross-couplings are also essential to 
produce desired behavior for the equation of state of pure neutron matter.  
Though, the effects are marginal, but, the quantitative agreement with the 
available empirical informations call for them \cite{pika12,Agrawal12}.
 Noticed that ERMF Lagrangian density in Eq. \eqref{eqq1} contains various 
coupling constants corresponding to different mesons. Thus, we can say that
each term has its own physical significance alongwith some limitations and 
if we omit any one of them then it will be very difficult to describe the 
Hamiltonian as a whole. So, to overcome the limitations all possible terms
are to be included and the model needs to be extended. For this, we strive 
for new parameter sets.

One may also consider the  contributions due to  the  couplings of the meson 
field gradients  to the nucleons as well as the tensor coupling of the mesons 
to the nucleons within the ERMF model \cite{estal01}. These additional 
couplings are required from the naturalness view point, but very often they 
are neglected. Only the parameterizations of the ERMF model in which the 
contributions from gradient and tensor couplings of mesons to the nucleons 
considered are the TM1$^*$, G1 and G2 \cite{estal01,furnstahl97}. However, 
these parameterizations display some disconcerting features. For instance,
the nuclear matter incompressibility and/or  the neutron-skin thickness
associated with the TM1$^*$, G1 and G2  parameter sets are  little too
large in view of their current estimates based on the measured values for
the isoscalar giant monopole and the isovector giant dipole resonances in
the $^{208}$Pb nucleus \cite{Garg, Rocamaza}.  The equation of state (EoS)
for the pure neutron matter at sub-saturation densities show noticeable
deviations with those calculated using realistic approaches.

Recently, the detection of gravitational waves from the binary neutron star 
GW170817 is a major breakthrough in astrophysics and was detected for the 
first time by the advanced Laser Interferometer Gravitational- wave Observatory 
(aLIGO) and advanced VIRGO detectors \cite{BNS}. This detection has certainly 
proved to be a valuable guidance to study matter under the most extreme 
conditions. In-spiraling and coalescing objects of a binary neutron star result 
in gravitational waves. Due to the merger, a compact remnant remains whose 
nature is decided by two factors: (i) the masses of the inspiraling objects
and (ii) the equation of state of the neutron star matter. For final
state, the formation of either a neutron star or a black hole depends
on the masses and stability of the objects. The chirp mass is measured
very precisely from data analysis  of GW170817 and it is found to be
1.188$^{+0.004}_{-0.002}M_\odot$ for the  90$\%$ credible intervals. It
is suggested that the total mass should be 2.74$^{+0.04}_{-0.01}M_\odot$ for
low-spin priors and 2.82$^{+0.47}_{-0.09}M_\odot$ for high-spin priors
\cite{BNS}. Moreover, the maximum mass of nonspinning neutron stars(NSs)
as a function of radius is observed with the highly precise measurements
of $M\approx 2.0M_\odot$. From the observations of gravitational waves,
we can extract information regarding the radii or tidal deformability of
the nonspinning and spinning NSs \cite{flan,tanja,tanja1}. Once we succeed
in getting this information, it is easy to get the neutron star matter
equation of state \cite{latt07,bharat}.

In this chapter, we constructed two set of new parameters G3, and the 
Institute of Physics Bhubaneswar-I (IOPB-I), using the simulated annealing 
method (SAM) \cite{Agrawal05,Agrawal06,kumar06} and explored the generic prediction of properties of finite 
nuclei, nuclear matter, and neutron stars within the ERMF formalism. Our new 
parameter set yields the considerable shell gap appearing at neutron numbers
N=172, 184 and 198 showing the magicity of these numbers. The behavior of the 
density-dependent symmetry energy of nuclear matter at low and high densities 
is examined in detail. The effects of the core EoS on the mass, radius, and 
tidal deformability of an NS are evaluated using the static $l=2$ perturbation 
of a Tolman-Oppenheimer-Volkoff solution.
\begin{table}
\begin{center}
\caption{The vector $v_0$ and $v_1$ contains the lower and upper limits of each of the components of the vector $v$ which used for implementing the SAM-based
algorithm for searching the global minimum of $\chi^2$. The vector d represents the maximum displacement allowed in a single step for the components of the vector $v$.}
\begin{tabular}{cccccccc}
%\hline
\hline
\multicolumn{1}{c}{Parameter}
&\multicolumn{1}{c}{$v$}
&\multicolumn{1}{c}{$v_0$}
&\multicolumn{1}{c}{$v_1$}
&\multicolumn{1}{c}{d}\\
\hline
$\mathcal{E}_{0}$  &  -16.02  & -16.30&-15.70&0.025 \\
$K_{\infty}$  &  230.0  & 210.0&245.0&1.0 \\
$\rho_{0}$  &  0.148  & 0.140&0.165&0.001 \\
$M^*/M$  &  0.525  & 0.5 &0.9&0.002\\
$J$  &  32.1  & 28.0 &35.0&0.08\\
$g_{\delta}$  &  2.0  & 0.0 & 15.0&0.2\\
$\eta_{1}$  &  0.410  & 0.4 &0.8&0.002\\
$\eta_{2}$  &  0.10  & 0.09 &0.12&0.002\\
$\eta_{\rho}$  &  0.590  & 0.1 &0.7&0.003\\
$\Lambda_{\omega}$  &  0.03  & 0.02 &0.09&0.002 \\
$\alpha_{1}$  &  1.73  & 1.0 &2.0&0.005\\
$\alpha_{2}$  &  -1.51  & -1.65 &-1.40&0.005\\
$\beta_\sigma$  &  -0.083  &  -0.09 &-0.08&0.00001\\
$\beta_\omega$  &  -0.55  &  -0.6 &-0.4&0.001\\
$\zeta_{0}$  &  1.01  & 1.01 &1.01&0.0\\
$f_\rho/4$  &  3.0  &  0.0 &6.0&0.03\\
$f_\omega/4$  &  0.4  &  0.0 &1.0&0.005\\
$m_{s}$  &  510.0  & 480.0 & 570.0&0.450\\
\hline
\end{tabular}
	\label{fitt}
\end{center}
\end{table}

\begin{table}[t]
	\begin{center}
		\caption{\small{The obtained new parameter sets G3, and IOPB-I along with 
NL3 \cite{lala97}, FSUGold2 \cite{fsu2}, FSUGarnet \cite{chai15} and 
G2 \cite{furnstahl97} sets are  listed. The nucleon mass $M$ is 939.0 MeV.  
All the coupling constants are dimensionless, except $k_3$ which is in 
		fm$^{-1}$. }}
\scalebox{1.}{
\begin{tabular}{cccccccccccc}
\Xhline{3\arrayrulewidth}
\multicolumn{1}{c}{}
&\multicolumn{1}{c}{NL3}
&\multicolumn{1}{c}{FSUGold2}
&\multicolumn{1}{c}{FSUGarnet}
&\multicolumn{1}{c}{G2}
&\multicolumn{1}{c}{G3}
&\multicolumn{1}{c}{IOPB-I}\\
\hline
	$m_{s}/M$  &  0.541  & 0.530 & 0.529&0.554  &    0.559 &0.533 \\
	$m_{\omega}/M$  &  0.833  & 0.833&0.833 & 0.832  &    0.832 &0.833 \\
	$m_{\rho}/M$  &  0.812 & 0.812 &0.812 & 0.820  &    0.820 &0.812 \\
	$m_{\delta}/M$   & 0.0  & 0.0 & 0.0& 0.0  &    1.043&0.0  \\
	$g_{s}/4 \pi$  &  0.813  &  0.827&0.837 & 0.835  &    0.782 &0.827 \\
	$g_{\omega}/4 \pi$  &  1.024  &  1.079&1.091 & 1.016  &    0.923&1.062 \\
	$g_{\rho}/4 \pi$  &  0.712  & 0.714 &1.105&  0.755  &   0.962 &0.885  \\
	$g_{\delta}/4 \pi$  &  0.0  & 0.0 & 0.0& 0.0  &    0.160 &0.0\\
	$k_{3} $   &  1.465  & 1.231 &1.368&  3.247  &  2.606 &1.496 \\
	$k_{4}$  &  -5.688  & -0.205 & -1.397& 0.632  &  1.694 &-2.932  \\
	$\zeta_{0}$  &  0.0  & 4.705 &4.410&  2.642  &  1.010  &3.103  \\
	$\eta_{1}$  &  0.0  & 0.0 &0.0&  0.650  &   0.424 &0.0  \\
	$\eta_{2}$  &  0.0  & 0.0 &0.0&  0.110  &   0.114 &0.0  \\
	$\eta_{\rho}$  &  0.0  & 0.0 &0.0&  0.390  &  0.645 &0.0  \\
	$\Lambda_{\omega}$  &  0.0  & 0.000823 &0.04337 &  0.0  &  0.038 &0.024  \\
	$\alpha_{1}$  &  0.0  & 0.0 &0.0&  1.723  &   2.000 &0.0 \\
	$\alpha_{2}$  &  0.0  & 0.0 &0.0&  -1.580  &  -1.468&0.0  \\
	$f_\omega/4$  &  0.0  &  0.0 &0.0& 0.173  &   0.220&0.0 \\
	$f_\rho/4$  &  0.0  &  0.0 &0.0& 0.962  &  1.239 &0.0\\
	$\beta_\sigma$  &  0.0  &  0.0 &0.0& -0.093  &-0.087 &0.0  \\
	$\beta_\omega$  &  0.0  &  0.0 &0.0& -0.460  &-0.484 &0.0  \\
\Xhline{3\arrayrulewidth}
\end{tabular}}
\label{table1}%}
	\end{center}
\end{table}

\begin{table}[t]
	\begin{center}
\caption{\small{The binding energy per nucleon B/A(MeV), charge radius $R_c$
(fm) and neutron skin thickness R$_{n}$-R$_{p}$ (fm) for some close shell 
nuclei compared with the NL3, FSUGold2, FSUGarnet, and G2 with experimental 
data \cite{audi12,angel13}.}}
\scalebox{0.9}{
\begin{tabular}{cccccccccc}
\Xhline{3\arrayrulewidth}
\multicolumn{1}{c}{Nucleus}&
\multicolumn{1}{c}{Obs.}&
\multicolumn{1}{c}{Expt.}&
\multicolumn{1}{c}{NL3}&
\multicolumn{1}{c}{FSUGold2}&
\multicolumn{1}{c}{FSUGarnet}&
\multicolumn{1}{c}{G2}&
\multicolumn{1}{c}{G3}&
\multicolumn{1}{c}{IOPB-I}\\
\hline
$^{16}$O&B/A & 7.976 &7.917&7.862&7.876 & 7.952&8.037&7.977  \\
 & R$_{c}$& 2.699 & 2.714&2.694&2.690   & 2.718&2.707&2.705  \\
 & R$_{n}$-R$_{p}$ & - & -0.026&-0.026&-0.028  & -0.028&-0.028&-0.027  \\
$^{40}$Ca & B/A  & 8.551 & 8.540&8.527&8.528  &8.529 &8.561&8.577  \\
 & R$_{c}$  & 3.478 &  3.466&3.444&3.438  & 3.453&3.459&3.458  \\
 & R$_{n}$-R$_{p}$  & - & -0.046&-0.047&-0.051  & -0.049&-0.049&-0.049  \\
$^{48}$Ca & B/A  & 8.666 & 8.636&8.616&8.609  & 8.668&8.671&8.638  \\
 & R$_{c}$  & 3.477 & 3.443&3.420&3.426   & 3.439&3.466&3.446  \\
 & R$_{n}$-R$_{p}$  & - &  0.229&0.235&0.169 & 0.213&0.174&0.202  \\
$^{68}$Ni & B/A  & 8.682 & 8.698&8.690&8.692  & 8.682&8.690&8.707  \\
 & R$_{c}$  & - & 3.870 & 3.846&3.861  &3.861&3.892&3.873  \\
 & R$_{n}$-R$_{p}$  & - &0.262 &0.268&0.184 &0.240&0.190&0.223  \\
$^{90}$Zr & B/A  & 8.709 & 8.695&8.685&8.693  & 8.684&8.699&8.691  \\
 & R$_{c}$  & 4.269 & 4.253& 4.230&4.231  & 4.240&4.276&4.253  \\
 & R$_{n}$-R$_{p}$  & - & 0.115&0.118&0.065   & 0.102&0.068&0.091 \\
$^{100}$Sn & B/A  & 8.258 & 8.301& 8.282&8.298   & 8.248&8.266&8.284 \\
 & R$_{c}$  & - & 4.469&4.453&4.426   &4.470&4.497&4.464  \\
 & R$_{n}$-R$_{p}$  & - & -0.073&-0.075&-0.078  & -0.079& -0.079&-0.077 \\
$^{132}$Sn & B/A  & 8.355 & 8.371&8.361&8.372  & 8.366&8.359&8.352 \\
 & R$_{c}$  & 4.709 & 4.697&4.679&4.687   & 4.690&4.732&4.706  \\
 & R$_{n}$-R$_{p}$  & - & 0.349&0.356&0.224   & 0.322&0.243&0.287 \\
$^{208}$Pb & B/A  & 7.867 &7.885& 7.881 &7.902 & 7.853&7.863&7.870 \\
 & R$_{c}$  &5.501  & 5.509&5.491&5.496 & 5.498&5.541&5.521  \\
 & R$_{n}$-R$_{p}$  & - &   0.283& 0.288&0.162 & 0.256& 0.180&0.221 \\
\Xhline{3\arrayrulewidth}
\end{tabular}
\label{table2}}
	\end{center}
\end{table}

\begin{table}
\begin{center}
	\caption{\small{The nuclear matter properties such as binding energy per nucleon, 
$\mathcal{E}_{0}$(MeV), saturation density $\rho_{0}$(fm$^{-3}$), 
incompressibility coefficient for symmetric nuclear matter $K$(MeV), 
effective mass ratio $M^*/M$, symmetry energy $J$(MeV), and linear density 
	dependence of the symmetry energy, $L$(MeV), at saturation.	}}
%\vspace{0.10 in}
\scalebox{0.9}{
\begin{tabular}{cccccccccc}
\Xhline{3\arrayrulewidth}
\multicolumn{1}{c}{}
&\multicolumn{1}{c}{NL3}
&\multicolumn{1}{c}{FSUGarnet}
&\multicolumn{1}{c}{G3}
&\multicolumn{1}{c}{IOPB-I}\\
\hline
	$\rho_{0}$ (fm$^{-3})$ &  0.148  &  0.153&  0.148&0.149  \\
$\mathcal{E}_{0}$(MeV)  &  -16.29  & -16.23 &  -16.02&-16.10  \\
$M^{*}/M$  &  0.595 & 0.578 &  0.699&0.593  \\
$J$(MeV)   & 37.43  &  30.95&   31.84&33.30  \\
$L$(MeV)  &  118.65  &  51.04 &  49.31&63.58 \\
$K_{sym}$(MeV)  &  101.34  & 59.36 & -106.07&-37.09 \\
$Q_{sym}$(MeV)  &  177.90  & 130.93&  915.47 &862.70  \\
$K$(MeV)  & 271.38  &  229.5&  243.96& 222.65 \\
$Q_{0} $(MeV)   &  211.94  & 15.76&   -466.61 &-101.37 \\
$K_{\tau}$(MeV)  &  -703.23  &  -250.41&-307.65 &-389.46  \\
$K_{asy}$(MeV)  & -610.56  & -246.89&  -401.97 &-418.58 \\
$K_{sat2}$(MeV)  & -703.23  &-250.41&  -307.65  &-389.46 \\
\Xhline{3\arrayrulewidth}
\end{tabular}
\label{table3}}
\end{center}
\end{table}

\begin{table}
	\begin{center}
\hspace{0.1 cm}
		\caption{\small{The binary neutron star masses ($m_1(M_\odot), m_2(M_\odot)$) and 
corresponding radii ($R_1$(km), $R_2$ (km)), tidal Love number 
($(k_2)_1$, $(k_2)_2$), and tidal deformabilities ($\lambda_1, \lambda_2$) 
in $1\times10^{36}$g cm$^2$ s$^2$ and dimensionless tidal deformabilities 
($\Lambda_1, \Lambda_2$). $\tilde{\Lambda}$, $\delta\tilde\Lambda$, $\cal{M}$$_{c}(M_\odot)$, and $\cal{R}$$_{c}(km)$ are the dimensionless tidal deformability,
		tidal correction, chirp mass, and radius of the binary neutron star, respectively.}}
\scalebox{0.6}{
\begin{tabular}{ccccccccccccccccccccccccc}
\Xhline{3\arrayrulewidth}
\multicolumn{15}{ c }{}\\
EoS &       $m_1(M_\odot)$   &       $m_2(M_\odot)$ &$R_1$(km)  &       
	$R_2$(km)       &       $(k_2)_1$        &       $(k_2)_2$ &  $\lambda_1$ & 
	$\lambda_2$ &$\Lambda_1$&     $\Lambda_2$&$\tilde{\Lambda}$&$\delta\tilde\Lambda$&$\cal{M}$$_c(M_\odot)$&$\cal{R}$$_{c}(km)$\\
\hline
	&&&&&&&&&&&&&&\\
NL3 
&      1.20&      1.20&    14.702&    14.702&    0.1139&    0.1139&     7.826&     7.826&   2983.15&   2983.15&   2983.15&     0.000&      1.04&    10.350 \\
&      1.50&      1.20&    14.736&    14.702&    0.0991&    0.1139&     6.889&     7.826&    854.06&   2983.15&   1608.40&   220.223&      1.17&    10.214 \\
&      1.25&      1.25&    14.708&    14.708&    0.1118&    0.1118&     7.962&     7.962&   2388.82&   2388.82&   2388.82&     0.000&      1.09&    10.313 \\
&      1.30&      1.30&    14.714&    14.714&    0.1094&    0.1094&     7.546&     7.546&   1923.71&   1923.71&   1923.71&     0.000&      1.13&    10.271 \\
&      1.35&      1.35&    14.720&    14.720&    0.1070&    0.1070&     7.393&     7.393&   1556.84&   1556.84&   1556.84&     0.000&      1.18&    10.224 \\
&      1.35&      1.25&    14.720&    14.708&    0.1070&    0.1118&     7.393&     7.962&   1556.84&   2388.82&   1930.02&    91.752&      1.13&    10.268 \\
&      1.37&      1.25&    14.722&    14.708&    0.1061&    0.1118&     7.339&     7.962&   1452.81&   2388.82&   1863.78&   100.532&      1.14&    10.271 \\
&      1.40&      1.20&    14.726&    14.702&    0.1044&    0.1139&     7.231&     7.826&   1267.07&   2983.15&   1950.08&   183.662&      1.13&    10.262 \\
&      1.40&      1.40&    14.726&    14.726&    0.1044&    0.1044&     7.231&     7.231&   1267.07&   1267.07&   1267.07&     0.000&      1.22&    10.174 \\
&      1.42&      1.29&    14.728&    14.712&    0.1031&    0.1099&     7.147&     7.572&   1145.72&   1994.02&   1515.18&    95.968&      1.18&    10.192 \\
&      1.44&      1.39&    14.730&    14.724&    0.1027&    0.1049&     7.120&     7.259&   1108.00&   1311.00&   1204.83&    19.212&      1.23&    10.179 \\
&      1.45&      1.45&    14.732&    14.732&    0.1018&    0.1018&     7.064&     7.064&   1037.13&   1037.13&   1037.13&     0.000&      1.26&    10.124 \\
&      1.54&      1.26&    14.740&    14.708&    0.0969&    0.1114&     6.741&     7.668&    729.95&   2303.95&   1308.91&   168.202&      1.21&    10.179 \\
&      1.60&      1.60&    14.746&    14.746&    0.0937&    0.0937&     6.532&     6.532&    589.92&    589.92&    589.92&     0.000&      1.39&     9.979 \\	
	&&&&&&&&&&&&&&\\
FSUGarnet 
&      1.20&      1.20&    12.944&    12.944&    0.1090&    0.1090&     3.961&     3.961&   1469.32&   1469.32&   1469.32&     0.000&      1.04&     8.983 \\
&      1.50&      1.20&    12.972&    12.944&    0.0893&    0.1090&     3.282&     3.961&    408.91&   1469.32&    784.09&   111.643&      1.17&     8.847 \\
&      1.25&      1.25&    12.958&    12.958&    0.1062&    0.1062&     3.880&     3.880&   1193.78&   1193.78&   1193.78&     0.000&      1.09&     8.977 \\
&      1.30&      1.30&    12.968&    12.968&    0.1030&    0.1030&     3.777&     3.777&    945.29&    945.29&    945.29&     0.000&      1.13&     8.910 \\
&      1.35&      1.35&    12.974&    12.974&    0.0998&    0.0998&     3.666&     3.666&    761.13&    761.13&    761.13&     0.000&      1.18&     8.860 \\
&      1.35&      1.25&    12.974&    12.958&    0.0998&    0.1062&     3.666&     3.880&    761.13&   1193.78&    955.00&    49.744&      1.13&     8.920 \\
&      1.37&      1.25&    12.976&    12.958&    0.0986&    0.1062&     3.629&     3.880&    710.62&   1193.78&    922.54&    53.853&      1.14&     8.924 \\
&      1.40&      1.20&    12.978&    12.944&    0.0965&    0.1090&     3.552&     3.961&    622.06&   1469.32&    959.22&    90.970&      1.13&     8.904 \\
&      1.40&      1.40&    12.978&    12.978&    0.0965&    0.0965&     3.552&     3.552&    622.06&    622.06&    622.06&     0.000&      1.22&     8.825 \\
&      1.42&      1.29&    12.978&    12.966&    0.0949&    0.1038&     3.495&     3.803&    565.47&   1001.18&    755.10&    50.492&      1.18&     8.867 \\
&      1.44&      1.39&    12.978&    12.978&    0.0939&    0.0973&     3.456&     3.582&    531.54&    653.60&    589.65&    14.148&      1.23&     8.823 \\
&      1.45&      1.45&    12.978&    12.978&    0.0931&    0.0931&     3.427&     3.427&    507.70&    507.70&    507.70&     0.000&      1.26&     8.776 \\
&      1.54&      1.26&    12.964&    12.960&    0.0862&    0.1057&     3.157&     3.864&    343.73&   1146.73&    638.35&    88.892&      1.21&     8.817 \\
&      1.60&      1.60&    12.944&    12.944&    0.0816&    0.0816&     2.964&     2.964&    266.20&    266.20&    266.20&     0.000&      1.39&     8.511 \\	
	&&&&&&&&&&&&&&\\
\Xhline{3\arrayrulewidth}
\end{tabular}}
\label{table4}
\end{center}
\end{table}

\begin{table}
	\begin{center}
\hspace{0.1 cm}
        \caption{Table \ref{table4} is continued \dots}
\scalebox{0.6}{
\begin{tabular}{ccccccccccccccccccccccccc}
\Xhline{3\arrayrulewidth}
\multicolumn{15}{ c }{}\\
EoS &       $m_1(M_\odot)$   &       $m_2(M_\odot)$ &$R_1$(km)  &
        $R_2$(km)       &       $(k_2)_1$        &       $(k_2)_2$ &  $\lambda_1$ &
        $\lambda_2$ &$\Lambda_1$&     $\Lambda_2$&$\tilde{\Lambda}$&$\delta\tilde\Lambda$&$\cal{M}$$_c(M_\odot)$&$\cal{R}$$_{c}(km)$\\
\hline
        &&&&&&&&&&&&&&\\
G3
&      1.20&      1.20&    12.466&    12.466&    0.1034&    0.1034&     3.114&     3.114&   1776.65&   1776.65&   1776.65&     0.000&      1.04&     9.331 \\
&      1.50&      1.20&   112.360&    12.466&    0.0800&    0.1034&     2.309&     3.114&    284.92&   1776.65&    803.43&   191.605&      1.17&     8.890 \\
&      1.25&      1.25&    12.460&    12.460&    0.1001&    0.1001&     3.007&     3.007&    939.79&    939.79&    939.79&     0.000&      1.09&     8.557 \\
&      1.30&      1.30&    12.448&    12.448&    0.0962&    0.0962&     2.875&     2.875&    728.07&    728.07&    728.07&     0.000&      1.13&     8.457 \\
&      1.35&      1.35&    12.434&    12.434&    0.0925&    0.0925&     2.750&     2.750&    582.26&    582.26&    582.26&     0.000&      1.18&     8.398 \\
&      1.35&      1.25&    12.434&    12.460&    0.0925&    0.1001&     2.750&     3.007&    582.26&    939.79&    742.29&    43.064&      1.13&     8.482 \\
&      1.37&      1.25&    12.428&    12.460&    0.0909&    0.1001&     2.696&     3.007&    530.66&    939.79&    709.72&    49.144&      1.14&     8.468 \\
&      1.40&      1.20&    12.416&    12.466&    0.0859&    0.1034&     2.613&     3.114&    461.03&   1776.65&    976.80&   183.274&      1.13&     8.937 \\
&      1.40&      1.40&    12.416&    12.416&    0.0859&    0.0859&     2.613&     2.613&    461.03&    461.03&    461.03&     0.000&      1.22&     8.312 \\
&      1.42&      1.29&    12.408&    12.450&    0.0868&    0.0972&     2.553&     2.905&    417.96&    772.17&    571.87&    43.226&      1.18&     8.387 \\
&      1.44&      1.39&    12.398&    12.420&    0.0854&    0.0894&     2.501&     2.643&    384.42&    484.90&    432.22&    12.671&      1.23&     8.292 \\
&      1.45&      1.45&    12.932&    12.392&    0.0846&    0.0846&     2.472&     2.472&    367.04&    367.04&    367.04&     0.000&      1.26&     8.225 \\
&      1.54&      1.26&    12.334&    12.458&    0.0769&    0.0992&     2.194&     2.976&    239.49&    883.46&    474.83&    75.175&      1.21&     8.311 \\
&      1.60&      1.60&    12.280&    12.280&    0.0716&    0.0716&     2.000&     2.000&    179.63&    179.63&    179.63&     0.000&      1.39&     7.867 \\
        &&&&&&&&&&&&&&\\
IOPB-I
&      1.20&      1.20&    13.222&    13.222&    0.1081&    0.1081&     4.369&     4.369&   1654.23&   1654.23&   1654.23&     0.000&      1.04&     9.199 \\
&      1.50&      1.20&    13.236&    13.222&    0.0894&    0.1081&     3.631&     4.369&    449.62&   1654.23&    875.35&   128.596&      1.17&     9.044 \\
&      1.25&      1.25&    13.230&    13.230&    0.1053&    0.1053&     4.268&     4.268&   1310.64&   1310.64&   1310.64&     0.000&      1.09&     9.146 \\
&      1.30&      1.30&    13.238&    13.238&    0.1024&    0.1024&     4.162&     4.162&   1053.07&   1053.07&   1053.07&     0.000&      1.13&     9.105 \\
&      1.35&      1.35&    13.240&    13.240&    0.0995&    0.0995&     4.050&     4.050&    857.53&    857.53&    857.53&     0.000&      1.18&     9.074 \\
&      1.35&      1.25&    13.240&    13.230&    0.0995&    0.1053&     4.050&     4.268&    857.53&   1310.64&   1060.81&    49.565&      1.13&     9.110 \\
&      1.37&      1.25&    13.242&    13.230&    0.0938&    0.1053&     4.004&     4.268&    791.92&   1310.64&   1019.60&    56.371&      1.14&     9.104 \\
&      1.40&      1.20&    13.242&    13.222&    0.0960&    0.1081&     3.911&     4.369&    680.79&   1654.23&   1067.64&   107.340&      1.13&     9.097 \\
&      1.40&      1.40&    13.242&    13.242&    0.0960&    0.0960&     3.911&     3.911&    680.79&    680.79&    680.79&     0.000&      1.22&     8.986 \\
&      1.42&      1.29&    13.242&    13.236&    0.0949&    0.1030&     3.864&     4.184&    632.31&   1099.78&    835.91&    52.836&      1.18&     9.049 \\
&      1.44&      1.39&    13.242&    13.242&    0.0935&    0.0969&     3.806&     3.946&    578.47&    719.80&    645.73&    17.094&      1.23&     8.985 \\
&      1.45&      1.45&    13.240&    13.240&    0.0927&    0.0927&     3.771&     3.771&    549.06&    549.06&    549.06&     0.000&      1.26&     8.915 \\
&      1.54&      1.26&    13.230&    13.232&    0.0868&    0.1047&     3.516&     4.247&    384.65&   1253.00&    703.58&    94.735&      1.21&     8.991 \\
&      1.60&      1.60&    13.212&    12.212&    0.0823&    0.0823&     3.314&     3.314&    296.81&    296.81&    296.81&     0.000&      1.39&     8.698 \\
\Xhline{3\arrayrulewidth}
\end{tabular}}
\label{table5}
	\end{center}
\end{table}

\section{Parameter Fitting}{\label{parameter}}
%\subsection{Parameter fitting}{\label{parameter}}
Having developed in chapter \ref{chapter2} most of the required formalism
related to the properties of finite nuclei, infinite nuclear matter, and the 
neutron star. We are now in a position to implement the calibration of a
new energy density functional by extending the model Lagrangian with addition
of relevant terms and determine the two new sets of force parameters using the 
SAM. The SAM is used to determine the 
parameters used in the Lagrangian density \cite{kirkpatrick83,press92}. The SAM is useful in the global minimization technique; i.e., it gives accurate 
results when there exists a global minimum within several local minima. 
Usually, this procedure is used in a system in which the number of parameters 
is more than the number of observables \cite{kirkpatrick84,ingber89,cohen94}.
In this simulation method, the system stabilizes when the temperature 
$T$ (a variable which controls the energy of the system) goes down 
\cite{Agrawal05,Agrawal06,kumar06}. Initially, the nuclear system is put at a 
high temperature (highly unstable) and then allowed to cool down slowly so 
that it is stabilized in a very smooth way and finally reaches the frozen 
temperature (stable or systematic system). The variation of $T$ should be very 
small near the stable state. The $\chi^2 =\chi^2(p_{1},\dots,p_{N})$ values of 
the considered systems are minimized (least-squares fit), which is governed by 
the model parameters $p_i$. The general expression of the $\chi^2$ can be 
given as:
 \begin{eqnarray}
\chi^{2}= \frac{1}{N_{d}-N_{p}} \sum_{i=1}^{N_{d}}
\left(\frac{M_{i}^{exp}-M_{i}^{th}}{\sigma_{i}}\right)^2.
\label{eq:chi2}\end{eqnarray}  
Here, $N_{d}$ and $N_{p}$ are the numbers of experimental data points and
fitting parameters, respectively. The experimental and theoretical
values of the observables are denoted by $M_{i}^{expt}$ and $M_{i}^{th}$,
respectively.  The $\sigma_{i}$'s are the adopted errors \cite{dob14}.
The adopted errors are composed of three components, namely, the
experimental, numerical, and theoretical errors \cite{dob14}. As the name
suggests, the experimental errors are associated with the measurements;
numerical and theoretical errors are associated with the  numerics and the 
shortcomings of the nuclear model employed, respectively. In principle, there 
exists some arbitrariness in choosing the values of $\sigma_i$, which is 
partially responsible for the proliferation of the mean-field models. The only 
guidance  available from the statistical analysis is that the $\chi^2$ per 
degree of freedom [Eq. \eqref{eq:chi2}] should be close to unity. In the 
present calculation, we used some selected fit data for binding energy and the 
root mean square radius of the charge distribution for some selected nuclei 
and the associated adopted errors on them \cite{klup09}.

In our calculations, we have built two new parameter sets G3, and IOPB-I and
analyzed its effects for finite and infinite nuclear systems. Thus,
we performed an overall fit with 13 parameters for G3, and 8 for IOPB-I, where 
the nucleons as well as the masses of the two vector mesons in free space are 
fixed at their experimental values, i.e., $M=939$ MeV, $m_{\omega}=782.5$ MeV, 
$m_{\rho}=763.0$ MeV, and $m_{\delta}=980.0$ MeV. The effective nucleon mass 
can be used as a nuclear matter constraint at the saturation density $\rho_0$ 
along with other empirical values like incompressibility, binding energy per 
nucleon, and asymmetric parameter $J$. While fitting the parameter, the values 
of effective nucleon mass $M^{*}/M$, nuclear matter incompressibility $K$, and
symmetry energy coefficient $J$ are constrained within 0.50--0.90, 210--245
MeV, and 28--35 MeV, respectively. 

We use SAM algorithm and follow the steps which are:
\begin{enumerate} 
\item We initiate the steps by taking a suitable guess value for the vector 
$v$ and then calculate $\chi^2$ (say, $\chi^{2}_{old}$) by the help of 
Eq. \eqref{eq:chi2} for particular set of experimental data and the corresponding ERMF results together with the theoretical errors.
\item By using the following steps we create a new set of ERMF parameters. We
choose a uniform random number for assigning a new value for the component 
$v_r$ as $v_{r}\rightarrow v_{r}+\eta d_{r}$, where $\eta$ is a uniform random
number that lies within the range -1 to +1. The above step is repeated 
again and again. We stop when the new value of $v_{r}$ comes out to be within
its allowed limits defined as $v_0$ and $v_{1}$. This modified $v$ is then 
used to generate new ERMF parameters. Thus, even a small change in the value 
of component of the vector $v$ will lead to changes in the values of several 
ERMF parameters. For example, a change in the value of $K_{\infty}$ will alter 
the values of the ERMF parameters $k_2$, $k_3$, and $M^*$.
\item We take the newly generated ERMF parameters and use in the Metropolis 
algorithm. We calculate the quantity $P(\chi^2)$ where
\end{enumerate}
\begin{equation}
	P(\chi^2)=e^{(\chi^{2}_{old}-\chi^{2}_{new})/T}.
	\end{equation}
We obtain the $\chi^{2}_{new}$ by taking the newly generated set of the ERMF
parameters. Also, the $T$ is a control parameter which we get from the Cauchy
annealing schedule given by 
\begin{equation}
	T(k)=T_{i}/(k+1);\;\;\;\;\; k=1,2,3,...
	\end{equation}
The newly generated ERMF parameter set is confirmed only when $P(\chi^2)>\beta$, where $\beta$ is an uniform random number that lies between 0 and 1.

%The minimum $\chi^{2}$ is obtained by the
%simulated annealing method \cite{ Agrawal05,Agrawal06,kumar06}  to fix
%the final parameters. 
The newly developed G3, and IOPB-I sets along with NL3
\cite{lala97}, FSUGold2 \cite{fsu2}, FSUGarnet \cite{chai15},  and G2 \cite{furnstahl97} are given for comparison in Table \ref{table1}. The calculated results of the binding energy and charge radius are compared with the known 
experimental data \cite{audi12,angel13}. It is to be noted that in the 
original ERMF parametrization, only five spherical nuclei were taken into
consideration while fitting the parameters with the binding energy, charge 
radius and single particle energy \cite{furnstahl97}. However, here, eight 
spherical nuclei are used for the fitting as listed in Table \ref{table2}.

\section{Results and Discussions}\label{result}

In this section we discuss our calculated results for finite nuclei, infinite 
nuclear matter and neutron stars. For finite nuclei, binding energy, rms radii 
for neutron and proton distributions, isotopic shift, two-neutron separation 
energy, and neutron-skin thickness are analyzed. Similarly, for infinite nuclear matter 
systems, the binding energies per particle for symmetric and asymmetric 
nuclear matter including pure neutron matter at both subsaturation and 
suprasaturation densities are compared with other theoretical results and 
experimental data. The parameter sets G3, and IOPB-I is also applied to study 
the structure of neutron stars using $\beta-$equilibrium and charge neutrality 
conditions.

\subsection{Finite Nuclei}

\begin{figure}[!b]
        \begin{center}
        \includegraphics[width=0.7\columnwidth]{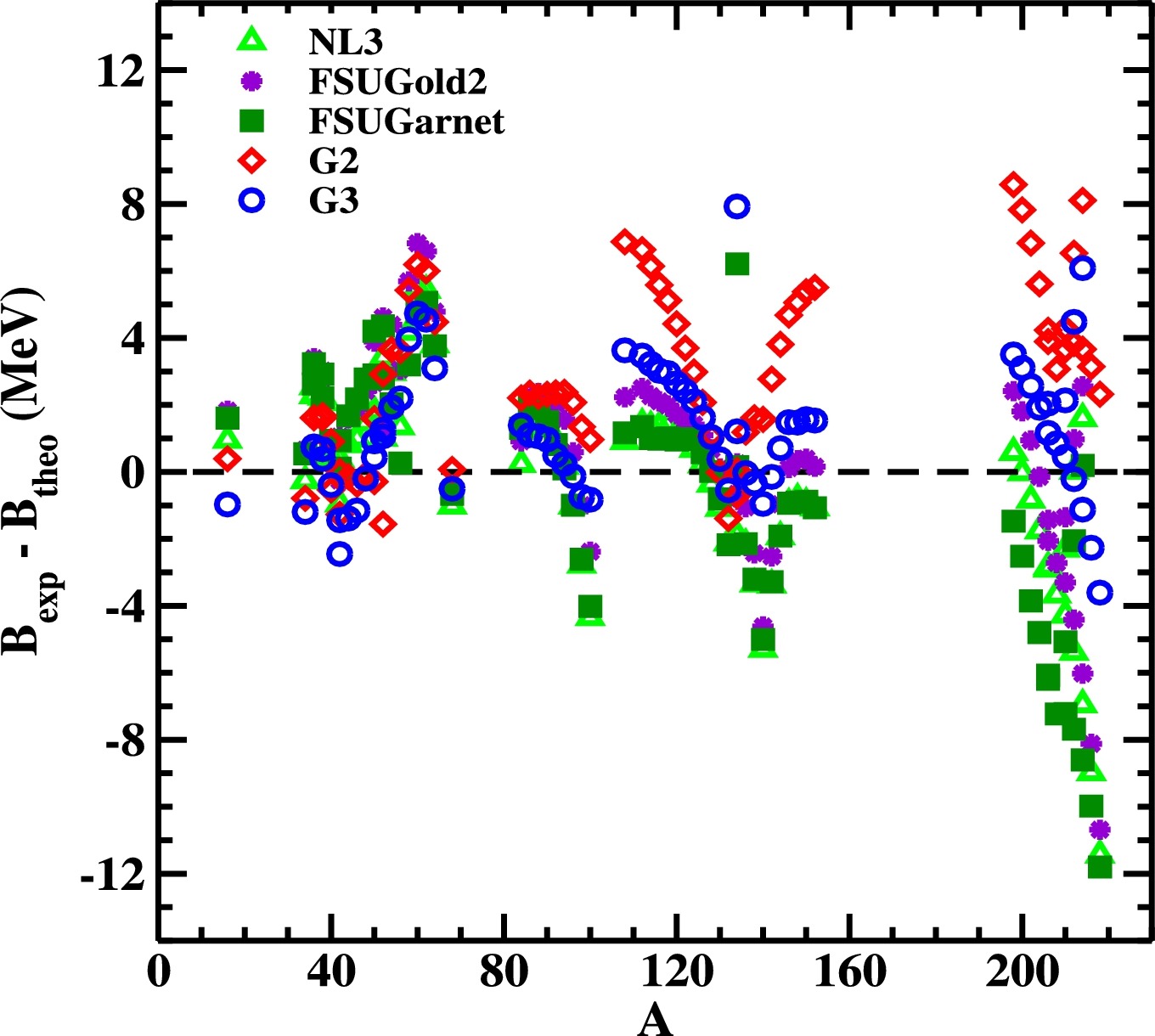}
\caption{ Difference between experimental and theoretical binding energies as a function of mass numbers for NL3 \cite{lala97}, FSUGold2 \cite{fsu2}, FSUGarnet \cite{chai15}, G2 \cite{furnstahl97} and G3 parameter sets. }
\label{be}
        \end{center}
\end{figure}

\begin{figure}[!b]
        \begin{center}
        \includegraphics[width=0.7\columnwidth]{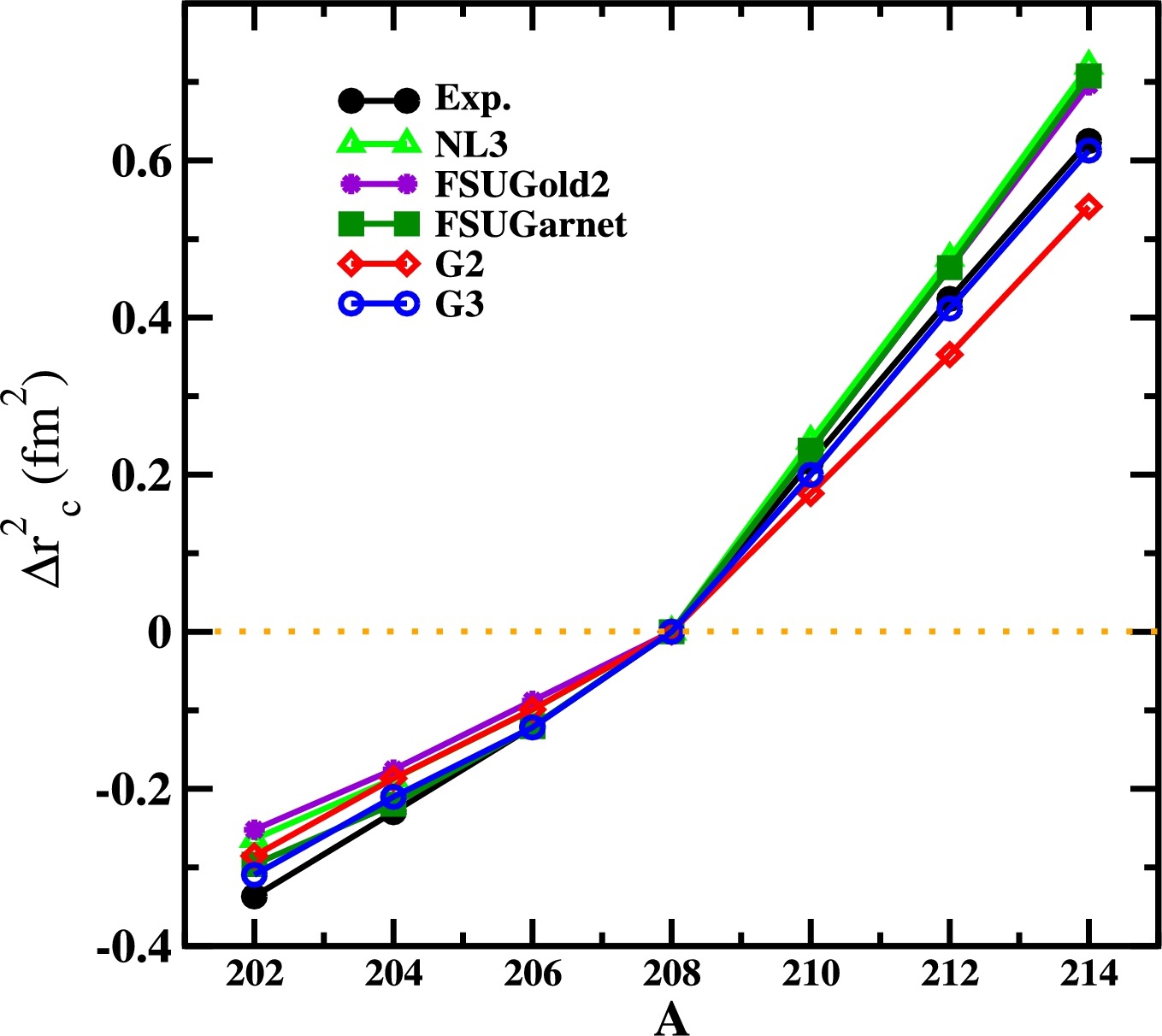}
        \caption{The isotopic shift $\Delta{r_c}^2=
R_{c}^2(208)-R_{c}^2(A)$ (fm$^2$) of Pb isotopes taking $R_{c}$ of $^{208}$Pb
as the standard value. Calculations with the NL3 \cite{lala97}, FSUGold2
\cite{fsu2}, FSUGarnet \cite{chai15}, G2 \cite{furnstahl97} and G3 parameter 
sets are compared.}
\label{shift}

\end{center}
\end{figure}

\begin{figure}[!b]
        \begin{center}
        \includegraphics[width=0.7\columnwidth]{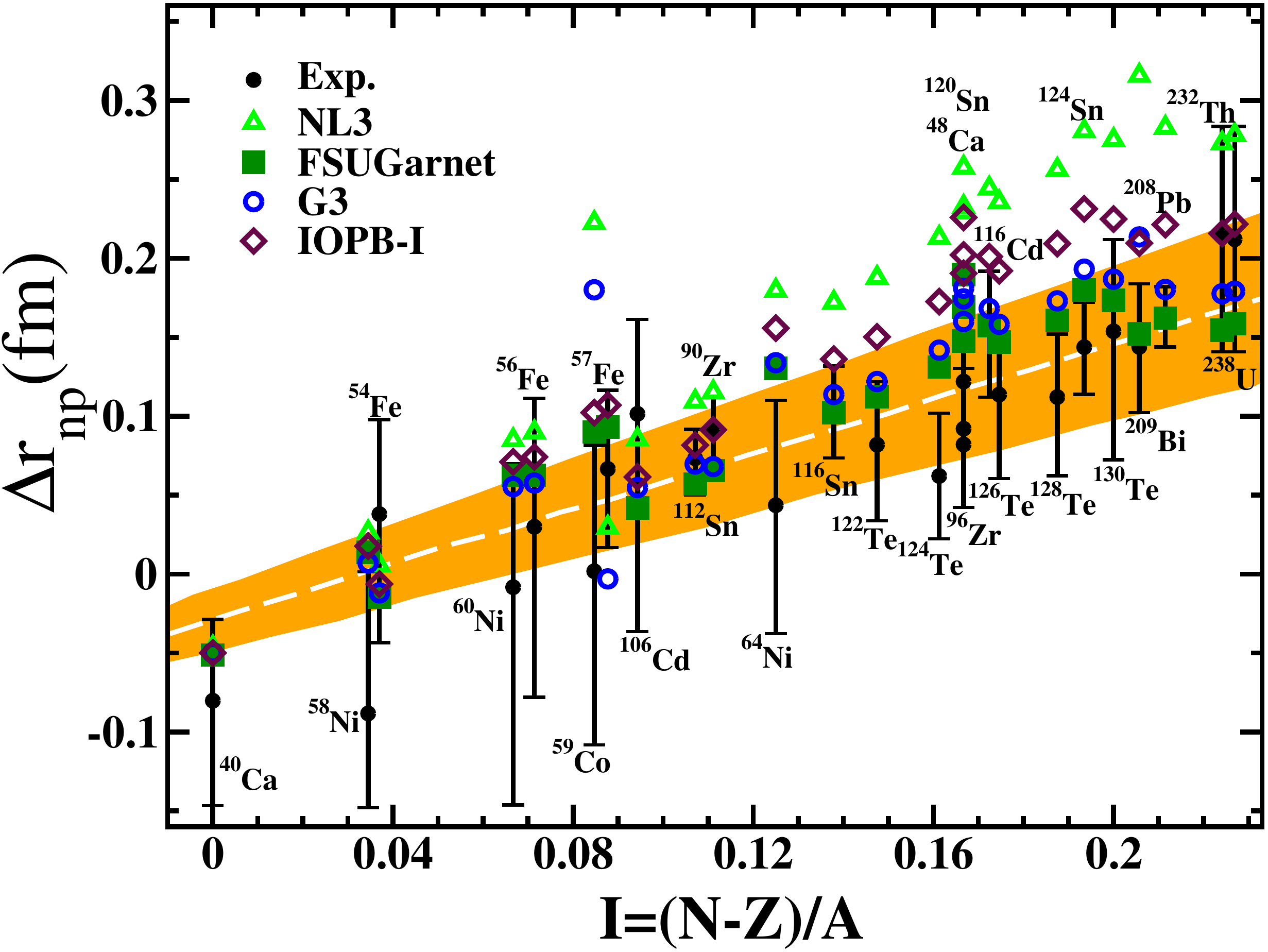}
\caption{The neutron-skin thickness as a function of the asymmetry parameter. 
Results obtained with the parameter sets G3, and IOPB-I are compared with 
those of the sets NL3 \cite{lala97}, FSUGarnet \cite{chai15}, and experimental 
values \cite{thick}. The shaded region is calculated using Eq. \eqref{eq:rnp}.}
        \label{thick}
        \end{center}
\end{figure}

\begin{center}
\bf (i) Binding energies, charge radii, isotopic shift and neutron-skin 
	thickness
\end{center}
We used eight spherical nuclei to fit the experimental ground-state binding 
energies and charge radii using the SAM. The calculated results are
listed in Table \ref{table2} and compared with other theoretical models
as well as experimental data \cite{audi12,angel13}. It can be seen
that the NL3 \cite{lala97}, FSUGold2 \cite{fsu2}, FSUGarnet \cite{chai15}, and 
G2 \cite{furnstahl97} models successfully reproduce the energies and charge 
radii as well. Even though the {\it "mean-field models are not expected to 
work well for the light nuclei," the results deviate only marginally for the 
ground-state properties for light nuclei} \cite{patra93}. We noticed that both 
the binding energy and charge radius of $^{16}$O are well produced by IOPB-I. 
However, the calculated charge radii of $^{40,48}$Ca slightly underestimate the data. We would like to emphasize that 
it is an open problem to mean-field models to predict the evolution of charge 
radii of $^{38-52}$Ca (see Fig. 3 in Ref. \cite{gar16}).

In Fig. \ref{be} we plot the differences between the calculated and 
experimental binding energies for 70 spherical nuclei \cite{klup09} obtained 
using different parameter sets.  The triangles, stars, squares, diamonds and 
circles are the results for the NL3, FSUGold2 , FSUGarnet, G2 and G3
parameterizations, respectively.  The above results affirm that G3 set
reproduces the experimental data better. The rms deviations  for the
binding energy as displayed in Fig. \ref{be} are  2.977, 3.062,
3.696, 3.827 and 2.308 MeV for NL3, FSUGold2, FSUGarnet, G2 and
G3 respectively. The rms error on the binding energy for G3 parameter
set is smaller in comparison to other parameter sets.

In Fig. \ref{shift}, the isotopic shift $\Delta{r}_c^2$ for Pb nucleus
is shown. The isotopic shift is defined as $\Delta{r}_c^2=R_{c}^2(A)-R_{c}^2(208)$
(fm$^2$), where $R_{c}^2(208)$ and $R_{c}^2(A)$ are the mean square radius
of $^{208}$Pb and  Pb isotopes having mass number A. From the figure,
one can see that $\Delta{r}_c^2$ increases with mass number monotonously
till A=208 ($\Delta{r}_c^2=0$ for $^{208}$Pb) and then gives a sudden
kink. It was first pointed by Sharma et al \cite{sharma93}, that the
non-relativistic parameterization fails to show this effect.  However,
this effect is well explained when a relativistic set like NL-SH
\cite{sharma93} is used. The NL3, FSUGold2 , FSUGarnet, G2 and G3
sets  also  appropriately predict this shift in Pb isotopes, but the
agreement with experimental data of the present parameter set G3  is
marginally better.

The excess of neutrons gives rise to a neutron-skin thickness. The neutron-skin 
thickness $\Delta r_{np}$ is defined as
\begin{equation} \Delta r_{np}=\langle r^2 \rangle^{1/2}_{n} -
\langle r^2 \rangle^{1/2}_{p} =R_n-R_p ,
 \end{equation} 
with $R_n$ and $R_p$ being the rms radii for the neutron and proton
distributions, respectively. 
The $\Delta r_{np}$, strongly correlated with the slope of the symmetry
energy \cite{pg10,brown00,maza11}, can probe the isovector  part of the
nuclear interaction. However, there is a large uncertainty in the
experimental measurement of the neutron distribution radius of the finite
nuclei.  The current values of neutron radius and neutron-skin thickness
of $^{208}$Pb are 5.78$^{+0.16}_{-0.18}$ and 0.33$^{+0.16}_{-0.18}$
fm, respectively \cite{abra12}. This error bar is too large to provide significant
constraints on the density-dependent symmetry energy. It is expected
that PREX-II result will give us the neutron radius of $^{208}$Pb within
$1\%$ accuracy. The inclusion of some isovector-dependent terms in the
Lagrangian density is needed, which would provide the freedom to refit
the coupling constants within the experimental data without compromising
the quality of fit.  The addition of $\omega$-$\rho$ cross-coupling into
the Lagrangian density controls the neutron-skin thickness of $^{208}$Pb
as well as that of other nuclei. In Fig. \ref{thick}, we show the neutron-skin
thickness $\Delta r_{np}$ for $^{40}$Ca to $^{238}$U nuclei as a function
of proton-neutron asymmetry $I=(N-Z)/A$. The calculated results of $\Delta
r_{np}$ for NL3, FSUGarnet, G3, and IOPB-I parameter sets are compared with
the corresponding experimental data \cite{thick}.  Experiments have been
done with antiprotons at CERN and  the $\Delta r_{np}$ are extracted 
for 26 stable nuclei ranging from $^{40}$Ca to $^{238}$U  as displayed in
the figure along with the error bars. The trend of the data points shows
approximately linear dependence of neutron-skin thickness on the relative
neutron excess $I$ of nucleus that can be fitted by \cite{thick,xavier14}:

\begin{eqnarray}
\Delta r_{np} = (0.90\pm0.15)I+(-0.03\pm0.02) \; {\text fm}.
\label{eq:rnp}
\end{eqnarray}

The values of  $\Delta r_{np}$ obtained with IOPB-I  for some of the
nuclei slightly deviate from the shaded region, as can be seen from Fig.
\ref{thick}.  This is because IOPB-I has a smaller strength of $\omega$-$\rho$ 
cross-coupling as compared to the FSUGarnet, and G3 sets. Recently, 
Fattoyev {\it et. al.} constrained the upper limit of $\Delta r_{np}\lesssim0.25$ fm for the $^{208}$Pb nucleus with the help of GW170817 observation data \cite{fatt17}. The calculated 
values of neutron-skin thickness for the $^{208}$Pb nucleus are 0.283, 0.162, 
0.180, and 0.221 fm for the NL3, FSUGarnet, G3, and IOPB-I parameter sets 
respectively. The proton elastic scattering experiment recently measured 
neutron-skin thickness $\Delta r_{np}=0.211^{+0.054}_{-0.063} $ fm for $^{208}$Pb
\cite{zeni10}.  Thus values of $\Delta r_{np} = 0.221$ for IOPB-I  are
consistent with the recent prediction of neutron-skin thickness.

\begin{figure}[!b]
        \begin{center}
        \includegraphics[width=0.7\columnwidth]{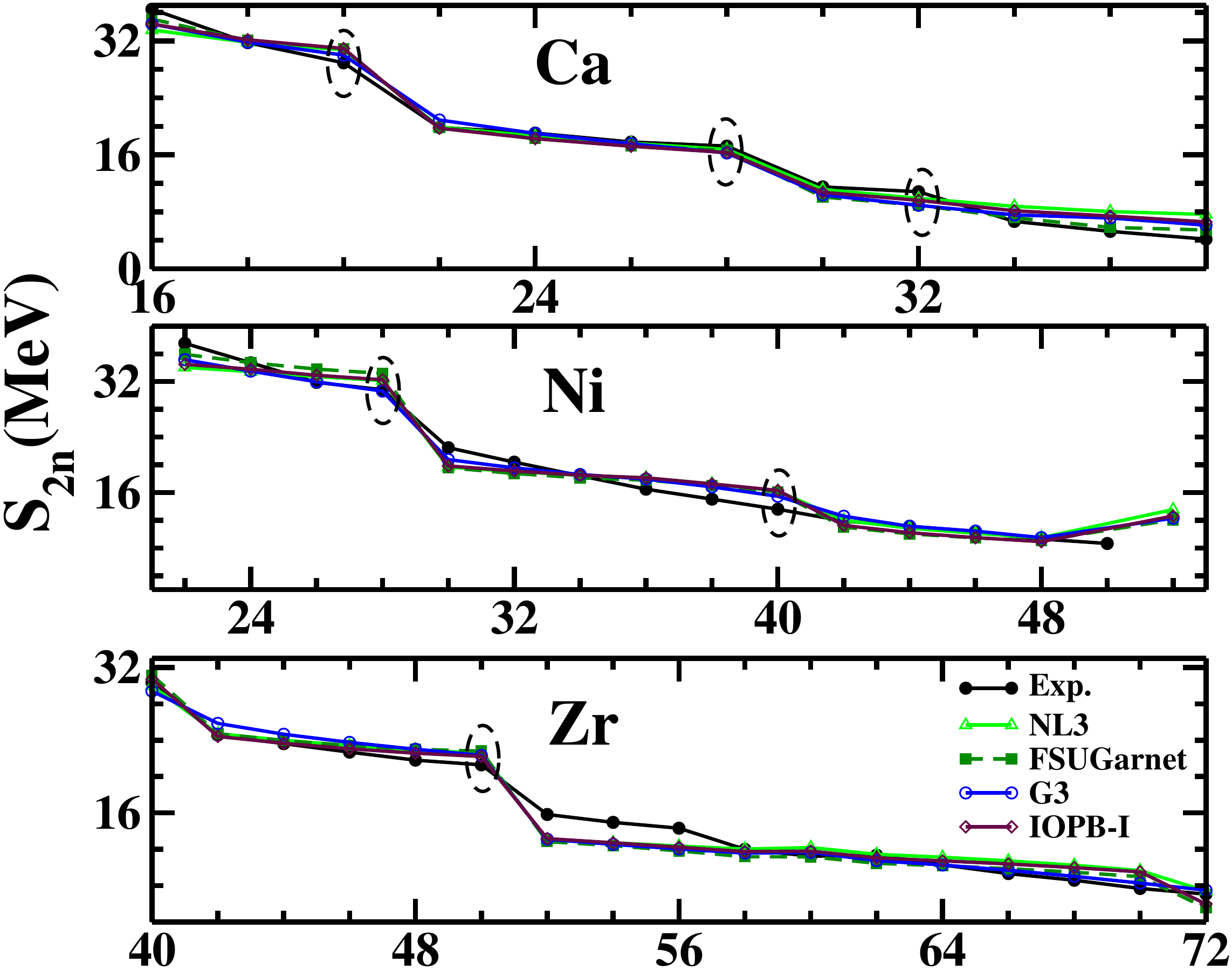}
        \includegraphics[width=0.7\columnwidth]{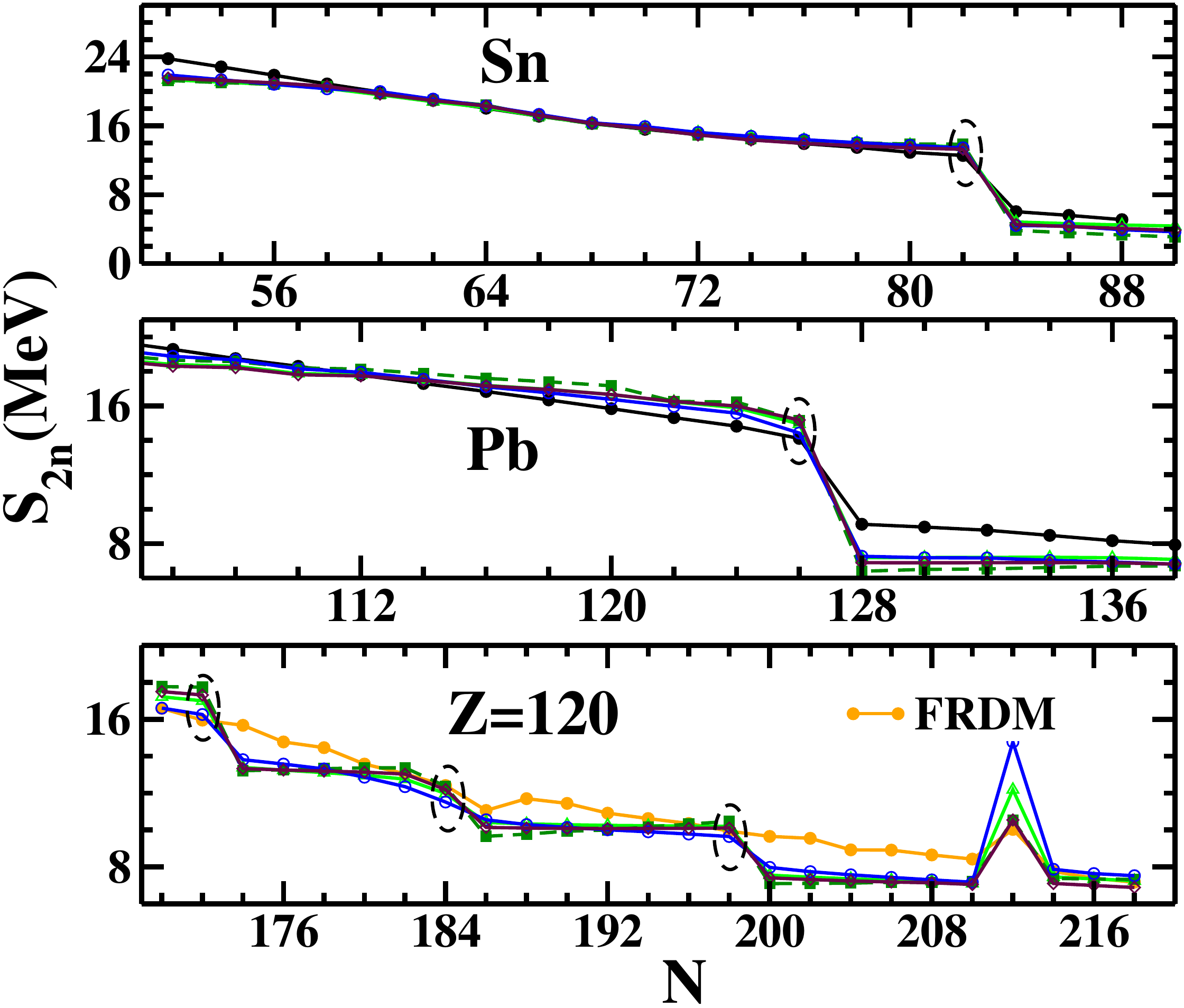}
\caption{The two-neutron separation energy as a function of neutron number for 
the isotopic series of  Ca, Ni, Zr, Sn, and Pb nuclei with NL3 \cite{lala97}, 
FSUGarnet \cite{chai15}, FRDM \cite{moller}, and experimental data \cite{audi12} whenever available. The dotted circle represents the 
magicity of the nuclei.}
        \label{s2n}
\end{center}
\end{figure}

\begin{center}
\bf (ii)  Two-neutron separation energy $S_{2n}(Z,N)$
\end{center}
The large shell gap in single-particle energy levels is an indication of
the magic number.  This is responsible for the extra stability for the
magic nuclei.  The extra stability for a particular  nucleon number
can be understood  from the sudden fall in the two-neutron separation
energy $S_{2n}$. The $S_{2n}$ can be estimated  by the difference in
ground state binding energies of two isotopes, i.e.,

\begin{equation}
S_{2n}(Z,N) = BE(Z,N) - BE(Z,N-2).
\end{equation}

In Fig. \ref{s2n}, we display results for the $S_{2n}$ as a function of
neutron numbers for  Ca, Ni, Zr, Sn, Pb, and Z=120 isotopic chains. The
calculated results are compared with the finite range droplet model (FRDM)
\cite{moller} and most recent experimental data \cite{audi12}. From the figure,
it is clear that there is an evolution of magicity as one moves from the valley
of stability to the drip line. In all cases, the $S_{2n}$ values
decrease gradually with increase in neutron number.  The experimental
manifestation of large shell gaps at neutron numbers N = 20, 28(Ca),
28(Ni), 50(Zr), 82(Sn), and 126(Pb) are reasonably well reproduced by the
four relativistic sets. 
%\st{The flat behavior of Ca isotopes in $S_{2n}$
%from N=30 to N=32 is recently confirmed with the mass measurement at
%TITAN/TRIUMF \cite{titan}.  It is shown that $^{52}$Ca is more bound
%by 1.74 MeV than $^{50}$Ca.} 
Figure \ref{s2n} shows that the experimental $S_{2n}$ of $^{50-52}$Ca are in 
good agreement with the prediction of the NL3 set. 
It is interesting to note that all sets predict the subshell closure
at N=40 for Ni isotopes.  Furthermore,  the two-neutron separation
energy for the isotopic chain of nuclei with  Z=120 is also displayed
in Fig. \ref{s2n}. For the isotopic chain of Z=120, no experimental information 
exits. The only comparison can be made with
theoretical models such as the FRDM \cite{audi12}. At N=172, 184 and
198 sharp falls in separation energy is seen for all forces,
which have been predicted by various theoretical models in the superheavy
mass region \cite{rutz,gupta,patra99,mehta}. It is to be noted that
the isotopes with Z=120 are shown to be spherical in their ground state
\cite{mehta}. 
In a detailed calculation, Bhuyan and Patra using both RMF and Skyrme-Hartree-Fock formalisms, predicted that Z=120 could be the next magic number after Z=82 in 
the superheavy region \cite{bunu12}. Thus, the deformation effects may not 
affect the results for Z=120. Therefore, a future mass measurement of 
$^{292,304,318}$120 would confirm a key test for the theory, as well as direct 
information about the closed-shell behavior at N=172, 184, and 198.

\begin{figure}
	\begin{center}
\includegraphics[width=0.7\columnwidth]{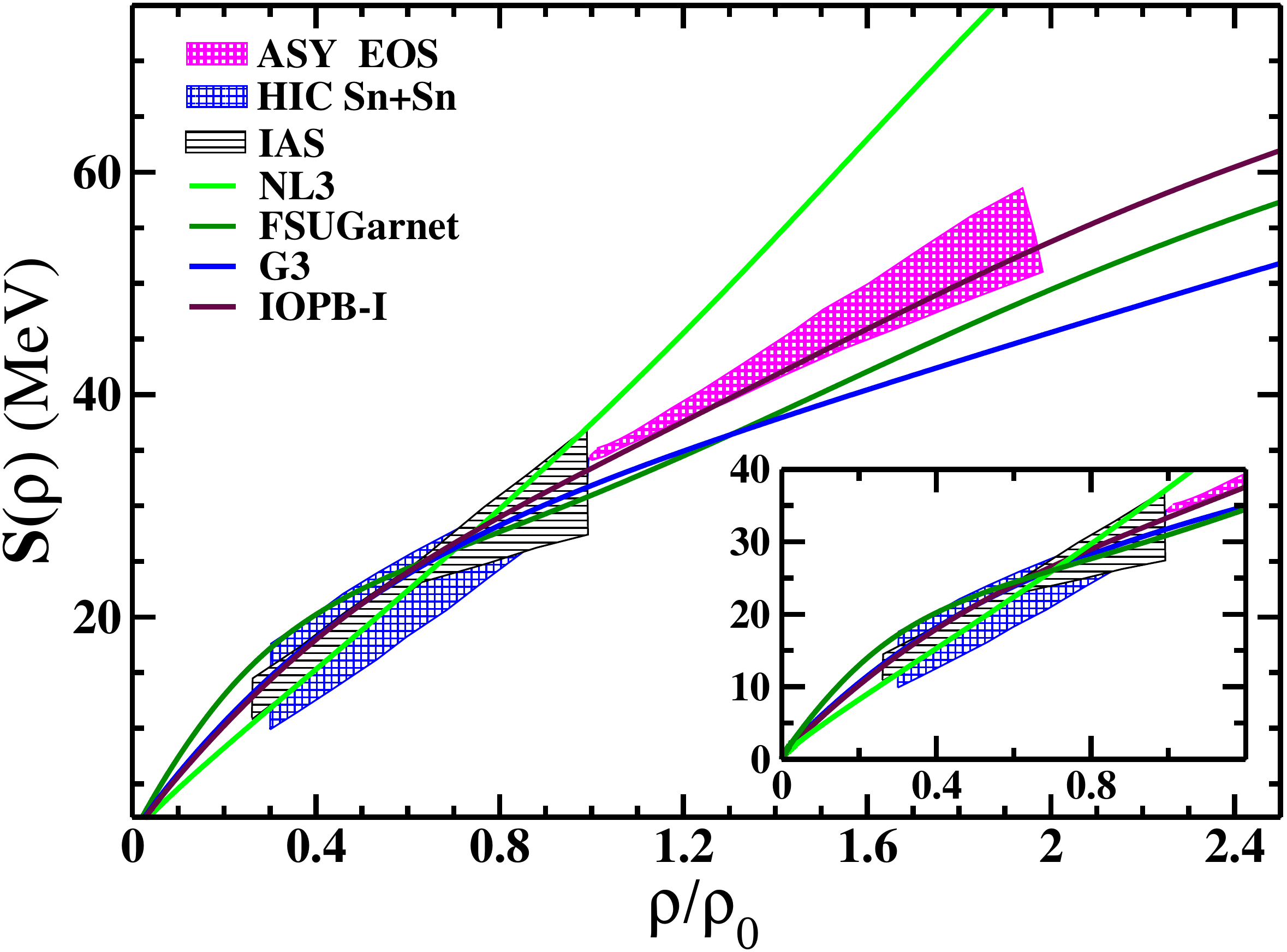}
\caption{Density-dependent symmetry energy from Eq. \eqref{eq30} with 
different ERMF parameter sets along with G3, and IOPB-I parametrizations. The 
shaded region is the symmetry energy from IAS \cite{ias}, HIC Sn+Sn \cite{snsn} and ASY-EoS experimental data \cite{asym}. The zoomed pattern of the symmetry energy 
at low densities is shown in the inset.}
        \label{sym}
		\end{center}
\end{figure}

\begin{figure}[!t]
        \begin{center}
\includegraphics[width=0.7\columnwidth]{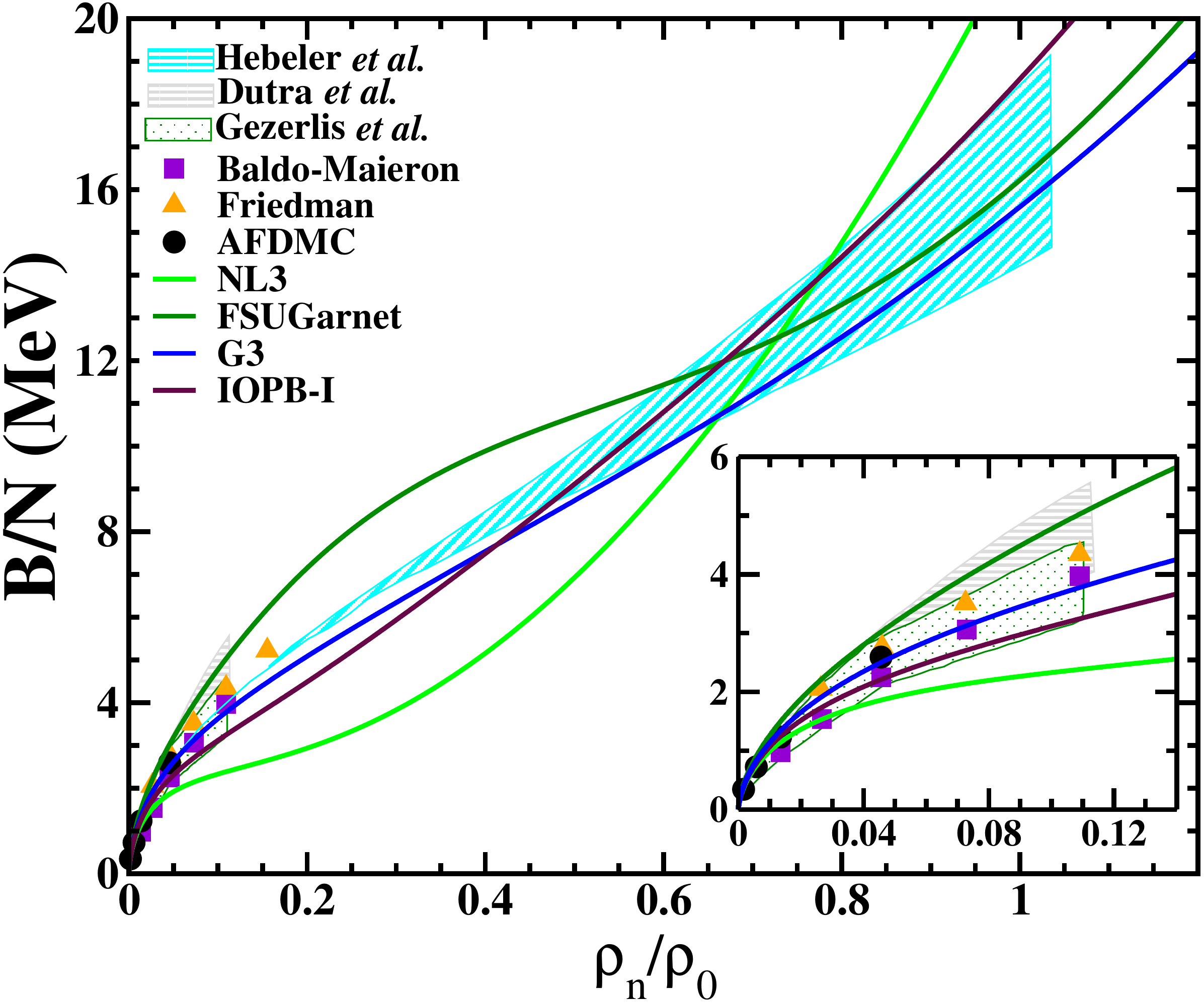}
\caption{The energy per neutron as a function of neutron density with NL3 \cite{lala97}, FSUGarnet \cite{chai15}, G3, and IOPB-I parameter sets. 
Other curves and shaded regions represent the results for various microscopic 
approaches such as Baldo-Maieron \cite{baldo}, Friedman \cite{friedman}, auxiliary-field diffusion Monte Carlo \cite{gando}, Dutra \cite{dutra12}, 
Gezerlis \cite{geze10} and Hebeler \cite{hebe13} methods.}
        \label{sat}
                \end{center}
\end{figure}

\begin{figure}
        \begin{center}
\includegraphics[width=0.7\columnwidth]{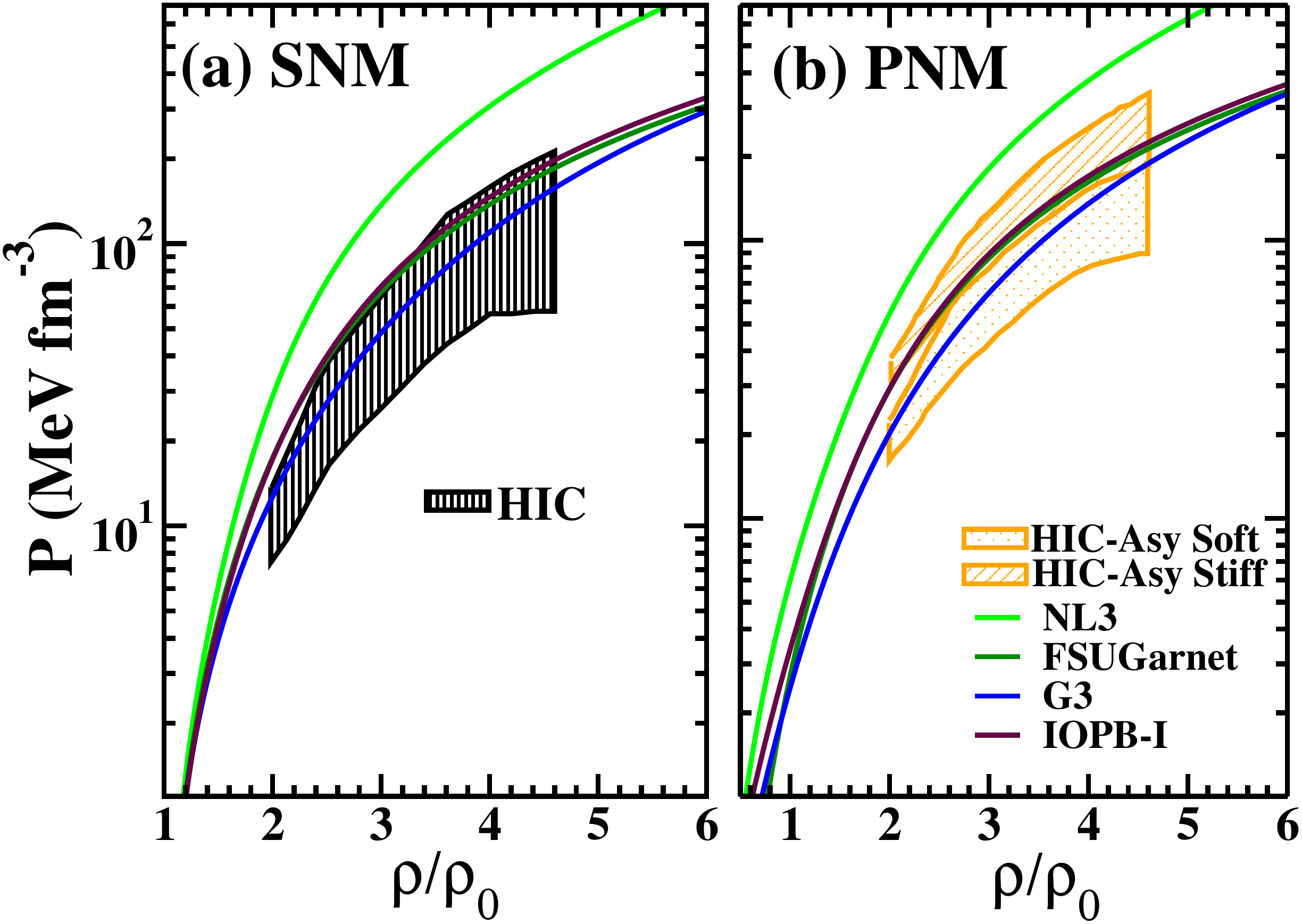}
\caption{Pressure as a function of baryon density for the G3, and IOPB-I 
forces. The results with NL3 \cite{lala97}, and FSUGarnet \cite{chai15} are 
compared with the EoS extracted from the analysis \cite{daniel02} for the (a) 
symmetric nuclear matter (SNM) and (b) pure neutron matter (PNM). }
        \label{eos}
                \end{center}
\end{figure}

\subsection{Infinite Nuclear  Matter}
The nuclear incompressibility $K$ determines the extent to which the nuclear 
matter can be compressed. This plays an important role in the nuclear EoS. 
Currently, the accepted value of $K=240\pm20$ MeV was determined from
isoscalar giant monopole resonance (ISGMR) for $^{90}$Zr and $^{208}$Pb nuclei 
\cite{colo14,piek14}. For our parameter set IOPB-I, we get $K=222.65$ MeV.
The density-dependent symmetry energy $S(\rho)$ is determined from 
Eq. \eqref{eq30} using IOPB-I along with three adopted models. The calculated 
results of the symmetry energy coefficient ($J$), the slope of symmetry energy 
($L$), and other saturation properties are listed in Table \ref{table3}. 
We find that in case of IOPB-I, $J=33.3$ MeV and $L=63.6$ MeV. These values 
are compatible with $J=31.6\pm2.66$ MeV and $L=58.9\pm16$ MeV obtained by 
various terrestrial experimental information and astrophysical observations 
\cite{li13}.

Another important constraint $K_\tau$ has been suggested which lies in the
range of -840 MeV to -350 MeV \cite{stone14,pear10,li10} by various 
experimental data on isoscalar giant monopole resonance, which we can calculate 
from Eq. \eqref{eq34}. It is to be noticed that the calculated values of 
$K_\tau$ are -703.23, -250.41, -307.65, and -389.46 MeV for NL3, FSUGarnet, 
G3, and IOPB-I parameter sets, respectively.  
The ISGMR measurement was investigated in a series of $^{112-124}$Sn isotopes, 
which extracted the value of $K_\tau=-395\pm40$ MeV \cite{garg07}. It is found 
that $K_\tau=-389.46$ MeV for the IOPB-I set is consistent with GMR 
measurement \cite{garg07}. In the absence of cross-coupling, $S(\rho)$ of NL3 
is stiffer at low and high density regimes as displayed in Fig. \ref{sym}.  
Alternatively, the presence of cross-coupling of $\rho$ mesons to the $\omega$ 
(in the case of FSUGarnet and IOPB-I) and $\sigma$ mesons (in case of G3) 
yields the softer symmetry energy at low density which is consistent with HIC 
Sn+Sn \cite{snsn} and IAS \cite{ias} data as shown in the figure. However, the 
IOPB-I set has softer $S(\rho)$ in comparison to the NL3 parameter set at 
higher density which lies inside the shaded region of ASY-EoS experimental data \cite{asym} .

 Next, we display in Fig. \ref{sat} the binding energy per neutron (B/N) as a 
function of the neutron density. Here, special attention is needed to build a 
nucleon-nucleon interaction to fit the data at subsaturation density.  For 
example, the EoS of pure neutron matter (PNM) at low density is obtained within 
the variational method, which is obtained with a Urbana $v14$ interaction \cite{friedman}. In this regard, the effective mean-field models also fulfill this 
demand to some extent \cite{chai15,mondal16,G3}. The cross-coupling 
$\omega$-$\rho$ plays an important role at low density of the PNM. The low 
density (zoomed pattern) nature of the FSUGarnet, G3 and IOPB-I sets are in 
harmony with the results obtained by microscopic calculations \cite{baldo,friedman,gando,dutra12,geze10}, while the results for NL3 deviate from the shaded 
region at low as well as high density regions. We also find a very good 
agreement for FSUGarnet, G3, and IOPB-I at higher densities, which have been 
obtained with chiral two-nucleon (NN) and three-nucleon (3N) interactions 
\cite{hebe13}.

In Fig. \ref{eos}, we show the calculated pressure $P$ for the SNM and PNM with 
the baryon density for the four ERMF models, which then are compared with the 
experimental flow data \cite{daniel02}. It is seen from Fig. \ref{eos}(a) that 
the SNM EoS for the G3 parameter set is in excellent agreement with the flow 
data for the entire density range. The SNM EoS for the  FSUGarnet and IOPB-I 
parameter sets are also compatible with the experimental HIC data  but they are 
stiffer relative to the EoS for the G3 parametrization.
In Fig. \ref{eos}(b), the bounds on the PNM EoS are divided into two categories 
(i) the upper one corresponds to a strong density dependence of symmetry energy 
$S(\rho)$ (HIC-Asy Stiff) and (ii) the lower one corresponds to the weakest 
$S(\rho)$ (HIC-Asy Soft) \cite{daniel02,prak88}. Our parameter set IOPB-I along 
with the G3 and FSUGarnet sets are reasonably in good agreement with experimental flow data. The PNM EoS for the IOPB-I model is quite stiffer than that of G3 at
high densities.

\begin{figure}
        \begin{center}
\includegraphics[width=0.7\columnwidth]{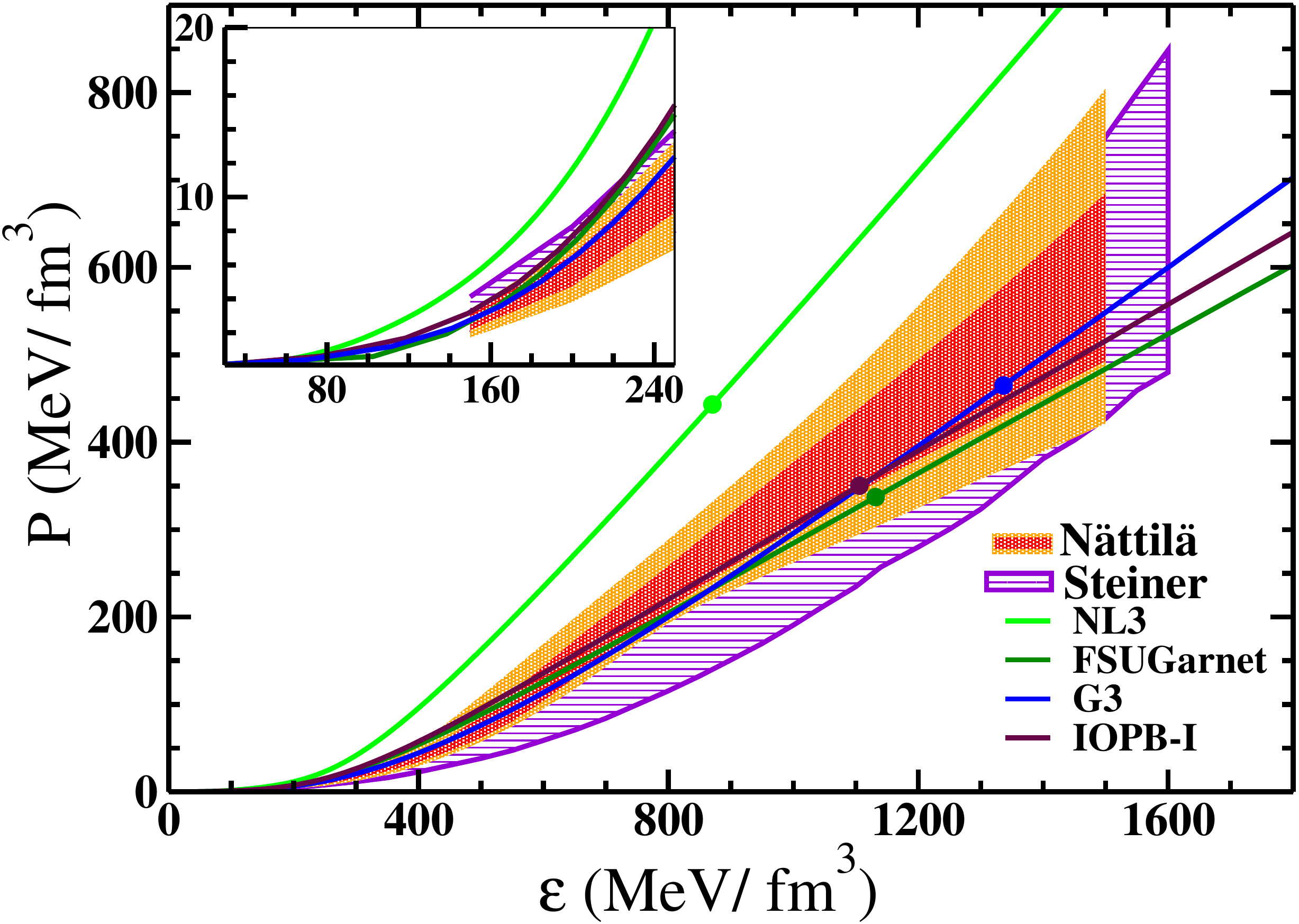}
\caption{The equations of state with NL3, FSUGarnet, G3, and IOPB-I sets for 
nuclear matter under charge neutrality as well as the $\beta$-equilibrium 
condition. The shaded region (violet) represents the observational constraint 
at $r_{ph} = R$ with uncertainty of 2$\sigma$ \cite{steiner10}. Here, $R$ and 
$r_{ph}$ are the neutron star radius and the photospheric radius, respectively. The other shaded region (red and orange) represents the QMC+Model A equation 
of state of cold dense matter with $95\%$ confidence limit \cite{joonas}. The 
region zoomed near the origin is shown in the inset.}
        \label{ep2}
                \end{center}
\end{figure}

\section{ Neutron Stars}
\begin{center}
	{\bf (i) \; Predicted equation of states}
\end{center}

We have solved Eqs. \eqref{eq20} and \eqref{eq21} for the energy density and 
pressure of the $\beta$-equilibrated charge neutral neutron star matter.
Figure \ref{ep2} displays the pressure as a function of energy density for the 
G3, and IOPB-I sets along with the NL3, and FSUGarnet sets. The solid circles 
are the central pressure and energy density corresponding to the maximum mass 
of the neutron star obtained from the above equations of state. The shaded 
region of the EoS can be divided into two parts as follows:

(i) N\"attli\"a {\it et. al.} applied the Bayesian cooling tail method to 
constraint (1$\sigma$ and 2$\sigma$ confidence limit) the EoS of cold dense 
matter inside the neutron stars \cite{joonas}.\\
(ii) Steiner {\it et. al.} determined an empirical dense matter EoS with a 
$95\%$ confidence limit from a heterogeneous data set containing PRE 
bursts and quiescent thermal emission from x-ray transients \cite{steiner10}.

From Fig. \ref{ep2}, it is clear that IOPB-I and FSUGarnet EoSs are similar at 
high density but they differ remarkably at low densities as shown in the zoomed 
area of the inset. The  NL3 set yields the stiffer EoS. Moreover, the IOPB-I set
shows the stiffest EoS up to energy densities $\mathcal{E}$$\lesssim$700
MeV fm$^{-3}$. It can be seen that the results of IOPB-I at very
high densities $\mathcal{E} \sim 400-1600$ MeV fm$^{-3}$ are consistent
with the EoS obtained by N\"attil\"a and Steiner {\it et al.}
\cite{joonas,steiner10}. However, the FSUGarnet set has a softer EoS at low energy
densities $\mathcal{E}$$\lesssim$200 MeV fm$^{-3}$ and stiffer EoS
at intermediate energy densities as compared to that for the G3 set. 
One can conclude from  Table \ref{table3} that the symmetry energy elements
$L$ and $K_{sym}$ are smaller in the G3 model compared to the IOPB-I, FSUGarnet, 
and NL3 sets. Hence, it yields the symmetry energy that is softer at higher
density.

\begin{figure}
        \begin{center}
\includegraphics[width=0.7\columnwidth]{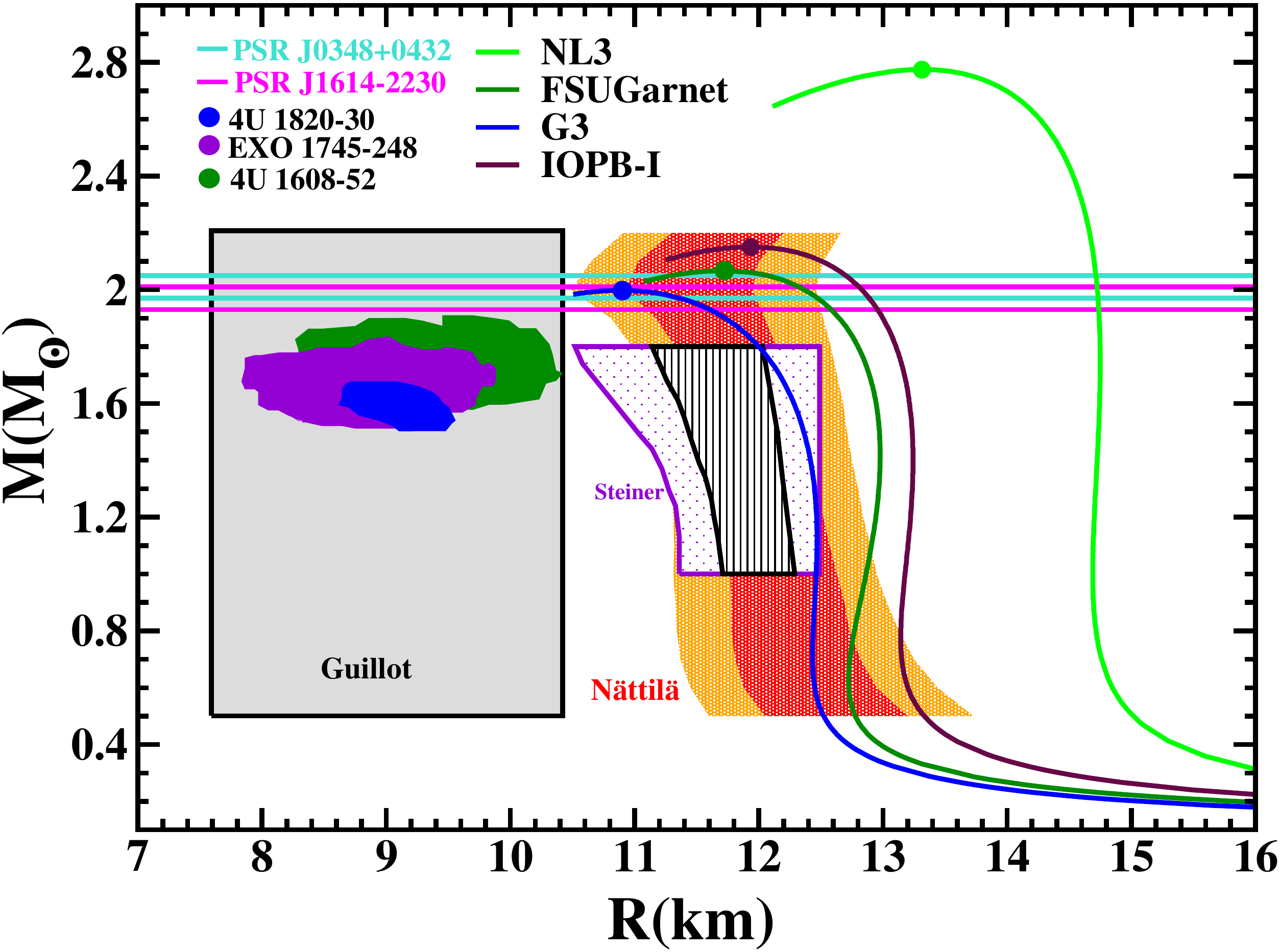}
\caption{The mass-radius profile predicted by NL3, FSUGarnet, G3, and IOPB-I 
parameter sets. The recent observational constraints on neutron-star 
masses \cite{demo10,antoni13} and radii \cite{ozel10,steiner10,sule11,joonas} 
are also shown. }
        \label{MR}
                \end{center}
\end{figure}

\begin{center}
\bf (ii) Mass-radius and tidal deformability of neutron star 
\end{center}
After fixing the equation of state for the various parameter sets, we extended 
our study to calculate the mass, radius, and tidal deformability of a nonrotating neutron star.  

In Fig. \ref{MR}, the horizontal bars in cyan and magenta include the results
from the precisely measured neutron stars masses, such as PSR J1614-2230
with mass $M=1.97\pm 0.04M_\odot$ \cite{demo10} and  PSR J0348+0432 with
$M=2.01\pm 0.04 M_\odot$ \cite{antoni13}.  These observations imply  that
the  maximum mass predicted by any theoretical model should reach the
limit $\sim 2.0 M_\odot$.  We also depict the $1\sigma$ and $2\sigma$
empirical mass-radius constraints for the cold densed matter inside the
NS which were obtained from a Bayesian analysis of type-I x-ray burst
observations \cite{joonas}. A similar approach was applied by Steiner {\it et
al.}, but they obtained the mass-radius from six sources, i.e.,
three from transient low-mass x-ray binaries and three from type-I x-ray
bursters with photospheric radius \cite{steiner10}.

The NL3 model of RMF theory suggests a larger and massive NS with mass
$2.77 M_\odot$ and the corresponding NS radius to be 13.314 km, which is
larger than the best observational radius estimates \cite{steiner10,joonas}. 
Hence it is clear that the new RMF was developed either through 
density-dependent couplings \cite{roca11} or higher order couplings 
\cite{G3,chai15}.  These models successfully reproduce the ground-state 
properties of finite nuclei, nuclear matter saturation properties, and also  
the maximum mass of the neutron stars. Another important advantage of these 
models is that they are consistent with the subsaturation density of
the pure neutron matter. Rezzolla {\it et al.} \cite{luci17} combined the recent 
gravitational-wave observation of a merging system of binary neutron stars via 
the event GW170817 with quasi-universal relations between the maximum mass of 
rotating and non-rotating NSs. It is found that the maximum mass for nonrotating 
NS should be in the range $2.01\pm0.04\lesssim M(M_\odot)$$\lesssim2.16\pm0.03$ 
\cite{luci17}, where the lower limit is
observed from massive pulsars in the binary system \cite{antoni13}. From the
results, we find that the maximum masses for IOPB-I along with FSUGarnet
and G3 EoS are consistent with the observed lower  bound on the
maximum NS mass. For the IOPB-I parametrization, 
the maximum mass of the NS is $2.15 M_\odot$ and  the radius (without
including crust) of the {\it canonical} mass is 13.242 km, which is
relatively larger as compared to the current x-ray observation radii of
range 10.5--12.8 km by N\"attil\"a {\it et. al.} \cite{joonas} and 11--12
km by Steiner {\it et. al.} \cite{steiner10}.  Similarly, FSUGarnet fails
to qualify radius constraint.  However, recently Annala {et al.} suggested that 
the radius of a $1.4M_\odot$ star should be in the range $11.1\leq R_{1.4M_\odot} \leq 13.4$ km \cite{eme17}, which is consistent with the IOPB-I and FSUGarnet 
sets. Furthermore, the G3 EoS is relatively softer at energy density 
$\cal{E}$$\gtrsim200$ MeV fm$^{-3}$ (see in Fig. \ref{ep2}), which is able to 
reproduce  the recent observational maximum mass of $2.0M_\odot$ as well as the 
radius of the {\it canonical} neutron star mass of 12.416 km.

\begin{figure}
        \begin{center}
        \includegraphics[width=0.7\columnwidth]{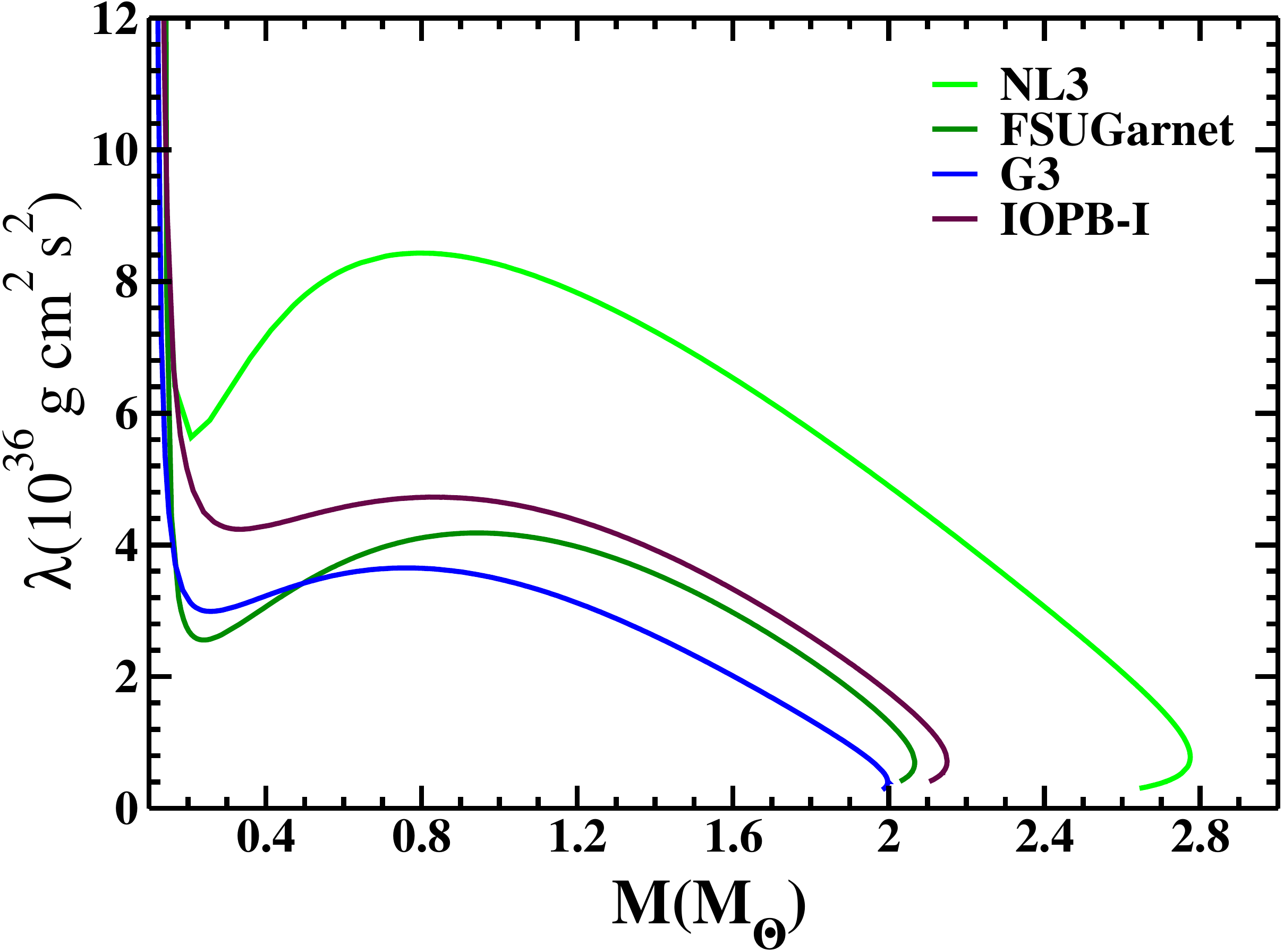}
\caption{The tidal deformability $\lambda$ as a function of neutron star mass 
with different EoS.}
        \label{lamb}
                \end{center}
\end{figure}

Now we move to results for the tidal deformability of the single neutron star 
as well as binary neutron stars (BNSs), which was recently discussed for 
GW170817 \cite{BNS}. 
Figure \ref{lamb} shows the tidal deformability as
a function of NS mass.  In particular, $\lambda$ takes a wide range
of values $\lambda\sim(1--8)\times10^{36}$ g cm$^{2}$ s$^{2}$ as shown
in Fig. \ref{lamb}. For the G3 parameter set, the tidal deformability
$\lambda$ is very low in the mass region $0.5--2.0M_\odot$ in comparison
with other sets. This is because the star exerts high central pressure
and energy density, resulting in the formation of a compact star which is
shown as solid dots in Fig. \ref{ep2}.  However, for the NL3 EoS case,
it turns out that, because of the stiffness of the EoS, the $\lambda$
value is increasing. The tidal deformabilities of the canonical NS
($1.4M_\odot$) of IOPB-I along with FSUGarnet and G3 EoSs are found to
be 3.191$\times 10^{36}$, 3.552$\times 10^{36}$, and 2.613 $\times 10^{36}$ g cm$^{2}$ s$^{2}$, respectively as shown in Tables \ref{table4} and \ref{table5}, 
which are consistent with the results obtained by Steiner {\it et al.} \cite{ste15}.

\begin{figure}
        \begin{center}
        \includegraphics[width=0.7\columnwidth]{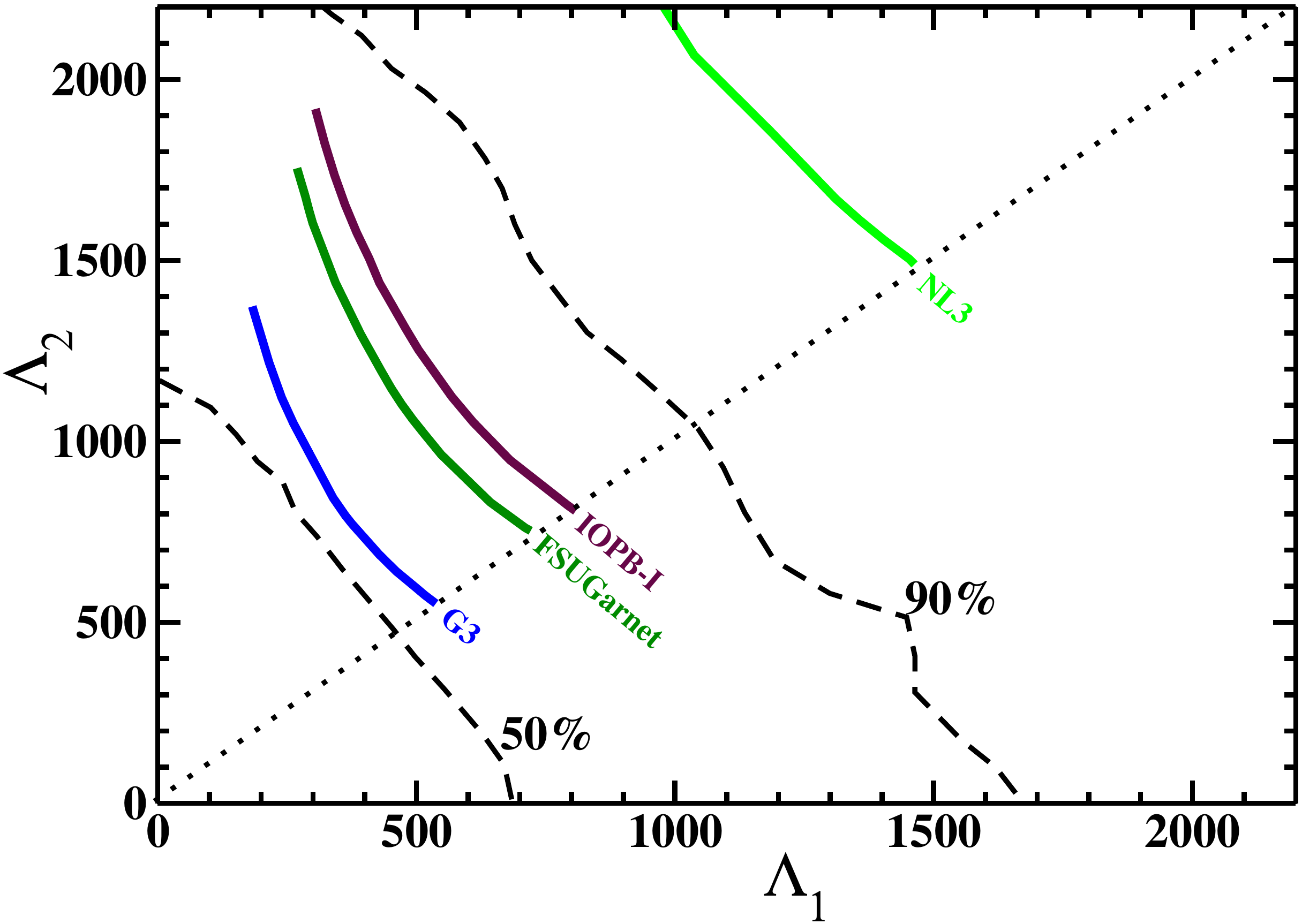}
\caption{Different values of $\Lambda$ generated by using G3, and IOPB-I along 
with NL3, and FSUGarnet EoS are compared with the $90\%$ and $50\%$ probability 
contour in the case of low spin, $|\chi|\leq 0.05$, as given in Fig. 5 of 
GW170817 \cite{BNS}.}
        \label{tidal}
                \end{center}
\end{figure}

Next, we discuss the weighted dimensionless tidal deformability of the 
BNS of mass $m_1$ and $m_2$ which is defined as \cite{BNS,fava,wade}
\begin{eqnarray}
	\tilde{\Lambda}=\frac{8}{13}\Bigg[(1+7\eta-31\eta^{2})(\Lambda_{1}
	\nonumber \\
	+\Lambda_{2})+\sqrt{1-4\eta}(1+9\eta-11\eta^{2})(\Lambda_{1}-
	\Lambda_{2})\Bigg]\; ,
	\label{lambar}
\end{eqnarray}
with tidal correction  
\begin{eqnarray}
	\delta{\tilde\Lambda}=\frac{1}{2}\Bigg[\sqrt{1-4\eta}\Bigg(1-\frac{13272}{1319}\eta+\frac{8944}{1319}\eta^{2}\Bigg)(\Lambda_{1}+\Lambda_{2})\nonumber
	\\
	+\Bigg(1-\frac{15910}{1319}\eta+\frac{32850}{1319}\eta^{2}+\frac{3380}
	{1319}\eta^{3}\Bigg)(\Lambda_{1}-\Lambda_{2})\Bigg].\;
\end{eqnarray}
Here, $\eta=m_{1} m_{2}/M^{2}$  is the symmetric mass ratio, $m_{1}$ and $m_{2}$ 
are the binary masses,  $M=m_{1}+m_{2}$ is the total mass, and $\Lambda_{1}$
and $\Lambda_{2}$ are the dimensionless tidal deformabilities of the BNS, for the
case $m_{1} \geq m_{2}$. Also, we have taken equal and unequal masses 
($m_1$ and $m_2$) BNS for the system as it has been done in Refs. \cite{BNS1,david17}. 
The calculated results for the $\Lambda_1$, $\Lambda_2$, and weighted tidal 
deformability $\tilde\Lambda$ of the present EoS are displayed in Tables 
\ref{table4} and \ref{table5}. 
In Fig. \ref{tidal}, we display the different dimensionless tidal
deformabilities corresponding to progenitor masses of the NS. It can be 
seen that the IOPB-I set along with FSUGarnet and G3 sets are in good agreement 
with the $90\%$ and $50\%$ probability contour of GW170817 \cite{BNS}.
Recently, aLIGO and VIRGO detectors measured a value of $\tilde\Lambda$ 
whose results are more precise than the  results found by considering the
individual values of $\Lambda_{1}$ and $\Lambda_{2}$ of the BNS \cite{BNS}.  
It is noticed that the values of $\tilde\Lambda\leq800$ in the low-spin
case and $\tilde\Lambda\leq700$ in the high-spin case are within the $90\%$
credible intervals which are consistent with the 680.79, 622.06, and
461.03 of the $1.4M_\odot$ NS binary for the IOPB-I, FSUGarnet, and G3
parameter sets, respectively (see Tables \ref{table4} and \ref{table5}
). We also find a reasonably good agreement in the $\tilde\Lambda$
value equal to 582.26 for $1.35M_\odot$ in the G3 EoS, which is
obtained using a Markov chain Monte Carlo simulation of a BNS with
$\tilde\Lambda\approx600$ at a signal-to-noise ratio of 30 in a
single aLIGO detector \cite{wade,fava}.  Finally, we close this
section with the discussion on chirp mass $\cal{M}$$_c$ and chirp radius
$\cal{R}$$_{c}$ of the BNS system  which are defined as 
\begin{eqnarray}
	{\cal{M}}_c = (m_1 m_2)^{3/5} (m_1+m_2)^{-1/5},\\ 
	{\cal{R}}_c =2 {\cal{M}}_c \;\tilde{\Lambda}^{1/5}.
\end{eqnarray} 
The precise mass measurements of the NSs were reported in Refs. \cite{demo10,antoni13}. However, until now no observation has been confirmed regarding the radius 
of the most massive NS. Recently,
aLIGO and VIRGO  measured a chirp mass of  $1.188^{+0.004}_{-0.002}M_\odot$ with
very good precision. With the help of this, we can easily calculate the
chirp radius $\cal{R}$$_{c}$ of the BNS system and we find that the chirp
radius is in the range $7.867 \leq \cal{R}$$_{c}\leq 10.350 \; $km for equal and 
unequal-mass BNS systems as shown in Tables \ref{table4} and \ref{table5}.

\section{Summary and Conclusions}{\label{summary}}

We have built two new relativistic effective interaction for finite nuclei, 
infinite nuclear matter, and neutron stars. The optimization was done
using experimental data for eight spherical nuclei such as binding energy and 
charge radius. The prediction of observables such as binding energies and radii 
with the new G3, and IOPB-I sets for finite nuclei are quite good.
The rms error on the total binding energy calculated with G3 set is noticeably 
smaller than the commonly used parameter sets NL3, FSUGold2, FSUGarnet, and G2. 
The Z=120 isotopic chain shows that the magicity appears at neutron numbers 
N=172, 184, and 198. Furthermore, we find that the IOPB-I set yields
slightly larger values for the neutron-skin thickness. This is due to the small 
strength of the $\omega$-$\rho$ cross-coupling. However, the $\Delta_{np}$ for
G3 parameterization calculated for nuclei over a wide range of masses are in 
harmony with the available experimental data. For infinite nuclear matter at subsaturation and suprasaturation densities, the results of our calculations 
agree well with the known experimental data. 
The nuclear matter properties obtained by this new parameter set IOPB-I are:
nuclear incompressibility $K = 222.65$ MeV, symmetry energy coefficient 
$J = 33.30$ MeV, symmetry energy slope  $L = 63.6$ MeV, and the asymmetry term 
of nuclear incompressibility $K_\tau = 389.46$ MeV, at saturation density 
$\rho_{0} = 0.149$ fm$^{-3}$. In general, all these values are consistent with 
current empirical data. The IOPB-I model satisfies the density dependence of 
the symmetry energy which is obtained from the different sets of experimental
data.  It also yields the NS maximum mass to be 2.15$M_\odot$, which is
consistent with the current GW170817 observational  constraint \cite{luci17}. 
The radius of the canonical neutron star is 13.24 km, compatible with the 
theoretical results in Ref. \cite{eme17}.  Similarly, the predicted values of 
dimensionless tidal deformabilities are in accordance with the GW170817 observational probability contour \cite{BNS}.

\setcounter{equation}{0}
\setcounter{figure}{0}

\newpage
\chapter {Summary and Conclusions}
\label{chapter7}
This thesis has touched several areas of active research in nuclear structure, 
nuclear matter, neutron star, and gravitational waves. Here, we have applied 
the effective field theory motivated RMF formalism which is very successful 
till now for studying the finite nuclei throughout the nuclear chart, infinite 
nuclear matter and neutron stars.  

After a brief introduction on various current phenomena for finite nuclei,
infinite nuclear matter in chapter \ref{chapter1}, chapter \ref{chapter2} 
covers mostly the mathematical derivations, which have been included in our 
work throughout. We began with the extended-RMF Lagrangian density 
by adding two extra terms {\it i.e.}, $\delta$ meson and $\omega-\rho$ 
cross-coupling which contain large number of terms with all types of self- and 
cross-coupling interactions. With the help of Euler-Lagrange equation we have 
reproduced the mean-field equations for different fields such as 
$\phi,\sigma,\omega,\rho$, and $\delta$, which are then solved 
self-consistently. The total energy of the nucleus comes from the energy 
contributions from nucleons and mesons. The temperature dependent BCS and 
Quasi-BCS pairing correlations for open-shell nuclei have been described in 
detail. Also, the equation of state is derived with the help of energy-momentum tensor and several expressions for the properties of the infinite nuclear 
matter are obtained in the RMF approximation.

In chapter \ref{chapter3}, on the basis of RMF formalism, we have studied
systematically the decay properties of recently predicted thermally fissile
Th and U isotopes. The potential energy surface is determined self-consistently
by employing the quadrupole constraint approach. It is found that $^{228-230}$Th
and $^{228-234}$U isotopes are showing three maxima in the PES. Then we 
have calculated fission barrier height of the experimentally known nuclei, which
agrees well with the available data. With the empirical estimation and quantum mechanical 
tunneling approach, we found that the neutron-rich isotopes of these thermally 
fissile nuclei such as $^{254}$Th and $^{256}$U are predicted to be stable 
against $\alpha$- and $cluster$-decays. Interestingly, these isotopes have 
shown the lower height and larger width which make the nucleus infinitely 
stable against spontaneous fission decay because of the decreasing penetrability. 
However, these nuclei are unstable against prompt neutron emission ($\beta$-decay) from the fragments at scission and has a half-life of the order of tens of 
seconds. These finite lifetime suggest that they could be very useful for 
energy production with the help of online synthesis of these nuclei by the 
nuclear reactor technology. Furthermore, $^{254}$Th and $^{256}$U are showing 
the close shell nature of the neutron magic number N=164. We expected that 
N=164 magic number may be identified by the next generation of experimental facilities such as Facility for Rare Isotope Beam (FRIB) at MSU and RIBF at RIKEN.

In chapter \ref{chapter4}, we have extended our study to calculate the 
relative binary mass distribution of $\beta$-stable nuclei $^{232}$Th and 
$^{236}$U and the neutron-rich thermally fissile nuclei $^{254}$Th and 
$^{250}$U within a statistical model.  
The level densities, the excitation energies, and the level density parameter
$a$ are calculated from the TRMF and FRDM formalisms. The excitation energies 
of the fragments are obtained from ground state single-particle energies
in the FRDM calculations. The TRMF model includes the thermal evolution's of 
the pairing gaps and deformation in a self-consistent manner while, these are 
ignored in the FRDM calculations. For $^{232}$Th and $^{236}$U, the 
binary fragments influence the most favorable yields at temperatures $T$ = 2 
and 3 MeV---whose mass distribution end up in the vicinity or exactly at the 
nucleon closed shell (N=82, 50 and Z = 28). However, the fragments of 
the neutron-rich thermally fissile nuclei $^{254}$Th and $^{250}$U have shown 
the neutron/proton close shell N=50,82, and 100 at T=2 and 3 MeV. Although, 
the TRMF and FRDM formalisms yield the closed shell nucleus as one of the 
favorable fragments, still the details of the relative binary fragmentation 
yields and probability of the mass distributions are quite different in these 
two formalisms.

In chapter \ref{chapter5}, the derivations of Newtonian and relativistic 
tides for neutron stars are presented. Using these derivations, we have 
thoroughly discussed the various types of Love numbers such as electric-type 
Love numbers $k_2, k_3, k_4$, magnetic-type 
Love number $j_2$, and shape Love number $h_l$ using the four EoSs for neutron 
and hyperon stars. These EoSs are calculated from an effective field theory 
motivated with RMF formalism under the $\beta$-equilibrium condition and we 
have compared the results with empirical astrophysical constraint. The 
inclusion of hyperon in our description which softens the EoS and there by 
reduces the maximum mass of the NS further. We find that the tidal 
deformability of the canonical neutron and hyperon stars either change 
slightly or remain unchanged with the addition of a hyperon in the NS. The 
recent observation GW170817 has also confirmed the tight constraint on the 
tidal deformability of the NS matter. We expect that the next generation of 
the gravitational wave detectors may provide obstacles on the tidal 
deformability of the hyperon star. Further, our calculations suggested that 
the higher order Love numbers $k_3$ and $k_4$ are very important for the 
detection of gravitational waves.

In chapter \ref{chapter6}, we improve the existing parameterizations of the 
ERMF model which includes couplings of the meson field gradients to the 
nucleons and the tensor couplings of the mesons to the nucleons in addition to 
the several self and cross-coupling terms. The nuclear matter 
incompressibility coefficient  and/or symmetry energy coefficient associated 
with earlier parameterizations of such ERMF model were a bit too large which 
has been taken care of in our new parameter set G3. The rms error on the total 
binding energy calculated from our parameter set is smaller than the commonly 
used parameter sets NL3, FSUGold2, FSUGarnet, and G2.  
The neutron-skin thickness $\Delta r_{np}$ for G3 set calculated for nuclei 
over a wide range of masses are in harmony with the available experimental 
data. However, IOPB-I gives slightly larger value of $\Delta r_{np}$,
due to the small strength of the $\omega-\rho$ cross-coupling. The neutron 
matter EoS at subsaturation densities for G3, and IOPB-I parameter sets show 
reasonable improvement over other parameter sets considered. Also, the maximum 
mass for the neutron star for G3 set $\sim 2.0M_\odot$, and $2.15M_\odot$ for 
IOPB-I set are compatible with the pulsar measurements, and GW170817 data. The 
radius of the neutron star with the canonical mass agree quite well with the 
two independent analyses of gravitational waves from the GW170817 neutron star 
merger \cite{fatt17,eme17}. The smallness of $R_{1.4}$ for G3 parameter set 
in comparison to those for the earlier parametrization of the RMF models, 
which are compatible with the observational constraint of $2M_\odot$, is a 
desirable feature. Motivated by recent astrophysical GW170817 observation
of the tight constraint on the dimensionless tidal deformability, we found that
the predictions of G3, and IOPB-I models to be in fairly good agreement with the
GW170817 probability contour in the low spin prior as given in Fig. 5 of Ref.
\cite{BNS}.\\

{\bf Future Prospects:} 
In my thesis work, I have tried to conjoint two major branches of physics viz., 
(i) nuclear structure physics and (ii) astrophysics and have tried to apply 
ERMF theory. Having worked in the ERMF theory and the newest calculations in 
nuclear structure and astrophysics, I can further apply the model to various 
applications on nuclear fission, neutron star and gravitational waves. A brief 
outline of my future work is as follows:
\begin{itemize}
\item[$\bullet$] As we have already discussed in the thesis that the 
neutron-rich thermally fissile nuclei may be used in the nuclear reactor so 
that we can make nuclear power. But we need more work in this direction. 
As we have pointed out that $\beta$-stable thermally fissile isotopes are 
found in the mass range A=230-240. Fortunately India has rich thorium deposits 
constituting roughly $25\%$ of the global reserves. As thorium isotopes too 
fall in the mass range A=230-240, so many tests can be done in this regard to 
find out other thermally fissile nuclei and our theoretical predictions can 
give input to the experimental findings.

\item[$\bullet$] In our upcoming work, we will perform a detailed covariance 
analysis for the G3, and IOPB-I models used in the present thesis and access the
uncertainties associated with the various parameters of the Lagrangian density. An appropriate covariance
analysis of our model requires a set of fitting data which include large
variety of nuclear and neutron star observables.
\item[$\bullet$]
The GW170817 observation has opened up a new era of gravitational astronomy
as well as nuclear physics. Still a lot of work is to be done in these
fields and several more events are expected to come up in near future for
extraction of NS properties such as the stars mass, spin, size, and shape,
which will finally lead to tighter constraints on the neutron star equation of state.
With the help of GW observations, we can connect the nuclear matter 
properties, in particular, the slope of the symmetry energy ($L$) 
with the structure properties of neutron star---which is an important task for
the nuclear physics community.
%Also, it can be tested whether the final state of neutron star merger will
%become a more massive NS or a black hole.

\end{itemize}

\setcounter{equation}{0}
\setcounter{figure}{0}

\begin{appendices}
\chapter{Derivation of the TOV equation}
Consider the general static, spherically symmetric metric \cite{misner,carrol}

\begin{eqnarray}
ds^2 = -e^{2 \nu(r)} dt^2 + e^{2 \lambda(r)} dr^2 + r^2 d\theta^2 + r^2 
sin^2 \theta d\phi^2 
\end{eqnarray}

Let's now take this metric and use Einstein's equation to solve the function
$\nu(r)$ and $\lambda(r)$. 

We are looking for non-vacuum solutions, so we turn to the full Einstein 
equation,

\begin{eqnarray}
G_{\mu \nu}= R_{\mu \nu}-\frac{1}{2} R g_{\mu \nu}=8 \pi   T_{\mu \nu}
\end{eqnarray}
 where $R_{\mu \nu}$ and $R$ are the Ricci tensor and Ricci scalar,
 respectively. $g_{\mu \nu}$ is the metric 
\[
g_{\mu \nu}=
\begin{bmatrix}
g_{tt}       & g_{tr} & g_{t\theta}  & g_{t\phi} \\
    g_{rt}       & g_{rr} & g_{r\theta}  & g_{r\phi} \\
    g_{\theta t}       & g_{\theta r} & g_{\theta \theta} & g_{\theta \phi}\\
g_{\phi t}       & g_{\phi r} & g_{\phi \theta} & g_{\phi \phi}
\end{bmatrix}
=
\begin{bmatrix}
-e^{2\nu(r)}      & 0 & 0  & 0 \\
   0       & e^{2\lambda(r)} & 0  & 0 \\
   0     & 0 & r^2 & 0\\
0       & 0 & 0 & r^2 sin^{2}\theta
\end{bmatrix}
\]
We begin by evaluating the christoffel (or affine connection or Levi-Civita
or Riemann connection)  symbols. If we use labels ($t,r,\theta,\phi$).

\begin{eqnarray}
\Gamma^{\lambda}_{\mu \nu}=\frac{1}{2} g^{\lambda \sigma} (\partial_\mu 
g_{\sigma \nu} + \partial_\nu  g_{\mu \sigma} - \partial_\sigma g_{\mu \nu})
\end{eqnarray}

\begin{eqnarray}
	\Gamma^{t}_{t r}&=&\frac{1}{2} g^{t \sigma} (\partial_t 
g_{\sigma r} + \partial_r  g_{t \sigma} - \partial_\sigma g_{t r})\nonumber\\
	&=&\frac{1}{2} g^{t \sigma} (0 + \partial_r  g_{t \sigma} -0))
\nonumber\\
	&=&\frac{1}{2} g^{t t} ( \partial_r  g_{t t})\nonumber\\
	&=&\frac{1}{2} (-e^{-2\nu(r)}) ( \partial_r (-e^{2\nu(r)}))
\nonumber\\
	&=&\frac{1}{2} (-e^{-2\nu(r)}) (-2 e^{2\nu(r)} \partial_r \nu)\nonumber\\
&=& \partial_r \nu
\end{eqnarray}

\begin{eqnarray}
	\Gamma^{r}_{t t}&=&\frac{1}{2} g^{r \sigma} (\partial_t 
g_{\sigma t} + \partial_t  g_{t \sigma} - \partial_\sigma g_{t t})\nonumber\\
	&=&\frac{1}{2} g^{r r} (\partial_t 
g_{r t} + \partial_t  g_{t r} - \partial_r g_{t t})\nonumber\\
	&=&\frac{1}{2} (e^{-2\lambda(r)}) (0 + 0 - \partial_r(-e^{2\nu(r)}))\nonumber\\
	&=&\frac{1}{2} (e^{-2\lambda(r)}) (2 e^{2\nu(r)}) \partial_r 
\nu\nonumber\\
&=& e^{2(\nu(r)-\lambda(r))}\partial_r \nu
\end{eqnarray}

\begin{eqnarray}
	\Gamma^{r}_{r r}&=&\frac{1}{2} g^{r \sigma} (\partial_r 
g_{\sigma r} + \partial_r  g_{r \sigma} - \partial_\sigma g_{r r})\nonumber\\
	&=&\frac{1}{2} g^{r r} (\partial_r g_{r r} + \partial_r  g_{r r} - 
\partial_r g_{r r})\nonumber\\
	&=&\frac{1}{2} g^{r r} (\partial_r g_{r r})\nonumber\\
	&=&\frac{1}{2} (e^{-2\lambda(r)}) \partial_r(e^{2\lambda(r)})\nonumber\\
	&=&\frac{1}{2} (e^{-2\lambda(r)}) (2) e^{2\lambda(r)} \partial_r \lambda\nonumber\\
&=& \partial_r \lambda
\end{eqnarray}

\begin{eqnarray}
	\Gamma^{\theta}_{r \theta}&=&\frac{1}{2} g^{\theta \sigma} (\partial_r 
g_{\sigma \theta} + \partial_\theta  g_{r \sigma} - \partial_\sigma g_{r \theta})\nonumber\\
	&=&\frac{1}{2} g^{\theta \theta} (\partial_r g_{\theta \theta} + \partial_\theta  g_{r \theta} - \partial_\theta g_{r \theta})\nonumber\\
	&=&\frac{1}{2} (r^{-2}) \partial_r (r^{2})=\frac{1}{2} (r^{-2}) (2 r)\nonumber\\
	\Gamma^{\theta}_{r \theta}&=&\frac{1}{r}
\end{eqnarray}

\begin{eqnarray}
	\Gamma^{r}_{\theta \theta}&=&\frac{1}{2} g^{r \sigma} (\partial_\theta 
g_{\sigma \theta} + \partial_\theta  g_{\theta \sigma} - \partial_\sigma g_{\theta \theta})\nonumber\\
	&=&\frac{1}{2} g^{r r} (\partial_\theta g_{r \theta} + \partial_\theta  g_{\theta r} - \partial_r g_{\theta \theta})\nonumber\\
	&=&\frac{1}{2} g^{r r} (0+0-\partial_r g_{\theta \theta})=\frac{1}{2} g^{r r}  (-\partial_r g_{\theta \theta})\nonumber\\
	&=&\frac{1}{2} (e^{-2\lambda(r)})(-\partial_r r^2)=-\frac{1}{2} (e^{-2\lambda(r)}) (2r)\nonumber\\
&=& -r e^{-2\lambda(r)}
\end{eqnarray}

\begin{eqnarray}
	\Gamma^{\phi}_{r \phi}&=&\frac{1}{2} g^{\phi \sigma} (\partial_r 
g_{\sigma \phi} + \partial_\phi  g_{r \sigma} - \partial_\sigma g_{r \phi})
\nonumber\\
	&=&\frac{1}{2} g^{\phi \phi} (\partial_r 
g_{\phi \phi} + \partial_\phi  g_{r \phi} - \partial_\phi g_{r \phi})
\nonumber\\
	&=&\frac{1}{2} g^{\phi \phi} (\partial_r g_{\phi \phi}+0-0)=\frac{1}{2} g^{\phi \phi} (\partial_r g_{\phi \phi})\nonumber\\
	&=&\frac{1}{2} (r^2 sin^2 \theta)^{-1} \partial_r (r^2 sin^2 \theta)\nonumber\\
	&=&\frac{1}{2} (r^2 sin^2 \theta)^{-1} (2 r sin^2 \theta)=\frac{1}{r}
\end{eqnarray}

\begin{eqnarray}
	\Gamma^{r}_{\phi \phi}&=&\frac{1}{2} g^{r \sigma} (\partial_\phi 
g_{\sigma \phi} + \partial_\phi  g_{\phi \sigma} - \partial_\sigma g_{\phi \phi})\nonumber\\
	&=&\frac{1}{2} g^{r r} (\partial_\phi g_{r \phi} + \partial_\phi  g_{\phi r} 
- \partial_r g_{\phi \phi})\nonumber\\
	&=&\frac{1}{2} g^{r r} (0+0-\partial_r g_{\phi \phi})=\frac{1}{2} g^{r r}(\partial_r g_{\phi \phi})\nonumber\\
	&=&\frac{1}{2} (e^{-2\lambda(r)})(-\partial_r (r^2 sin^2\theta))\nonumber\\
	&=&\frac{1}{2} (e^{-2\lambda(r)}) (-2r sin^2\theta)\nonumber\\
&=&-r sin^2\theta e^{-2\lambda(r)}
\end{eqnarray}

\begin{eqnarray}
	\Gamma^{\theta}_{\phi \phi}&=&\frac{1}{2} g^{\theta \sigma} (\partial_\phi 
g_{\sigma \phi} + \partial_\phi  g_{\phi \sigma} - \partial_\sigma g_{\phi \phi})\nonumber\\
	&=&\frac{1}{2} g^{\theta \theta} (\partial_\phi g_{\theta \phi} + \partial_\phi  
g_{\phi \theta} - \partial_\theta g_{\phi \phi})\nonumber\\
	&=&\frac{1}{2} g^{\theta \theta} (0+0-\partial_\theta g_{\phi \phi})\nonumber\\
	&=&\frac{1}{2}(r^2)^{-1}(-\partial_\theta(r^2 sin^2\theta))\nonumber\\
&=&-sin{\theta} cos\theta
\end{eqnarray}

\begin{eqnarray}
	\Gamma^{\phi}_{\theta \phi}&=&\frac{1}{2} g^{\phi \sigma} (\partial_\theta 
g_{\sigma \phi} + \partial_\phi  g_{\theta \sigma} - \partial_\sigma g_{\theta \phi})\nonumber\\
	&=&\frac{1}{2} g^{\phi \phi} (\partial_\theta g_{\phi \phi} + \partial_\phi  
g_{\theta \phi} - \partial_\phi g_{\theta \phi})\nonumber\\
	&=&\frac{1}{2} g^{\phi \phi}(\partial_\theta g_{\phi \phi}+0-0)\nonumber\\
	&=&\frac{1}{2}(r^2 sin^2\theta)^{-1}\partial_\theta(r^2 sin^2\theta)\nonumber\\
	&=&\frac{1}{2}(r^2 sin^2\theta)^{-1}(2r^2sin{\theta}cos\theta)\nonumber\\
&=&\frac{cos\theta}{sin\theta}
\end{eqnarray}

We know that Riemann tensor
\begin{eqnarray}
R^{\rho}_{\sigma \mu \nu}=\partial_\mu \Gamma^{\rho}_{\nu \sigma}-\partial_\nu 
\Gamma^{\rho}_{\mu \sigma}+\Gamma^{\rho}_{\mu \lambda} \Gamma^{\lambda}_{\nu \sigma} - \Gamma^{\rho}_{\nu \lambda} \Gamma^{\lambda}_{\mu \sigma}
\end{eqnarray}
Now, We calculate Ricci tensor
\begin{eqnarray}
	R_{\mu \nu}&=& R^{\lambda}_{\mu \lambda \nu}\nonumber\\
	R_{t t}&=& R^{\lambda}_{t \lambda t}=R^{r}_{t r t}+R^{\theta}_{t \theta t}
+R^{\phi}_{t \phi t}
\end{eqnarray}
\begin{eqnarray}
	R^{r}_{t r t}&=& \partial_r \Gamma^{r}_{t t}-\partial_t \Gamma^{r}_{r t}+
\Gamma^{r}_{r \lambda} \Gamma^{\lambda}_{t t} - \Gamma^{r}_{t \lambda} 
\Gamma^{\lambda}_{r t}\nonumber\\
	&=&\partial_r \Gamma^{r}_{t t}-0+\Gamma^{r}_{r r} \Gamma^{r}_{t t}
+\Gamma^{r}_{r \theta} \Gamma^{\theta}_{t t}+\Gamma^{r}_{r \phi} 
\Gamma^{\phi}_{t t}- \Gamma^{r}_{t t} \Gamma^{t}_{r t}- \Gamma^{r}_{t r} 
\Gamma^{r}_{r t}- \Gamma^{r}_{t \theta} \Gamma^{\theta}_{r t}- 
\Gamma^{r}_{t \phi}\Gamma^{\phi}_{r t}\nonumber\\
	&=&\partial_r(e^{2(\nu-\lambda)} \partial_r\nu)+\partial_r {\lambda}  
e^{2(\nu-\lambda)} \partial_r \nu+0+0-e^{2(\nu-\lambda)} \partial_r 
\nu \partial_r \nu -0-0-0 \nonumber\\  
	&=&\partial^{2}_r{\nu} e^{2(\nu-\lambda)}+\partial_r{\nu} \partial_r 
(e^{2(\nu-\lambda)})+\partial_r{\nu} \partial_r{\lambda} e^{2(\nu-\lambda)}
-e^{2(\nu-\lambda)} (\partial_r \nu)^2\nonumber\\
	&=&\partial^{2}_r{\nu} e^{2(\nu-\lambda)} + (2 e^{2(\nu-\lambda)} 
\partial_r{\nu} -2 e^{2(\nu-\lambda)} \partial_r{\lambda})\partial_r{\nu}
+\partial_r{\nu} \partial_r{\lambda} e^{2(\nu-\lambda)} -e^{2(\nu-\lambda)} (\partial_r \nu)^2\nonumber\\
	&=&e^{2(\nu-\lambda)}[\partial^{2}_r{\nu}+2(\partial_r{\nu})^2-2\partial_r \nu \partial_r \lambda+\partial_r \nu \partial_r \lambda-(\partial_r{\nu})^2]\nonumber\\
&=& e^{2(\nu-\lambda)}[\partial^{2}_r{\nu}+(\partial_r{\nu})^2-\partial_r \nu \partial_r \lambda] 
\end{eqnarray}

\begin{eqnarray}
	R^{\theta}_{t \theta t}&=& \partial_\theta \Gamma^{\theta}_{t t}-\partial_t \Gamma^{\theta}_{\theta t}+ \Gamma^{\theta}_{\theta \lambda} \Gamma^{\lambda}_{t t} - 
\Gamma^{\theta}_{t \lambda} \Gamma^{\lambda}_{\theta t}\nonumber\\
	&=& 0+0+ \Gamma^{\theta}_{\theta t} \Gamma^{t}_{t t}+\Gamma^{\theta}_{\theta r} 
\Gamma^{r}_{t t}+\Gamma^{\theta}_{\theta \theta} \Gamma^{\theta}_{t t} 
+\Gamma^{\theta}_{\theta \phi} \Gamma^{\phi}_{t t}- 
\Gamma^{\theta}_{t t} \Gamma^{t}_{\theta t}-\Gamma^{\theta}_{t r} 
\Gamma^{r}_{\theta t}-\Gamma^{\theta}_{t \theta} \Gamma^{\theta}_{\theta t}-
\Gamma^{\theta}_{t \phi} \Gamma^{\phi}_{\theta t}\nonumber\\
	&=&0+0+0+\Gamma^{\theta}_{\theta r} \Gamma^{r}_{t t}+0+0-0-0-0-0\nonumber\\
&=&\frac{1}{r} e^{2(\nu-\lambda)}\partial_r\nu
\end{eqnarray}

\begin{eqnarray}
	R^{\phi}_{t \phi t}&=& \partial_\phi \Gamma^{\phi}_{t t}-\partial_t \Gamma^{\phi}_{\phi t}+ \Gamma^{\phi}_{\phi \lambda} \Gamma^{\lambda}_{t t} - \Gamma^{\phi}_{t \lambda} \Gamma^{\lambda}_{\phi t}\nonumber\\
	&=&0+0+ \Gamma^{\phi}_{\phi \lambda} \Gamma^{\lambda}_{t t} - \Gamma^{\phi}_{t \lambda} \Gamma^{\lambda}_{\phi t}\nonumber\\
	&=&\Gamma^{\phi}_{\phi t} \Gamma^{t}_{t t}+\Gamma^{\phi}_{\phi r} \Gamma^{r}_{t t}
+\Gamma^{\phi}_{\phi \theta} \Gamma^{\theta}_{t t}+\Gamma^{\phi}_{\phi \phi} 
\Gamma^{\phi}_{t t}-\Gamma^{\phi}_{t t} \Gamma^{t}_{\phi t}-\Gamma^{\phi}_{t r} \Gamma^{r}_{\phi t}-\Gamma^{\phi}_{t \theta} \Gamma^{\theta}_{\phi t}-\Gamma^{\phi}_{t \phi} \Gamma^{\phi}_{\phi t}\nonumber\\
	&=&0+\Gamma^{\phi}_{\phi r} \Gamma^{r}_{t t}+0+0-0-0-0-0\nonumber\\
&=&\frac{1}{r} e^{2(\nu-\lambda)}\partial_r\nu
\end{eqnarray}
From eq.(A.15) can be written as:
\begin{eqnarray}
R_{t t}&=& R^{\lambda}_{t \lambda t}=R^{r}_{t r t}+R^{\theta}_{t \theta t}
+R^{\phi}_{t \phi t}\nonumber \\
&=&e^{2(\nu-\lambda)}[\partial^{2}_r{\nu}+(\partial_r{\nu})^2-\partial_r \nu \partial_r \lambda]+\frac{2}{r} e^{2(\nu-\lambda)}\partial_r\nu\nonumber\\
	&=& e^{2(\nu-\lambda)}[\partial^{2}_r{\nu}+(\partial_r{\nu})^2-\partial_r \nu \partial_r \lambda+\frac{2}{r} \partial_r\nu]
\end{eqnarray}

\begin{eqnarray}
R_{r r}&=& R^{\lambda}_{r\lambda r}=R^{t}_{r t r}+R^{\theta}_{r \theta r}
+R^{\phi}_{r \phi r}\nonumber
\end{eqnarray}

\begin{eqnarray}
	R^{t}_{r t r}&=& \partial_t \Gamma^{t}_{r r}-\partial_r \Gamma^{t}_{t r}+ 
\Gamma^{t}_{t \lambda} \Gamma^{\lambda}_{r r} - \Gamma^{t}_{r \lambda} 
\Gamma^{\lambda}_{t r}\nonumber\\
	&=&0-\partial_r \Gamma^{t}_{t r}+\Gamma^{t}_{t t} \Gamma^{t}_{r r}+\Gamma^{t}_{t r} \Gamma^{r}_{r r}+\Gamma^{t}_{t \theta} \Gamma^{\theta}_{r r}+\Gamma^{t}_{t 
\phi} \Gamma^{\phi}_{r r}-\Gamma^{t}_{r t} \Gamma^{t}_{t r}-\Gamma^{t}_{r 
r}\Gamma^{r}_{t r}-\Gamma^{t}_{r \theta} \Gamma^{\theta}_{t r}-\Gamma^{t}_{r 
\phi}\Gamma^{\phi}_{t r}\nonumber\\
	&=&-\partial_r(\partial_r \nu)+0+\partial_r \nu \partial_r \lambda+0+0-\partial_r \nu \partial_r \nu -0-0-0\nonumber\\
&=&\partial_r \nu \partial_r \lambda-\partial^{2}_r \nu-(\partial_r \nu)^2
\end{eqnarray}

\begin{eqnarray}
	R^{\theta}_{r \theta r}&=& \partial_\theta \Gamma^{\theta}_{r r}-\partial_r 
\Gamma^{\theta}_{\theta r}+ \Gamma^{\theta}_{\theta \lambda} \Gamma^{\lambda}_{r r} - \Gamma^{\theta}_{r \lambda} \Gamma^{\lambda}_{\theta r}\nonumber\\
	&=&0-\partial_r (\frac{1}{r})+\Gamma^{\theta}_{\theta t} \Gamma^{t}_{r r} +
\Gamma^{\theta}_{\theta r} \Gamma^{r}_{r r}+\Gamma^{\theta}_{\theta \theta} 
\Gamma^{\theta}_{r r}+\Gamma^{\theta}_{\theta \phi} \Gamma^{\phi}_{r r}-
\Gamma^{\theta}_{r t} \Gamma^{t}_{\theta r}-\Gamma^{\theta}_{r r} 
\Gamma^{r}_{\theta r}- \Gamma^{\theta}_{r \theta} \Gamma^{\theta}_{\theta r}
-\Gamma^{\theta}_{r \phi} \Gamma^{\phi}_{\theta r}\nonumber\\
	&=&\frac{1}{r^2}+0+\frac{1}{r}\partial_r \lambda+0-\frac{1}{r}\frac{1}{r}-0-0-0\nonumber\\
&=&\frac{1}{r}\partial_r \lambda 
\end{eqnarray}

\begin{eqnarray}
	R^{\phi}_{r \phi r}&=& \partial_\phi \Gamma^{\phi}_{r r}-\partial_r 
\Gamma^{\phi}_{\phi r}+ \Gamma^{\phi}_{\phi \lambda} \Gamma^{\lambda}_{r r} - \Gamma^{\phi}_{r \lambda} \Gamma^{\lambda}_{\phi r}\nonumber\\
	&=&0-\partial_r (\frac{1}{r})+\Gamma^{\phi}_{\phi t} \Gamma^{t}_{r r} +
\Gamma^{\phi}_{\phi r} \Gamma^{r}_{r r}+\Gamma^{\phi}_{\phi \theta} \Gamma^{\theta}_{r r}+\Gamma^{\phi}_{\phi \phi} \Gamma^{\phi}_{r r}-\Gamma^{\phi}_{r t} \Gamma^{t}_{\phi r}-\Gamma^{\phi}_{r r} \Gamma^{r}_{\phi r} - \Gamma^{\phi}_{r \theta} \Gamma^{\theta}_{\phi r}-\Gamma^{\phi}_{r \phi} \Gamma^{\phi}_{\phi r}\nonumber\\
	&=&\frac{1}{r^2}+0+\frac{1}{r}\partial_r \lambda+0-0-0-0-\frac{1}{r}\frac{1}{r}
\nonumber\\
	&=& \frac{1}{r}\partial_r \lambda 
\end{eqnarray}

\begin{eqnarray}
	R_{rr}&=&\partial_r \nu \partial_r \lambda-\partial^{2}_r \nu-(\partial_r \nu)^2+\frac{1}{r}\partial_r \lambda+\frac{1}{r}\partial_r \lambda\nonumber\\
	&=&\partial_r \nu \partial_r \lambda-\partial^{2}_r \nu-(\partial_r \nu)^2+\frac{2}{r}\partial_r \lambda
\end{eqnarray}

\begin{eqnarray}
	R_{\theta \theta}&=& R^{\lambda}_{\theta \lambda \theta}=R^{t}_{\theta t \theta}
+R^{r}_{\theta r \theta} +R^{\phi}_{\theta \phi \theta}\nonumber
\end{eqnarray}

\begin{eqnarray}
	R^{t}_{\theta t \theta}&=&\partial_t \Gamma^{t}_{\theta \theta}-\partial_\theta
\Gamma^{t}_{t \theta}+ \Gamma^{t}_{t \lambda} \Gamma^{\lambda}_{\theta \theta} 
- \Gamma^{t}_{\theta \lambda} \Gamma^{\lambda}_{t \theta}\nonumber\\
	&=&0-0+\Gamma^{t}_{t t} \Gamma^{t}_{\theta \theta}+\Gamma^{t}_{t r} \Gamma^{r}_{\theta \theta}+\Gamma^{t}_{t \theta} \Gamma^{\theta}_{\theta \theta}+
\Gamma^{t}_{t \phi} \Gamma^{\phi}_{\theta \theta}-\Gamma^{t}_{\theta t} \Gamma^{t}_{t \theta}-\Gamma^{t}_{\theta r} \Gamma^{r}_{t \theta}-\Gamma^{t}_{\theta \theta} \Gamma^{\theta}_{t \theta}-\Gamma^{t}_{\theta \phi} \Gamma^{\phi}_{t \theta}\nonumber\\
	&=&0+\partial_r \nu (-r e^{-2\lambda})+0+0-0-0-0-0\nonumber\\
	&=&-r e^{-2\lambda} \partial_r \nu \;\;\;\;\;\;\;
\end{eqnarray}

\begin{eqnarray}
	R^{r}_{\theta r \theta}&=&\partial_r \Gamma^{r}_{\theta \theta}-\partial_\theta
\Gamma^{r}_{r \theta}+ \Gamma^{r}_{r \lambda} \Gamma^{\lambda}_{\theta \theta} 
- \Gamma^{r}_{\theta \lambda} \Gamma^{\lambda}_{r \theta}\nonumber\\
	&=&\partial_r(-r e^{-2\lambda})-0+\Gamma^{r}_{r t} \Gamma^{t}_{\theta \theta}+\Gamma^{r}_{r r} \Gamma^{r}_{\theta \theta}+\Gamma^{r}_{r \theta} \Gamma^{\theta}_{\theta \theta}+
\Gamma^{r}_{r \phi} \Gamma^{\phi}_{\theta \theta}-\Gamma^{r}_{\theta t} 
\Gamma^{t}_{r \theta}-\Gamma^{r}_{\theta r} \Gamma^{r}_{r \theta}-\Gamma^{r}_{\theta \theta} \Gamma^{\theta}_{r \theta}-\Gamma^{r}_{\theta \phi} \Gamma^{\phi}_{r \theta}\nonumber\\
	&=&2r e^{-2\lambda} \partial_r \lambda-e^{-2\lambda}+0+\partial_r\lambda(-re^{-2\lambda})+0+0-0-0-(-r e^{-2\lambda})\frac{1}{r}-0\nonumber\\
	&=&2r e^{-2\lambda} \partial_r \lambda-e^{-2\lambda}+\partial_r\lambda(-re^{-2\lambda})+e^{-2\lambda}\nonumber\\
	&=& r e^{-2\lambda} \partial_r \lambda 
\end{eqnarray}

\begin{eqnarray}
	R^{\phi}_{\theta \phi \theta}&=&\partial_\phi \Gamma^{\phi}_{\theta \theta}-\partial_\theta \Gamma^{\phi}_{\phi \theta}+ \Gamma^{\phi}_{\phi \lambda} 
\Gamma^{\lambda}_{\theta \theta} - \Gamma^{\phi}_{\theta \lambda} 
\Gamma^{\lambda}_{\phi \theta}\nonumber\\
	&=&0-\partial_\theta \Gamma^{\phi}_{\phi \theta}+ \Gamma^{\phi}_{\phi t} \Gamma^{t}_{\theta \theta}+\Gamma^{\phi}_{\phi r} \Gamma^{r}_{\theta \theta}+\Gamma^{\phi}_{\phi \theta} \Gamma^{\theta}_{\theta \theta}+\Gamma^{\phi}_{\phi \phi} \Gamma^{\phi}_{\theta \theta}-\Gamma^{\phi}_{\theta t} \Gamma^{t}_{\phi \theta}-\Gamma^{\phi}_{\theta r} \Gamma^{r}_{\phi \theta}-\Gamma^{\phi}_{\theta \theta} \Gamma^{\theta}_{\phi \theta}-\Gamma^{\phi}_{\theta \phi} \Gamma^{\phi}_{\phi \theta}\nonumber\\
	&=&-\partial_\theta(cot \theta)+0+\frac{1}{r}(-r e^{-2\lambda})+0-0-0-0-cot^2{\theta}\nonumber\\
	&=&(1-e^{-2\lambda})
\end{eqnarray}

\begin{eqnarray}
	R_{\theta \theta}&=&-r e^{-2\lambda} \partial_r\nu+r e^{-2\lambda} \partial_r\lambda
+(1-e^{-2\lambda})\nonumber\\
	&=&e^{-2\lambda}[r(\partial_r\lambda-\partial_r\nu)-1]+1
\end{eqnarray}

\begin{eqnarray}
	R_{\phi \phi}&=&R^{\lambda}_{\phi\lambda\phi}=R^{t}_{\phi t\phi}+R^{r}_{\phi r\phi}+R^{\theta}_{\phi\theta\phi}\nonumber\\
	&=&-r e^{-2\lambda} sin^2{\theta} \partial_r\nu+r e^{-2\lambda} sin^2{\theta} \partial_r\lambda+(1-e^{-2\lambda})sin^2{\theta}\nonumber\\
	&=&\{e^{-2\lambda}[r (\partial_r\lambda-\partial_r\nu)-1]+1\}sin^2{\theta}\nonumber\\
	&=& sin^2{\theta} R_{\theta \theta}
\end{eqnarray}
Note: $\partial^{2}_r \nu=\nu'', \partial_r \nu=\nu',\partial_r  \lambda=\lambda'$\\
Curvature scalar or Ricci scalar is
\begin{eqnarray}
	R&=&g^{\mu \nu} R_{\mu \nu}=g^{t t} R_{t t}+g^{r r} R_{r r}+g^{\theta \theta} R_{\theta \theta}+g^{\phi \phi} R_{\phi \phi}\nonumber\\
	&=&-e^{-2\nu} e^{2(\nu-\lambda)}[\nu''-\nu'^2-\nu'\lambda'+\frac{2}{r}
\nu']+e^{-2\lambda}[-\nu''-\nu'^2+\nu'\lambda'+\frac{2}{r}\lambda']+
	\nonumber\\
	&&\frac{1}{r^2}\{e^{-2\lambda}[r(\lambda'-\nu')-1]+1\}
+\frac{1}{r^2 sin^2\theta}\{e^{-2\lambda}[r(\lambda'-\nu')-1]+1\}sin^2\theta\nonumber\\
	&=&e^{-2\lambda}[-\nu''-\nu'^2+\nu'\lambda'-\frac{2}{r}\nu'-\nu''-\nu'^2+\nu'\lambda'+\frac{2}{r}\lambda']\nonumber\\
	&&	+\frac{2}{r^2}\{e^{-2\lambda}[r(\lambda'-\nu')-1]+1\}\nonumber\\
	&=&2e^{-2\lambda}[-\nu''-\nu'^2+\nu'\lambda'-\frac{1}{r}(\nu'-\lambda')]+2e^{-2\lambda}[\frac{1}{r}(\lambda'-\nu')-\frac{1}{r^2}]+\frac{2}{r^2}\nonumber\\
	&=&2e^{-2\lambda}[-\nu''-\nu'^2+\nu'\lambda'-\frac{2}{r}(\nu'-\lambda')-\frac{1}{r^2}]+\frac{2}{r^2}\nonumber\\
	&=&-2e^{-2\lambda}[\nu''+\nu'^2-\nu'\lambda'+\frac{2}{r}(\nu'-\lambda')+\frac{1}{r^2}(1-e^{2\lambda})]\nonumber\\
	&=&-2e^{-2\lambda}[\nu''+\nu'^2-\nu'\lambda'+\frac{2}{r}(\nu'-\lambda')+\frac{1}{r^2}(1-e^{2\lambda})] 
\end{eqnarray}
 Now, calculate full Einstein's equation
\begin{eqnarray}
	G_{\mu \nu}&=&R_{\mu \nu}-\frac{1}{2}g_{\mu \nu} R
\end{eqnarray}
$tt$-component:
\begin{eqnarray}
	G_{tt}&=&R_{tt}-\frac{1}{2}g_{tt} R\nonumber\\
	&=&e^{2(\nu-\lambda)}[\nu''+\nu'^2-\nu'\lambda'+\frac{2}{r}\nu']-\frac{1}{
2}(-e^{2\nu})(-2e^{-2\lambda})[\nu''+\nu'\nonumber\\
&&-\nu'\lambda'+\frac{2}{r}(\nu'-\lambda')+\frac{1}{r^2}(1-e^{2\lambda}]\nonumber\\
	&=&e^{2(\nu-\lambda)}[\nu''+\nu'^2-\nu'\lambda'+\frac{2}{r}\nu'-\nu''-\nu'^2+\nu'\lambda'\nonumber\\
	&&-\frac{2}{r}(\nu'-\lambda')-\frac{1}{r^2}(1-e^{2\lambda})]\nonumber\\
	&=&e^{2(\nu-\lambda)}[\frac{2}{r}\lambda'-\frac{1}{r^2}(1-e^{2\lambda})]\nonumber\\
	&=&\frac{1}{r^2}e^{2(\nu-\lambda)}[2\lambda'r-(1-e^{2\lambda})]\nonumber\\
&=&\frac{1}{r^2}e^{2(\nu-\lambda)}[2r\lambda'-1+e^{2\lambda}]
\end{eqnarray}
$rr$-component:
\begin{eqnarray}
	G_{rr}&=&R_{rr}-\frac{1}{2}g_{rr}R\nonumber\\
	&=&-\nu''-\nu'^2+\nu'\lambda'+\frac{2}{r}\lambda'-\frac{1}{2}e^{2\lambda}(-2e^{-2\lambda}) [\nu''+\nu'^2\nonumber\\
&&-\nu'\lambda'+\frac{2}{r}(\nu'-\lambda')+\frac{1}{r^2}(1-e^{2\lambda})] \nonumber\\
	&=&\frac{2}{r}\nu'+\frac{1}{r^2}(1-e^{2\lambda})\nonumber\\
	&=&\frac{1}{r^2}[2r\nu'+1-e^{2\lambda}] 
\end{eqnarray}

\begin{eqnarray}
	G_{\theta \theta}&=&R_{\theta \theta}-\frac{1}{2}g_{\theta\theta}R\nonumber\\
	&=&e^{-2\lambda}[r(\lambda'-\nu')-1]+1-\frac{1}{r^2}(-2e^{-2\lambda})[\nu''+\nu'^2\nonumber\\
&&-\nu'\lambda'+\frac{2}{r}(\nu'-\lambda')+\frac{1}{r^2}(1-e^{2\lambda})]\nonumber\\
	&=&r^2e^{-2\lambda}[\nu''+\nu'^2-\nu'\lambda'+\frac{1}{r}(\nu'-\lambda')] 
\end{eqnarray}

\begin{eqnarray}
	G_{\phi \phi}&=&R_{\phi \phi}-\frac{1}{2}g_{\phi \phi}R\nonumber\\
	&=&sin^2\theta\{e^{-2\lambda}[r(\lambda'-\nu')-1]+1\}-\frac{1}{2}r^2sin^2\theta (-2e^{-2\lambda})[\nu''+\nu'^2\nonumber\\
	&&-\nu'\lambda'+\frac{2}{r}(\nu'-\lambda')+\frac{1}{r^2}(1-e^{2\lambda}]\nonumber\\
	&=&e^{-2\lambda}sin^2\theta[r(\lambda'-\nu')-1+e^{2\lambda}]+r^2sin^2\theta e^{-2\lambda}[\nu''+\nu'^2\nonumber\\
&&-\nu'\lambda'+\frac{2}{r}(\nu'-\lambda')+\frac{1}{r^2}(1-e^{2\lambda})]\nonumber\\
&=&e^{-2\lambda}sin^2\theta[r\lambda'-r\nu'-1+e^{2\lambda}+r^2\nu''+r^2\nu'^2
	\nonumber\\
&&-r^2\nu'\lambda'+2r\nu'-2r\lambda'+1-e^{2\lambda}]\nonumber\\
&=&e^{-2\lambda}sin^2\theta[r\nu'-r\lambda'+r^2\nu''+r^2\nu'^2-r^2\nu'\lambda']\nonumber\\
	&=&sin^2\theta G_{\theta \theta}
\end{eqnarray}

The energy-momentum tensor of the star itself as a perfect fluid is given by
\begin{eqnarray}
	T_{\mu \nu}&=&(\mathcal{E}+P)u_{\mu}u_{\nu}+P g_{\mu\nu}
	\label{Tun}
\end{eqnarray}
Here, the energy density $(\mathcal{E})$ and pressure density ($P$) is a 
function of $r$ alone. Since, the fluid is static, there are no other 
contributions to the fluid velocity except time-like components ($\mu_0$). 
The velocity is normalized so that $u_{\mu}u^{\mu}=-1$. Then, 
it becomes
\begin{eqnarray}
	u_{\mu}&=&(e^\nu,0,0,0)
\end{eqnarray}
So, the component of energy momentum tensor will be  

\[
	T=
\begin{bmatrix}
	-e^{2\nu(r)}\mathcal{E}      & 0 & 0  & 0 \\
   0       & e^{2\lambda(r)} P& 0  & 0 \\
   0     & 0 & r^2 P & 0\\
0       & 0 & 0 & r^2 P sin^{2}\theta
\end{bmatrix}
\]

Einstein tensor:
\begin{eqnarray}
G_{\mu \nu}=R_{\mu \nu}-\frac{1}{2}g_{\mu \nu}R=8\pi  T_{\mu\nu}
	\label{EIN}
\end{eqnarray} 
Using all previous results we can find that the components of the Einstein
tensor are:\\
The $tt$-component:
\begin{eqnarray}
	\frac{1}{r^2}e^{2(\nu-\lambda)}(2r\lambda'-1+e^{2\lambda})&=&8\pi  e^{2\nu}\mathcal{E}\nonumber\\
	\frac{1}{r^2}e^{-2\lambda}(2r\lambda'-1+e^{2\lambda})&=&8\pi  \mathcal{E}
\end{eqnarray}

The $rr$-component:

\begin{eqnarray}
	\frac{1}{r^2}(2r\nu'+1-e^{2\lambda})&=&8\pi  e^{2\lambda}P\nonumber\\
	\frac{1}{r^2}e^{-2\lambda}(2r\nu'+1-e^{2\lambda})&=&8\pi  P
\end{eqnarray}

The $\theta \theta$-component:
\begin{eqnarray}
	r^2 e^{-2\lambda}[\nu''+\nu'^2-\nu'\lambda'+\frac{1}{r}(\nu'-\lambda')]&=&8\pi  r^2 P\nonumber\\
	e^{-2\lambda}[\nu''+\nu'^2-\nu'\lambda'+\frac{1}{r}(\nu'-\lambda')]&=&8\pi  P
\end{eqnarray}

The $\phi\phi$-component is proportional to the $\theta \theta$-equation, so there
is no need to consider separately. So, the tt-component of the Einstein equation gives:
\begin{eqnarray}
	\frac{1}{r^2}e^{-2\lambda}(2r\lambda'-1+e^{2\lambda})&=&8\pi  \mathcal{E}\nonumber\\
	e^{-2\lambda}(2r\lambda'-1)+1&=&8\pi  \mathcal{E} r^2\nonumber\\
	-\frac{d}{dr}\{r(e^{-2\lambda}-1)\}&=&8\pi  \mathcal{E} r^2\nonumber\\
	\frac{d}{dr}\{r(e^{-2\lambda}-1)\}&=&-k \mathcal{E} r^2\nonumber\\
	d\{r(e^{-2\lambda}-1)\}&=&-k \mathcal{E} r^2 dr
\end{eqnarray}
Where, $k=-8\pi$ and previous equation can be integrated 

\begin{eqnarray}
	e^{-2\lambda}=1+\frac{k}{r}\int^{R}_{0} \mathcal{E}(r) r^2 dr
\end{eqnarray}
Let us define
\begin{eqnarray}
	m(r)&=&4\pi\int^{R}_{0} \mathcal{E}(r) r^2 dr\nonumber\\
	e^{-2\lambda}&=&1-\frac{8\pi G}{4\pi r}m(r)\nonumber\\
	e^{-2\lambda}&=&\big(1-\frac{2 G m(r)}{r}\big)\nonumber\\
	e^{2\lambda}&=&\big(1-\frac{2 G m(r)}{r}\big)^{-1}
\end{eqnarray}
So that, the metric will be
\begin{eqnarray}
ds^2=-e^{2\nu(r)} dt^2+\big(1-\frac{2 G m(r)}{r}\big)^{-1} dr^2+r^2 d\theta^2
+r^2 sin^2\theta d\phi^2
\end{eqnarray}

The gravitational mass of the neutron stars is given by
\begin{eqnarray}
M_B&=&4\pi\int^{R}_{0} \rho_{B}(r) r^2 e^{2\lambda} dr\nonumber\\
M_B&=&4\pi\int^{R}_{0} \frac{\rho_{B}(r) r^2}{(1-\frac{2 G m(r)}{r})}dr
\end{eqnarray}
The binding energy due to the internal gravitational attraction of the fluid
elements in the star, which is given by 
\begin{eqnarray}
E_B=M-M_B<0
\end{eqnarray}
The binding energy is the amount of energy that would be required to disperse 
the matter in the star to be infinity. The $rr$-component can be written as
\begin{eqnarray}
	\frac{1}{r^2}e^{-2\lambda}(2r\frac{d\nu}{dr}+1-e^{-2\lambda})&=&8\pi  p\nonumber\\
	2r\frac{d\nu}{dr}+1-e^{2\lambda}&=&8\pi  P r^2 e^{2\lambda}\nonumber\\
	2r\frac{d\nu}{dr}&=&8\pi  P r^2 e^{2\lambda}+e^{2\lambda}-1\nonumber\\
	&=&\frac{(8\pi  P r^2+1)}{(1-\frac{2  m(r)}{r})}-1\nonumber\\
	&=&\frac{(8\pi  P r^3+r)}{(r-2 m(r))}-1\nonumber\\
	&=&\frac{8\pi  P r^3+r-r+2 m(r)}{(r-2 m(r))}\nonumber\\
	&=&\frac{8\pi  P r^3+2 m(r)}{(r-2 m(r))}\nonumber\\
	\frac{d\nu}{dr}&=&\frac{(4\pi  P r^3+ m(r))}{r(r-2 m(r))}
\end{eqnarray}
Now, we have to calculate $\frac{dP}{dr}=?$.
Again, solve the tt-component equation
\begin{eqnarray}
(2r\lambda'-1+e^{2\lambda})&=&8\pi  \epsilon r^2 e^{2\lambda}\nonumber\\
	2r\lambda'&=&(8\pi  \mathcal{E} r^2 e^{2\lambda}-1)e^{2\lambda}+1
\end{eqnarray}
\begin{eqnarray}
	(2r\nu'+1-e^{2\lambda})&=&8\pi  P r^2 e^{2\lambda}\nonumber\\
	2r\nu'&=&(1+8\pi  P r^2 e^{2\lambda})e^{2\lambda}-1
\end{eqnarray}
Take the derivative of the equation (A.48) then
\begin{eqnarray}
	2\nu'+2r\nu''&=&(8\pi  P' r^2 +16 \pi  P r)e^{2\lambda}+2\lambda'
(1+8\pi  P r^2)e^{2\lambda}\nonumber\\
	2r\nu'+2r^2\nu''&=&[2r\lambda'(1+8\pi  r^2 P)+(16\pi  r^2 P + 8\pi  r^3 P')] e^{2\lambda}\nonumber\\
2r^2\nu''&=&[2r\lambda'(1+8\pi  r^2 P)+(16\pi  r^2 P + 8\pi  r^3 P')] e
^{2\lambda}-2r\nu'
\end{eqnarray}
Put the values of $2r\nu'$ and $2r\lambda'$ from eqs.(A.47) and (A.48), then 

\begin{eqnarray}
	2 r^2 \nu''&=&1+(16\pi  r^2 P+8\pi  r^3 P')e^{2\lambda}-(1+8\pi r^2 P)(1-8\pi  r^2 \mathcal{E})e^{4\lambda}
\end{eqnarray}
Square eq.(A.50) to obtain the result
\begin{eqnarray}
2r^2\nu'^2&=&\frac{1}{2}(1+8\pi  r^2 P)^2 e^{4\lambda}-(1+8\pi  r^2 P)e^{2\lambda}-\frac{1}{2}
\end{eqnarray}
 Now, we have the expressions for $\nu',\nu'', \nu'^{2}$, and $\lambda'$
in terms of $P,P',\mathcal{E},$ and $e^{2\lambda}$. Hence, equation (A.51) can be 
written as:
\begin{eqnarray}
	(\mathcal{E}+P)\frac{d\nu}{dr}&=&-\frac{dP}{dr}\nonumber\\
\frac{dP}{dr}&=&-\frac{(\mathcal{E}+P)( m(r)+4\pi  r^3 P)}{r(r-2  m(r))}
\end{eqnarray}
Summary: the TOV equations is written as \cite{glen20}
\begin{eqnarray}
\frac{dP}{dr}&=&-\frac{(\mathcal{E}+P)( m(r)+4\pi  r^3 P)}{r(r-2  m(r))}\\
	\frac{dm}{dr}&=&4\pi r^2 \mathcal{E}
\end{eqnarray} 
\label{tov}
\end{appendices}

\setcounter{equation}{0}
\setcounter{figure}{0}

%\include{appendixB}
%\setcounter{equation}{0}
%\setcounter{figure}{0}

%\newpage
%\include{summ}
%\setcounter{equation}{0}
%\setcounter{figure}{0}

\newpage
\addcontentsline{toc}{chapter}{REFERENCES}
\renewcommand{\bibname}{REFERENCES}
\bibliographystyle{utcaps}
\bibliography{mythesis}

\end{document}